\documentclass[twocolumn,preprintnumbers,amsmath,amssymb,rmp,floatfix]{revtex4}

\usepackage{graphicx}
\usepackage{dcolumn}
\usepackage{bm}
\usepackage{hyperref}
\hypersetup{backref=true, pagebackref=true, hyperindex=true, breaklinks=true,colorlinks=true,urlcolor=blue, linkcolor=black,  citecolor=blue,pagecolor=red, bookmarks=true, bookmarksopen=true}

\begin{document}

\title{LECTURE NOTES ON THE FORMATION AND \\ EARLY EVOLUTION OF PLANETARY SYSTEMS\footnote{{\em Astrophysics 
of Planet Formation} \citep{armitage10} is a graduate level textbook based on earlier versions of these notes. 
I plan to continue updating these notes as an open access resource.}}

\author{Philip J. Armitage}
\affiliation{JILA, University of Colorado, Boulder}

\begin{abstract}
These notes provide an introduction to the 
theory of the formation and early evolution of planetary systems. Topics 
covered include the structure, evolution and dispersal of protoplanetary 
disks; the formation of planetesimals, terrestrial and gas giant planets; 
and orbital evolution due to gas disk migration, planetesimal scattering,
planet-planet interactions, and tides.
\end{abstract}

\maketitle

\tableofcontents

\section{Introduction}
The theoretical study of planet formation has a long history. Many of the 
fundamental ideas in the theory of terrestrial planet formation were laid 
out by \citet{safronov69} in his monograph ``Evolution of the 
Protoplanetary Cloud and Formation of the Earth and the Planets". The 
core accretion theory for gas giant formation was discussed by 
Cameron in the early 1970's \citep{perri73} and had been developed in 
recognizable detail by 1980 \citep{mizuno80}. The data that motivated and 
tested these theories, however, was relatively meagre and limited to the Solar System.
The last twenty-five years have seen a wealth of new observations, 
including imaging and spectroscopy  
of protoplanetary disks, the discovery of the Solar System's Kuiper 
Belt, and the detection and characterization of extrasolar planetary systems. 
Many of these observations have revealed unexpected properties of disks 
and planetary systems, highlighting not so much gaps in our theoretical knowledge as 
a lack of understanding of how known physical processes combine to form 
the planetary systems.

The goal of these notes is to introduce the concepts underlying planet formation, via  
a mix of worked-through derivations and (necessarily incomplete) references to 
the literature. The main questions we hope to answer are,
\begin{itemize}
\item
How small solid particles grown to macroscopic dimensions within the environment of 
protoplanetary disks.
\item
How terrestrial and giant planets form.
\item
What processes determine the final architecture of planetary systems, and might 
explain the astounding diversity of observed extrasolar planets.
\end{itemize}
First though,  we briefly review observational properties of the Solar System 
and extrasolar planetary systems that we might hope a theory of planet formation 
would explain.

\subsection{Critical Solar System observations}
\subsubsection{Architecture}

\begin{table*}
\label{origins_t1}
\caption{Basic properties of planets in the Solar System, the semi-major axis $a$, eccentricity $e$, orbital inclination $i$,  
mass $M_p$, and mean radius $R_p$.}
\begin{ruledtabular}
\begin{tabular}{lccccc}
 & $a / {\rm AU}$ & $e$ & $i$ & $M_p / {\rm g}$ & $R_p / {\rm km}$ \\
\hline
Mercury & 0.387 & 0.206 & 7.0$^\circ$ & $3.3 \times 10^{26}$ & $2.4 \times 10^3$ \\
Venus   & 0.723 & 0.007 & 3.4$^\circ$ & $4.9 \times 10^{27}$ & $6.1 \times 10^3$ \\
Earth   & 1.000 & 0.017 & 0.0$^\circ$ & $6.0 \times 10^{27}$ & $6.4 \times 10^3$ \\
Mars    & 1.524 & 0.093 & 1.9$^\circ$ & $6.4 \times 10^{26}$ & $3.4 \times 10^3$ \\
Jupiter & 5.203 & 0.048 & 1.3$^\circ$ & $1.9 \times 10^{30}$ & $7.1 \times 10^4$ \\
Saturn  & 9.537 & 0.054 & 2.5$^\circ$ & $5.7 \times 10^{29}$ & $6.0 \times 10^4$ \\
Uranus  & 19.191 & 0.047 & 0.8$^\circ$ & $8.7 \times 10^{28}$ & $2.6 \times 10^4$ \\
Neptune & 30.069 & 0.009 & 1.8$^\circ$ & $1.0 \times 10^{29}$ & $2.5 \times 10^4$ \\
\end{tabular}
\end{ruledtabular}
\end{table*}

The orbital properties, masses and radii of the Solar System's planets are listed in Table~I. The dominant
planets in the Solar System are our two 
{\em gas giants}, Jupiter and Saturn. These planets 
are composed primarily of hydrogen and helium -- like the Sun -- 
though they have a higher abundance of heavier elements as compared to Solar composition. Saturn is known to have a 
substantial core. Descending in mass there are two {\em ice giants} 
(Uranus and Neptune) composed of water, ammonia, methane, silicates 
and metals, plus low mass hydrogen / helium atmospheres; two large 
{\em terrestrial planets} (Earth and Venus) plus two smaller 
terrestrial planets (Mercury and Mars).  Apart from Mercury, all 
of the planets have low eccentricities and orbital inclinations. They 
orbit in a plane that is approximately, but not exactly, perpendicular to the Solar 
rotation axis (the misalignment angle is about $7^\circ$).

In the Solar System the giant and terrestrial planets are clearly 
segregated in orbital radius, with the inner zone occupied by the 
terrestrial planets being separated from the outer giant planet 
region by the main asteroid belt. The orbital radii of the giant planets 
coincide with where we expect the 
protoplanetary disk to have been cool enough for ices to have been 
present. This is a significant observation in the classical theory 
of giant planet formation, since in that theory the time scale for 
giant planet formation depends upon the mass of 
condensable materials. One would therefore expect faster growth to 
occur in the outer ice-rich part of the protoplanetary disk.

\subsubsection{Mass and angular momentum}

The mass of the Sun is $M_\odot = 1.989 \times 10^{33} \ {\rm g}$, made up 
of hydrogen (fraction by mass $X = 0.73$), helium ($Y=0.25$) and ``metals" 
(which includes everything else, $Z=0.02$). One observes immediately that,
\begin{equation}
 Z M_\odot \gg \sum M_p,
\end{equation} 
i.e. most of the heavy elements 
in the Solar System {\em are found in the Sun} rather than in the planets. 
If most of the mass in the Sun passed through a disk 
during star formation the planet formation process 
need not be very efficient. 

The angular momentum budget for the Solar System is dominated by the 
orbital angular momentum of the planets. The angular momentum in the 
Solar rotation is,
\begin{equation}
 L_\odot \simeq k^2 M_\odot R_\odot^2 \Omega,
\end{equation} 
assuming for simplicity solid body rotation. Taking $\Omega = 2.9 \times 10^{-6} 
\ {\rm s}^{-1}$ and adopting $k^2 = 0.1$ (roughly appropriate for a star 
with a radiative core), $L_\odot \simeq 3 \times 10^{48} \ {\rm g \ cm}^{2} {\rm \ s}^{-1}$. 
By comparison the orbital angular momentum of Jupiter is,
\begin{equation}
 L_J = M_J \sqrt{ GM_\odot a } = 2 \times 10^{50} \ {\rm g \ cm}^{2} \ {\rm s}^{-1}.
\end{equation}
This result implies that substantial 
segregation of mass and angular momentum must have taken place 
during (and subsequent to) the star formation process. We will 
look into how such segregation arises during disk accretion later. 

\subsubsection{Minimum mass Solar Nebula}

We can use the observed masses and compositions of the planets to derive 
a {\em lower limit} to the amount of gas that must have been present when the 
planets formed. This is called the Minimum Mass Solar Nebula 
\cite{weidenschilling77}. The procedure is:
\begin{itemize}
\item[1.]
Start from the known mass of heavy elements (say iron) in each planet, 
and augment this mass with enough hydrogen and helium to bring the 
mixture to Solar composition. This is a mild augmentation for Jupiter, 
but a lot more for the Earth.
\item[2.]
Divide the Solar System into annuli, with one planet per annulus. 
Distribute the augmented mass for each planet uniformly across the 
annuli, to yield a characteristic gas surface density $\Sigma$ (units ${\rm g \ cm}^{-2}$) 
at the location of each planet.
\end{itemize}

The result is that between Venus and Neptune (and ignoring the asteroid 
belt) $\Sigma \propto r^{-3/2}$. The precise normalization is mostly a 
matter of convention, but if one needs a 
specific number the most common value used is that due to \citet{hayashi81},
\begin{equation}
 \Sigma = 1.7 \times 10^3 \left( \frac{r}{\rm AU} \right)^{-3/2} \ {\rm g \ cm}^{-2}.
\end{equation}
Integrating out to 30~AU the enclosed mass is
around 0.01~$M_\odot$, which is in the same ball park as estimates 
of protoplanetary disk masses observed around other stars.

As the name should remind you this is a {\em minimum} mass. It is not an 
estimate of the disk mass at the time the Sun formed, nor is the $\Sigma \propto r^{-3/2}$ scaling necessarily the 
actual surface density profile for a protoplanetary disk. Theoretical 
models of disks based on the $\alpha$-prescription
 predict a shallower slope more akin to  $\Sigma \propto r^{-1}$ \citep{bell97}, 
while models based on first-principles calculations of disk angular momentum 
transport suggest a complex $\Sigma$ profile that is not well-described by a 
single power-law. Observations of protoplanetary disks around other stars do not 
directly probe the planet-forming region at a few~AU, although on larger scales (beyond 
20~AU) sub-mm images are consistent with a median profile $\Sigma \propto r^{-0.9}$ 
\citep{andrews09}. 

\subsubsection{Resonances}

A {\em resonance} occurs when there is a near-exact relation between 
characteristic frequencies of two bodies. For example, a mean-motion 
resonance between two planets with orbital periods $P_1$ and $P_2$ occurs 
when,
\begin{equation}
 \frac{P_1}{P_2} \simeq \frac{i}{j},
\end{equation}
with $i$, $j$ integers (the resonance is typically important if $i$ and $j$, or their 
difference,  
are {\em small} integers). The ``approximately equal to" sign in this expression 
reflects the fact that resonances have a finite width, which varies with the 
particular resonance and with the eccentricities of the bodies involved. 
Resonant widths can be calculated precisely, though the methods needed to do so are beyond the 
scope of these notes \citep[a standard reference is][]{murray99}. 
In the Solar System Neptune and Pluto (along with 
many other Kuiper Belt objects) are in a 3:2 resonance, while Jupiter and 
Saturn are close to but outside a 5:2 mean-motion resonance\footnote{A delightful 
account of how it was recognized that this proximity influences the motion of 
Jupiter and Saturn is in \citet{lovett95}.}. There are many resonant pairs 
among planetary moons. Jupiter's satellites Io, Europa and Ganymede, for 
example, form a resonant chain in which Io is in 2:1 resonance with Europa, 
which itself is in a 2:1 resonance with Ganymede. In the Saturnian system, 
the small moons Prometheus and Pandora occupy a 121:118 resonance. 
If planetary (or satellite) orbits were distributed randomly, subject only to 
the requirement that they be stable for long periods, then the chances that 
two bodies would find themselves in a resonance is low. Seeing a resonance is thus 
strong circumstantial 
evidence that dissipative processes (tides being the prototypical example) resulted in orbital 
evolution and trapping into resonance at some point in the past history of the system \citep{goldreich65}.  

Although there are no mean-motion resonances today between the Solar System's major planets, 
other resonances are dynamically important. In particular, secular resonances, which occur 
when the {\em precession} frequencies of two bodies match, couple the dynamics of the 
giant planets to that of the asteroid belt and inner Solar System. The $\nu_6$ resonance, 
for example, which roughly speaking corresponds to the precession rate 
of Saturn's orbit, defines the inner edge of the asteroid belt. It is important for the 
delivery of meteorites and Near Earth Asteroids to the Earth \citep{scholl91}.

\subsubsection{Minor bodies}

As a rough generalization the Solar System is dynamically full, in 
that most locations where test particle orbits would be stable for 
5~Gyr are, in fact, occupied by minor bodies \citep[e.g., for the outer Solar System see][]{holman93}. 
In the inner and middle 
Solar System the main asteroid belt is the largest reservoir of minor 
bodies. The asteroid belt displays considerable structure, most 
notably in the form of sharp decreases in the number of asteroids in 
the {\em Kirkwood gaps}. The existence of these gaps 
provides a striking illustration of the importance of 
resonances (in this case with Jupiter) in influencing dynamics. 
The asteroid belt also preserves radial gradients in composition, 
with the water-rich bodies that are the source of meteorites known as carbonaceous chondrites 
residing in the outer belt, while the inner belt is dominated by water-poor 
asteroids that source the enstatite chondrites \citep{morbi00}.  

Beyond Neptune orbit Kuiper Belt Objects (KBOs), with sizes 
ranging up to a few thousand km \citep{jewitt93}. The differential size distribution, deduced indirectly from the measured 
luminosity function, is roughly a power-law for large bodies 
with diameters $D \gtrsim 100 \ {\rm km}$ \citep{trujillo01}. A 
determination by \citet{fraser09} infers a power-law slope $q \simeq 4.8$ for large 
bodies together with a break to a much shallower slope at small sizes.The dynamical structure of the 
Kuiper Belt is extraordinarily rich, and this motivates a dynamical classification 
of KBOs into several classes \citep{chiang06},
\begin{itemize}
\item[1.]
{\em Resonant KBOs} are in mean-motion resonances with Neptune. This class includes Pluto and the 
other ``Plutinos"  in Neptune's exterior 3:2 resonance, and provided some of the original empirical 
motivation for the idea of giant planet migration in the Solar System \citep{malhotra93}. 
\item[2.]
{\em Classical KBOs} are objects whose orbits do not, and will not, cross the orbit of Neptune 
given the current configuration of the outer Solar System. Many classical KBOs have  
low inclinations, and hence these bodies may have suffered relatively little in the way of 
dynamical excitation during the past history of the Solar System. 
\item[3.]
{\em Scattered disk KBOs} are objects, also with perihelion distances beyond Neptune, that have 
typically high eccentricities and inclinations. These can also be described as a ``hot" Classical population. 
\end{itemize}
The total mass in the observed Kuiper Belt populations today is low \citep[$M \sim 0.01 \ M_\oplus$;][]{fraser14}, 
though it is commonly suggested to have been many orders of magnitude higher in the past.
The rich dynamical structure of the Kuiper Belt preserves information about the early dynamical 
history of the Solar System, and is our best hope when it comes to distinguishing between 
models for the formation and migration of the giant planets.
We will discuss some of the popular models later, 
but for now just direct the reader to a handful of representative models that give a 
flavor of the physical considerations \citep{dawson12,batygin11,levison08,hahn05}.

The Classical KBOs have an apparent edge to their radial distribution at about 
50~AU \citep{trujillo01}. There are, however, a handful of known objects at 
larger distances, including some with perihelia large enough that they are 
dynamically detached from Neptune and the current outer Solar System. 
{\em Sedna}, a large body with semi-major axis $a = 480 \pm 40 \ {\rm AU}$ and 
eccentricity $e = 0.84 \pm 0.01$, falls in this class \citep{brown04}. The orbital 
elements of the detached objects do not appear to be randomly distributed, 
a result which could imply the existence of a planetary perturber 
\citep{trujillo14} or of a massive planetesimal disk \citep{madigan16} at 
very large radii in the Solar System. The most developed model is that of \citet{batygin16}, 
who find that a planet with a mass of $\approx 10 M_\oplus$, semi-major axis 
$a \approx 600 \ {\rm AU}$, eccentricity $e \approx 0.5$ and inclination 
$i \approx 30^\circ$ would be consistent with the observations. The possible 
positions of this hypothetical planet, which would be bright enough to potentially 
detect in the near-term, are constrained but not excluded by 
more direct observations, for example ranging data to the {\em Cassini} 
spacecraft around Saturn \citep{fienga16}. From a theoretical perspective, 
the existence of {\em Sedna} demonstrates that dynamical perturbations other 
than those of the known planets are or were operative in the outer Solar System, 
and it is certainly possible to imagine that an additional ice giant was ejected from the 
region of planet formation and captured into a high perihelion orbit due to 
perturbations from other stars in the Sun's birth cluster \citep[reviewed, e.g., by][]{adams10}.

The discovery of large numbers of extrasolar planetary systems with short period super-Earth or 
ice giant planets raises the question of why there are no Solar System bodies interior to 
Mercury. Dynamically, an annulus of orbits between about 0.1~AU and 0.2~AU would be 
stable \citep{evans99}. An inner asteroid belt would, however, be subject to severe 
collisional and radiative depletion \citep{stern00}, so while it may be a puzzle why there are 
no {\em planets} interior to Mercury the lack of a large population of Vulcanoid asteroids is 
less surprising.

\subsubsection{Ages}

Radioactive dating of meteorites provides an absolute 
age of the Solar System, together with constraints on the 
time scales of some phases of planet formation. The details are 
an important topic that is not part of these lectures. Typical numbers quoted are 
a Solar System age of 4.57~Gyr, a time scale for the formation 
of large bodies within the asteroid belt of $< 5$~Myr \citep{wadhwa06}, and a 
time scale for final assembly of the Earth of $\sim 100$~Myr.

\subsubsection{Satellites}

Most of the planets possess satellite systems, some of which are 
very extensive. Their observed properties, and by inference their 
origins, are heterogeneous. All four giant planets possess systems of 
{\em regular} satellites that have prograde 
orbits approximately coincident with the equatorial plane of the 
planet. The regular satellites orbit relatively close to their 
planets (in one definition, regular satellites orbit less than  
0.05~Hill radii away from their planet, where the Hill radius is 
defined as $r_H \equiv (M_p/3M_\odot)^{1/3} a$). The {\em irregular} 
satellites orbit further out and exhibit a large range of 
eccentricities and inclinations. Finally, the Earth's Moon and Pluto's 
companion Charon are so anomalously massive as 
to suggest that they belong to a third class.

There is a consensus that the Moon formed as a consequence of a giant impact 
event late in the final assembly of the Earth \citep{benz86,canup04}. The 
probability of a suitable collision is moderately high ---  
of the order of 10\% \citep{elser11} --- and it is well-established that 
an impact can eject debris that would rapidly cool and coagulate to form a 
satellite \citep{kokubo00}. The principle quantitative challenge for 
giant impact models is to 
explain the extremely close match between the composition of the Earth and the 
Moon, measured for example in terms of lunar and terrestrial oxygen 
isotope ratios. This is a problem\footnote{Amusingly, early discussions of the 
giant impact hypothesis stress the gross compositional properties of the Moon 
as motivation for the model \citep{hartmann75}.} because simulations of an 
impact that is just large enough to produce the Moon predict that the 
disk is preferentially composed of material from the impactor, which would 
have formed in at least a slightly different environment within the 
protoplanetary disk. A variety of ideas have been advanced to explain 
the observed compositional similarity, including strong  
turbulent mixing between the Earth and the initially molten 
Moon-forming disk \citep{pahlevan07}, or a larger impact that 
generated a disk with an excess of angular momentum that was 
subsequently lost \citep{canup12,cuk12,cuk16}.

The orbits of the irregular satellites suggest that they 
were captured from heliocentric orbits \citep{jewitt07}. Under restricted 
3-body gravitational dynamics (the Sun, the planet, and a massless test 
particle), however, permanent capture is impossible. Several mechanisms 
have been advanced to evade this restriction, including collisions of small 
bodies close to the planet, tidal disruption of small body binaries 
\citep{agnor06,kobayashi12} and capture facilitated by 
planetary perturbations during giant planet migration \citep{nesvorny07}. 

The dynamically cold orbits of the regular satellite systems make it tempting to 
regard them as miniature planetary systems, with an analogous formation 
mechanism \citep{lunine82}. The compositional gradient of Jupiter's Galilean satellites, 
which become increasingly ice-rich with distance from the planet, is 
consistent with such a scenario, and all models for regular satellite 
formation are based upon growth in a sub-nebular disk 
\citep[for a review, see e.g.][]{estrada09}. Recent examples of 
models for the feeding and structure of such disks include 
\citet{tanigawa12} and \citet{martin11}. It is important, however, to 
recognize that satellite formation involves significantly 
different physics and is, in some respects, even more uncertain. In 
addition to well-understood differences in the dynamics, 
neither the initial conditions for the gaseous disk component (which is 
at least initially derived from the protoplanetary disk), nor for the 
solid component (which at the late epoch of satellite formation 
is expected to be highly evolved), are very well known. Different authors have 
considered qualitatively distinct satellite formation models. 
\citet{canup02,canup08} described a satellite formation scenario (the ``gas-starved" model) 
within a disk whose physics closely parallels standard actively accreting 
protoplanetary disk models. Aspects of this model have been further 
developed by \citet{sasaki10} and \citet{ogihara12}. A different scenario (the 
``solids-enhanced minimum mass disk") has been advanced by 
\citet{estrada03a,estrada03b}. In this model the regular satellites form 
within a disk that is (at most) weakly turbulent, and hence almost static. 

\subsection{Extrasolar planet search methods}
The first extrasolar planetary system was discovered by \citet{frail92} around the millisecond 
pulsar PSR1257+12. High precision timing of the radio pulses from the neutron star was used 
to infer the reflex motion caused by the orbiting planets. Shortly afterwards the first generally 
accepted detection of an extrasolar planet orbiting a main-sequence star, 51~Peg~b, was 
announced by \citet{mayor95}. The detection method was conceptually identical --- high 
precision spectroscopy was used to measure the time-dependent radial velocity shifts 
that the planet induces on the star. 51~Peg~b, a gas giant with a 4.2~day orbital period, 
is unlike any Solar System object and is the prototype for the ``hot Jupiter" class of extrasolar 
planets.

\begin{figure}
\includegraphics[width=0.75\columnwidth,angle=-90]{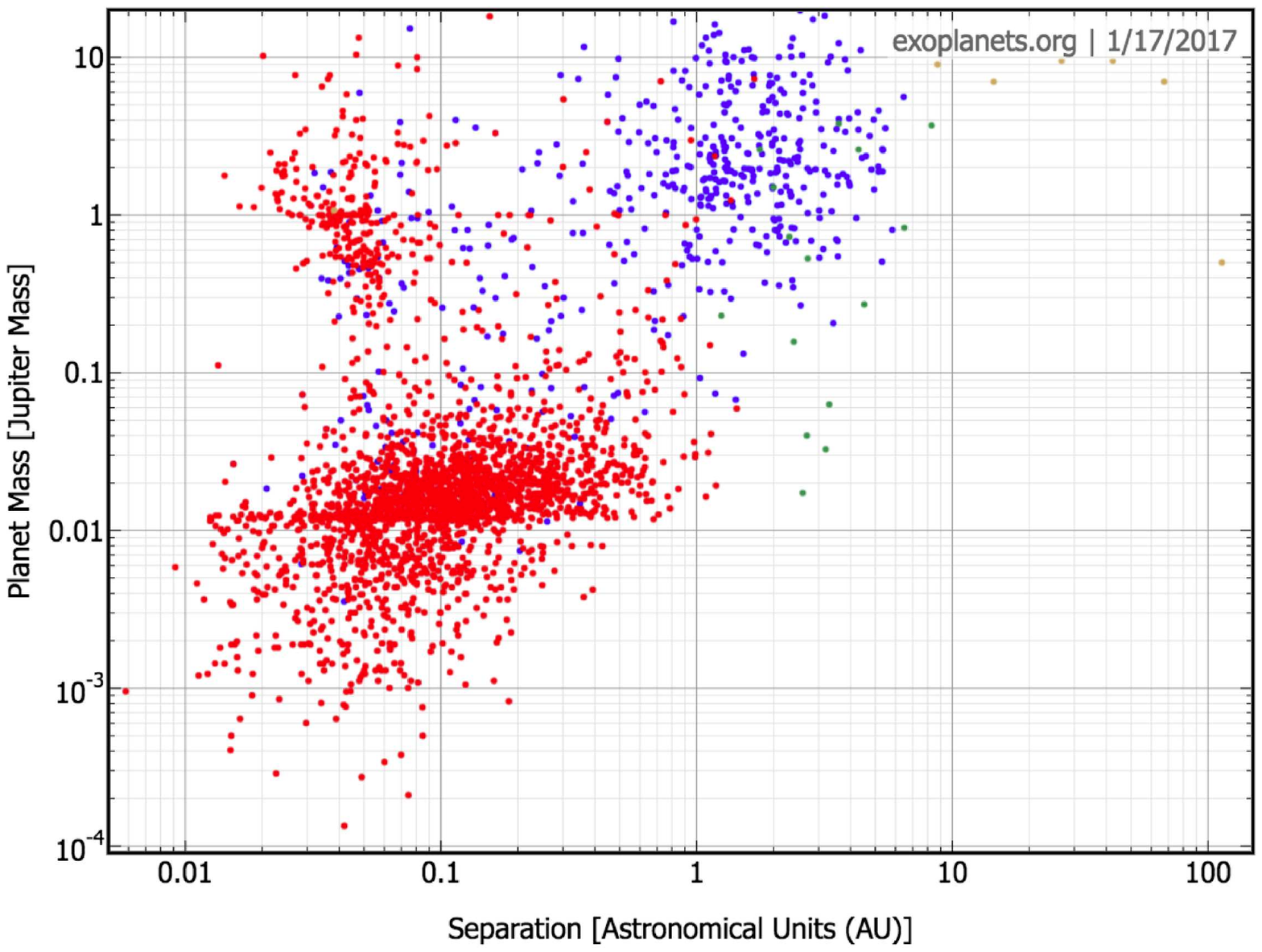}
\caption{The masses and orbital radii of many of the confirmed extrasolar planets, as of 
early 2017 (this plot was generated from {\tt exoplanets.org}). The color coding shows the 
discovery technique: radial velocity (blue), transit (red), microlensing (green) and direct 
imaging (yellow).}
\label{fig_exoplanets}
\end{figure}

Figure~\ref{fig_exoplanets} shows the distribution of a sample of extrasolar planets 
as a function of mass and orbital radius. Several thousand planets have been discovered 
from radial velocity surveys and transit searches, with NASA's {\em Kepler} 
mission contributing the largest numbers. Direct imaging and microlensing 
searches have found smaller numbers of systems, but among them are some of particular 
interest for constraining planet formation theory. Despite this bonanza, 
it is clear from Figure~\ref{fig_exoplanets} that large regions of parameter space remain 
to be explored. There is, to give one example, no current method that can find an extrasolar 
analog of Saturn, which plays a significant role in Solar System dynamics.

\subsubsection{Radial velocity searches}

\begin{figure}
\includegraphics[width=\columnwidth]{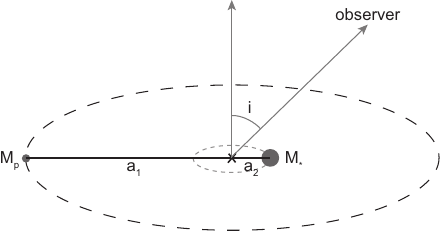}
\caption{A planet of mass $M_p$ orbits the common center 
of mass at distance $a_1$, while the star of mass $M_*$ orbits 
at distance $a_2$. The system is observed at inclination angle $i$.}
\label{fig_geometry}
\end{figure}

The observable in a radial velocity search for extrasolar planets is the time dependence of the 
radial velocity of a star due to the presence of an orbiting planet. 
For a planet on a circular orbit the geometry is shown in 
Figure~\ref{fig_geometry}. The star orbits the center of mass 
with a velocity,
\begin{equation}
 v_* \simeq \left( \frac{M_p}{M_*} \right) \sqrt \frac{GM_*}{a}.
\end{equation} 
Observing the system at an inclination angle $i$, we see the 
radial velocity vary with a semi-amplitude $K = v_* \sin i$,
\begin{equation}
 K \propto M_p \sin i a^{-1/2}.
\end{equation} 
If the inclination is unknown, what we measure ($K$) determines a 
{\em lower limit} to the planet mass $M_p$. Note that $M_*$ is 
not determined from the radial velocity curve, but must instead be  
determined from the stellar spectral properties. If the planet 
has an eccentric orbit, $e$ can be determined by fitting the 
non-sinusoidal radial velocity curve.

The noise sources for radial velocity surveys comprise photon 
noise, intrinsic jitter in the star (e.g. from convection or 
stellar oscillations), and instrumental effects. The magnitude 
of these effects vary (sometimes dramatically) from star 
to star. However, if we imagine an idealized survey for which 
the noise per observation was a constant, then the selection 
limit would be defined by,
\begin{equation} 
 M_p \sin i \vert_{\rm minimum} = C a^{1/2},
\end{equation}
with $C$ a constant. Planets with masses below this threshold 
would be undetectable, as would planets with orbital periods 
exceeding the duration of the survey (since orbital solutions 
are poorly constrained when only part of an orbit 
is observed unless the signal to noise of the observations is 
very high). The effect of such a selection boundary is 
evident in the distribution of the blue points in Figure~\ref{fig_exoplanets}. It 
favors the detection of low mass planets at small orbital radii, and 
has a relatively sharp cutoff beyond about 5~AU.

\begin{figure}
\includegraphics[width=\columnwidth]{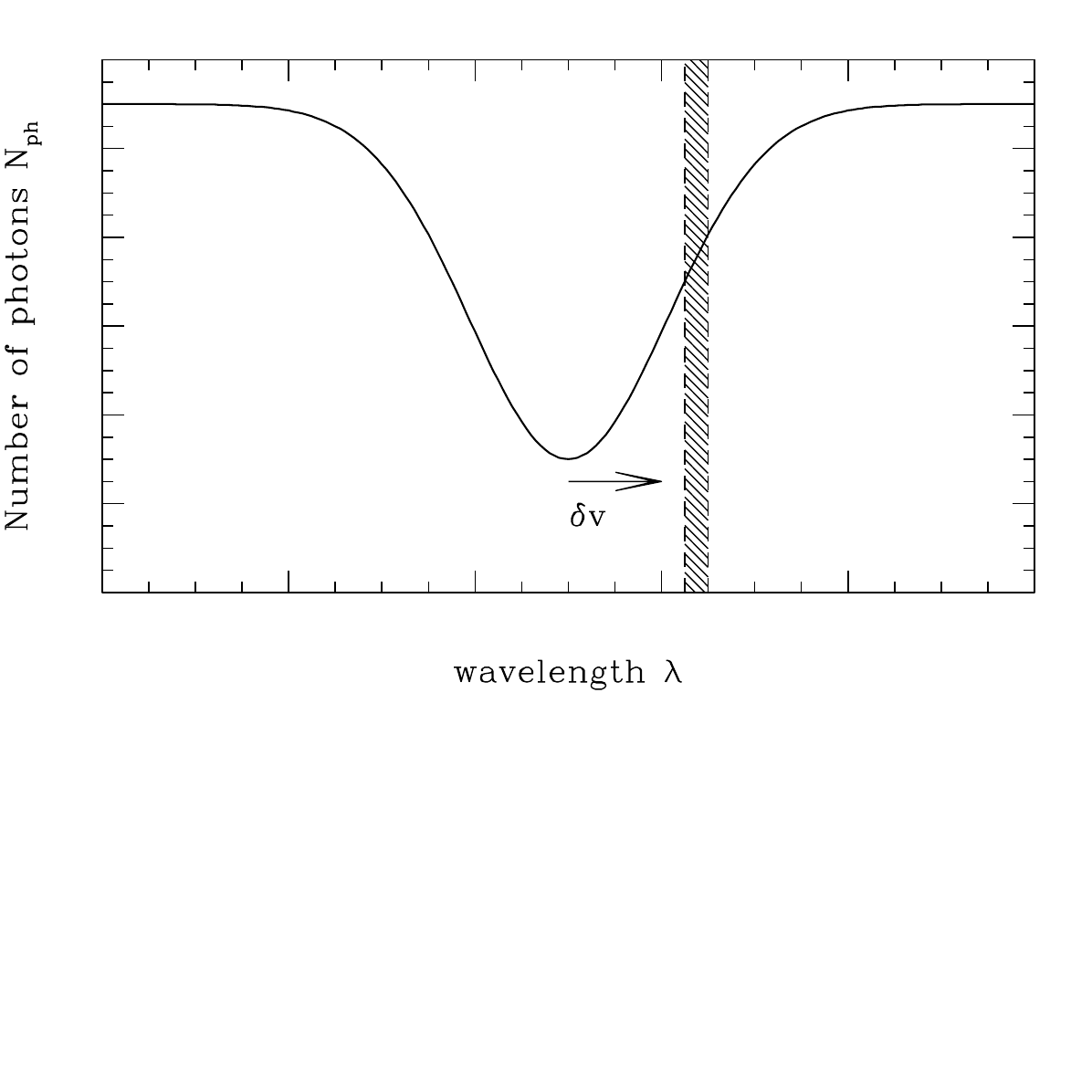}
\vspace{-3.5cm}
\caption{Schematic spectrum in the vicinity of a single spectral line of the 
host star. The wavelength range that corresponds to a single pixel in the 
observed spectrum is shown as the vertical shaded band. If the spectrum 
shifts by a velocity $\delta v$ the number of photons detected at that 
pixel will vary by an amount that depends upon the local slope of the spectrum.}
\label{figure_C1_rvprecision}
\end{figure}

Extremely accurate radial velocity measurements are a prerequisite 
for discovering planets via this technique. For the Solar System,
\begin{eqnarray}
 v_* & \approx & 12 \ {\rm ms}^{-1} \,\,\, {\rm (Jupiter)} \nonumber \\
 v_* & \approx & 0.1 \ {\rm ms}^{-1} \,\,\, {\rm (Earth)}.
\end{eqnarray}
Given that astronomical spectrographs have a resolving power of the 
order of $10^5$ (which corresponds, in velocity units, to a precision 
of the order of {\em kilometers} per second) it 
might seem impossible to find planets with such small radial 
velocity signatures. To appreciate how detection of small 
(sub-pixel) shifts is possible, it is useful to consider the 
precision that is possible against the background of shot noise 
(i.e. uncertainty in the number of photons due purely 
to counting statistics). An estimate of the photon noise limit can be derived 
by considering a very simple problem: 
how accurately can velocity shifts be estimated given measurement 
of the flux in a single pixel on the detector? To do this, we follow the 
basic approach of \citet{butler96} and consider 
the spectrum in the vicinity of a spectral line, as shown in 
Figure~\ref{figure_C1_rvprecision}. Assume that, in an observation of 
some given duration, $N_{\rm ph}$ photons are detected in the 
wavelength interval corresponding to the shaded vertical band. If we 
now imagine displacing the spectrum by an amount (in velocity units) 
$\delta v$ the change in the mean number of photons is,
\begin{equation} 
 \delta N_{\rm ph} = \frac{ {\rm d}N_{\rm ph}}{ {\rm d}v } \delta v.
\end{equation}
Since a 1$\sigma$ detection of the shift requires that $\delta N_{\rm ph} 
\approx N_{\rm ph}^{1/2}$, the minimum velocity displacement that is 
detectable is,
\begin{equation}
 \delta v_{\rm min} \approx \frac{N_{\rm ph}^{1/2}}{ {\rm d}N_{\rm ph} / {\rm d}v}.
\label{eq_C1_vmin} 
\end{equation}
This formula makes intuitive sense -- regions of the spectrum that are 
flat are useless for measuring $\delta v$ while sharp spectral features are 
good. For Solar-type stars with photospheric temperatures $T_{\rm eff} \approx 
6000 \ {\rm K}$ the sound speed at the photosphere is around 10~kms$^{-1}$. Taking 
this as an estimate of the thermal broadening of spectral lines, the slope 
of the spectrum is at most,
\begin{equation} 
 \frac{1}{N_{\rm ph}} \frac{ {\rm d}N_{\rm ph} }{ {\rm d}v } 
 \sim \frac{1}{10 \ {\rm kms}^{-1} } \sim 10^{-4} \ {\rm m}^{-1}{\rm s}.
\label{eq_C1_slope}
\end{equation} 
Combining Equations~(\ref{eq_C1_vmin}) and (\ref{eq_C1_slope}) allows us to estimate 
the photon-limited radial velocity precision. For example, if the 
spectrum has a signal to noise ratio of 100 (and there are no other 
noise sources) then each pixel receives $N_{\rm ph} \sim 10^4$ photons 
and $\delta v_{\rm min} \sim 100 \ {\rm ms}^{-1}$. If the spectrum 
contains $N_{\rm pix}$ such pixels the combined limit to the radial 
velocity precision is,
\begin{equation} 
 \delta v_{\rm shot} = \frac{\delta v_{\rm min}}{N_{\rm pix}^{1/2}} \sim 
 \frac{100 \ {\rm ms}^{-1}}{N_{\rm pix}^{1/2}}.
\end{equation}
Obviously this discussion ignores many aspects that are practically 
important in searching for planets from radial velocity data. However, 
it suffices to reveal the key feature: given a high signal 
to noise spectrum and stable wavelength calibration, photon noise is 
small enough that a radial velocity measurement with the ms$^{-1}$ 
precision needed to detect extrasolar planets is feasible.

Records for the smallest amplitude radial velocity signal that can 
be extracted from the noise have improved dramatically over the years. 
Planets have now been detected for which $K$ is 
as small as about 0.5~${\rm ms}^{-1}$ \citep{pepe11}, and there are 
plans (e.g. the {\em ESPRESSO} instrument on ESO's VLT) for 
next-generation instruments able to reach the 0.1~${\rm ms}^{-1}$ 
precision needed to find Earth analogs.
It is important to remember that these 
are best-case values -- many stars are not stable enough to 
allow anything like such high precision and 
complete samples of extrasolar planets 
that are suitable for statistical studies only exist for much 
larger $K$.

Detailed modeling is necessary in order to assess whether a particular 
survey has a selection bias in eccentricity. Naively you can argue it 
either way -- an eccentric planet produces a larger perturbation at 
closest stellar approach, but most of the time the planet is further 
out and the radial velocity is smaller. A good starting point for 
studying these issues is the explicit calculation for the Keck Planet 
Search reported by \citet{cumming08}. These authors find that the 
Keck search is complete for sufficiently massive planets 
(and thus trivially unbiased) for $e \lesssim 0.6$.

\subsubsection{Transit searches}

\begin{figure}
\includegraphics[width=\columnwidth]{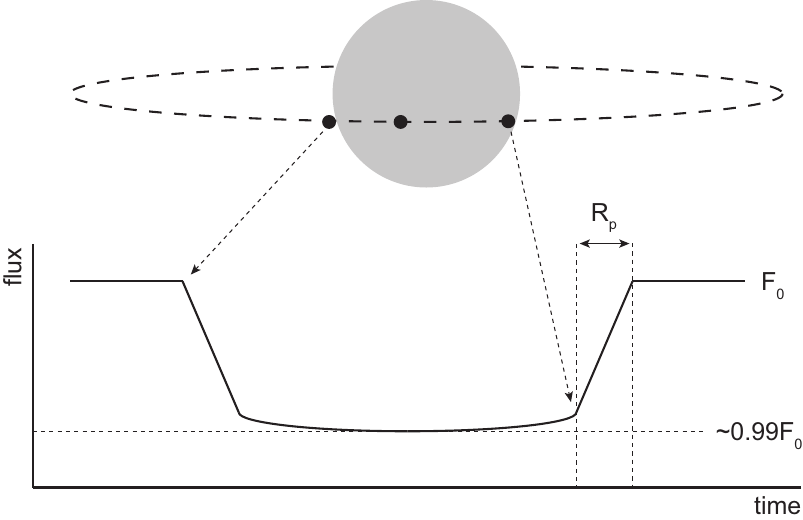}
\caption{Illustration of the light curve expected for the transit of a gas giant planet across a 
Solar-type star.}
\label{fig_transit}
\end{figure}

The observable for transit surveys is the stellar flux as a function of time. Planets emit 
very little flux in the visible, so to a good approximation a transiting planet 
produces the ``U-shaped" light curve that would result from a perfectly obscuring disk moving 
across the stellar surface as seen from Earth. Simple geometrical considerations, illustrated in 
Figure~\ref{fig_transit}, allow us to deduce two important facts. The {\em transit depth} (the 
fraction of the stellar flux that is blocked by the planet) is,
\begin{equation}
 \frac{\Delta F}{F} = \left( \frac{R_p}{R_*} \right)^2,
\end{equation}
where $R_p$ and $R_*$ are the planetary and stellar radii. For giant planets the depth 
is of the order of 1\%, while for the Earth around a Solar type star $\Delta F / F \simeq 8.4 \times 10^{-5}$.  
To see a transit requires a favorable, almost edge-on, orbital alignment. For a planet at orbital 
radius $a$, in a system observed at inclination angle $i$, some part of the planet will touch 
the stellar disk provided that $\cos(i) \le (R_p + R_*) / a$. Given random inclinations, the 
{\em probability of transit} is then,
\begin{equation}
 P_{\rm transit} = \frac{R_p + R_*}{a}.
\end{equation}
For an Earth analog this is about 0.5\%. As with radial velocity surveys, transit searches are 
thus strongly biased toward small orbital radii. Once planets are observed to transit, the 
measurable properties are the orbital period and the ratio of the planetary to stellar radius. 
The semi-major axis and planetary radius follow {\em provided} that the stellar mass and 
radius are known to good precision.

Transit searches have to contend with both noise and false positives --- astronomical events 
unrelated to planets that masquerade as transit signals (eclipsing binaries whose light is blended 
with an unrelated third star are a major source of the latter). For ground-based transit searches the 
dominant noise component is atmospheric fluctuations, which make it hard to measure stellar 
fluxes to a fractional precision better than around the $10^{-3}$ level. For Solar-type stars 
this restricts ground-based detections to the regime of gas or ice giants. \citep[Low-mass stars' 
smaller radii allow the detection of smaller planets, with GJ1214b having $R_p \simeq 2.7 \ R_\oplus$;][]{charbonneau09}. 
From space, depending on the aperture of the telescope and the brightness of the target, some combination 
of photon noise and intrinsic stellar variability dominates the noise budget. Analyses of {\em Kepler} 
data by \citet{gilliland12} and \citet{basri13} come to somewhat different conclusions, but are 
consistent with the broad-brush statement that the Sun's noise level is somewhere between typical and 
moderately quiescent as compared to other Solar-type stars. The measured stellar noise levels 
(when added to photometric and instrumental noise sources) allowed {\em Kepler} to 
discover large numbers of small planets, though the realized precision and limited lifetime of the 
original mission proved to be marginal 
for the original goal of measuring the frequency of Earth-like planets at 1~AU around Solar-type 
stars.

The information yielded by transit detections can be increased in various special circumstances. 
The observation of {\em multiple transit signals} for a single target star provides, first, near-certainty 
that the photometric signal is genuinely caused by a planet rather than being a spurious false 
positive \citep[because the probability of multiple false positive signals, with different periods, 
afflicting one star is very small;][]{lissauer12}. Second, if the planets producing the multiple 
transit signals are relatively closely spaced, their mutual gravitational perturbations may give 
rise to measurable Transit Timing Variations (TTVs) \citep{agol05,holman05}. The strength 
of TTVs is a (complex) function of the planets' masses and orbital elements, but in a useful 
subset of cases enough information is available to constrain the planets' masses using transit 
data alone \citep{ford12}. This is particularly important for the {\em Kepler} systems around 
faint hosts, where precision radial velocity follow-up is difficult and time-consuming.

\begin{figure*}
\includegraphics[width=0.75\columnwidth,angle=90]{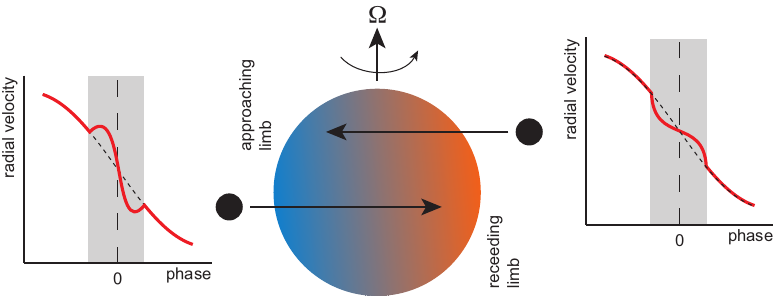}
\caption{Illustration of how radial velocity measurements {\em during transit} can constrain the 
degree of alignment between the planetary orbital axis and the stellar spin. A planet whose 
orbit is aligned with the spin first obscures a fraction of the stellar disk that is rotating towards 
us (blue shifted). A counter-rotating planet, on the other hand, first obscures a red shifted 
piece of the stellar photosphere. The shape of the radial velocity perturbation caused by this 
effect constrains the sky-projected angle between 
the orbital and spin vectors.}
\label{fig_rmeffect}
\end{figure*}

When radial velocity data {\em is} available for a star with one or more transiting planets it 
is immediately possible to estimate the true mass and density of the planets. Less obviously, 
with sufficiently precise radial velocity data it is possible to determine whether the transiting 
planet orbits within the plane defined by the rotating star's equator. This is possible because, 
as shown in Figure~\ref{fig_rmeffect}, an extra radial velocity perturbation is produced as the 
planet obscures rotationally red-shifted or blue-shifted portions of the stellar photosphere.
When this effect, known as the Rossiter-McLaughlin 
effect \citep{rossiter24,mclaughlin24}\footnote{The physical principles at work here 
long precede the detection of extrasolar planets. Detections of the 
``rotational effect" (as it was then called) in eclipsing binaries were published 
by Richard Rossiter (as part of his Ph.D. studying 
the beta Lyrae system), and by Dean McLaughlin (who studied Algol). Frank 
Schlesinger, and possibly others, may have seen similar effects in binaries.}, can be measured, 
it is possible to determine the sky-projected angle between the orbital angular momentum 
vector of the planet and the spin vector of the star. Although this is not the true inclination 
angle of the orbit, it nonetheless provides very useful information that can be used to 
test theories for the formation of close-in planetary systems. 

\subsubsection{Other exoplanet search methods}
Several other search techniques, although less important for our current understanding of the 
exoplanet population, have either furnished unique information or have significant future discovery 
potential. 

{\em Gravitational microlensing}, which works by detecting the planetary perturbation to the light curve 
of a distant star lensed by a foreground planet host, is the ground-based technique with the best 
sensitivity to low-mass planets. A planet with a mass of roughly $5 \ M_\oplus$ was found with this 
technique more that a decade ago \citep{beaulieu06}. The method is most sensitive to planets orbiting 
near the Einstein ring radius (the radius at which light from the background star traverses the lens 
system en route to us) which, interestingly, is at about the radius of the snow line (a few AU). A 
review of the method and results can be found in \citet{gaudi10}. NASA's proposed {\em WFIRST} 
mission would be able to detect a large number of low-mass planets via this technique.

{\em Direct imaging} is presently not competitive as a means of discovering planets that would be 
analogs of the Solar System's terrestrial or giant planets, but is sensitive enough to detect 
massive planets at larger orbital radii. From a theoretical viewpoint, by far the most interesting 
system seen to date is that surrounding HR~8799 \citep{marois08,marois10}. The system has four very 
massive planets orbiting at projected radii that extend out to 70~AU. As we will discuss later, it is 
hard to see how such a system could form {\em in situ}. Existing survey results show that systems 
similar to HR~8799 are moderately rare \citep[occurring with a frequency of the order of 1\%;][]{galicher16}, 
but the error bars are large. An improvement is expected with results from surveys using newer 
instruments, including the {\em Gemini Planet 
Imager} and {\em VLT Sphere}.

{\em Astrometry} works in conceptually exactly the same way as radial velocity surveys, except that 
the observable is the variation of the two-dimensional position of the star in the plane of the sky rather than the 
one-dimensional line of sight velocity. The {\em GAIA} mission, currently flying, is expected to discover a 
large number of planets via this technique.

\subsection{Exoplanet properties}
The time has long since passed when a few pages could summarize what is known observationally about 
extrasolar planetary systems. Here, we summarize some of their basic properties and highlight a few 
of the open issues that seem especially relevant to planet formation theory.

\subsubsection{Planetary masses and radii}
The mass distribution of extrasolar planets has been well-constrained by radial velocity surveys across the 
range of masses associated with ice and gas giants.  An analysis of data from 2,500 stars targeted as 
part of the Lick / Keck / AAT survey identified 250 planets, distributed in mass and radius as \citep{marcy08},
\begin{eqnarray}
 \frac{ {\rm d}N }{ {\rm d}M_p } \propto M_p^{-1.1} \\
 \frac{ {\rm d}N }{ {\rm d}\log a } \propto a^{0.4}.
\end{eqnarray}
Relatively few planets with orbital radii beyond 5~AU are known, but with that caveat the observed mass 
distribution between about 5~$M_\oplus$ and 10~$M_J$ can be considered reliably determined \citep[compare the 
above analysis, for example, to earlier work by][]{tabachnik02}.  A relatively modest extrapolation suggests that 
around 20\% of Solar-type stars are orbited by giant planets with semi-major axis less than 20~AU \citep{marcy08}. 
Most of these planets are {\em not} part of the hot Jupiter systems that were the first to be discovered, but rather 
orbit at larger distances from their hosts.

The most surprising result from the {\em Kepler} mission has been the discovery of a 
very large population of small planets in short-period orbits. For periods $P < 50 \ {\rm days}$ and 
$R_p \ge 0.5 R_\oplus$, for example, \citet{youdin11} estimate the number of planets per star to 
be around unity ($N \simeq 0.7-1.4$). As with the radial velocity sample, these planets are smoothly distributed 
in size with a distribution that increases steeply toward small radii. \citet{howard12} found that for 
planets interior to 0.25~AU the size distribution followed,
\begin{equation}
 \frac{ {\rm d}N }{ {\rm d} \log R_p} \propto R_p^{-1.9},
\end{equation}
down to radii $R_p \simeq 2 \ R_\oplus$. Intriguingly, the data does not display a bimodal distribution 
of sizes, as might be expected based on the clear separation between the radii of Solar System 
terrestrial and giant planets. Below $2 \ R_\oplus$ there is a decrease in the slope of the 
size distribution, which may be roughly flat between $1-2 \ R_\oplus$ \cite{petigura13}. 

Masses (and hence mean densities) are only available for the small subset of the {\em Kepler} 
sample that have precision Doppler measurements or useful transit timing variation constraints. 
It is clear, however, that the ``mid-sized" {\em Kepler} planets form a heterogeneous sample 
containing both ``super-Earths" (rocky planets with masses and radii greater than the Earth) 
and ``mini-Neptunes" (planets with cores but also substantial gaseous envelopes). The 
Kepler-36 system, for example, contains two planets in adjacent orbits, one with a mass 
of $4.5 \ M_\oplus$ and a density of $7.5 \ {\rm g \ cm}^{-3}$, and the other with a mass of 
$8 \ M_\oplus$ and a density of $0.9 \ {\rm g \ cm}^{-3}$ \citep{carter12}. Analysis and follow-up 
of the {\em Kepler} data is ongoing, but current work is consistent with a picture where planets 
with $R_p \lesssim 1.5 \ R_\oplus$ are predominantly super-Earths, while samples of larger planets 
contain a rising population of mini-Neptunes \citep{weiss14,marcy14}.

\subsubsection{Orbital properties}
\begin{figure}
\includegraphics[width=\columnwidth]{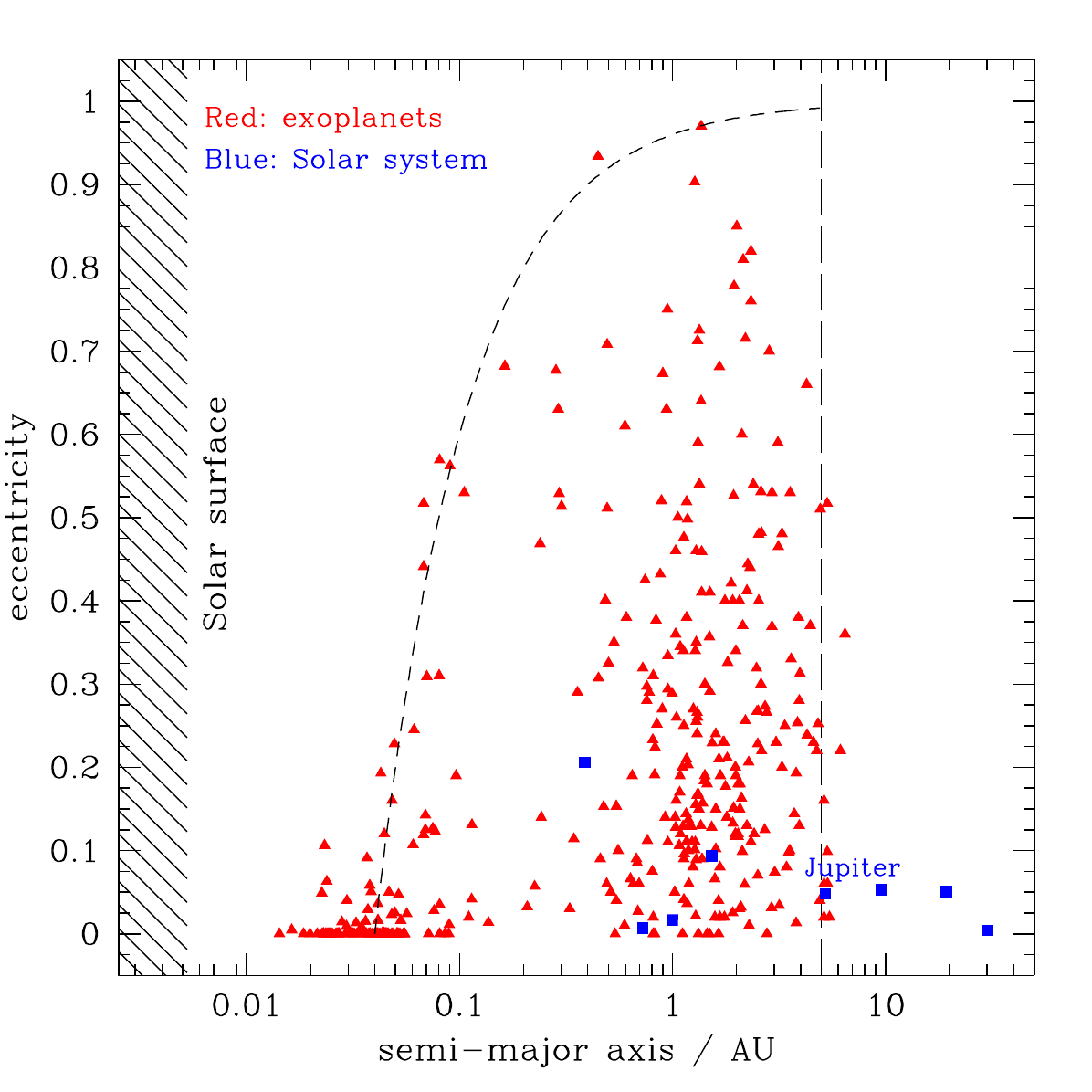}
\caption{The distribution of a sample of extrasolar planets in semi-major 
axis and eccentricity (red triangles). Solar System planets are 
shown for comparison as the blue squares. The dashed curve denotes a line of 
constant periastron distance. The figure uses data from {\tt exoplanets.org} and includes planets that have $1 \ M_J \leq M_p \sin i \leq 10 \ M_J$.}
\label{fig_data1}
\end{figure}

The distribution of {\em giant} planets in the $a$-$e$ plane is shown in Figure~\ref{fig_data1}, 
using a sample of data taken from the {\tt exoplanets.org} database. The 
closest-in hot Jupiters have circular orbits, due to tidal dissipation in the star and planet\footnote{Using a 
tidal model \citet{hansen10} fits a circularization period of about 3~days to similar data.}. At larger radii, 
however, the observed sample of exoplanets shows a striking spread in eccentricity. The median 
eccentricity is $\langle e \rangle \simeq 0.28$, and some extremely eccentric planets exist with 
$e > 0.8$. One should bear in mind that most of the detected planets are at smaller orbital radius 
than any of the gas giants in the Solar System, and many are more massive. Nonetheless, these 
large eccentricities are strikingly unlike the near-circular orbits that we are familiar with.

Several properties of observed giant planet systems are considered to furnish clues 
to the origin of eccentricity and hot Jupiters. One is the fact that ``hot Jupiters are (almost always) alone". 
Around stars that {\em do not} have a hot Jupiter, detections of multiple giant planets are reasonably 
common, with \citet{hartman14} quoting an abundance of 22\% (this number is evidently affected by 
many selection effects, so its absolute value is not important). In contrast, those systems with a hot 
Jupiter (defined as $P < 10 \ {\rm days}$) have an abundance of detected companions that is 
only around 3\%. A qualitatively similar result holds true for lower mass companions to hot 
Jupiters \citep{steffen12}. This paucity of nearby companions suggests that the formation 
process of hot Jupiters is most often inconsistent with the formation or survival of another close-in planet. 

An independent clue to hot Jupiter origins comes from measurements of the Rossiter-McLaughlin effect 
for transiting hot Jupiters. \citet{winn12} found that hot Jupiters orbiting stars with effective 
temperatures $T_{\rm eft} \gtrsim 6250 \ {\rm K}$ showed a broad distribution of projected 
obliquities, including some systems with polar and retrograde orbits. Cooler stars, on the 
other hand, showed a greater preponderance of aligned planetary orbits. The current 
obliquities may well be affected by tidal evolution --- complicating quantitative comparisons --- 
but the existence of some highly misaligned hot Jupiters certainly suggests that the formation 
process knew nothing about the spin axis of the star.

The prevalence of resonant planetary systems is also of interest. An early example was the 
GJ~876 system, which contains two massive planets in a 2:1 mean motion resonance. Unfortunately, 
an iron-clad determination of resonant behavior in an exoplanet system requires detailed observations 
that are not available for all known multi-planet systems. Among the best-characterized multi-planet 
systems containing gas giants, however, resonant configurations appear to be common. \citet{wright11}, 
for example, estimate a resonant fraction of around a third. As in the Solar System, the existence of 
these resonances is taken as evidence for dissipative processes occurring during the evolution of the 
system \citep{lee02}.

The above discussion of resonances applies to giant planet systems discovered via radial velocity 
surveys. The period ratios observed in multiple planet {\em Kepler} systems show a subtle, but 
even more intriguing structure. Most {\em Kepler} multiple systems are non-resonant, but there is a 
significant excess of pairs that are {\em just outside} first order MMRs such as the 2:1 and 3:2 \citep{fabrycky14}. 
This result is not easy to interpret, as it seems to imply simultaneously that these planets are influenced 
by resonant effects while avoiding the large-scale trapping into resonance that would be the simplest 
prediction of gas disk migration models. The short assembly time scale of planets in close-in orbits 
means that the effects of gas disk migration are likely significant, and hence one idea is that 
a higher fraction of primordial resonances has been subsequently disrupted. A broad range of 
theoretical ideas have been studied, but there is no consensus as to the most important physical 
processes responsible for the observed  {\em Kepler} systems. Paper that 
discuss various aspects of the problem include \citet{petrovich13}, \citet{goldreich14},  
\citet{hands14}, \citet{chatterjee15}, \citet{pu15} and \citet{coleman16}.

\subsubsection{Host properties}

\begin{figure}
\includegraphics[width=\columnwidth]{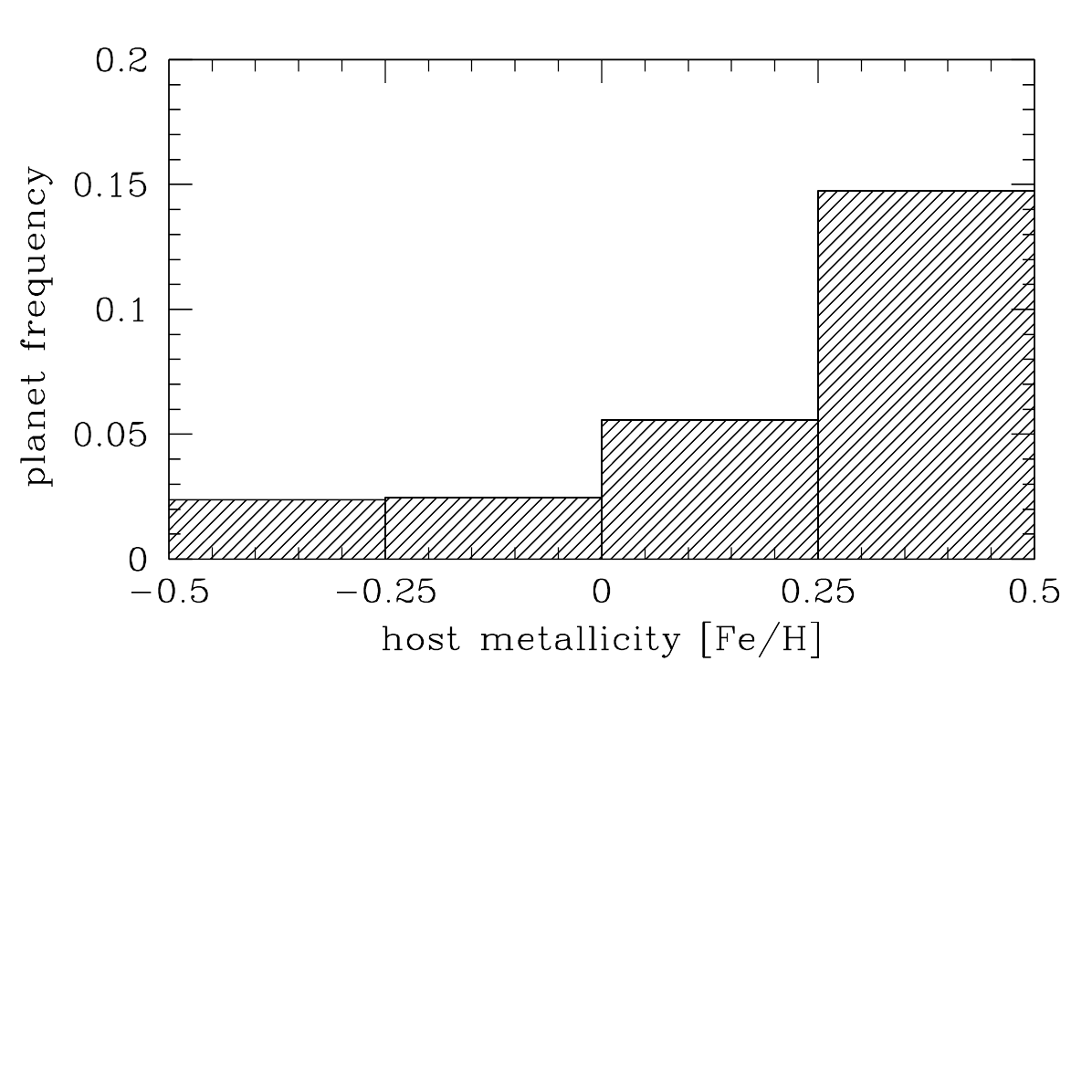}
\vspace{-1.4truein}
\caption{The fraction of stars that host giant extrasolar planets is 
plotted as a function of the stellar metallicity, from data (their Figure~4) 
reported by \citet{fischer05}.}
\label{fig_feh}
\end{figure}

The dependence of giant planet frequency with stellar metallicity is shown in Figure~\ref{fig_feh}, 
using data from the paper by \citet{fischer05}. A strong trend is evident. Changes in metallicity 
of a factor of a few lead to large variations in the incidence of detected giant planets. This is not 
surprising. Within the core accretion model for giant planet formation, a prerequisite for forming a 
gas giant is the ability to assemble a solid core of $5-10 \ M_\oplus$ during the few million year 
lifetime of the gas disk, and this is evidently easier to fulfill if the total inventory of disk solids is boosted. 
The same is not true of lower mass planets. \citet{sousa08} found that the abundance of Neptune 
analogs is not a strong function of host metallicity, and \citet{everett13}, and other groups, find that 
the same is true of the smaller planets in the {\em Kepler} sample. These results suggest that 
even if critical stages of planet formation --- such as the formation of planetesimals --- require 
threshold levels of metallicity \citep[as suggested by, e.g.,][]{johansen09}, it is still possible for 
stars with moderately sub-Solar metallicity to form systems of lower-mass planets.

The frequency of relatively close-in planets has been measured as a function of stellar 
type from {\em Kepler} data. \citet{howard12} find that planets with radii of $2-4 \ R_\oplus$ 
are substantially (by a factor of 7) more abundant around the coolest stars ($T \simeq 4000 \ {\rm K}$) 
than around stars with $T \simeq 7000 \ {\rm K}$. I am not aware of a simple explanation for 
this trend.

{\em Kepler} data has also identified a small number of {\em circumbinary} planets \citep{doyle11,welsh12}, 
whose properties are consistent with low mass gas giants. Estimates suggest that of the order of 1\% of 
tight binaries have such gas giants in almost coplanar orbits, so these are not particularly rare systems. 
They are particularly interesting for planet formation because gravitational perturbations from the binary would have 
increased the collision velocities of planetesimals above the values seen around single stars, making it 
harder for cores to grow in situ \citep[][and references therein]{lines14}.

\subsubsection{Planetary structure}
Empirical determination of the planetary mass-radius relation (from a combination of transit measurements 
of the radius, and radial velocity or TTV determinations of the mass) provides a test of models for planetary 
structure. To leading order the expectation for gas giants is that the mass-radius relation ought to be flat, 
with $R_p \simeq R_J$ being a decent approximation for sub-Jovian to several Jupiter mass planets. 
Actual transit data, however, shows that hot Jupiter radii scatter substantially above and below the 
expected values. The undersized 
gas giants are interesting, but pose no special theoretical 
conundrum. To first order, the radius of a gas giant of a given 
mass varies with the total mass of heavy elements it contains\footnote{Whether 
those heavy elements are distributed evenly within the planet or concentrated 
at the center in a core also affects the radius, but at a more subtle 
level.}; hence a plausible explanation for any small planet is that 
it has an above-average heavy element content. The measured radius of the 
Saturn mass planet orbiting HD~149026, for example, is generally 
interpreted as providing evidence for approximately $70 M_\oplus$ 
of heavy elements in the interior \citep{sato05}. The inflated 
planets, on the other hand, are more mysterious, since some (examples include TrES-4 and WASP-12b) 
are too large even when compared to pure hydrogen / helium models. Explaining their 
radii requires an additional source of heat. 

The origin of the heat source needed to explain inflated hot Jupiter radii is not fully 
understood, and may not be unique. Empirically it is observed that the 
prevalence of inflated radii increases with the degree of stellar irradiation \citep[see, e.g., plots in][]{demory11,spiegel13}, 
suggesting that in at least some cases stellar heating can couple into the convective 
interior efficiently enough to impact the radius. Suggestions for how this 
coupling might be realized physically include substantial changes to atmospheric 
opacities \citep{burrows07}, waves that connect the radiative and convective 
regions \citep{guillot02}, and magnetic fields that generate Ohmic heating 
of the interior \citep{batygin10,ginsburg16}. \citet{spiegel13} provide a much more 
extensive list of references to proposed mechanisms. 

The composition of lower mass planets is plausibly much more diverse. In the Solar System we 
have have only the terrestrial planets (dominated by rock, with atmospheres that are negligible 
from a mass-radius perspective) and the ice giants, but the results from {\em Kepler} show 
that the Solar System gap between these classes is not a general outcome of planet formation. 
Taking generality to its extreme limit, we might then consider the structure of low-mass planets 
composed of arbitrary mixtures of iron, silicates, ices and H/He. This approach yields instructive 
limits: an inferred density higher than that of a pure iron planet is unphysical, while a density 
lower than that of a pure silicate world implies the existence of an atmosphere of volatiles. 
An analysis by \citet{rogers15} suggests that most {\em Kepler} planets larger than  
$1.6 \ M_\oplus$ have transit radii that are determined by their atmospheres or envelopes. 
It is clear, of course, that there are normally too many variables to admit a unique 
determination of the composition given only measurements of the mass and radius, and 
other constraints are needed to break degeneracies. Such constraints could come from 
additional observations (e.g. of the atmospheric composition) or from theoretical priors 
(e.g. a 100\% water planet is hard to construct outside of science fiction).

\subsubsection{Habitability}
The primary long-term goal of observational exoplanet research is to identify low-mass 
planets and characterize their atmospheres via either transmission or emission 
spectroscopy. We currently know almost nothing about the diversity of terrestrial 
planet atmospheres, so such an exercise is certain to be scientifically interesting. 
Moreover, it is possible that we might identify 
one or more {\em biomarkers} --- atmospheric constituents that have a biological 
origin on Earth and which would be removed from the atmosphere by abiotic 
processes on a short time scale. Oxygen is much the most important of these. 
If it proves possible both to measure one or more biomarkers, and to robustly 
exclude non-biological interpretations, we will have discovered evidence for 
life elsewhere.

\begin{figure}
\includegraphics[width=\columnwidth]{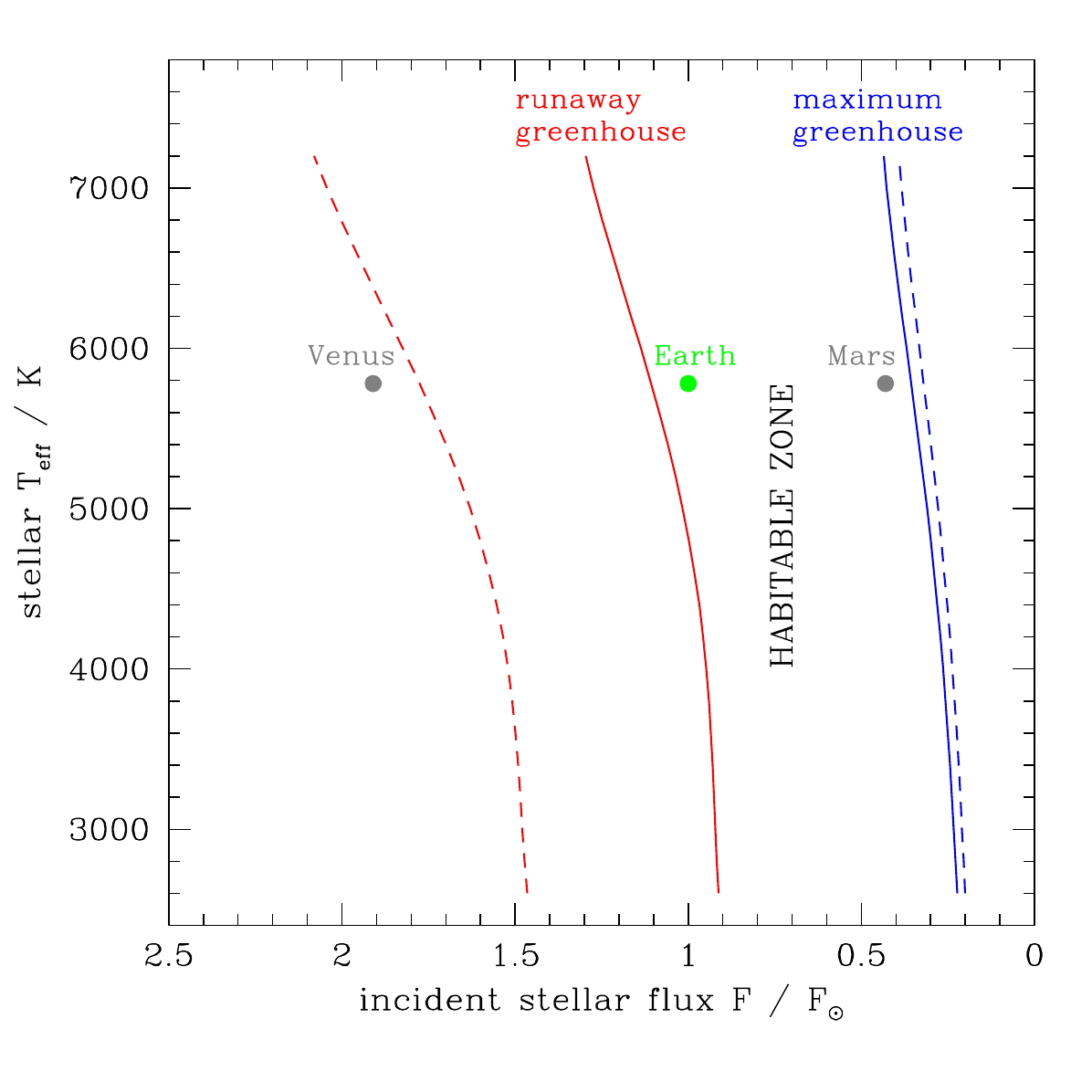}
\caption{The predicted width of the habitable zone for stars with different effective 
temperatures $T_{\rm eff}$, here plotted in terms of the incident stellar flux on the 
planet relative to that for the present-day Earth. The solid red and blue curves 
show theoretical estimates for the location of the inner and outer boundaries of the 
habitable zone, based on one-dimensional planetary atmosphere models that 
incorporate the warming effects of greenhouse gases \citep{kopparapu13}. The dashed curves show 
empirical estimates based on the idea that relatively recent Venus and early Mars may both 
have been habitable.}
\label{fig_HZ}
\end{figure}

In the current absence of such empirical evidence, the best we can do is to make educated 
guesses as to which extrasolar planets have the best chance of being habitable. 
Habitability is not a precisely defined concept, and discussion of it invites speculation as to which planetary properties 
are either essential or favorable for life. On the Earth, for example, we owe the long-term stability 
of the climate to the negative feedback of the carbonate-silicate cycle, by which the volcanic 
outgassing of greenhouse gases is balanced against the temperature-dependent weathering 
of silicate rocks \citep{walker81}. The operation of this cycle requires plate tectonics, which then 
may (or may not) be a prerequisite for habitability. Similarly, magnetic fields --- which reduce the 
rate of atmospheric erosion due to high energy radiation --- and a stable obliquity have been 
suggested to contribute to the Earth's benign environment. 

The prospects for remotely measuring {\em all} of the properties that might impact habitability 
are slim. We can, however, plausibly identify planets 
whose temperatures and pressures could support liquid water on their surfaces. 
The presence of liquid water on the surface is probably neither a necessary nor a sufficient 
condition for a planet to be habitable (note that in the Solar System, there is interest in moons 
such as Europa that likely support sub-surface oceans), but by convention the range of orbital 
radii across which planets on circular orbits could maintain surface water is called the 
{\em habitable zone} \citep{kasting93}. The habitable zone varies with both time \citep[the young Sun was as much as 
30\% fainter than it is today;][]{sagan72} and planetary mass \citep{kopparapu14}. The uncertain 
role of greenhouse gases means that even this simplest element of habitability is not easy to calculate accurately. Let us first 
consider a planet devoid of any atmosphere. Balancing incoming stellar radiation 
$\pi R_p^2 L_* / (4 \pi a^2)$ against outgoing thermal radiation $4 \pi R_p^2 \sigma T_{\rm s}^4$ 
gives, for a planetary albedo $A$, 
\begin{equation}
 T_{\rm s} = 255 \left( \frac{1 - A}{0.7} \right)^{1/4} \left( \frac{L_*}{L_\odot} \right)^{1/4} 
 				\left( \frac{a}{\rm 1 \ AU} \right)^{-1/2} {\rm K}.
\end{equation}				
This estimate gives a temperature significantly lower than the actual temperature when 
applied to the Earth (taking $A=0.3$). Clearly, an accounting for the warming effects of the atmosphere is essential. 

Two approaches have been used to estimate the extent of the habitable zone. The theoretical 
approach, pioneering by \citet{kasting93}, uses planetary atmosphere models to bracket 
the conditions under which a greenhouse gas atmosphere can sustain liquid water on the 
surface. The inner edge of the habitable zone is set by the onset of a {\em runaway greenhouse}, 
in which increased surface temperatures lead to increased evaporation of surface water (itself 
a greenhouse gas), so that the entire ocean inventory of water ultimately ends up in the atmosphere. 
The outer edge is set by a {\em maximum greenhouse} condition. Although a volcanic 
planet can outgas very large quantities of CO$_2$, the maximum atmospheric content (and consequently 
the maximum extent of warming) is limited by the onset of CO$_2$ condensation. Figure~\ref{fig_HZ} 
shows the width of the habitable zone defined theoretically by these physical considerations \citep{kopparapu13}.

The theoretical habitable zone is not very broad. For current Solar conditions the inner edge is not far 
inside 1~AU, while the outer extent would not stretch to encompass the orbit of Mars given the faintness  
of the young Sun. As discussed in the review by \citet{gudel14} it is likely that the true habitable zone 
differs from the idealized theoretical one, due to known simplifications (e.g. using one dimensional 
atmosphere models) and, possibly, neglected physical effects. It is then useful to consider an 
empirical habitable zone defined, not by theory, but rather by Solar System observations. There is 
both {\em in situ} and geomorphological evidence that liquid water flowed on Mars around 4~Gyr 
ago, suggesting but not proving that Mars lay inside the outer edge of the habitable zone despite the lower 
Solar flux at that time. Less securely, there are suggestions that Venus may have been 
habitable in the relatively recent past, even though it is well inside the theoretical inner 
boundary of the habitable zone. From these considerations we can define an empirical 
habitable zone for the Solar System and, by appropriate scaling, for other stellar types. 
These limits, which are shown as the dashed lines in Figure~\ref{fig_HZ}, can be 
regarded as optimistic inner and outer bounds.

\section{Protoplanetary Disks}

A more extensive review of protoplanetary disk physics can be found in ``Physical processes in protoplanetary disks" \citep{armitage15}. The reader whose main interests lie in disks may want to start there.

\subsection{The star formation context}

Stars form in the Galaxy today from the small fraction of gas that 
exists in dense molecular clouds. Molecular clouds are observed in one 
or more molecular tracers -- examples include CO, $^{13}$CO and NH$_3$ -- 
which can be used both to probe different regimes of column density and 
to furnish kinematic information that can give clues as to the presence 
of rotation, infall and outflows. Observations of the dense, small scale 
{\em cores} within molecular clouds (with scales of the order of 
0.1~pc) that are the immediate precursors of star formation show 
velocity gradients that are of the order of $1 \ {\rm km \ s}^{-1} \ {\rm pc}^{-1}$. 
Even if all of such a gradient is attributed to rotation, the parameter,
\begin{equation}
 \beta \equiv \frac{E_{\rm rot}}{\vert E_{\rm grav} \vert}
\end{equation}
is small -- often of the order of 0.01. Hence rotation is dynamically 
unimportant during the early stages of collapse. The angular momentum, 
on the other hand, is large, with a ballpark figure being 
$J_{\rm core} \sim 10^{54} \ {\rm g \ cm}^{2} \ {\rm s}^{-1}$. This is 
much larger than the angular momentum in the Solar System, never mind 
that of the Sun, a discrepancy that is described as the {\em angular 
momentum problem} of star formation. The problem has multiple 
solutions. Many stars are part of binary systems with large amounts 
of orbital angular momentum. For the single stars, magnetic flux 
that is approximately conserved within the collapsing gas can remove 
angular momentum from the system \citep[this process can be {\em too} 
efficient, resulting in a ``magnetic braking catastrophe" that precludes 
disk formation;][]{li14}. For our 
purposes, it suffices to note that the specific angular momentum 
of gas in molecular cloud cores would typically match the specific 
angular momentum of gas in Keplerian orbit around a Solar mass star 
at a radius of $\sim 10 - 10^2$~AU. 

The observed properties of  
molecular cloud cores are thus 
consistent with the formation of large disks -- of the size of the 
Solar System and above -- around newly formed stars. At least 
initially, those disks could be quite massive. One would also expect 
the disks to retain some net magnetic field that is a residual of the 
complex fields that likely threaded the molecular cloud core.

\begin{figure}
\includegraphics[width=0.8\columnwidth]{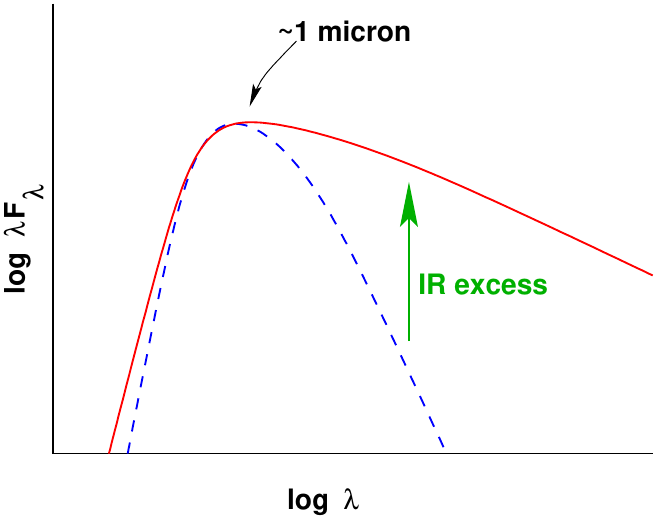}
\caption{Schematic depiction of the Spectral Energy 
Distribution of a young star surrounded by a disk. The presence 
of a disk is inferred from an infra-red excess (above the 
expected photospheric value) at wavelengths longward of 
around $1 \ \mu {\rm m}$. An ultra-violet excess is also 
commonly detected, and this is attributed to gas accretion 
on to the stellar surface producing hot spots.}
\label{fig_sed}
\end{figure}

Young Stellar Objects (YSOs) are classified observationally according 
to the shape of their Spectral Energy Distribution $\lambda F_\lambda (\lambda)$ 
in the infra-red. As shown schematically in Figure \ref{fig_sed}, YSOs 
often display,
\begin{itemize}
\item[1.]
An infra-red excess (over the stellar photospheric contribution) that is 
attributed to hot dust in the disk near the star.
\item[2.]
An ultra-violet excess, which is ascribed to high temperature regions 
(probably hot spots) on the stellar surface where gas from the disk 
is being accreted.
\end{itemize}
To quantify the magnitude of the IR excess, it is useful to define 
a measure of the slope of the IR SED,
\begin{equation} 
 \alpha_{\rm IR} = \frac{\Delta \log (\lambda F_\lambda)}{\Delta \log \lambda}
\end{equation}
between the near-IR and the mid-IR. Conventions vary, but for 
illustration we can assume that the slope is measured between 
the K band (at 2.2$\mu {\rm m}$) and 
the N band (at 10$\mu {\rm m}$). We can then classify YSOs as,
\begin{itemize}
\item
{\bf Class 0}: SED peaks in the far-IR or mm part of the spectrum 
($\sim 100 \ \mu{\rm m}$), with no flux being detectable in the 
near-IR.
\item
{\bf Class I}: approximately flat or rising SED into mid-IR ($\alpha_{\rm IR} > 0$).
\item
{\bf Class II}: falling SED into mid-IR ($-1.5 < \alpha_{\rm IR} < 0$). 
These objects are also called ``Classical T Tauri stars".
\item
{\bf Class III}: pre-main-sequence stars with little or no excess 
in the IR. These are the ``Weak lined T Tauri stars" (note that 
although WTTs are defined via 
the equivalent width of the H$\alpha$ line, this is an accretion 
signature that correlates well with the presence of an IR excess).
\end{itemize}
This observational classification scheme is theoretically interpreted, 
in part, as an evolutionary sequence \citep{adams87}. In particular, 
clearly objects in Classes 0 through II eventually lose their disks 
and become Class III sources. Observational estimates for the 
duration of the gas disk phase are typically a few Myr \citep{haisch01}. 
While the gas is present, however, viewing angle may well play a 
role in determining whether a given source is observed as a Class I 
or Class II object.

\subsection{Passive circumstellar disks}
An important physical distinction needs to be drawn between 
{\em passive} circumstellar disks, which derive most of their luminosity 
from reprocessed starlight, and {\em active} disks, which are instead 
powered by the release of gravitational potential energy as gas flows 
inward. For a disk with an accretion rate $\dot{M}$, surrounding a star 
with luminosity $L_\odot$ and radius $R_* = 2 R_\odot$, the critical accretion 
rate below which the accretion energy can be neglected may be estimated 
as,
\begin{equation}
 \frac{1}{4} L_\odot = \frac{GM_* \dot{M}}{2 R_*},
\end{equation}
where we have anticipated the result, derived below, that a flat disk 
intercepts one quarter of the stellar flux. Numerically,
\begin{equation}  
 \dot{M} \approx 3 \times 10^{-8} \ M_\odot {\rm yr}^{-1}.
\end{equation} 
Measured accretion rates of Classical T Tauri stars \citep{gullbring98} 
range from an order of magnitude above this critical rate to two orders 
of magnitude below, so it is oversimplifying to assume that protoplanetary disks 
are either always passive or always active. Rather, the dominant source 
of energy for a disk is a function of both time and radius. 
We expect internal heating to dominate at early epochs and / or small 
orbital radii, while at late times and at large radii reprocessing dominates.

\subsubsection{Vertical structure}

\begin{figure}
\includegraphics[width=\columnwidth]{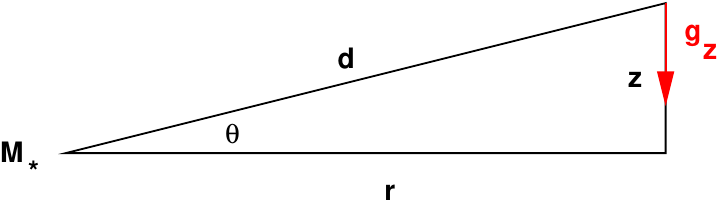}
\caption{Geometry for calculation of the vertical hydrostatic 
equilibrium of a circumstellar disk.}
\label{fig_gravity}
\end{figure}

The vertical structure of a geometrically thin disk (either passive 
or active) is derived by considering vertical hydrostatic equilibrium 
(Figure \ref{fig_gravity}). The pressure gradient is, 
\begin{equation}
 \frac{ {\rm d}P }{ {\rm d}z } = -\rho g_z,
\label{eq_hydrostatic}
\end{equation}
where $\rho$ is the gas density. Ignoring any contribution to 
the gravitational force from the disk (this is justified provided that 
the disk is not too massive), 
the vertical component of gravity seen by a parcel of gas at 
cylindrical radius $r$ and height above the midplane $z$ is,
\begin{equation} 
 g_z = \frac{GM_*}{d^2} \sin \theta = \frac{GM_*}{d^3} z.
\end{equation} 
For a thin disk $z \ll r$, so
\begin{equation} 
 g_z \simeq \Omega^2 z
\end{equation}
where $\Omega \equiv \sqrt{GM_* / r^3}$ is the Keplerian angular 
velocity. If we assume for simplicity that the disk is vertically 
isothermal (this will be a decent approximation for a passive 
disk, less so for an active disk) then the equation of state 
is $P = \rho c_s^2$, where $c_s$ is the (constant) sound speed. The 
equation of hydrostatic equilibrium (equation \ref{eq_hydrostatic}) 
then becomes,
\begin{equation}
 c_s^2 \frac{ {\rm d} \rho }{ {\rm d}z } = -\Omega^2 \rho z.
\end{equation} 
The solution is,
\begin{equation} 
 \rho = \rho_0 e^{-z^2 / 2 h^2}
\label{eq_disk_gaussian} 
\end{equation}
where $\rho_0 = \rho(z=0)$ and $h$, the vertical scale height, is given by,
\begin{equation}
 h = \frac{c_s}{\Omega}.
\end{equation}  
Integrating equation~(\ref{eq_disk_gaussian}) over $z$, we can write the 
mid-plane density $\rho_0$ in terms of the surface density and vertical 
scale height,
\begin{equation}
 \rho_0 = \frac{1}{\sqrt{2 \pi}} \frac{\Sigma}{h}.
\end{equation} 
We can also compare the disk thickness to the radius,
\begin{equation} 
 \frac{h}{r} = \frac{c_s}{v_\phi}
\end{equation}
where $v_\phi$ is the local orbital velocity. We see that the aspect 
ratio of the disk $h/r$ is inversely proportional to the Mach number 
of the flow. 

The {\em shape} of the disk depends upon $h(r)/r$. If we parameterize the 
radial variation of the sound speed via,
\begin{equation} 
 c_s \propto r^{-\beta}
\end{equation}
then the aspect ratio varies as,
\begin{equation} 
 \frac{h}{r} \propto r^{-\beta + 1/2}.
\label{eq_aspect} 
\end{equation}   
The disk will {\em flare} -- i.e. $h/r$ will increase with radius giving 
the disk a bowl-like shape -- if $\beta < 1/2$. This requires a 
temperature profile $T(r) \propto r^{-1}$ or shallower. As we will 
show shortly, flaring disks are expected to be the norm.

\subsubsection{Radial temperature profile}
The physics of the calculation of the radial temperature profile of a 
passive disk is described in papers by \citet{adams86}, \citet{kenyon87} 
and \citet{chiang97}. We begin by considering the absolute simplest model: 
a flat thin disk in the equatorial plane that absorbs all incident 
stellar radiation and re-emits it as a single temperature blackbody. 
The back-warming of the star by the disk is neglected.

\begin{figure}
\includegraphics[width=\columnwidth]{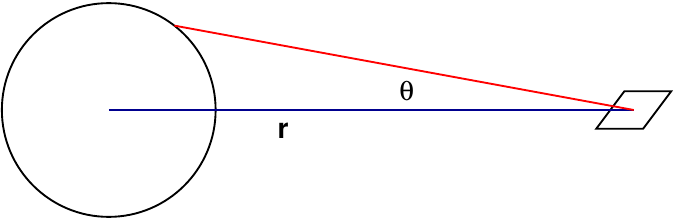}
\caption{Geometry for calculating the temperature profile of a flat, 
passive disk. We consider unit surface area in the disk plane at distance 
$r$ from a star of radius $R_*$. The axis of spherical polar co-ordinates 
is the line between the surface and the center of the star, with $\phi = 0$ 
in the direction of the stellar pole.}
\label{fig_passive}
\end{figure}

We consider a surface in the plane of the disk at distance $r$ from a star 
of radius $R_*$. The star is assumed to be a sphere of constant brightness $I_*$. 
Setting up spherical polar co-ordinates, as shown in Figure \ref{fig_passive}, 
the stellar flux passing through this surface is,
\begin{equation} 
 F = \int I_* \sin \theta \cos \phi d \Omega.
\end{equation}
We count the flux coming from the top half of the star only (and 
to be consistent equate that to radiation from only the top surface 
of the disk), so the limits on the integral are,
\begin{eqnarray}
 -\pi / 2 < & \phi & \leq \pi / 2 \nonumber \\
 0 < & \theta & < \sin^{-1} \left( \frac{R_*}{r} \right).
\end{eqnarray} 
Substituting $d\Omega = \sin \theta d\theta d\phi$, the integral for the 
flux is,
\begin{equation}
 F = I_* \int_{-\pi/2}^{\pi/2} \cos \phi d\phi 
 \int_0^{\sin^{-1}(R_*/r)} \sin^2 \theta d\theta,
\end{equation} 
which evaluates to,
\begin{equation}
 F = I_* \left[ \sin^{-1} \left( \frac{R_*}{r} \right) - 
 \left( \frac{R_*}{r} \right) \sqrt{1 - \left( \frac{R_*}{r} \right)^2} \right].
\end{equation} 
For a star with effective temperature $T_*$, the brightness $I_* = (1 / \pi) 
\sigma T_*^4$, with $\sigma$ the Stefan-Boltzmann constant \citep{rybicki79}. 
Equating $F$ to the one-sided disk emission $\sigma T_{\rm disk}^4$ we obtain a radial 
temperature profile,
\begin{equation}
 \left( \frac{T_{\rm disk}}{T_*} \right)^4 = \frac{1}{\pi} 
 \left[ \sin^{-1} \left( \frac{R_*}{r} \right) - 
 \left( \frac{R_*}{r} \right) \sqrt{1 - \left( \frac{R_*}{r} \right)^2} \right].
\label{eq_tpassive} 
\end{equation}
Integrating over radii, we obtain the total disk flux,
\begin{eqnarray}
 F_{\rm disk} & = & 2 \times \int_{R_*}^\infty 2 \pi r \sigma T_{\rm disk}^4 dr \nonumber \\
              & = & \frac{1}{4} F_*.
\end{eqnarray}
We conclude that a flat passive disk extending all the way to the stellar 
equator intercepts a quarter of the stellar flux. The ratio of the 
observed bolometric luminosity of such a disk to the stellar luminosity 
will vary with viewing angle, but clearly a flat passive disk is predicted 
to be less luminous than the star. 

The form of the temperature profile given by equation (\ref{eq_tpassive}) is 
not very transparent. Expanding the right hand side in a Taylor series, 
assuming that $(R_* / r) \ll 1$ (i.e. far from the stellar surface), we 
obtain,
\begin{equation}
 T_{\rm disk} \propto r^{-3/4},
\label{eq_tpassive2} 
\end{equation} 
as the limiting temperature profile of a thin, flat, passive disk. For 
fixed molecular weight $\mu$ this in turn implies a sound speed profile,
\begin{equation}
 c_s \propto r^{-3/8}.
\end{equation} 
Assuming vertical isothermality, the aspect ratio given by 
equation (\ref{eq_aspect}) is,
\begin{equation}
 \frac{h}{r} \propto r^{1/8},
\end{equation} 
and we predict that the disk ought to flare modestly to larger 
radii. If the disk does flare, then the outer regions intercept a 
larger fraction of stellar photons, leading to a higher temperature. 
As a consequence, a temperature profile $T_{\rm disk} \propto r^{-3/4}$ 
is probably the steepest profile we would expect to obtain for a 
passive disk.

\subsubsection{Spectral energy distribution (SED)}

\begin{figure}
\includegraphics[width=0.8\columnwidth]{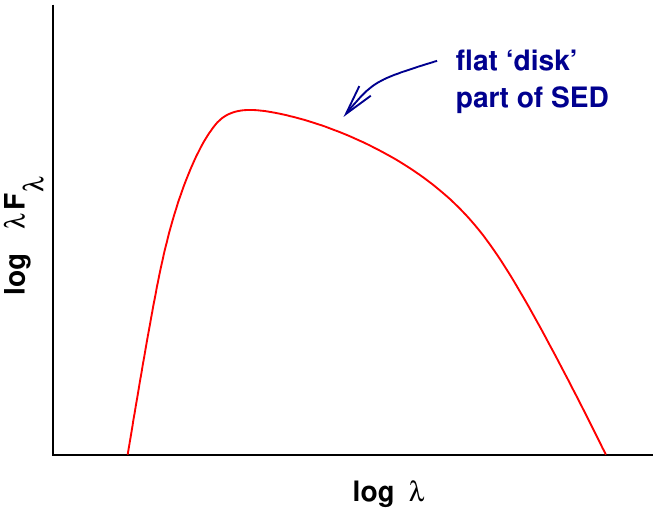}
\caption{Schematic disk spectrum. At short wavelengths, we see an 
exponential cut-off corresponding to the highest temperature 
annulus in the disk (normally close to or at the inner edge). At 
long wavelengths, there is a Rayleigh-Jeans tail reflecting the 
coldest material in the outer disk. At intermediate wavelengths, there 
is a flatter portion of the spectrum, so that the overall SED 
resembles a stretched blackbody.}
\label{fig_disk_sed}
\end{figure}

Suppose that each annulus in the disk radiates as a blackbody at the 
local temperature $T_{\rm disk} (r)$. If the disk extends from 
$r_{\rm in}$ to $r_{\rm out}$, the disk spectrum is just the 
sum of these blackbodies weighted by the disk area,
\begin{equation}
 F_\lambda \propto \int_{r_{\rm in}}^{r_{\rm out}} 
 2 \pi r B_\lambda [T(r)]dr
\label{eq_sed} 
\end{equation}
where $B_\lambda$ is the Planck function,
\begin{equation}
 B_\lambda (T) = \frac{2 h c^2}{\lambda^5}
 \frac{1}{e^{hc/\lambda k T} - 1}.
\end{equation}
The behavior of the spectrum implied by equation (\ref{eq_sed})  
is easy to derive. At long wavelengths $\lambda \gg hc / kT(r_{\rm out})$ 
we recover the Rayleigh-Jeans form,
\begin{equation} 
 \lambda F_\lambda \propto \lambda^{-3}
\end{equation} 
while at short wavelengths $\lambda \ll hc / kT(r_{\rm in})$ there is 
an exponential cut-off that matches that of the hottest annulus in the 
disk,
\begin{equation}
 \lambda F_\lambda \propto \lambda^{-4} e^{-hc/\lambda k T(r_{\rm in})}.
\end{equation} 
For intermediate wavelengths,
\begin{equation} 
 \frac{hc}{kT(r_{\rm in})} \ll \lambda \ll \frac{hc}{kT(r_{\rm out})}
\end{equation}
the form of the spectrum can be found by substituting,
\begin{equation}
 x \equiv \frac{hc}{\lambda k T(r_{\rm in})} \left( \frac{r}{r_{\rm in}} \right)^{3/4}
\end{equation} 
into equation (\ref{eq_sed}). We then have, approximately,
\begin{equation}
 F_\lambda \propto \lambda^{-7/3} \int_0^\infty \frac{x^{5/3} dx}{e^x - 1} \propto \lambda^{-7/3}
\end{equation}
and so
\begin{equation} 
 \lambda F_\lambda \propto \lambda^{-4/3}.
\end{equation} 
The overall spectrum, shown schematically in Figure \ref{fig_disk_sed}, is that 
of a ``stretched" blackbody \citep{lyndenbell69}. 

The SED predicted by this simple model generates an IR-excess, but with a 
declining SED in the mid-IR. This is too steep to match the observations of 
even most Class II sources.

\subsubsection{Sketch of more complete models}
Two additional pieces of physics need to be included when computing 
detailed models of the SEDs of passive disks. First, as already noted 
above, all reasonable disk models flare toward large $r$, and as a 
consequence intercept and reprocess a larger fraction of the stellar flux. 
At large radii, \citet{kenyon87} find that consistent flared disk 
models approach a temperature profile,
\begin{equation} 
 T_{\rm disk} \propto r^{-1/2},
\end{equation}
which is much flatter than the profile derived previously. Second, the 
assumption that the emission from the disk can be approximated as a 
single blackbody is too simple. In fact, dust in the surface layers of the 
disk radiates at a significantly higher temperature because the dust is 
more efficient at absorbing short-wavelength stellar radiation than it 
is at emitting in the IR \citep{shlosman89}. Dust particles of size $a$ 
absorb radiation efficiently for $\lambda < 2 \pi a$, but are inefficient 
absorbers and emitters for $\lambda > 2 \pi a$ (i.e. the opacity is a 
declining function of wavelength). As a result, the disk absorbs 
stellar radiation close to the surface (where $\tau_{1 \mu {\rm m}} \sim 1$), 
where the optical depth to emission at longer IR wavelengths $\tau_{\rm IR} \ll 1$. The 
surface emission comes from low optical depth, and is not at the 
blackbody temperature previously derived. \citet{chiang97} showed that a 
relatively simple disk model made up of,
\begin{itemize}
\item[1.]
A hot surface dust layer that directly re-radiates half of the 
stellar flux
\item[2.]
A cooler disk interior that reprocesses the other half of the stellar 
flux and re-emits it as thermal radiation
\end{itemize} 
can, when combined with a flaring geometry, reproduce most SEDs quite well. 
A review of recent disk modeling work is given by \citet{dullemond06}.

The above considerations are largely sufficient to understand the 
structure and SEDs of Class II sources. For Class I sources, however, 
the possible presence of an envelope (usually envisaged to comprise 
dust and gas that is still infalling toward the star-disk system) 
also needs to be considered. The reader is directed to \citet{eisner05} 
for one example of how modeling of such systems can be used 
to try and constrain their physical properties and evolutionary 
state.

\subsection{Actively accreting disks}
The radial force balance in a passive disk includes contributions 
from gravity, centrifugal force, and radial pressure gradients. The 
equation reads,
\begin{equation} 
 \frac{v_\phi^2}{r} = \frac{GM_*}{r^2} + \frac{1}{\rho} \frac{{\rm d}P}{{\rm  d}r},
\end{equation}
where $v_\phi$ is the orbital velocity of the gas and $P$ is the pressure. 
To estimate the magnitude of the pressure gradient term we note that,
\begin{eqnarray}
   \frac{1}{\rho} \frac{{\rm d}P}{{\rm  d}r} & \sim & - \frac{1}{\rho} \frac{P}{r} \nonumber \\
   & \sim & - \frac{1}{\rho} \frac{\rho c_s^2}{r}  \nonumber \\
   & \sim & - \frac{GM_*}{r^2} \left( \frac{h}{r} \right)^2, 
\end{eqnarray}   
where for the final step we have made use of the relation $h = c_s / \Omega$. 
If $v_K$ is the Keplerian velocity at radius $r$, we then have that,
\begin{equation}
 v_\phi^2 = v_K^2 \left[ 1 - {\cal{O}} \left( \frac{h}{r} \right)^2 \right],
\label{eq_vphivk} 
\end{equation} 
i.e pressure gradients make a negligible contribution to the rotation curve 
of gas in a geometrically thin $(h/r \ll 1)$ disk\footnote{This is not to 
say that pressure gradients are unimportant -- as we will see later the 
small difference between $v_\phi$ and $v_K$ is of critical importance 
for the dynamics of small rocks within the disk.}. To a good approximation, 
the specific angular momentum of the gas within the disk is just that of 
a Keplerian orbit,
\begin{equation} 
 l = r^2 \Omega = \sqrt{GM_*r},
\end{equation}
which is an {\bf increasing function} of radius. To accrete on to the star, 
gas in a disk must lose angular momentum, either,
\begin{itemize}
\item[1.] 
Via redistribution of angular momentum within the disk (normally described as 
being due to ``viscosity", though this is a loaded term, best avoided where 
possible).
\item[2.]
Via loss of angular momentum from the star-disk system, for example in a 
magnetically driven disk wind.
\end{itemize}
Aspects of models in the second class have been studied for a long time -- 
the famous disk wind solution of \citet{blandford82}, for example, describes 
how a wind can carry away angular momentum from an underlying disk. 
Observationally, it is not known whether magnetic winds are launched from 
protoplanetary disks on $1-100$~AU scales (jets, of course, are observed, 
but these are probably launched closer to the star), and hence the question 
of whether winds are important for the large-scale evolution of disks 
remains open. An review of the theory of disk winds as applied 
to protostellar systems is given by \citet{konigl10}, while \citet{bai16} 
present wind models, motivated by recent simulations, that incorporate 
both magnetic and thermal driving. To get started though, we'll initially 
assume that winds are {\em not} the dominant driver of evolution, and 
derive the equation for the time evolution of the 
surface density for a thin, viscous disk \citep{lyndenbell74,shakura73}. 
Clear reviews of the fundamentals of accretion disk theory can be 
found in \citet{pringle81} and in \citet{frank02}.

\subsubsection{Diffusive evolution equation}
Let the disk have surface density $\Sigma(r,t)$ and radial velocity 
$v_r(r,t)$ (defined such that $v_r < 0$ for inflow). The potential is 
assumed fixed so that the angular velocity $\Omega = \Omega(r)$ only. 
In cylindrical co-ordinates, the continuity equation for an axisymmetric 
flow gives (see e.g. \citet{pringle81} for an elementary derivation),
\begin{equation}
 r \frac{\partial \Sigma}{\partial t} + \frac{\partial}{\partial r} 
 \left( r \Sigma v_r \right) = 0.
\label{eq_mass} 
\end{equation} 
Similarly, conservation of angular momentum yields,
\begin{equation}
 r \frac{\partial \left( \Sigma r^2 \Omega \right)}{\partial t} + 
 \frac{\partial}{\partial r} \left( r \Sigma v_r \cdot r^2 \Omega \right) =
 \frac{1}{2 \pi} \frac{\partial G}{\partial r},
\label{eq_angular}
\end{equation}
where the term on the right-hand side represents the net torque 
acting on the fluid due to viscous stresses. From fluid dynamics 
\citep{pringle81}, $G$ is given in terms of the kinematic viscosity 
$\nu$ by the expression,
\begin{equation} 
 G = 2 \pi r \cdot \nu \Sigma r \frac{{\rm d}\Omega}{{\rm d}r} \cdot r
\end{equation}
where the right-hand side is the product of the circumference, 
the viscous force per unit length, and the level arm $r$. 
If we substitute for $G$, eliminate $v_r$ between equation (\ref{eq_mass}) 
and equation (\ref{eq_angular}), and specialize to a Keplerian 
potential with $\Omega \propto r^{-3/2}$, we obtain the evolution 
equation for the surface density of a thin accretion disk in 
its normal form,
\begin{equation}
 \frac{\partial \Sigma}{\partial t} = \frac{3}{r} \frac{\partial}{\partial r}
 \left[ r^{1/2} \frac{\partial}{\partial r} \left( 
 \nu \Sigma r^{1/2} \right) \right].
\label{eq_viscdisk}
\end{equation}  
This partial differential equation for the evolution of the surface 
density $\Sigma$ has the form of a diffusion equation. To make that 
explicit, we change variables to,
\begin{eqnarray} 
 X & \equiv & 2 r^{1/2} \nonumber \\
 f & \equiv & \frac{3}{2} \Sigma X. 
\end{eqnarray} 
For a constant $\nu$, equation (\ref{eq_viscdisk}) then takes the 
prototypical form for a diffusion equation,
\begin{equation}
 \frac{\partial f}{\partial t} = D \frac{\partial^2 f}{\partial X^2}
\label{eq_diffusion} 
\end{equation}
with a diffusion coefficient,
\begin{equation} 
 D = \frac{12 \nu}{X^2}.
\end{equation}  
The characteristic diffusion time scale implied by equation (\ref{eq_diffusion}) 
is $X^2 / D$. Converting back to the physical variables, we find that the 
evolution time scale for a disk of scale $r$ with kinematic viscosity $\nu$ is,
\begin{equation} 
 \tau \simeq \frac{r^2}{\nu}.
\end{equation}
Observations of disk evolution (for example determinations of the time scale 
for the secular decline in the accretion rate) can therefore  
be combined with estimates of the disk size to yield an estimate of the 
effective viscosity in the disk \citep{hartmann98}.

\subsubsection{Solutions}

In general $\nu$ is expected to be some function of the local 
conditions within the disk (surface density, radius, temperature, 
ionization fraction etc). If $\nu$ depends on $\Sigma$, then 
equation (\ref{eq_viscdisk}) becomes a non-linear equation 
with no analytic solution (except in some special cases), while 
if there is a more complex 
dependence on the local conditions then the surface density 
evolution equation will often need to be solved simultaneously with 
an evolution equation for the central temperature \citep{pringle86}. 
Analytic solutions {\em are} possible, however, if $\nu$ can be written as 
a power-law in radius \citep{lyndenbell74}, and these suffice to illustrate 
the essential behavior implied by equation (\ref{eq_viscdisk}).

\begin{figure}
\includegraphics[width=\columnwidth]{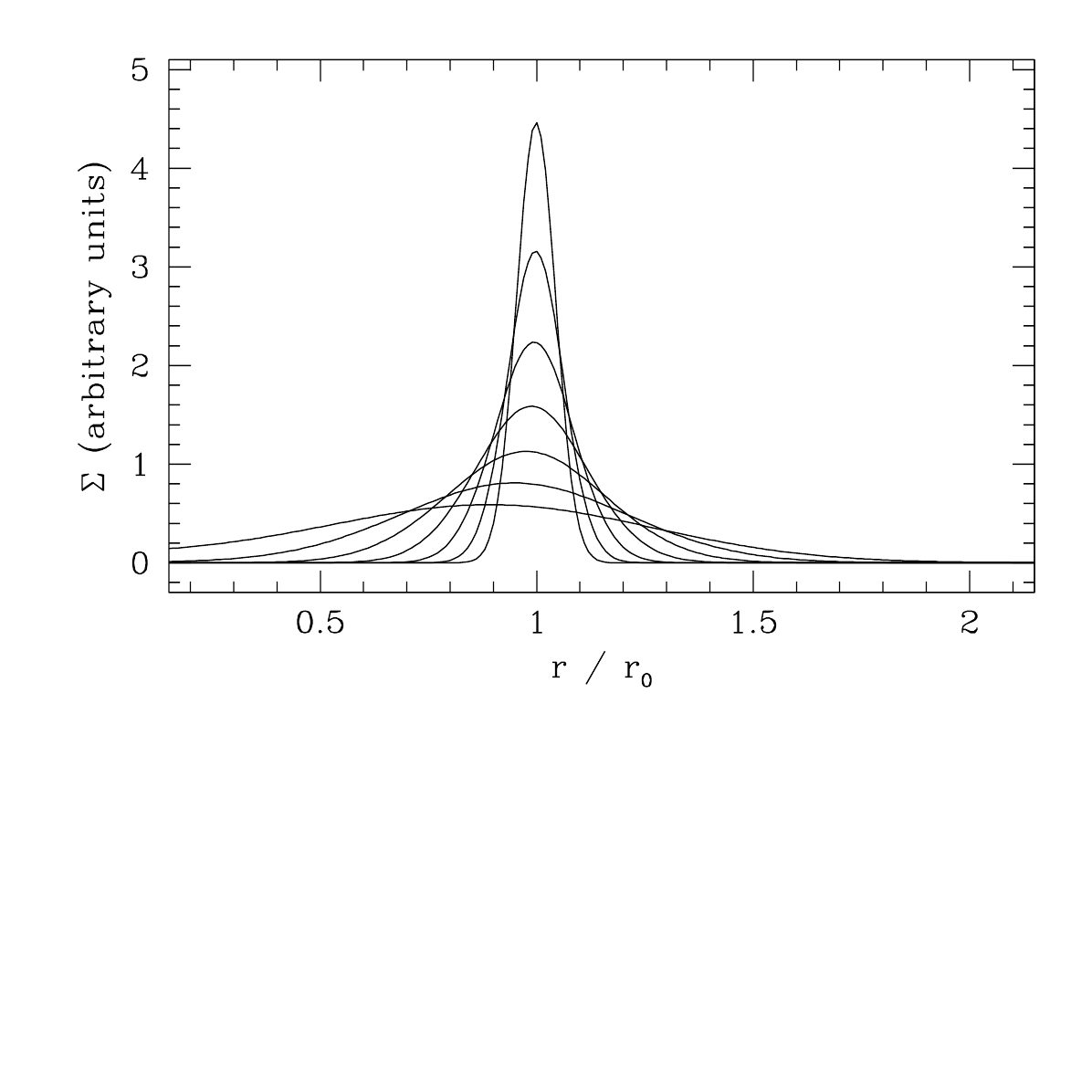}
\vspace{-3.5cm}
\caption{The Green's function solution to the disk evolution equation with 
$\nu = {\rm constant}$, showing the spreading of a ring of mass 
initially orbiting at $r = r_0$. From top down the curves show the 
behavior as a function of the scaled time variable $\tau=12 \nu r_0^{-2} t$, 
for $\tau = 0.004$, $\tau=0.008$, $\tau=0.016$, $\tau=0.032$, $\tau=0.064$, 
$\tau=0.128$, and $\tau=0.256$.}
\label{fig_sigma_solution}
\end{figure}

First, we describe a Green's function solution to equation (\ref{eq_viscdisk}) 
for the case $\nu = {\rm constant}$. Suppose that at $t=0$, all of the gas 
lies in a thin ring of mass $m$ at radius $r_0$,
\begin{equation}
 \Sigma(r,t=0) = \frac{m}{2 \pi r_0} \delta (r-r_0).
\end{equation}
One can show that the solution is then,
\begin{equation}
 \Sigma (x,\tau) = \frac{m}{\pi r_0^2} \frac{1}{\tau} 
 x^{-1/4} e^{-(1+x^2) / \tau} I_{1/4} \left( \frac{2x}{\tau} \right),
\end{equation}
where we have written the solution in terms of dimensionless variables 
$x \equiv r / r_0$, $\tau \equiv 12 \nu r_0^{-2} t$, and $I_{1/4}$ is a 
modified Bessel function of the first kind.

Unless you have a special affinity for Bessel functions, this Green's 
function solution is not terribly transparent. The evolution it implies 
is shown in Figure \ref{fig_sigma_solution}. The most important features 
of the solution are that, as $t \rightarrow \infty$,
\begin{itemize}
\item 
The {\bf mass} flows to $r = 0$.
\item
The {\bf angular momentum}, carried by a negligible fraction of the 
mass, flows toward $r = \infty$.
\end{itemize}
This segregation of mass and angular momentum is a generic 
feature of viscous disk evolution, and is obviously relevant 
to the angular momentum problem of star formation. 

\begin{figure}
\includegraphics[width=\columnwidth]{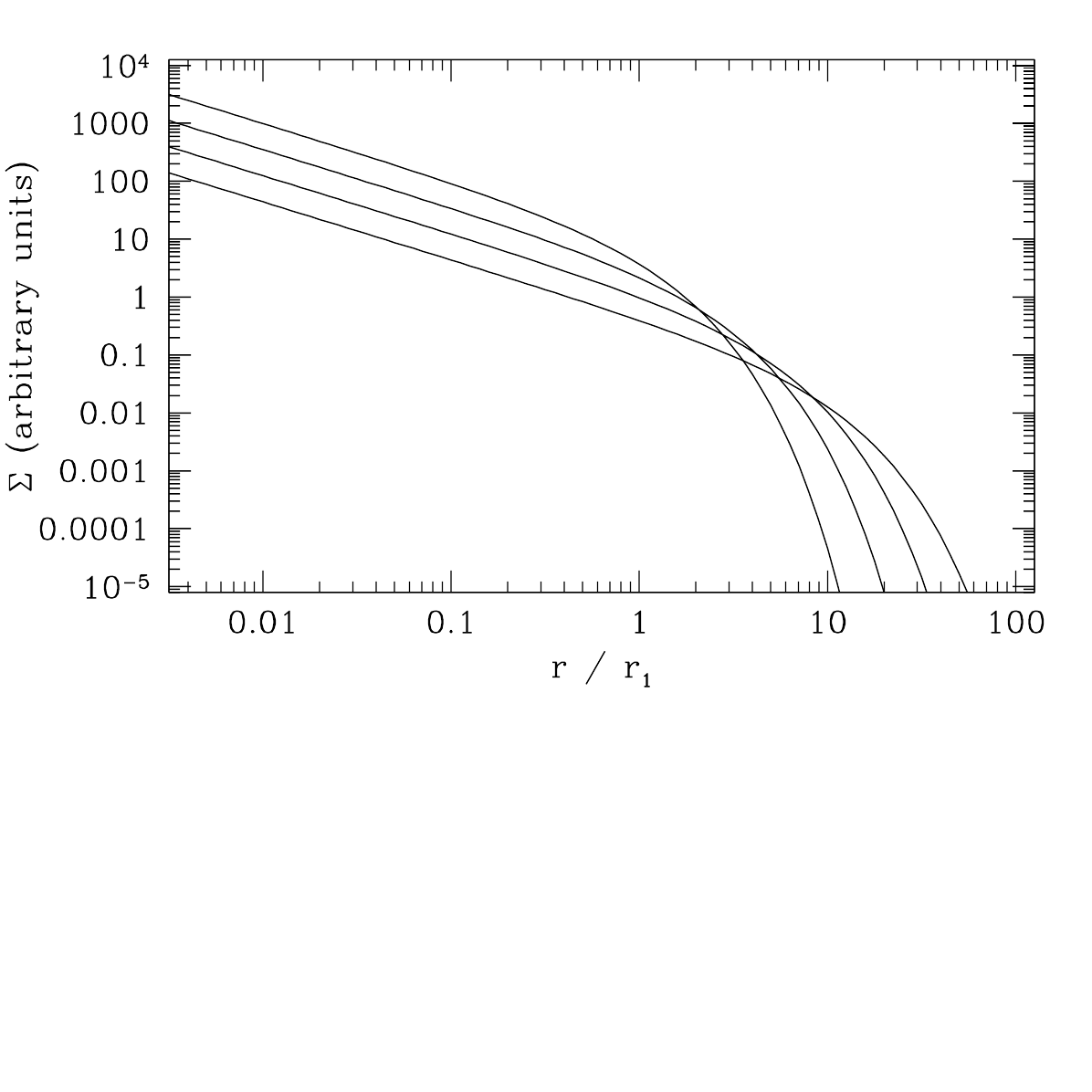}
\vspace{-3.5cm}
\caption{The self-similar solution to the disk evolution equation is plotted 
for a viscosity $\nu \propto r$. The initial surface density tracks the 
profile for a steady-state disk (with $\Sigma \propto r^{-1}$) 
at small radius, before cutting off exponentially beyond $r = r_1$. The curves 
show the surface density at the initial value of the scaled time $T=1$, and at subsequent 
times $T=2$, $T=4$ and $T=8$.}
\label{fig_selfsimilar}
\end{figure}

Of greater practical utility is the self-similar solution also derived 
by \citet{lyndenbell74}. Consider a disk in which the viscosity can 
be approximated as a power-law in radius,
\begin{equation} 
 \nu \propto r^\gamma.
\end{equation}
Suppose that the disk at time $t=0$ has the surface density profile 
corresponding to a steady-state solution (with this viscosity law) 
out to $r = r_1$, with an exponential cut-off at larger radii. As 
we will shortly show, the initial surface density then has the form,
\begin{equation} 
 \Sigma (t=0) = \frac{C}{3 \pi \nu_1 \tilde{r}^\gamma} 
 \exp \left[ -\tilde{r}^{(2-\gamma)} \right],
\end{equation}
where $C$ is a normalization constant, $\tilde{r} \equiv r / r_1$, 
and $\nu_1 \equiv \nu(r_1)$. The self-similar solution is then,
\begin{equation}
 \Sigma(\tilde{r},T) = \frac{C}{3 \pi \nu_1 \tilde{r}^\gamma} 
 T^{-(5/2-\gamma)/(2-\gamma)}  \exp \left[ - \frac{\tilde{r}^{{(2-\gamma)}}}{T} \right],
\label{eq_selfsimilar}
\end{equation}
where,
\begin{eqnarray}
 T & \equiv & \frac{t}{t_s} + 1 \nonumber \\
 t_s & \equiv & \frac{1}{3(2-\gamma)^2} \frac{r_1^2}{\nu_1}. 
\end{eqnarray}  
This solution is plotted in Figure \ref{fig_selfsimilar}. Over 
time, the disk mass decreases while the characteristic scale 
of the disk (initially $r_1$) expands to conserve angular 
momentum. This solution is quite useful both for studying 
evolving disks analytically, and for comparing observations 
of disk masses, accretion rates or radii with theory 
\citep{hartmann98}.  

\begin{figure}
\includegraphics[width=0.8\columnwidth]{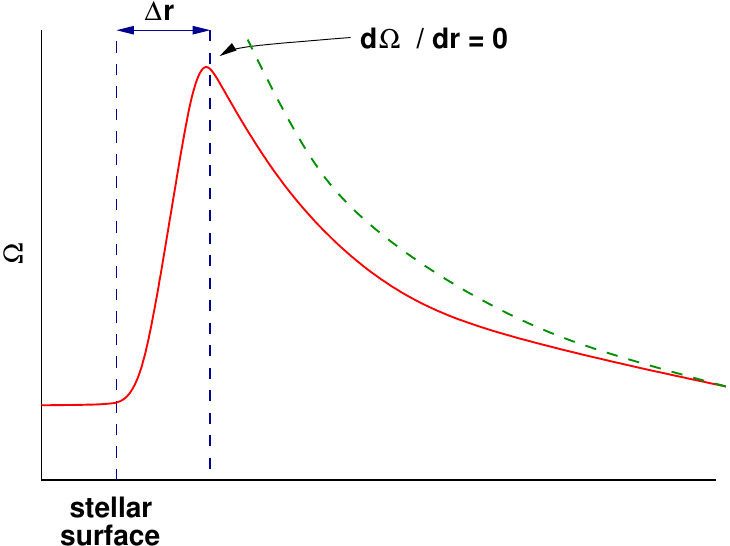}
\caption{Schematic depiction of the angular velocity $\Omega(r)$ 
for a slowly rotating star surrounded by a thin accretion disk that 
extends to the stellar equator. At large radii in the disk, the 
angular velocity has the normal Keplerian form $\Omega^{-3/2}$, 
shown as the dashed green curve. To match smoothly on to the star, 
the angular velocity must turn over at smaller radii in a transition 
zone known as the boundary layer. The existence of a boundary 
layer implies that at some radius ${\rm d} \Omega / {\rm d}r = 0$,  
at which point the viscous stress vanishes.}
\label{fig_bl}
\end{figure}

A steady-state solution for the radial dependence of the surface 
density can be derived by setting $\partial / \partial t = 0$ and 
integrating the angular momentum conservation equation (\ref{eq_angular}). 
This yields,
\begin{equation}
 \Sigma r^3 \Omega v_r = \nu \Sigma r^3 \frac{{\rm d}\Omega}{{\rm d}r} + {\rm constant}.
\end{equation}
Noting that the mass accretion rate $\dot{M} = - 2 \pi r \Sigma v_r$ we have,
\begin{equation}
 - \frac{\dot{M}}{2 \pi} r^2 \Omega = \nu \Sigma r^3 \frac{{\rm d}\Omega}{{\rm d}r} + {\rm constant}.
\label{eq_bl_integral} 
\end{equation}  
To determine the constant of integration, we note that the torque within 
the disk vanishes if ${{\rm d}\Omega} / {{\rm d}r} = 0$. At such a 
location, the constant can be evaluated and is just proportional to the 
local flux of angular momentum
\begin{equation}
 {\rm constant} \propto \dot{M} r^2 \Omega.
\end{equation} 
Usually this is determined at the inner boundary. A particularly simple 
example is the case of a disk that extends to the equator of a slowly 
rotating star. This case is illustrated in Figure \ref{fig_bl}. In 
order for there to be a transition between the Keplerian angular velocity profile 
in the disk and the much smaller angular velocity at the stellar surface there 
must be a maximum in $\Omega$ at some radius $r_* + \Delta r$. Elementary 
arguments \citep{pringle77} -- which may fail at the very high accretion rates 
of FU Orionis objects \citep{popham93} but which are probably reliable otherwise -- 
suggest that $\Delta r \ll r_*$, so that the transition occurs in a narrow 
{\em boundary layer} close to the stellar surface. The constant can then 
be evaluated as,
\begin{equation} 
 {\rm constant} \approx - \frac{\dot{M}}{2 \pi} r_*^2 \sqrt{\frac{GM_*}{r_*^3}},
\end{equation}
and equation (\ref{eq_bl_integral}) becomes,
\begin{equation} 
 \nu \Sigma = \frac{\dot{M}}{3 \pi} \left( 1 - \sqrt{\frac{r_*}{r}} \right).
\label{eq_steady} 
\end{equation}
Given a viscosity, this equation defines the steady-state surface density 
profile for a disk with an accretion rate $\dot{M}$. Away from the boundaries, 
$\Sigma(r) \propto \nu^{-1}$. 

The origin of angular momentum transport within the  
boundary layer itself presents interesting complications, since the boundary 
layer is a region of strong shear that is stable against the magnetorotational 
instabilities that we will argue later are critical for transporting angular 
momentum within disks. As a consequence, magnetic field evolution is 
qualitatively different within the boundary layer as compared to the 
Keplerian disk \citep{pringle89,armitage02b}. Analytic and simulation 
work by \citet{belyaev13} shows that acoustic waves provide the dominant 
source of transport in the boundary layer region.

The inner boundary condition which leads to equation (\ref{eq_steady}) is described as 
a {\em zero-torque} boundary condition. As noted, zero-torque conditions are 
physically realized in the case where there is a boundary layer between the 
star and its disk. This is not, however, the case in most Classical T Tauri 
stars. Observational evidence suggests \citep{bouvier06} that in accreting 
T Tauri stars the stellar magnetosphere disrupts the inner accretion disk, 
leading to a {\em magnetospheric} mode of accretion in which gas becomes 
tied to stellar field lines and falls ballistically on to the stellar 
surface \citep{konigl91}. The magnetic coupling between the star and 
its disk allows for angular momentum exchange, modifies the steady-state 
surface density profile close to the inner truncation radius, and may 
allow the star to rotate more slowly than would otherwise be the case 
\citep{acc93,armitage96}. Whether such ``disk-locking" actually regulates 
the spin of young stars remains a matter of debate, however, and both theoretical 
and observational studies have returned somewhat ambiguous results \citep{matt05,herbst05,rebull06}. 

\subsubsection{Temperature profile}
Following \citet{frank02}, we derive the radial dependence 
of the effective temperature of an actively accreting disk by considering the 
net torque on a ring of width $\Delta r$. This torque -- $(\partial G / \partial r)
\Delta r$ -- does work at a rate,
\begin{equation}
 \Omega \frac{\partial G}{\partial r} \Delta r \equiv 
 \left[ \frac{\partial}{\partial r} \left( G \Omega \right) - G \Omega^\prime \right] 
 \Delta r
\end{equation}
where $\Omega^\prime = {\rm d}\Omega / {\rm d}r$. Written this way, we note 
that if we consider the whole disk (by integrating over $r$) the first term 
on the right-hand-side is determined solely by the boundary values of 
$G\Omega$. We therefore identify this term with the {\em transport} of 
energy, associated with the viscous torque, through the annulus. The second 
term represents the rate of loss of energy to the gas. 
We assume that this is ultimately converted into heat and radiated, so that 
the dissipation rate per unit surface area of the disk (allowing that the 
disk has two sides) is,
\begin{equation} 
 D(r) = \frac{G\Omega^\prime}{4 \pi r} = \frac{9}{8} \nu \Sigma \Omega^2,
\end{equation}
where we have assumed a Keplerian angular velocity profile. For 
blackbody emission $D(r) = \sigma T_{\rm disk}^4$. Substituting for 
$\Omega$, and for $\nu \Sigma$ using the steady-state solution given by 
equation (\ref{eq_steady}), we obtain,
\begin{equation} 
 T_{\rm disk}^4 = \frac{3 G M_* \dot{M}}{8 \pi \sigma r^3} 
 \left( 1 - \sqrt{\frac{r_*}{r}} \right).
\end{equation}       
We note that,
\begin{itemize}
\item
Away from the boundaries $(r \gg r_*)$, the temperature profile of an 
actively accreting disk is $T_{\rm disk} \propto r^{-3/4}$. This has 
the same form as for a passive disk given by equation (\ref{eq_tpassive2}).
\item
The temperature profile does {\em not} depend upon the viscosity. This 
is an attractive feature of the theory given uncertainties regarding the 
origin and efficiency of disk angular momentum transport. On the flip 
side, it eliminates many possible routes to learning about the physics underlying 
$\nu$ via observations of steady-disks.
\end{itemize}
Substituting a representative value for the accretion rate of 
$\dot{M} = 10^{-7} \ M_\odot {\rm yr}^{-1}$, we obtain for a Solar 
mass star at 1~AU an effective temperature $T_{\rm disk} = 150 \ {\rm K}$. 
This is the {\em surface} temperature, as we will show shortly the 
central temperature is predicted to be substantially higher.

\subsubsection{Shakura-Sunyaev disks}
Molecular viscosity is negligible in protoplanetary disks. For a gas 
in which the mean free path is $\lambda$, the viscosity
\begin{equation}
 \nu \sim \lambda c_s
\end{equation}
where $c_s$ is the sound speed. In turn, the mean free path is 
given by $\lambda = 1 / n\sigma$, where $n$ is the number density 
of molecules with cross-section for collision $\sigma$. These 
quantities are readily estimated. For example, consider a 
protoplanetary disk with $\Sigma = 10^3 \ {\rm g \ cm}^{-2}$ and 
$h/r = 0.05$ at 1~AU. The midplane density is of the order 
of $n \sim \Sigma / 2 m_H h \sim 4 \times 10^{14} \ {\rm cm}^{-3}$, 
while the sound speed implied by the specified $h/r$ is 
$c_s \approx 1.5 \times 10^5 \ {\rm cm \ s}^{-1}$. The collision cross-section 
of a hydrogen molecule is of the order of \cite{chapman70},
\begin{equation}
 \sigma \sim 2 \times 10^{-15} \ {\rm cm}^2,
\end{equation} 
and hence we estimate,
\begin{eqnarray} 
 \lambda & \sim & 1 \ {\rm cm} \nonumber \\
 \nu & \sim & 2 \times 10^5 \ {\rm cm}^2 \ {\rm s}^{-1}. 
\end{eqnarray}  
The implied disk evolution time scale $\tau \simeq r^2 / \nu$ then 
works out to be of the order of $10^{13}$~yr -- at least $10^6$ times 
too slow to account for observed disk evolution.

In a classic paper \citet{shakura73} noted that turbulence within the 
disk can provide an effective viscosity that greatly exceeds molecular 
viscosity. For isotropic turbulence, the maximum scale of turbulent 
cells within the disk will be of the same order as the vertical 
scale height $h$, while the maximum velocity of turbulent motions 
relative to the mean flow is comparable to the sound speed $c_s$ 
(any larger velocity would lead to shocks and rapid dissipation of 
turbulent kinetic energy into heat). Such considerations motivate a 
parameterization, 
\begin{equation}
 \nu = \alpha c_s h, 
\label{eq_SS}
\end{equation}
where $\alpha$ is a dimensionless parameter that measures how efficient 
the turbulence is at creating angular momentum transport. We note at the 
outset that the existence of turbulence within the disk does not, a 
priori, guarantee that the outward angular momentum transport necessary 
to drive accretion will occur. 

In the standard theory of so-called ``$\alpha$-disks", $\alpha$ is treated as 
a constant. If this is done, it is possible to solve analytically for the 
approximate vertical structure of an actively accreting disk and derive 
a scaling for $\nu$ as a function of $r$, $\Sigma$ and $\alpha$. Textbook discussions 
of this procedure can be found in \citet{frank02}, \citet{armitage10}, and 
many other places. Combining the 
known functional form for $\nu$ with the disk evolution equation (\ref{eq_viscdisk}) 
then yields a full theory for the predicted time dependence of the disk, 
with the only unknown being the appropriate value for $\alpha$. This is 
all very well, but there is no physical reason to assume that $\alpha$ is 
a constant, and if instead $\alpha$ is regarded as a free {\em function} then 
much of the beguiling simplicity of the theory is lost. $\alpha$-disk models 
should therefore be regarded as illustrative rather than definitive predictions for 
the evolution of the disk.

It is straightforward to estimate how large $\alpha$ must be to account for 
the observed evolution of protoplanetary disks. Suppose, for example, that 
the evolution time scale at 50~AU is 1~Myr. Then starting from the $\alpha$-prescription 
(equation \ref{eq_SS}), and noting that $c_s \simeq h \Omega$, the evolution 
time scale becomes,
\begin{equation}
 \tau = \frac{r^2}{\nu} = \left( \frac{h}{r} \right)^{-2} \frac{1}{\alpha \Omega}.
\end{equation}
Substituting for $\tau$ and $r$, and assuming again that $h/r = 0.05$, we 
obtain an estimate for $\alpha \simeq 0.02$. This is fairly typical -- 
observational attempts to constrain $\alpha$ on large scales in protoplanetary 
disks (none of which are much more sophisticated than our crude estimate) tend 
to result in estimates that are around $10^{-2}$ \cite{hartmann98}\footnote{An 
important exception is modeling of the large-amplitude eruptive events known as FU 
Orionis outbursts \cite{hartmann95}, which, if interpreted as self-regulated thermal instabilities, 
require small values of $\alpha$ of the order of $10^{-3}$ or less \cite{bell94}. 
My own opinion is that these values are unreasonably small, and that FU Orionis 
events are instead triggered by instabilities that arise at larger radii \citep{armitage01,zhu09,martin11}.}. 
These values are an order of magnitude smaller 
than the values of $\alpha$ derived from the modeling of dwarf nova outbursts 
that occur in accretion disks around white dwarfs \cite{cannizzo93,king07}. 
Of course the disks around white dwarfs, and around 
other compact objects, are invariably more highly ionized than protoplanetary disks.

\subsection{Angular momentum transport}
Significant uncertainties persist as to 
the physical origin and properties of angular momentum transport within 
protoplanetary disks. The Reynolds number of the flow in the disk,
\begin{equation}
 {\rm Re} \equiv \frac{UL}{\nu}
\end{equation}
where $U$ is a characteristic velocity and $L$ a characteristic size 
scale, is extremely large (of the order of $10^{14}$ using the parameters 
that we previously estimated when considering the magnitude of molecular 
viscosity). Terrestrial flows typically develop turbulence above a 
critical Reynolds number of the order of $10^4$, so one's intuition  
would suggest that disk flows would surely be highly turbulent due to 
purely hydrodynamic effects. Detailed 
studies, however, do {\em not} support this conclusion. We first note that 
the condition for linear hydrodynamic stability in a differentially 
rotating fluid (the Rayleigh criterion) is that the specific angular 
momentum increase outward,
\begin{equation} 
 \frac{\rm d}{{\rm d}r} \left( r^2 \Omega \right) > 0.
\label{eq_rayleigh} 
\end{equation}
In a Keplerian disk, $r^2 \Omega \propto r^{1/2}$, so the flow is always 
linearly stable.

Many authors have investigated the possibility of non-linear 
instabilities that might lead to turbulence within accretion disks. To date 
there is no compelling evidence that astrophysically relevant instabilities exist. 
At least in the model system of an isothermal unstratified disk, hydrodynamic 
perturbations are found to have the potential to exhibit transient growth 
\cite{ioannou01,afshordi05} but ultimately decay \cite{balbus96,shen06,balbus06}. 
Experiments yield a similar result \cite{ji06,schartman12}. In stratified disks 
it has been suggested that vortices may be able to replicate themselves and 
give rise to turbulence \citep{marcus15}, though more work is needed both 
to clarify numerical aspects of this instability and the regime in which it 
would operate in protoplanetary disks \citep{lesur16}.

In the absence of simple hydrodynamic instabilities, the standard model 
appeals to angular momentum transport by self-gravity (at early times) 
and magnetorotational instability \citep{balbus91}. As we discuss below, the 
action of the magnetorotational instability, and more generally {\em any} 
magnetohydrodynamic process, will be strongly modified by the low ionization 
fraction of protoplanetary disks. The poor coupling between magnetic fields 
and fluid that results leads to the qualitative concept of a ``dead zone" \citep{gammie96}. 
There are also known hydrodynamic instabilities that act, not on the stable radial 
shear, but rather on either the {\em vertical} shear or the radial {\em entropy} 
gradient. The effect these hydrodynamic processes have on the qualitative picture 
of protoplanetary disks remains rather murky.

\subsubsection{Magnetorotational instability}
The hydrodynamic stability 
condition given by equation (\ref{eq_rayleigh}) is dramatically 
altered by the presence of a weak magnetic field. Whereas a 
hydrodynamic flow is stable provided only that the specific 
angular momentum increase outward, a magnetohydrodynamic (MHD) 
flow requires that the angular velocity itself be an increasing 
function of radius,
\begin{equation}  
 \frac{\rm d}{{\rm d}r} \left( \Omega^2 \right) > 0, 
\end{equation}
in order to be stable \cite{chandra61,balbus91,velikhov59}\footnote{The 
significance of Chandrasekhar's result for the origin of 
turbulence within the protoplanetary disk was appreciated by 
\citet{safronov69}, who noted that the MHD stability criterion 
does not reduce to the Rayleigh criterion as the magnetic field 
tends toward zero, and that ``for a weak magnetic field the cloud 
should be less stable than we found earlier in the absence of 
the field". Safronov then, however, dismisses the MRI on 
the (incorrect) grounds that the instability requires that the 
viscosity and diffusivity are identically zero. The importance 
of the MRI for accretion disks was only appreciated more than 
20 years later by Balbus \& Hawley.}. This condition 
is {\em not} satisfied in Keplerian disks. As a consequence, 
in ideal (zero diffusivity) MHD an arbitrarily weak magnetic field 
renders a Keplerian disk linearly unstable, with 
perturbations growing exponentially on a dynamical time scale. 

A comprehensive review of the physics of this instability -- called 
the magnetorotational (MRI) instability -- is 
given by \citet{balbus98}. The MRI is a linear instability that 
leads to self-sustaining turbulence within sufficiently well-ionized 
accretion disks \cite{axel95,stone96b}. It transports angular 
momentum outward, as is required to allow mass to flow inward and 
liberate gravitational potential energy. The magnitude of the 
effective $\alpha$, generated by the MRI under ideal MHD conditions, 
has been estimated from local simulations to be of the order of 
$\alpha \sim 10^{-2}$ \citep{davis10,simon12}. This may appear to be in 
encouraging agreement with the values inferred empirically for 
protoplanetary disks, but (as discussed below) ideal MHD is a poor approximation across much 
of the radial extent of real disks. Accordingly, although the MRI 
is generally accepted to solve the problem of angular momentum 
transport in well-ionized disks around black holes and compact 
objects, it remains possible that hydrodynamic processes play an 
important role in the protoplanetary context.

\subsubsection{Hydrodynamic transport processes}
Hydrodynamic transport is known to occur if disk self-gravity is important. 
A sufficiently massive disk is unstable \cite{toomre64} to the 
development of trailing spiral arms, which act to transport angular 
momentum outward. We will discuss the physics underlying this instability 
later in the context of models for planetesimal and giant planet 
formation, but for now we note that instability occurs when, roughly, 
\begin{equation} 
 \frac{M_{\rm disk}}{M_*} > \frac{h}{r}.
\end{equation}  
Self-gravity could therefore play a role in protoplanetary disks at 
early epochs -- when the disk may well be massive enough -- but will 
not be important at late times when $M_{\rm disk} \ll M_*$. Models 
for when self-gravity is important, and for the long-term evolution 
of disks evolving under the action of self-gravity, have been 
calculated by several authors \citep{clarke09,rafikov09,rice09}. 
The basic conclusion of such models is that -- {\em if other 
sources of angular momentum transport are weak or non-existent} -- 
then gas in the disk will settle into a stable self-gravitating 
state out to $\sim 10^2$~AU. Such disks are necessarily massive, 
and have a steep surface density profile.

In non-self-gravitating disks linear hydrodynamic instability is possible 
if the vertical entropy profile is unstable to convection. For many years 
it was thought that convection in disks transports angular momentum 
{\em inward} but simulations by \citet{lesur10} demonstrate 
that convection does yield a positive 
value of $\alpha$. Nonetheless, the difficulty of sustaining a 
sufficiently unstable vertical entropy profile means that convection 
is not considered likely to be an important process in protoplanetary 
disks.

Linear instability is also possible if there is a sufficiently 
strong {\em vertical} shear. In the presence of both radial and 
vertical gradients of specific angular momentum $l$, the stability 
criterion involves both these quantities and the gradients of 
specific entropy $S$. Specifically, stability requires,
\begin{equation}
 \frac{\partial l^2}{\partial r} \frac{\partial S}{\partial z} - 
 \frac{\partial l^2}{\partial z} \frac{\partial S}{\partial r} > 0.
\end{equation} 
A non-zero vertical shear can thus destabilize the disk. This 
``vertical shear instability" has been analyzed in detail by 
\citet{nelson13}, who show that it can give rise to significant levels 
of turbulence and transport in the upper layers of disks where the 
radiative cooling time scale is relatively short. It is closely related to 
the Goldreich-Schubert-Fricke instability of differentially rotating 
stars \citep{goldreich67,fricke68}. \citet{lin15} discuss where and 
when the VSI might operate in protoplanetary disks.

If we instead consider purely radial displacements, 
the condition for a rotating flow to 
be stable to linear axisymmetric disturbances in the presence 
of an entropy gradient is known as the Solberg-Ho\"iland 
criterion. For a Keplerian disk it can be written as,
\begin{equation}
 N^2 + \Omega^2 > 0,
\end{equation}
where $N$, the Brunt-V\"ais\"al\"a 
frequency, is,
\begin{equation}
 N^2 = - \frac{1}{\gamma \rho} \frac{\partial P}{\partial r} 
 \frac{\partial}{\partial r} \ln \left( \frac{P}{\rho^\gamma} \right),
\end{equation}
with $\gamma$ the adiabatic index. Protoplanetary disks are 
stable to radial convection by this criterion. They can, 
however, be unstable to a local, finite amplitude instability 
that is driven by the radial entropy gradient. This instability, 
called the {\em subcritical baroclinic instability} \citep{petersen07,lesur10b}, 
is present when,
\begin{equation}
 N^2 < 0,
\end{equation}
(i.e. when the disk is {\em Schwarzschild unstable}), and there is either 
significant thermal diffusion or a thermal balance set by irradiation 
and radiative cooling. The subcritical baroclinic instability results in the 
formation of vortices \citep{klahr03} -- hydrodynamic structures of the 
type exemplified by Jupiter's Great Red Spot with non-zero vorticity ${\bf \omega} \equiv 
\nabla \times {\bf v}$. Vortices are of particular 
interest because they can both transport angular momentum and,  
by trapping dust within their cores, accelerate 
the formation of larger solid bodies \cite{barge95}. 
How efficient they are at accomplishing these tasks is 
quite hard to assess, because in three dimensional disks 
vortices are subject to disruptive instabilities 
\cite{barranco05,shen06,lesur09,lithwick09}\footnote{In two dimensions, 
on the other hand, vortices are known to be long lived and 
quite effective agents of angular momentum transport \cite{godon99,johnson05}.}. 
The population of 
vortices present in a disk will reflect a balance between mechanisms that 
generate vorticity and instabilities that destroy it. 

\subsubsection{Simple dead zone models}
\label{sec_dead_zone}
Returning now to magnetohydrodynamic transport processes, a critical 
complication arises because the low ionization fraction in protoplanetary disks 
leads to a finite conductivity. Resistivity (and other 
departures from ideal MHD due to ambipolar diffusion and 
the Hall effect) can then damp or modify the MRI and 
suppress turbulence and resulting angular momentum 
transport. The linear physics in this regime has been 
analyzed in numerous papers, including works by \citet{blaes94}, \citet{desch04} 
and \citet{salmeron05}. Reviews by \citet{balbus09}, \citet{armitage11} and \citet{turner14} 
discuss the physics of the MRI in the non-ideal regime, which is highly complex. A 
good place to start though is with a toy model, in which the MRI is modified 
from the ideal MHD limit by Ohmic diffusion alone. This is the ``dead zone" or ``layered 
accretion" model proposed by \citet{gammie96}.

Following \citet{gammie96}, we begin by noting that in the 
presence of resistivity the magnetic field obeys the usual 
induction equation,
\begin{equation}
 \frac{\partial {\bf B}}{\partial t} = \nabla \times 
 \left( {\bf v} \times {\bf B} \right) - 
 \nabla \times \left( \eta \nabla \times {\bf B} \right), 
\end{equation}
where $\eta$ is the magnetic diffusivity. In turn, $\eta$ can 
be written in terms of the electron fraction $x \equiv n_e / n_H$ 
via,
\begin{equation}
 \eta = 6.5 \times 10^{-3} x^{-1} \ {\rm cm^2 s^{-1}}.
\end{equation}
Our goal is to determine the minimum $x$ for which the MRI will 
be able to operate despite the damping caused by the diffusivity. 
To do this, we note that resistivity damps small scales most easily. We 
therefore consider the largest disk scale $l = h$, and equate the 
MRI growth time scale \cite{balbus98},
\begin{equation} 
 \tau_{\rm MRI} \sim \frac{h}{v_A}
\end{equation} 
where $v_A = \sqrt{B^2 / (4 \pi \rho)}$ is the Alfv\'en speed, 
with the damping time scale,
\begin{equation} 
 \tau_{\rm damp} \sim \frac{h^2}{\eta}.
\end{equation}
This yields a simple criterion for the MRI to operate:
\begin{equation}
  \eta < h v_A.
\end{equation} 
It remains to estimate appropriate values for $h$ and $v_A$. 
For a crude estimate, we can guess that at 1~AU $h \sim 10^{12} \ 
{\rm cm}$ and that $v_A \sim c_s \sim 10^5 \ {\rm cm \ s^{-1}}$ 
(more accurately, $v_A \sim \alpha^{1/2} c_s$ in MRI turbulence 
that yields an effective Shakura-Sunyaev $\alpha$). In that 
case the limit becomes $\eta < 10^{17} \ {\rm cm^2 \ s^{-1}}$ 
which translates into a minimum electron fraction,
\begin{equation} 
 x > 10^{-13},
\end{equation}
which is more or less the ``right" value derived from more 
rigorous analyses \cite{gammie96,balbus98}. The most important 
thing to note is that this is an 
{\em extremely} small electron fraction! The linear MRI growth 
rate is so large that a tiny electron fraction couples the 
gas to the magnetic field well enough that the MRI can 
overcome the stabilizing influence of diffusion.  

Although only a small degree of ionization is required for the MRI 
to work, there may be regions in the protoplanetary disk where even 
$x \sim 10^{-13}$ is not attained. Considering first thermal 
ionization processes, calculations of collisional ionization by \citet{umebayashi83} 
show that ionization of the alkali metals suffices to drive 
$x > 10^{-13}$. This, however, requires temperatures $T \approx 10^3 \ {\rm K}$ 
and above. Only the {\em very} innermost disk -- within a few tenths 
of an AU of the star -- will therefore be able to sustain the MRI as a 
result of purely thermal ionization. 

At larger disk radii the ionization fraction will be determined by a 
balance between non-thermal ionization processes and recombination. 
Various sources of ionization are potentially important, 
\begin{itemize}
\item
Ionization by stellar X-rays. T Tauri stars are observed to be 
strong X-ray sources \cite{feigelson07}, and the harder components 
of the emission are penetrating enough to ionize a 
fraction of the column through the disk \citep{ercolano13}.
\item
Ionization by cosmic rays. Cosmic rays have a stopping length 
that is of the order of $\Sigma_{\rm layer} = 100 \ {\rm g \ cm^{-2}}$ \cite{umebayashi81}. 
{\em If present} they are therefore likely to be more penetrating and 
important than X-rays. The disk may, however, be screened from the interstellar 
cosmic ray flux by the magnetized plasma flowing away from the system in a wind. 
\citet{cleeves15} present evidence for such screening derived from observations 
of molecular line emission from the disk in the TW~Hya system.
\item
Ionization by far ultraviolet radiation \citep{perezbecker11}, which yields a relatively high 
ionization fraction ($x \sim 10^{-5}$, coming from elements such as carbon) 
within a very thin skin on the surface of the disk ($\Sigma_{\rm layer} \sim 10^{-2} - 10^{-1} 
\ {\rm g \ cm^{-2}}$).
\item
Radioactive decay, primarily from short-lived radionuclides such as $^{26}{\rm Al}$, 
which provides a minimum level of ionization independent of the other external 
ionizing agents.
\end{itemize}
The degree of ionization that results from these processes also depends on the 
efficiency of recombination, which is a sensitive function of the abundance of 
metal ions and dust particles. Detailed calculations, however, show that at radii 
where the disk is simultaneously too cool to be collisionally ionized, and dense 
enough that the interior is shielded from non-thermal ionization, non-ideal MHD 
effects will be very important. In the case of Ohmic dissipation, as originally 
considered by \citet{gammie96}, the prediction is that MHD turbulence in the 
mid-plane ought to be strongly damped. Accretion in that case would occur 
primarily through an ionized surface layer, with the interior forming a ``dead zone".

\subsubsection{Non-ideal MHD transport processes}

\begin{figure}
\includegraphics[width=\columnwidth]{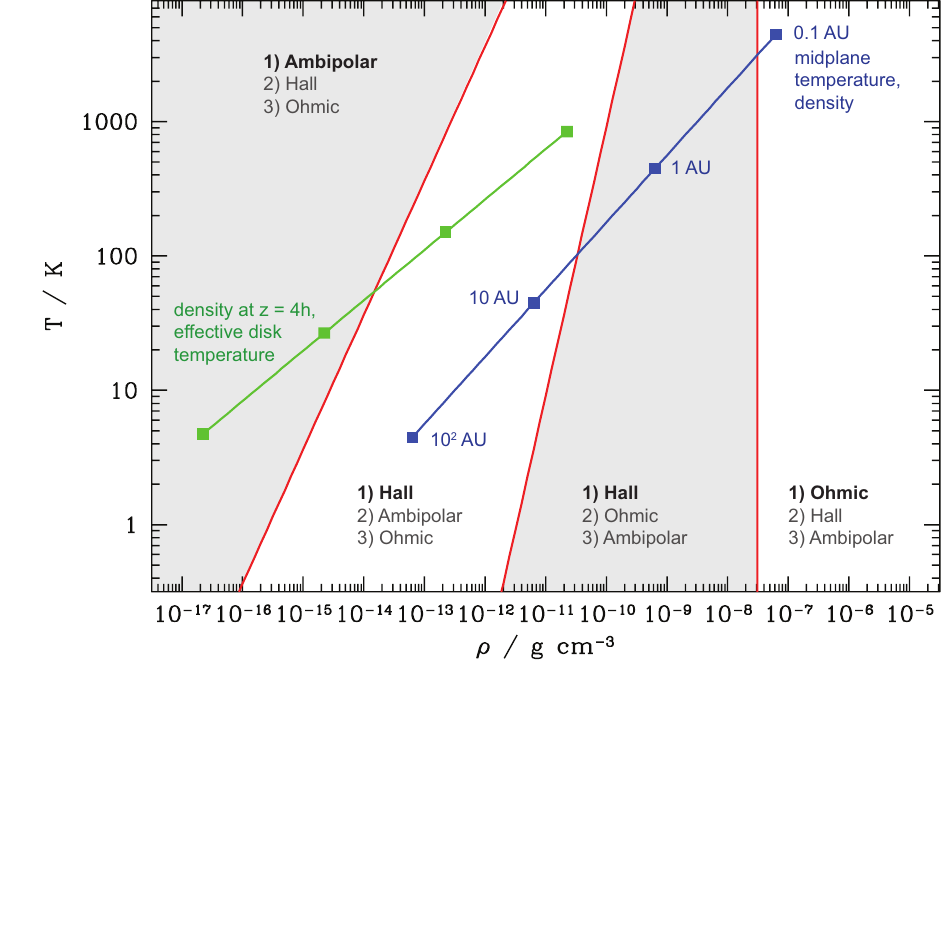}
\vspace{-1.0truein}
\caption{The relative ordering of non-ideal MHD terms plotted in the 
$(\rho,T)$ plane \citep[this version from a review by][]{armitage11}, for a magnetic field 
with $v_A / c_s = 0.1$ and a dust-free disk model. Also shown are approximate tracks 
of the radial variation of density and temperature at the mid-plane, and near 
the disk surface. The mid-plane conditions are appropriate 
for a disk around a star with $\Sigma = 10^3 (r / 1~{\rm AU})^{-1} \ 
{\rm g \ cm}^{-2}$, 
and $(h/r) = 0.04$. The surface conditions assume the density 
at $z = \pm 4 h$ (using a gaussian density profile), and a  
temperature given by the effective temperature for a steady disk accreting 
at $\dot{M} = 10^{-7} \ M_\odot {\rm yr}^{-1}$.}
\label{fig_regimes}
\end{figure}

In reality, of course, Ohmic diffusion is {\em not} the only non-ideal process that can 
affect the evolution of the MRI. The full non-ideal MHD induction equation reads,
\begin{eqnarray}
 \frac{\partial {\bf B}}{\partial t} = \nabla \times 
 \left[ {\bf v} \times {\bf B} 
 - \eta \nabla \times {\bf B} \right. \nonumber \\
 \left. - \frac{ {\bf J} \times {\bf B} }{e n_e} 
 + \frac{ ( {\bf J} \times {\bf B} ) \times {\bf B} }{c \gamma \rho \rho_I} \right].
\label{eq_full_induction} 
\end{eqnarray}
Here the current  ${\bf J} = ({c}/{4 \pi}) \nabla \times {\bf B}$, and the third and 
fourth terms on the right-hand-side describe the Hall effect and ambipolar diffusion. 
The Hall effect depends upon the electron number density $n_e$, 
while ambipolar diffusion depends upon the ion density $\rho_I$ and on the 
drag co-efficient $\gamma$ describing the collisional coupling between ions and neutrals.

The absolute strength of the non-ideal effects depends upon the ionization state of the 
disk, and can be calculated with a chemical model. These models are generally 
complex, and subject to significant uncertainties arising, for example, from the unknown 
abundance of small dust grains that play an important role in recombination. Somewhat simpler 
is an estimate of the {\em relative} strengths of the different non-ideal terms, which follows 
from a dimensional analysis of equation~(\ref{eq_full_induction}) given assumptions as to 
the strength and structure of the magnetic field \citep{balbus01,kunz04}. Such an 
analysis is shown in Figure~\ref{fig_regimes}, 
which shows the estimated ordering of the importance of the different non-ideal terms as 
a function of the disk density and temperature. Ohmic diffusion is important 
at high densities (in the inner disk) whereas ambipolar diffusion dominates in disk 
regions where the density is very low (at large radii, and in the disk atmosphere). 
The Hall effect is most important at intermediate densities.

The nature of accretion that is driven by MHD processes in non-ideal disks has been 
studied with both local and global numerical simulations. Local or ``shearing-box" 
simulations model a small co-rotating patch of disk with a linearized shear profile 
and radially periodic boundary conditions that are modified to account for the shear.  
Local simulations with reasonably realistic ionization models have been used to 
study protoplanetary accretion at radii between 1 and 100~AU:
\begin{itemize}
\item
In the {\bf outer disk} (30-100~AU, and beyond) ambipolar diffusion is the dominant 
non-ideal effect. An approach similar to that used for Ohmic diffusion in \S\ref{sec_dead_zone} 
can be used to estimate when ambipolar diffusion damps the MRI, and the general 
expectations of such an estimate are borne out by simulations \citep{bai11}. In 
the absence of any net magnetic field ambipolar diffusion damps turbulence to such 
an extent that it is difficult to achieve accretion rates that match those observed 
much closer to the star \citep{simon13}. Accretion rates that are of the right order 
of magnitude are only recovered if the disk is threaded with a weak net vertical 
field, with a ratio of mid-plane gas to magnetic pressure (in the vertical component) 
$\beta_z = P_{\rm gas} / (B_z^2 / 8 \pi) \sim 10^4$ \citep{simon13b}. This result 
suggests that net fields---which are in any event an almost unavoidable consequence 
of star formation from magnetized molecular cloud cores---are also critical drivers of 
disk accretion.
\item
In the {\bf inner disk} (1-10~AU) the physics is more complex. Ohmic and ambipolar 
diffusion damp turbulence throughout most of the disk column, though in the 
presence of a net field accretion could still occur via angular momentum loss in a 
disk wind \citep{bai13}. Accretion in this region can also be driven by the Hall 
effect, as a result of the ``Hall shear instability" \citep{kunz08} which can amplify 
magnetic fields via a mechanism that is essentially independent of the usual MRI. 
The novel feature of the Hall effect is that it operates differently depending upon the 
{\em sign} of the net field relative to the angular momentum vector of the disk 
rotation. An aligned field leads to a substantial, largely laminar magnetic stress, 
while an anti-aligned field supports only very weak stress \citep{lesur14,bai14}.
\end{itemize}
Taken together the results of local simulations suggest that the strength and 
evolution of net disk fields are major players in disk evolution, and that the 
nature of accretion in the inner disk depends upon not just the strength 
but also the orientation of the field. Simulations by \citet{simon15} indicate 
that the angular momentum transport efficiency at 1~AU could differ by 
1-2 orders of magnitude depending upon whether the net field in this 
region is aligned or anti-aligned to the rotation axis.

Local simulations are unsuitable tools for studies of disk winds, and so the 
fact that local models point to winds being important for protoplanetary disk 
accretion is more than a little concerning. Global simulations that include 
Ohmic and ambipolar diffusion have been reported by \citet{gressel15}, 
while \citet{bethune17} present simulations including all three non-ideal effects. 
The global results are rather hard to summarize, but they confirm the likely 
importance of disk winds and suggest a weaker but still significant polarity 
dependence of angular momentum transport as a consequence the Hall effect. 
They also show the development of small-scale radial structure in the net 
field, via a process of self-organization that may have observable consequences 
for the distribution of dust within disks.

\begin{figure}
\includegraphics[width=\columnwidth]{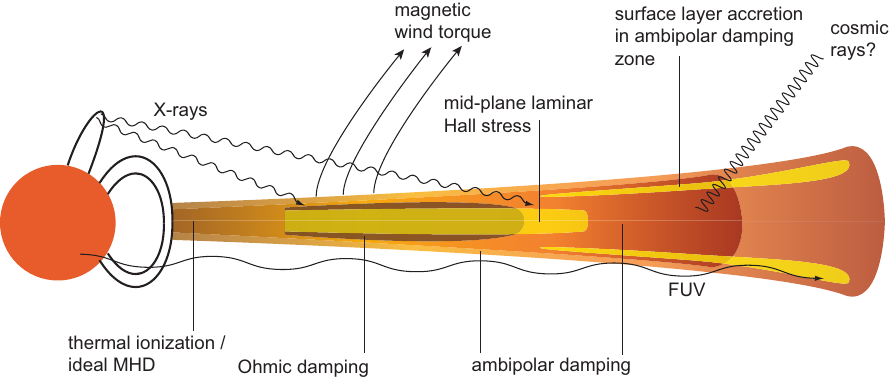}
\caption{Schematic illustration of the dead zone model for 
protoplanetary disks, originally proposed (in somewhat different 
form) by \citet{gammie96}. In this model, the innermost regions of the 
disk are hot enough that thermal ionization suffices to couple the 
magnetic field to the gas and allow the MRI to operate. At larger 
radii ionization is provided by a combination of non-thermal processes, 
stellar X-rays, far ultraviolet radiation, cosmic rays, and radioactive 
decay. The MRI is damped by Ohmic and ambipolar diffusion, 
while the Hall effect leads to a laminar magnetic stress at 
intermediate radii. There may also be magnetic torques exerted on the 
surface of the disk due to a wind.}
\label{fig_layered}
\end{figure}

Figure~\ref{fig_layered} shows a schematic depiction of our ``best guess" for what 
these non-ideal MHD simulation results imply for the global structure of protoplanetary disks. 
Three main regions are indicated. The innermost disk is hot enough to be thermally 
ionized, ideal MHD is a reasonable approximation, and the MRI results in turbulent 
transport of angular momentum. Further out turbulence is damped by a combination 
of Ohmic diffusion (near the mid-plane) and ambipolar diffusion (in the atmosphere). 
Angular momentum transport here may occur via largely laminar magnetic stresses, 
and there may also be angular momentum loss via disk winds. Finally, 
in the outer disk ambipolar diffusion damps the MRI strongly, with accretion occurring 
primarily in thin surface layers.

The paradigm of protoplanetary accretion described above, developed from theoretical 
considerations, is testable via several avenues. Furthest along are efforts to measure the 
turbulent broadening of molecular lines, which probe different regions of the vertical 
column depending upon their optical thickness (CO $3 \rightarrow 2$ transitions, for example, 
are probes of the disk atmosphere, whereas HCO$^+$ traces the mid-plane). To date, 
\citet{flaherty15} report upper limits of a few percent of the sound speed on turbulent 
broadening of CO lines in the HD~163296 system, while \citet{teague16} find results 
consistent with turbulence in the TW~Hya disk at a level of 20-40 percent of the sound 
speed. These results may be compared to theoretical predictions for HD~163296 
which imply turbulent velocities of at least tens of percent of the sound speed in 
the layer of the disk probed by CO \citep{simon15b}. There is an interesting 
conflict here between observations and theory, whose consequences remain to be 
understood.

\subsubsection{Disk dispersal}
Loss of the gaseous component of protoplanetary disks sets a 
time limit for the completion of gas giant formation, and will 
affect the environment for terrestrial planet formation as well. 
If only accretion is involved then the self-similar solution for a viscous disk 
(equation \ref{eq_selfsimilar}) predicts that the surface density 
and accretion rate decline as power-laws at late times, and hence 
that the transition between disk and diskless states should be 
very gradual. This will be modified if winds remove significant 
amounts of either mass or angular momentum from the disk. The 
best-developed models for dispersal focus on mass loss 
via thermal winds (photoevaporation), but mass or angular 
momentum loss in MHD flows may very well play a role as 
well if the disk retains magnetic flux throughout its lifetime \citep{armitage13,bai16b}.

The original motivation for considering photoevaporarion came from  
HST observations of low mass stars exposed to the strong ionizing 
flux produced by massive stars in the core of the Orion Nebula's 
Trapezium cluster \cite{odell93}. The images reveal tadpole-shaped 
nebulae surrounding young stars with circumstellar disks, which are 
interpreted as the signature of {\em photoevaporation} and escape of 
disk gas as a result of illumination by external ionizing radiation 
\cite{johnstone98}. The physics underlying this process is relatively 
simple \cite{bally82,shu93,hollenbach94}, and is closely related  
to the well-studied problem of Compton heated winds from accretion 
disks in Active Galactic Nuclei \cite{begelman83}. Extreme ultraviolet (EUV) 
photons with $E > 13.6 \ {\rm eV}$ ionize and heat a surface layer 
of the disk, raising it to a temperature $T \simeq 10^4 \ {\rm K}$ 
characteristic of an HII region. The sound speed in the photoionized 
gas is $c_s \simeq 10 \ {\rm kms}^{-1}$. Outside a critical radius 
$r_g$, given by,
\begin{equation}  
 r_g = \frac{GM_*}{c_s^2}
\label{eq_rg} 
\end{equation}
the sound speed in the hot gas exceeds the local Keplerian speed. 
The gas is then unbound, and flows away from the disk as a 
thermal wind. For a Solar mass star, $r_g$ as estimated by 
equation (\ref{eq_rg}) is at 9~AU.

The same basic process can occur regardless of whether the EUV flux 
arises from an external source, such as a massive star in a cluster, 
or from the central star itself. In the typical star formation 
environment \cite{lada03}, however, most low mass stars receive too 
low a dose of EUV radiation from external sources to destroy their 
disks \cite{adams06}. Photoevaporation due to radiation from the 
central star is therefore likely to be necessary for 
disk dispersal. In this regime, \citet{hollenbach94} derived an 
estimate for the mass loss rate due to photoevaporation,
\begin{equation}
 \dot{M}_{\rm wind} \simeq 4 \times 10^{-10} 
 \left( \frac{\Phi}{10^{41} \ {\rm s}^{-1}} \right)^{1/2} 
 \left( \frac{M_*}{M_\odot} \right)^{1/2} \ 
 M_\odot {\rm yr}^{-1}
\end{equation} 
where $\Phi$ is the stellar ionizing flux. Most of the wind 
mass loss is predicted to originate close to $r_g$, with a 
radial dependence of the mass loss given by $\dot{\Sigma} 
\propto r^{-5/2}$. Numerical hydrodynamic simulations by 
\citet{font04} largely confirm this basic picture, although 
in detail it is found both that mass is lost for radii 
$r < r_g$ and that the integrated mass loss is a factor 
of a few smaller than that predicted by the above equation.

\begin{figure}
\includegraphics[width=\columnwidth,angle=90]{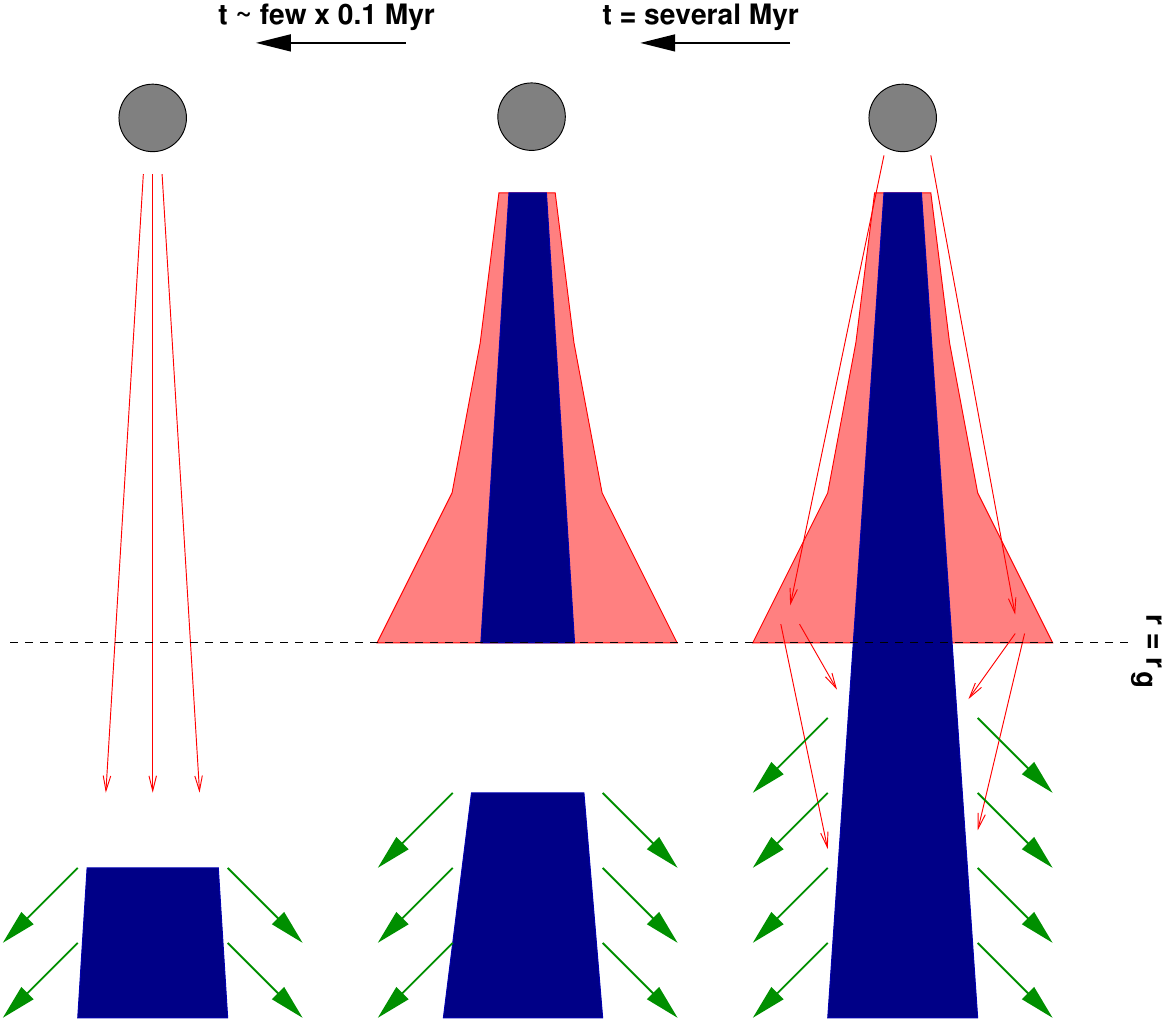}
\caption{Schematic depiction of how photoevaporation driven by a 
central source of UV radiation is predicted to disperse the 
protoplanetary disk. In the initial phase, UV radiation (shown 
as the red arrows) ionizes 
the surface of the disk, producing a vertically extended bound 
atmosphere for $r < r_g$ and mass loss in a thermal wind for 
$r > r_g$. The ionizing flux that photoevaporates the outer disk 
arises primarily from stellar photons scattered by the atmosphere 
at small radii (the `diffuse field'). After several Myr, the disk accretion rate drops 
to a value that is of the same order as the wind mass loss rate. 
At this point, the wind opens up a gap in the disk close to $r_g$, 
cutting off the inner disk from resupply by the disk further out. 
The inner disk then drains rapidly on to the star -- producing an 
inner hole -- and the direct UV flux from the star photoevaporates 
the outer region.}
\label{fig_disperse}
\end{figure}

The combination of a photoevaporative wind and viscous disk evolution 
leads to rapid disk dispersal \cite{clarke01}. Calculations by 
\citet{alexander06b} suggest a three-stage scenario depicted schematically 
in Figure~\ref{fig_disperse},
\begin{itemize}
\item
Initially $\dot{M} \gg \dot{M}_{\rm wind}$. The wind mass loss has 
negligible effect on the disk, which evolves in a similar way to an 
ordinary viscous model. The mass accretion rate and surface density 
gradually drop on the viscous time scale of the entire disk (determined 
at large radii), which is of the order of a Myr.
\item
After a few Myr, the accretion rate drops sufficiently so that 
$\dot{M} \sim \dot{M}_{\rm wind}$. The wind is then strong enough 
to dominate the disk surface density evolution near $r_g$, opening up 
a gap in the disk and cutting off the inner disk from resupply by 
gas flowing in from the reservoir at larger radii. The inner disk 
then drains on to the star on its own (short) viscous time scale, 
which can be of the order of $10^5$~yr or less.
\item
Once the inner disk has drained, the remaining gas in the outer 
disk is directly illuminated by UV radiation from the star (previously, 
the dominant flux was photons scattered on to the outer disk from a 
bound atmosphere at smaller radii). This radiation rapidly burns 
through the outer disk removing all remaining gas.
\end{itemize}
The primary source of uncertainty in these models is the origin and 
magnitude of the stellar ionizing flux. There are few constraints 
on $\Phi$ for Solar mass T Tauri stars \cite{alexander05}, and 
essentially no information on any scaling with stellar mass.

The 
physics of the EUV-ionized gas flows described above is particularly easy to calculate. 
Qualitatively similar flows, however, can be driven by softer FUV 
radiation ($6 \ {\rm eV} < E < 13.6 \ {\rm eV}$), which 
suffices to dissociate H$_2$ molecules and drives  
evaporative flow from the outer disk where the escape velocity is 
smaller. The detailed physics of such flows -- which resemble 
photodissociation regions rather than HII regions -- is harder to 
calculate because the temperature of the heated gas is determined by 
a balance between grain photoelectric heating and cooling by both 
atomic and molecular lines \cite{adams06,gorti09}. Harder X-ray 
photons can also be important, and  
X-rays may in fact dominate the total photoevaporative mass loss 
rate for protoplanetary disks \cite{ercolano09}. \citet{alexander14} provide a 
review of both the theory of these flows and some of the observational 
constraints on photoevaporation models. 

\subsection{The condensation sequence}
In an actively accreting disk, there must be a temperature gradient 
in $z$ in order for energy to be transported from the dense midplane 
where it is probably liberated to the photosphere where it is 
radiated (note that for a thin disk with $h/r \ll 1$ gradients in 
$z$ will dominate over radial gradients, which can consistently 
be ignored). A simple application of the theory of radiative 
transport in plane-parallel media \cite{rybicki79} allows us 
to derive the relation between the central disk temperature $T_{\rm c}$ 
and the effective disk temperature $T_{\rm disk}$.

To proceed, we define the optical depth to the disk midplane,
\begin{equation} 
 \tau = \frac{1}{2} \kappa_R \Sigma,
\end{equation}
where $\kappa_R$ is the Rosseland mean opacity and $\Sigma$ is the 
disk surface density. The vertical density profile of the disk 
is $\rho(z)$. If the vertical energy transport occurs 
via radiative diffusion (in some regions convection may also be 
important), then for $\tau \gg 1$ the vertical energy flux $F(z)$ 
is given by the equation of radiative diffusion,
\begin{equation} 
 F(z) = - \frac{16 \sigma T^3}{3 \kappa_R \rho} \frac{{\rm d}T}{{\rm d}z}.
\end{equation} 
Let us assume for simplicity that {\em all} the energy dissipation 
occurs at $z=0$\footnote{In magnetized disks this is certainly 
not true, and both ideal \citep{miller00} and non-ideal \citep{hirose11} MHD 
simulations suggest that an interesting fraction of the dissipation may 
occur at low optical depths.}. In that case $F(z) = \sigma T_{\rm disk}^4$ 
is a constant with height. Integrating assuming that the 
opacity is a constant,
\begin{eqnarray}
 - \frac{16 \sigma}{3 \kappa_R} \int_{T_{\rm c}}^{T_{\rm disk}} T^3 dT & = & 
 \sigma T_{\rm disk}^4 \int_0^z \rho ( z^\prime ) dz^\prime \nonumber \\
 - \frac{16 \sigma}{3 \kappa_R} \left[ \frac{T^4}{4} \right]_{T_{\rm c}}^{T_{\rm disk}} 
 & = & T_{\rm disk}^4 \frac{\Sigma}{2},
\end{eqnarray} 
where for the final equality we have used the fact that for $\tau \gg 1$ 
almost all of the disk gas lies below the photosphere. For large $\tau$ 
we expect that $T_{\rm c}^4 \gg  T_{\rm disk}^4$, and the equation 
simplifies to,
\begin{equation} 
 \frac{T_{\rm c}^4}{T_{\rm disk}^4} \simeq \frac{3}{4} \tau.
\end{equation}
The implication of this result is that active disks with large 
optical depths are substantially hotter at the midplane than at 
the surface. For example, if $\tau = 10^2$ to the thermal radiation 
emitted by the disk at some radius then 
$T_{\rm c} \approx 3 T_{\rm disk}$. This is important since it 
is the {\em central} temperature that determines the 
sound speed that enters into the viscosity (equation \ref{eq_SS}),  
and it is also the central temperature that determines which ices 
or minerals can be present. Relatively modest 
levels of accretion can thus affect the thermal structure of the 
disk substantially.

For both terrestrial planet formation, and gas giant planet formation 
if it occurs via the core accretion mechanism, the evolution of the 
trace solid component of the disk is of great interest. The gas that 
forms the protoplanetary disk will contain interstellar dust grains 
made up of a mixture of silicates, graphite and polycyclic aromatic 
hydrocarbons (PAHs). In the interstellar medium, measurements of the 
wavelength dependence of extinction can be fit by assuming that the 
dust grains follow a power-law size distribution \citep{mathis77},
\begin{equation} 
 n(a) \propto a^{-3.5}
\end{equation}
where $a$ is the grain size (assumed to be spherical) and the distribution 
extends from 0.005~$\mu$m to about 1~$\mu$m. This distribution is 
generally assumed to be the starting point for further evolution within the 
denser conditions prevailing within the disk. In the hottest, inner regions 
of the disk the central temperature can be high enough to destroy the 
grains (1000~K to 2000~K, depending on whether the grains are made of carbon 
or silicate). The resulting absence of dust very close to the star constitutes 
one of the main arguments against an in situ origin for hot Jupiters, but the 
dust destruction radius is sufficiently close in (normally substantially less 
than 1~AU) that it rarely impacts either terrestrial or, especially, giant 
planet formation. It is, however, important observationally, since once the 
dust is destroyed the remaining gas phase opacity is greatly reduced. There 
will therefore be an opacity ``hole" in the inner disk even if gas is present 
there.

\begin{table}
\label{origins_t2}
\caption{Condensation temperatures for selected materials}
\begin{ruledtabular}
\begin{tabular}{cc}
 $T$ & Material \\
\hline
1680~K & Al$_2$O$_3$ \\
1590~K & CaTiO$_3$ \\
1400~K & MgAl$_2$O$_4$ \\
1350~K & Mg$_2$SiO$_4$, iron alloys \\
370~K  & Fe$_3$O$_4$ \\
180~K & water ice \\
130~K & NH$_3$ $\cdot$ H$_2$O \\
40~K -- 80~K & methane, methane ices \\
50~K & argon \\
\end{tabular}
\end{ruledtabular}
\end{table}

If the gas that makes up the protoplanetary disk has a known 
elemental composition (for example that of the Sun), then it is 
a well defined problem (for a chemist!) to calculate the {\em 
most thermodynamically stable} mix of chemical species at any 
given pressure and temperature. The abundance of different minerals 
and ices within the disk will follow this condensation sequence 
provided that there is sufficient time for chemical reactions to 
reach equilibrium. This may be a reasonable assumption in the 
hot inner disk but deviations will occur due to slow chemical 
reactions in the cool outer disk and radial drift of both gas and 
solids. The equilibrium mix depends more strongly on temperature than 
on pressure, so we can roughly map the condensation sequence into a 
predicted variation of disk composition with radius. Table II,  
adapted from \citet{lodders03}, lists characteristic temperatures 
below which different ices and minerals are predicted to be 
dominant. Of these, by far the most important is the temperature 
below which water ice can be present -- this is 180~K at a 
pressure of 10$^{-4}$ bar (though for the conditions in the 
protoplanetary disk, water ice requires moderately cooler conditions 
with $T \approx 150 \ {\rm K}$). For a Solar mix of elements, 
the surface density of condensable materials rises dramatically 
once water ice forms,
\begin{equation} 
 \Sigma ({\rm ices + rock}) \simeq 4 \Sigma ({\rm rock})
\end{equation}
though the ratio depends upon still uncertain determinations 
of the exact Solar composition \cite{lodders03}. It is tempting -- and 
extremely plausible -- to associate changes in the efficiency or 
outcome of planet formation (in particular the division between 
terrestrial and gas giant planets in the Solar System) with the large 
change in the predicted surface density of solids that occurs at this radius.

The radius in the protoplanetary disk beyond which water ice can 
be present is called the {\em snow line}. In the Solar System, 
water-rich asteroids are found only in the outer asteroid belt \cite{morbi00}, 
which suggests that the snow line in the Solar Nebula fell 
at around 3~AU. Passive protoplanetary disks are predicted to 
have snow lines at substantially smaller radii -- in some cases interior to 1~AU -- 
though accretion rates within the plausible range for Classical T Tauri disks
suffice to push the snow line out to the inferred location of 3~AU \cite{lecar06,garaud06}.

\section{Planet Formation}
The formation of planets from sub-micron size dust particles requires growth through 
at least 12 orders of magnitude in spatial scale. It is useful to consider different 
size regimes in which the interaction between the solid component and the gas is 
qualitatively distinct:
\begin{itemize}
\item
{\bf Dust} -- small particles ranging from sub-micron to cm in scale. These 
particles are well-coupled to the gas, but there can be slow drift 
either vertically or radially. Growth occurs through physical collisions 
leading to agglomeration.
\item
{\bf ``Rocks"} -- objects of meter scale. These particles have 
increasingly weak coupling to the gas, and so it can be useful to 
approximate their dynamics as being a combination of Keplerian orbits 
plus aerodynamic drag. Growth through this size regime is deduced to 
be rapid but the mechanism remains uncertain.
\item
{\bf Planetesimals} -- bodies of km scale and above. Planetesimals 
are massive enough that their dynamics is largely decoupled from 
that of the gas. A population of bodies of this size is often considered 
as the initial condition for subsequent stage of planet formation, 
since the evolution of such a population is a well-posed N-body problem 
involving primarily purely 
gravitational forces (though for smaller planetesimals, questions 
regarding the bodies material strength can also be pertinent). In the 
classical scenario for planet formation we ignore dust and rocks once 
planetesimals have formed, but in fact aerodynamically assisted accretion 
of small bodies may play a substantial role in protoplanetary growth. 
This is ``pebble accretion".
\item
{\bf Earth mass} planets or progenitors of the giant planet cores. Once 
growing planets reach masses of the order of that of Earth, they again 
become coupled to the gas disk, though this time via gravitational 
rather than aerodynamic interactions. We will discuss this coupling 
later in the context of different regimes of planetary {\em migration}. 
For terrestrial planet formation it is possible that the formation of 
Earth mass bodies occurs {\em after} the gas disk has been dispersed 
(in which case this coupling is moot), but for growing giant planet 
cores it is inevitable that interaction will take place.
\item
{\bf Planetary cores} with masses of the order of $10 \ M_\oplus$. 
At around this mass, there is a transition from a quasi-hydrostatic 
core + envelope structure to a regime of rapid gas accretion.
\end{itemize}
Although it predates the discovery of extrasolar planetary systems, 
the review by \citet{lissauer93} remains an excellent, readable 
summary of much of the physics that we will address in this section.

\subsection{From dust to planetesimals}
A spherical particle of radius $a$, moving relative to the gas 
at velocity $v$, experiences an aerodynamic drag force $F_D$ that opposes 
its motion,
\begin{equation}
 F_D = - \frac{1}{2} C_D \cdot \pi a^2 \cdot \rho v^2
\end{equation}
where $C_D$ is the {\em drag coefficient}. The form of the drag 
coefficient depends upon the size of the particle compared to the 
mean free path $\lambda$ of molecules in the gas \cite{weidenschilling77b,whipple72}. 
For small particles (typically of cm size or less) for which, 
\begin{equation}
 a < \frac{9}{4} \lambda,
\end{equation}
the drag coefficient is,
\begin{equation}
 C_D = \frac{8}{3} \frac{\bar{v}}{v},
\label{eq_epstein} 
\end{equation}
where $\bar{v} = ({8/\pi})^{1/2} c_s$ is the mean thermal velocity in the 
gas. This is called the {\em Epstein regime} of drag. For larger particles 
the {\em Stokes} drag law is valid. Defining the Reynolds number via,
\begin{equation}
 {\rm Re} = \frac{2 a v}{\nu}
\end{equation}
where $\nu$ is the microscopic (molecular) viscosity in the gas, the drag 
coefficient can be expressed as a piecewise function,
\begin{eqnarray}
 C_D & = & 24 {\rm Re}^{-1} \,\,\,\,\,\,\, {\rm Re} < 1 \nonumber \\
 C_D & = & 24 {\rm Re}^{-0.6} \,\,\, 1 < {\rm Re} < 800 \nonumber \\  
 C_D & = & 0.44 \,\,\,\,\,\,\,\,\,\,\,\,\,\,\,\, {\rm Re} > 800.
\label{eq_stokes} 
\end{eqnarray} 
We will apply these drag laws to consider both the vertical distribution and 
radial drift of small solid bodies within the gas disk.

\subsubsection{Dust settling}
Dust particles are strongly coupled to the gas via drag forces. For a particle 
of mass $m$, the {\em friction time scale} is defined as,
\begin{equation}
 t_{\rm fric} = \frac{mv}{|F_D|}.
\label{eq_friction}
\end{equation}
It is the time scale on which drag will lead to order unity changes in the 
relative velocity between the particle and the gas. Writing the particle 
mass $m = (4/3)\pi a^3 \rho_d$ in terms of the material density $\rho_d$, 
the friction time scale has a simple form in the Epstein regime,
\begin{equation} 
 t_{\rm fric} = \frac{\rho_d}{\rho} \frac{a}{\bar{v}}.
\end{equation} 
Adopting conditions appropriate to 1~AU within the disk, $\rho = 5 \times 10^{-10} \ 
{\rm g \ cm}^{-3}$, $\bar{v} = 2.4 \times 10^5 \ {\rm cms}^{-1}$ and 
$\rho_d = 3 \ {\rm g \ cm}^{-3}$ we obtain $t_{\rm fric} \approx 2.5 \ {\rm s}$. 
Small particles are thus very tightly coupled to the gas.

Consider a thin, vertically isothermal gas disk with surface density $\Sigma$ 
and scale height $h = c_s / \Omega_K$. The vertical density profile is,
\begin{equation}
 \rho(z) = \frac{\Sigma}{h \sqrt{2 \pi}} e^{-z^2 / 2 h^2}.
\end{equation} 
To start with, let us ignore the effects of turbulence and assume that the 
disk is entirely quiescent. In this case the important forces acting on a 
particle at height $z$ above the midplane are the vertical component 
of gravity and gas drag, given by,
\begin{eqnarray}
 \vert F_{\rm grav} \vert & = & m \Omega_K^2 z \nonumber \\
 \vert F_D \vert & = & \frac{4}{3} \pi a^2 \bar{v} \rho v.
\end{eqnarray} 
Given the strong coupling expected for dust particles terminal 
velocity will rapidly be attained, so we equate these to obtain the 
settling speed,
\begin{equation}
 v_{\rm settle} = \left( \frac{\Omega_K^2}{\bar{v}} \right) 
 \frac{\rho_d}{\rho} a z.
\end{equation} 
Settling is more rapid at higher $z$ (where the gas density is 
lower and the vertical component of gravity stronger), and for 
larger grains. For example, for micron sized dust particles at 
$z=h$ at 1~AU the settling velocity is $v_{\rm settle} \approx 0.1 \ 
{\rm cms}^{-1}$ and the settling time scale,
\begin{equation}
 t_{\rm settle} = \frac{z}{|v_{\rm settle}|} \sim 2 \times 10^5 \ {\rm yr}.
\end{equation}
In the absence of turbulence, then, we expect micron sized dust particles 
to sediment out of the upper layers of the disk on a time scale that is 
short compared to the disk lifetime, while for particles with sizes 
$< 0.1 \ \mu{\rm m}$ the time scale is marginal.

Only the density in the equation for the settling time scale is a 
function of height. Inserting the expression for the vertical density 
profile the general expression for the settling time scale becomes,
\begin{equation}
 t_{\rm settle} = \frac{2}{\pi} \frac{\Sigma}{\Omega_K \rho_d a} 
 e^{-z^2 / 2 h^2}.
\label{eq_tsettle} 
\end{equation}
The strong $z$-dependence implies that dust will settle out of the 
upper regions of the disk rather rapidly in the absence of 
turbulence. This is of some interest since scattered light images 
of protoplanetary disks \cite{burrows96} are sensitive to dust well away from 
the midplane.  

\subsubsection{Settling in the presence of turbulence}
The conditions necessary for turbulence to stir up the dust enough to 
oppose vertical settling can be estimated by comparing the settling 
time (equation~\ref{eq_tsettle}) with the time scale on which 
diffusion will erase spatial gradients in the particle concentration. 
To diffuse vertically across a scale $z$ requires a time scale,
\begin{equation}
 t_{\rm diffuse} = \frac{z^2}{D},
\end{equation} 
where $D$ is an anomalous (i.e. turbulent) diffusion co-efficient. Equating the settling and 
diffusion time scales at $z=h$ we find that turbulence will inhibit 
the formation of a particle layer with a thickness less than $h$ 
provided that,
\begin{equation}
 D \gtrsim \frac{\pi e^{1/2}}{2} \frac{\rho_m a h^2 \Omega_K}{\Sigma}.
\end{equation} 
This result is not terribly transparent. We can cast it into a more 
interesting form if we assume that the turbulence stirring up the 
particles is the same turbulence responsible for angular momentum 
transport within the disk. In that case it is plausible that the anomalous 
diffusion co-efficient has the same magnitude and scaling as the 
anomalous viscosity, which motivates us to write,
\begin{equation}
 D \sim \nu = \frac{\alpha c_s^2}{\Omega_K}.
\end{equation} 
With this form for $D$ the minimum value of $\alpha$ required for 
turbulence to oppose settling becomes,
\begin{equation}
 \alpha \gtrsim \frac{\pi e^{1/2}}{2} \frac{\rho_d a}{\Sigma},
\end{equation} 
which is roughly the ratio between the column density through a 
single solid particle and that of the whole gas disk. For small 
particles this critical value of $\alpha$ is extremely small. 
If we take $\Sigma = 10^2 \ {\rm g} \ {\rm cm}^{-2}$, 
$\rho_d = 3 \ {\rm g} \ {\rm cm}^{-3}$ and $a = 1 \ \mu {\rm m}$, 
for example, we obtain $\alpha \gtrsim 10^{-5}$. This implies that small particles of dust 
will remain suspended throughout much of the vertical extent 
of the disk in the presence of turbulence with any plausible 
strength. For larger particles the result is different. If we 
consider particles of radius 1~mm -- a size that we know from observations is present 
within disks -- we find that the critical value of $\alpha \sim 10^{-2}$. 
This value is comparable to most large scale estimates of 
$\alpha$ for protoplanetary disks. 
Particles of this size and above will therefore not have the same vertical 
distribution as the gas in the disk.

To proceed more formally we can consider the solid particles as a 
separate fluid that is subject to the competing influence of 
settling and turbulent diffusion. If the ``dust" fluid with 
density $\rho_p$ can be treated as a trace species within the 
disk (i.e. that $\rho_p / \rho \ll 1$) then it evolves according to 
an advection-diffusion equation of the form \citep{fromang06,dubrulle95},
\begin{equation}
 \frac{\partial \rho_p}{\partial t} = 
 D \frac{\partial}{\partial z} \left[ \rho 
 \frac{\partial}{\partial z} \left( \frac{\rho_p}{\rho} \right) \right] 
 + \frac{\partial}{\partial z} \left( \Omega_K^2 t_{\rm fric} \rho_p z \right).
\label{eq_C4_advdiffz} 
\end{equation}
Simple steady-state solutions to this equation can be found in the 
case where the dust layer is thin enough that the {\em gas} density 
varies little across the dust scale height. In that limit the 
dimensionless friction time $\Omega t_{\rm fric}$ is independent of 
$z$ and we obtain,
\begin{equation}
 \frac{\rho_p}{\rho} = \left( \frac{\rho_p}{\rho} \right)_{z=0} 
 \exp \left[-\frac{z^2}{2 h_d^2} \right],
\end{equation}
where $h_d$, the scale height describing the vertical distribution 
of the particle concentration $\rho_p / \rho$, is,
\begin{equation}
 h_d = \sqrt{ \frac{D}{\Omega_K^2 t_{\rm fric}} }.
\end{equation}
If, as previously, we assume that $D \sim \nu$, we can write a 
compact expression for the ratio of the concentration scale height 
to the usual gas scale height,
\begin{equation}
 \frac{h_d}{h} \simeq \sqrt{ \frac{\alpha}{\Omega_K t_{\rm fric}} }.
\label{eq_C4_hdd} 
\end{equation}   
The condition for solid particles to become strongly concentrated 
toward the disk midplane is then that the dimensionless friction 
time is substantially greater than $\alpha$. For any reasonable 
value of $\alpha$ this implies that substantial particle growth 
is required before settling takes place. 

Our discussion of dust settling in the presence of turbulence sweeps a number 
of tricky issues under the carpet. More careful treatments need to consider:
\begin{itemize}
\item
Whether plausible sources of disk turbulence {\em really} generate an 
effective turbulent diffusivity that is related to the effective turbulent 
viscosity. There has been a great deal of work on this issue over the 
years -- a good modern starting point is \citet{zhu15}.
\item
The relationship between the effective diffusivity of the gas and that 
of particles aerodynamically coupled to it. This is non-trivial once 
particles become large enough that $\Omega_K t_{\rm fric} \sim 1$ 
\citep{youdin07}.
\end{itemize}

\subsubsection{Settling with coagulation}
In addition to being affected by turbulence, settling is also coupled to 
coagulation and particle growth. The settling velocity 
increases with the particle size, so coagulation hastens the 
collapse of the dust toward the disk midplane.

To estimate how fast particles could grow during sedimentation we appeal 
to a simple single particle growth model \citep{safronov69,dullemond05}. Imagine that a single ``large" particle, of radius 
$a$ and mass $m = (4/3) \pi a^3 \rho_d$, is settling toward the disk midplane 
at velocity $v_{\rm settle}$ through a background of much smaller solid 
particles. By virtue of their small size, the settling of the small particles 
can be neglected. If every collision leads to coagulation, the large particle 
grows in mass at a rate that reflects the amount of solid material in the 
volume swept out by its geometric cross-section,
\begin{equation}
 \frac{{\rm d}m}{{\rm d}t} = \pi a^2 | v_{\rm settle} | f \rho (z),
\end{equation}
where $f$ is the dust to gas ratio in the disk. Substituting for the 
settling velocity one finds,
\begin{equation}
 \frac{{\rm d}m}{{\rm d}t} = \frac{3}{4} \frac{\Omega^2 f}{\bar{v}} z m.
\label{eq_C4_sc1} 
\end{equation} 
Since $z = z(t)$ this Equation cannot generally be integrated immediately\footnote{Note 
however that if the particle grows rapidly (i.e. more rapidly than it sediments) then 
the form of the equation implies exponential growth of $m$ with time.}, but rather 
must be solved in concert with the equation for the height of the particle 
above the midplane,
\begin{equation}
 \frac{{\rm d}z}{{\rm d}t} = - \frac{\rho_d}{\rho} \frac{a}{\bar{v}} \Omega^2 z.
\label{eq_C4_sc2} 
\end{equation}
Solutions to these equations provide a very simple model for particle growth and 
sedimentation in a non-turbulent disk. 

\begin{figure}
\includegraphics[width=\columnwidth]{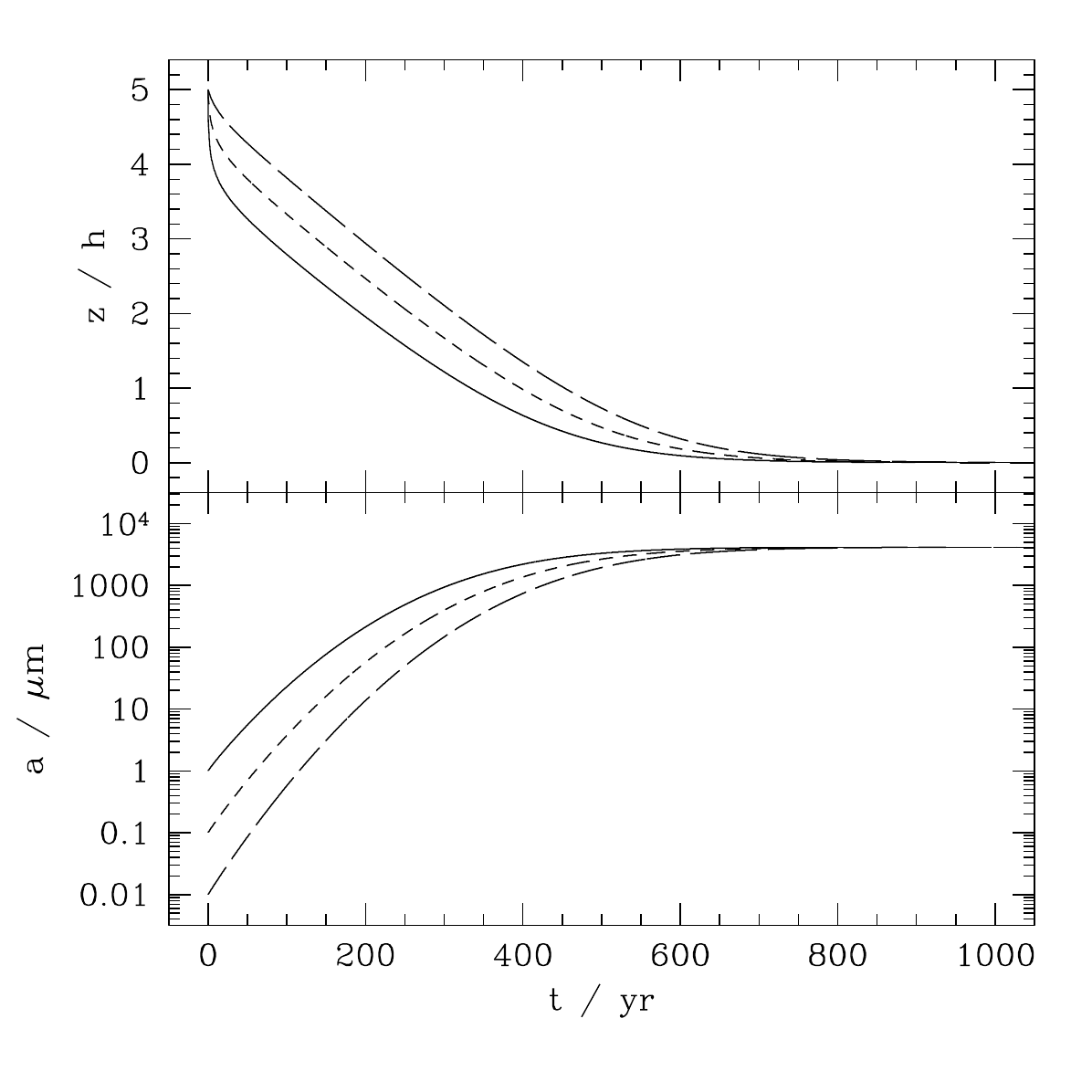}
\caption{The settling and growth of a single particle in a laminar 
(non-turbulent) protoplanetary disk. The model assumes that a single 
particle (with initial size $a = 1 \ \mu {\rm m}$ (solid line), 0.1~$\mu$m 
(dashed line), or 0.01~$\mu$m (long dashed line) accretes all smaller 
particles it encounters as it settles toward the disk midplane. The 
smaller particles are assumed to be at rest. The upper panel shows the 
height above the midplane as a function of time, the lower panel the 
particle radius $a$. For this example the disk parameters adopted are: 
orbital radius $r = 1 \ {\rm AU}$, scale height $h = 3 \times 10^{11} \ {\rm cm}$, 
surface density $\Sigma = 10^3 \ {\rm g} \ {\rm cm}^{-2}$, dust to gas ratio 
$f = 10^{-2}$, and mean thermal speed $\bar{v} = 10^5 \ {\rm cm} \ {\rm s}^{-1}$. 
The dust particle is taken to have a material density $\rho_d = 3 \ {\rm g} \ {\rm cm}^{-3}$ 
and to start settling from a height $z_0 = 5 h$.}
\label{figure_C4_settle}
\end{figure}

Figure~\ref{figure_C4_settle} shows solutions to 
equations~(\ref{eq_C4_sc1}) and (\ref{eq_C4_sc2}) for initial particle 
sizes of 0.01~$\mu$m, 0.1~$\mu$m and 1~$\mu$m. The particles settle from 
an initial height $z_0 = 5 h$ through a disk whose parameters are chosen 
to be roughly appropriate to a (laminar) Solar Nebula model at 1~AU from the 
Sun. Both particle growth and vertical settling are extremely rapid. 
With the inclusion of coagulation, particles settle to the disk midplane 
on a time scale of the order of $10^3$~yr -- more than two orders of 
magnitude faster than the equivalent time scale in the absence of particle 
growth. By the time that the particles reach the midplane they have grown 
to a final size of a few~mm, irrespective of their initial radius.

The single particle model described above is very simple, both in its 
neglect of turbulence and because it assumes that the only reason that 
particle-particle collisions occur is because the particles have different 
vertical settling velocities. Other drivers of collisions include Brownian 
motion, turbulence, and differential {\em radial} velocities. The basic 
result, however, is confirmed by more sophisticated models \citep{dullemond05}, 
which show that if collisions lead to particle adhesion 
growth from sub-micron scales up to small macroscopic scales (of the order 
of a mm) occurs rapidly. There are no time scale problems involved with the 
very earliest phases of particle growth. Indeed, what is more problematic 
is to understand how the population of small grains -- which are unquestionably 
present given the IR excesses characteristic of Classical T~Tauri star -- 
survive to late times. The likely solution to this quandary involves the 
inclusion of particle {\em fragmentation} in sufficiently energetic collisions, 
which allows a broad distribution of particle sizes to survive out to late 
times. Fragmentation is not likely given collisions at relative velocities of 
the order of a cm~s$^{-1}$ -- values typical of settling for micron-sized 
particles -- but becomes more probable for collisions at velocities of 
a m~s$^{-1}$ or higher. 

\subsubsection{Radial drift of particles}
Previously we showed (equation \ref{eq_vphivk}) that the azimuthal 
velocity of gas within a geometrically thin disk is close to the 
Keplerian velocity. That it is not identical, however, turns out 
to have important consequences for the evolution of small solid 
bodies within the disk \cite{weidenschilling77b}. We can 
distinguish two regimes,
\begin{itemize}
\item
{\bf Small particles} ($a < {\rm cm}$) are well-coupled to the gas. To a 
first approximation we can imagine that they orbit with the 
{\em gas} velocity. Since they don't experience the same radial 
pressure gradient as the gas, however, this means that they feel 
a net inward force and drift inward at their radial terminal 
velocity.
\item
{\bf Rocks} ($a > {\rm m}$) are less strongly coupled to the gas. To a 
first approximation we can imagine that they orbit with the 
{\em Keplerian} velocity. This is faster than the gas velocity, 
so the rocks see a headwind that saps their angular momentum 
and causes them to spiral in toward the star.
\end{itemize}
To quantify these effects, we first compute the magnitude of the 
deviation between the gas and Keplerian orbital velocities. 
Starting from the radial component of the momentum equation, 
\begin{equation} 
 \frac{v_{\phi,{\rm gas}}^2}{r} = \frac{GM_*}{r^2} + \frac{1}{\rho} \frac{{\rm d}P}{{\rm  d}r},
\end{equation}
we write the variation of the midplane pressure with radius as a 
power-law near radius $r_0$,
\begin{equation}
 P = P_0 \left( \frac{r}{r_0} \right)^{-n}
\label{eq_def_n} 
\end{equation}
where $P_0 = \rho_0 c_s^2$. Substituting, we find,
\begin{equation}
 v_{\phi,{\rm gas}} = v_K \left( 1 - \eta \right)^{1/2}
\label{eq_dv} 
\end{equation}
where
\begin{equation}
 \eta = n \frac{c_s^2}{v_K^2}.
\label{eq_eta_dependence} 
\end{equation}  
Typically $n$ is positive (i.e. the pressure decreases outward), 
so the gas orbits slightly slower than the local Keplerian 
velocity. For example, for a disk of constant $h(r)/r = 0.05$ and 
surface density profile $\Sigma \propto r^{-1}$ we have $n=3$ 
and,
\begin{equation}
 v_{\phi,{\rm gas}} \simeq 0.996 v_K.
\end{equation}
The fractional difference between the gas and Keplerian velocities is small 
indeed! However, at 1~AU even this small fractional difference amounts to 
a relative velocity of the order of 100~ms$^{-1}$. Large rocks will 
then experience a substantial, albeit subsonic, headwind.

The effect of the drag force on the dynamics of particles of arbitrary 
sizes has been calculated by \citet{weidenschilling77b}. Here, we 
adopt the approach of \citet{takeuchi02} and proceed by considering 
the radial and azimuthal equations of motion for the particle\footnote{Although 
this calculation is straightforward, it's easy to confuse the three 
different azimuthal velocities that are involved -- that of the 
particle, that of the gas, and the Kepler speed. Be careful!},
\begin{eqnarray}
 \frac{{\rm d}v_r}{{\rm d}t} & = & \frac{v_\phi^2}{r} 
 - \Omega_K^2 r - \frac{1}{t_{\rm fric}} \left( v_r - v_{r,{\rm gas}} \right) \nonumber \\
 \frac{\rm d}{{\rm d}t} \left( r v_\phi \right) & = & 
 - \frac{r}{t_{\rm fric}} \left( v_\phi - v_{\phi,{\rm gas}} \right).
\end{eqnarray}
We simplify the azimuthal equation by noting that the specific angular 
momentum always remains close to Keplerian (i.e. the particle spirals 
in through a succession of almost circular, almost Keplerian orbits),
\begin{equation} 
 \frac{\rm d}{{\rm d}t} \left( r v_\phi \right) \simeq 
 v_r \frac{\rm d}{{\rm d}r} \left( r v_K \right) = 
 \frac{1}{2} v_r v_K.
\end{equation}
This yields,
\begin{equation}
 v_\phi - v_{\phi,{\rm gas}} \simeq - \frac{1}{2} \frac{t_{\rm fric}v_r v_K}{r}.
\label{eq_101} 
\end{equation}   
Turning now to the radial equation, we substitute for $\Omega_K$ using 
equation (\ref{eq_dv}). Retaining only the lowest order terms,
\begin{equation}
 \frac{{\rm d}v_r}{{\rm d}t} = - \eta \frac{v_K^2}{r} + 
 \frac{2 v_K}{r} \left( v_\phi - v_{\phi,{\rm gas}} \right) 
 - \frac{1}{t_{\rm fric}} \left( v_r - v_{r,{\rm gas}} \right).
\label{eq_102}
\end{equation}
The ${{\rm d}v_r}/{{\rm d}t}$ term is negligible, and for 
simplicity we also assume that $v_{r,{\rm gas}} \ll v_r$, which  
will be true for those particles experiencing the most rapid 
orbital decay. Eliminating $(v_\phi - v_{\phi,{\rm gas}})$ 
between equations (\ref{eq_101}) and (\ref{eq_102}) we obtain,
\begin{equation}
 \frac{v_r}{v_K} = \frac{-\eta}{\frac{v_K}{r} t_{\rm fric} + 
 \frac{r}{v_K} t_{\rm fric}^{-1}}.
\end{equation} 
This result can be cast into a more intuitive form by 
defining a dimensionless stopping time,
\begin{equation}
 \tau_{\rm fric} \equiv t_{\rm fric} \Omega_K,
\end{equation} 
in terms of which the particle radial velocity is,
\begin{equation}
 \frac{v_r}{v_K} = \frac{-\eta}{\tau_{\rm fric} + 
 \tau_{\rm fric}^{-1}}.
\label{eq_vrrate} 
\end{equation}
The peak radial velocity is attained when $\tau_{\rm fric} = 1$ 
(i.e. when the friction time scale equals $\Omega_K^{-1}$), and 
equals $\eta v_K / 2$ independent of the disk properties.

\begin{figure}
\includegraphics[width=\columnwidth]{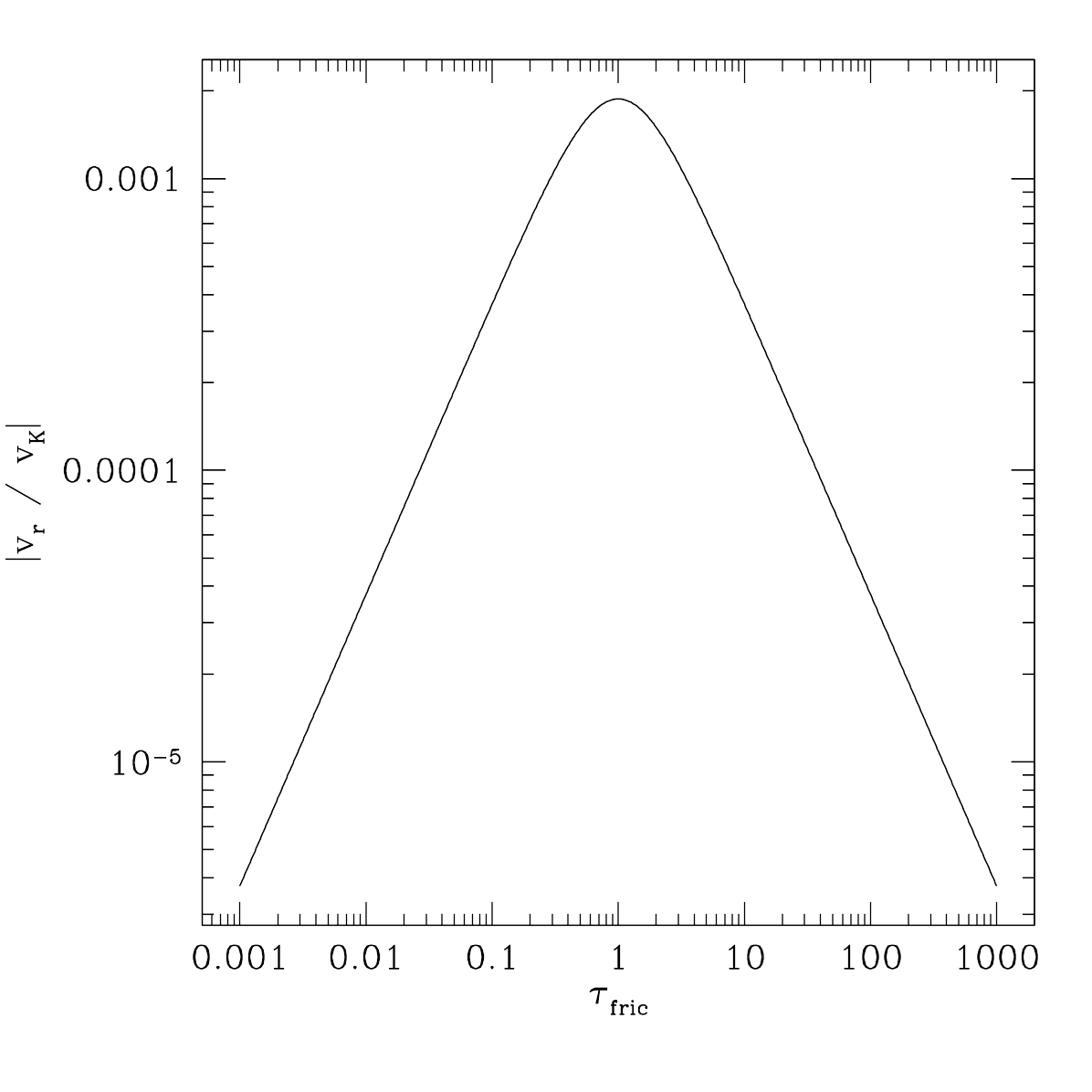}
\caption{Radial drift velocity of particles at the midplane of a 
protoplanetary disk with $h/r = 0.05$, plotted as a function of 
the dimensionless stopping time $\tau_{\rm fric}$. The radial 
velocity of the gas has been set to zero. The most rapid 
inward drift occurs for a physical stopping time $\Omega_K^{-1}$, 
which for typical disk models translates to a particle size in 
the 10~cm to m range. At 1~AU, the peak inward velocity is around 
60~ms$^{-1}$, which implies a decay time of less than 100~yr.}
\label{fig_vr}
\end{figure}

Figure \ref{fig_vr} plots $v_r / v_K$ as a function of the 
dimensionless stopping time for a fiducial disk with $h/r = 0.05$. 
Using equations (\ref{eq_epstein}) and (\ref{eq_stokes}), one can 
associate a particular $\tau_{\rm fric}$ with a unique particle 
size $a$ given known conditions in the protoplanetary disk. Generically, 
one finds that at radii of a few AU the peak inspiral rate is 
attained for particles with size of the order of 10~cm to a few m. 
The minimum inspiral time scale at a given orbital radius depends 
only on $\eta$ -- at 1~AU it is of the order of 100~yr. The inescapable 
conclusion is that the {\bf radial drift time scale $\ll$ disk lifetime 
for meter-scale bodies} in the protoplanetary disk.

The above analysis assumes that the density of solid particles is low enough 
(compared to the gas density) that there is no back-reaction of the solids on 
the gas. In some important circumstances (particularly when considering 
models for planetesimal formation) this criterion will be violated. 
\citet{nakagawa86} have calculated models of settling and radial drift 
that are valid in the more general case where the solid and gas phases 
can have comparable densities. 

As we noted earlier, the fact that most of the heavy elements in the 
Solar System are found in the Sun means that we can tolerate some 
loss of planetary raw material during planet formation. However, 
radial drift time scales as short as 100~yr would clearly lead to 
a catastrophic loss of mass into the star unless, in fact, growth 
through the meter-scale size regime is very fast. The most important 
conclusion from this analysis is, therefore, that {\em planetesimal 
formation must be a rapid process}. This is a robust inference since 
it derives directly from the unavoidable existence of a velocity differential 
between the gas disk and solid bodies orbiting within it. 

The radial drift velocities given by equation (\ref{eq_vrrate}) imply significant 
radial migration over the lifetime of the disk -- not just for particles at the 
most vulnerable meter-scale size range but also for substantially 
smaller and larger bodies. This means that we should expect substantial 
changes in the local ratio of solids to gas as a function of time and 
radius in the disk \cite{takeuchi05}. Under some circumstances, radial 
drift may allow solids to pileup within the inner disk, potentially 
improving the chances of forming planetesimals there \cite{youdin04}.

\begin{figure}
\includegraphics[width=\columnwidth]{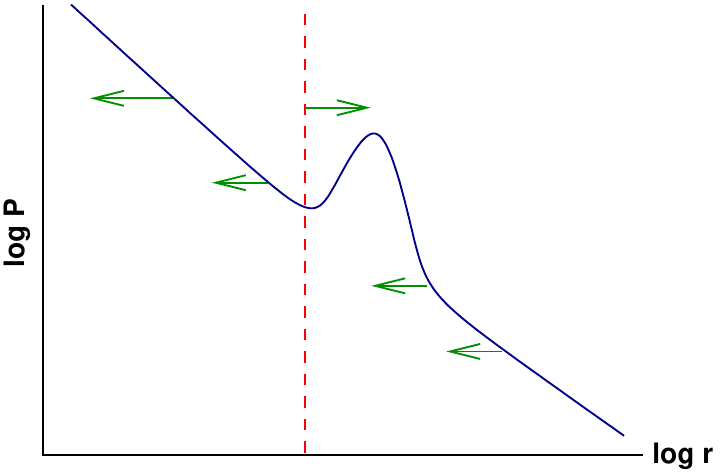}
\caption{Illustration of how local pressure maxima within a disk 
could concentrate solid bodies, forming a ring in this idealized 
axisymmetric example. Local pressure maxima might arise as a 
consequence of turbulence within the disk.}
\label{fig_concentrate}
\end{figure}

Radial drift can be slowed or locally reversed if the gas disk has a 
non-monotonic radial pressure profile. The inward motion of 
solid bodies embedded within the disk occurs 
as a consequence of a gas pressure gradient that leads to 
sub-Keplerian gas orbital velocities. In general, radial drift 
drives particles toward {\em pressure maxima}, so in a disk where the 
mid-plane pressure declines smoothly the motion 
is typically inward. If, on the other hand, it were possible to create 
local pressure maxima these would also act as sites where solids 
concentrate. This possibility was recognized in a prescient 
paper by \citet{whipple72}, whose Figure~1 is more or less 
reproduced here as Figure~\ref{fig_concentrate}. If 
the perturbation to the pressure occurs on a scale 
$\Delta r$, then to obtain a maximum we require that the local pressure gradient $\sim P / \Delta r$ 
exceed the global gradient $\sim P / r$. The time scale to 
concentrate solids locally is then faster than the global inspiral 
time by a factor $\sim (\Delta r / r)^2$. 

Several physical processes can create pressure maxima. 
Persistent local pressure maxima 
could form in a disk at particular locations (for example, at the 
inner edge of dead zones), while transient maxima could 
occur due to the large-scale effects of disk turbulence. 
\citet{rice04} and \citet{durisen05} discussed how 
self-gravitating turbulence might concentrate particles within 
spiral arms or axisymmetric gas rings. \citet{johansen09b} 
showed that ``zonal flows"---axisymmetric local perturbations 
to the pressure that are maintained by variations in $v_\phi$---could 
form within MRI turbulent disks. The edges of gaps carved by 
massive planets are another obvious location where a pressure 
maxima may be expected.

\subsubsection{Planetesimal formation via coagulation}
The growth of micron-sized dust particles up to small macroscopic dimensions 
(of the order of a mm) is driven by pairwise collisions that lead to sticking and 
particle growth. (Simultaneously, high velocity impacts may lead to fragmentation.) 
The most economical hypothesis for planetesimal formation is that the 
same process continues uninterrupted up to 
the planetesimal size scale \citep[for an early calculation, see e.g.][]{weidenschilling80}. 
A coagulation model for planetesimal formation, however, faces two independent 
challenges. First, the {\em material properties} of colliding bodies with a realistic 
velocity distribution must permit growth (rather than bouncing or fragmentation) 
across the full range of sizes between dust particles and planetesimals. 
Second, the {\em rate of growth} must be high enough to form planetesimals 
before the material is lost into the star via aerodynamic drift. These 
constraints are not easily satisfied (or summarized), but neither are they 
obviously insurmountable. 

The outcome of collisions between micron to cm-sized bodies can be studied 
experimentally, ideally under low-pressure microgravity conditions. Good introductions 
to the extensive literature on experimental results are given by \citet{blum08}, 
\citet{guttler10} and \citet{testi14}. The bulk of the experimental work has been 
performed using silicate particles and aggregates, representative of materials 
found interior to the snow line. The most basic result is the critical velocity below 
which individual grains stick together. For $1 \ \mu {\rm m}$ silicate monomers 
this velocity is about 1~${\rm m} \ {\rm s}^{-1}$ \citep{poppe00}. For similarly 
sized water ice monomers the velocity is approximately 10~${\rm m} \ {\rm s}^{-1}$ 
\citep{gundlach15}. This order of magnitude difference presages the likelihood 
of differences in particle growth outcomes interior to and outside the water snow line.

Going beyond monomers, it is generally assumed that the 
colliding bodies are aggregates made up these smaller sub-units. The porosity 
of the aggregates represents an extra dimension that must be taken into 
account in modeling growth. A porous body can dissipate energy on collision 
through rearrangement of its structures (and hence can be ``stronger" in some 
sense than a single particle), but can also be compactified by the action of 
multiple collisions. Theoretical models suggest that aggregates growing 
outside the snow line can, in some situations, be {\em extremely} porous, 
with internal densities as low as $10^{-4} \ {\rm g \ cm}^{-3}$ \citep{okuzumi12}.

Models for particle growth need to account consistently for how the size 
distribution evolves given the predicted collision speeds between particles 
\citep{weidenschilling93,ormel07}, the experimentally measured or 
theoretically predicted collision outcomes, and the local change in 
solid density due to radial drift. It's a difficult problem whose solution 
is uncertain. Generally it appears that:
\begin{itemize}
\item
There are no material barriers preventing monomers from growing into at 
least mm-sized particles anywhere in the disk. The size distribution of 
particles in the range between $\mu {\rm m}$ and $mm-cm$ 
is likely set by a balance 
of coagulation and fragmentation processes \citep{birnstiel11}.
\item
Interior to the snow line, the onset of bouncing creates a barrier to 
growth at around mm or cm sizes \citep{zsom10}. Growth likely 
continues beyond these sizes, but is more severely limited by 
fragmentation and is not fast enough to form planetesimals in 
the presence of radial drift.
\item
Outside the snow line, growth to larger sizes is in principle allowed 
because icy particles stick at higher velocities and are more 
resistant to fragmentation (which may not occur until collision 
velocities reach several tens of meters per second). Growth may 
be limited by radial drift itself, leading to a ``drift-limited growth" 
scenario in which the particle size as a function of radius is 
determined by the condition that that the growth time roughly 
equals the drift time \citep{birnstiel12}. Growth to larger sizes 
is possible if the disk structure supports persistent or transient 
particle traps that slow radial drift \citep{pinilla12}.
\item
In vapor-rich regions adjacent to ice lines \citep{stevenson88} growth via condensation 
of vapor on to pre-existing particles may play a role \citep{ros13}.
\end{itemize}
Based upon these results most workers infer that a planetesimal 
formation mechanism distinct from simple collisional growth is needed 
to explain how planetesimals can form across a broad range of radii 
both interior to and outside the snow line.

\subsubsection{The Goldreich-Ward mechanism}
The alternate hypothesis for planetesimal formation holds that planetesimals 
form from gravitational fragmentation of dense clumps of particles. In its 
simplest form, particles might settle vertically so strongly that a dense 
sub-disk of solids at the mid-plane becomes vulnerable to collapse 
\citep{goldreich73}\footnote{Similar 
considerations are discussed in \citet{safronov69}, who in turn 
quotes earlier work by Gurevich \& Lebedinskii from as early as 
1950.}. As we will discuss this specific model 
does not work, because it turns out to be very difficult to 
settle a layer of small dust particles to densities high enough 
for gravitational instability. Nonetheless the basic idea remains 
attractive since it forms planetesimals while entirely bypassing the size 
scales that are most vulnerable to radial drift, and it useful to discuss 
the original idea before considering contemporary theories for planetesimal 
formation that likewise invoke collective instabilities followed by 
fragmentation.

The basic idea of the  
Goldreich-Ward (1973) mechanism for planetesimal formation is that 
vertical settling and radial drift results in the formation of a dense dust sub-disk 
within which the solid density exceeds the local gas density 
(this obviously requires a very thin sub-disk if the local 
ratio of gas to dust surface density is comparable to the 
fiducial global value of 100). The solid sub-disk then 
becomes gravitationally unstable, and fragments into 
bound clumps of solid particles that subsequently dissipate 
energy via physical collisions and collapse to form 
planetesimals.

Gravitational instability requires that the disk be massive 
(high surface density) and / or dynamically cold (low velocity 
dispersion). The classic analysis of the conditions for 
gravitational instability is that of \citet{toomre64}. Here, 
we consider the stability of a rotating {\em fluid} sheet -- 
this is somewhat easier than the collisionless calculation, 
gives the same answer to a small numerical factor when 
the gas sound speed is identified with the particle 
velocity dispersion, and carries over to the instability 
of a gas disk that we will discuss later. The simplest 
system to analyze is that of a uniformly rotating sheet -- 
in what follows I follow the notation and approach of 
\citet{binney87}.

\begin{figure}
\includegraphics[width=\columnwidth]{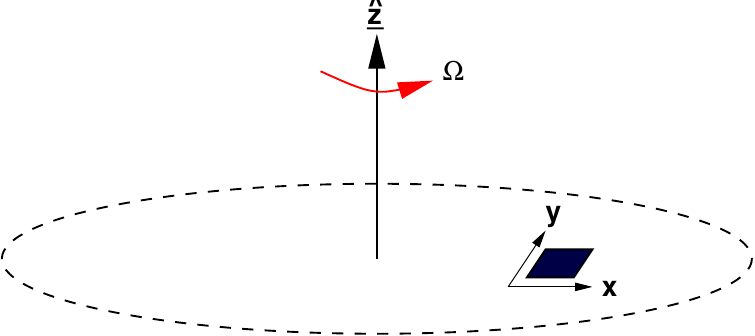}
\caption{Geometry for the calculation of the stability of a 
uniformly rotating sheet.}
\label{fig_gw_setup}
\end{figure}

The setup for the calculation is as shown in Figure \ref{fig_gw_setup}. 
We consider a sheet of negligible thickness in the $z=0$ plane, 
with constant surface density $\Sigma_0$ and angular velocity 
${\bf \Omega} = \Omega \hat{\bf z}$. Our aim is to calculate the 
stability of the sheet to in-plane perturbations. Working in a 
frame that corotates with the (unperturbed) angular velocity 
$\Omega$, the fluid equations are,
\begin{eqnarray}
 \frac{\partial \Sigma}{\partial t} + \nabla \cdot 
 \left( \Sigma {\bf v} \right) & = & 0 \\
 \frac{\partial {\bf v}}{\partial t} + \left( 
 {\bf v} \cdot \nabla \right) {\bf v} & = & 
 - \frac{\nabla p}{\Sigma} - \nabla \Phi 
 -2 {\bf \Omega} \times {\bf v} \nonumber \\ & & + \Omega^2 
 \left( x \hat{\bf x} + y \hat{\bf y} \right)
\end{eqnarray} 
where the momentum equation picks up terms for the Coriolis and 
centrifugal forces in the rotating frame. These equations apply 
in the $z=0$ plane only. The gravitational potential $\Phi$ is 
given by the Poisson equation,
\begin{equation}  
 \nabla^2 \Phi = 4 \pi G \Sigma \delta(z)
\end{equation}
which describes $\Phi$ in all space. In these equations, 
${\bf v} = v_x \hat{\bf x} + v_y \hat{\bf y}$ is the 
velocity {\em in the rotating frame}, $\Sigma$ is the 
surface density, and $p = p(\Sigma)$ is the vertically 
integrated pressure. The sound speed is defined via,
\begin{equation} 
 c_s^2 \equiv \left. \frac{{\rm d}p}{{\rm d}\Sigma} \right\vert_{\Sigma = \Sigma_0}.
\end{equation}  
In the unperturbed state, $\Sigma = \Sigma_0$, $\Phi = \Phi_0$, ${\bf v} = 0$ and 
$p = p_0 = p(\Sigma_0)$. Substituting these values into the 
momentum equation yields $\nabla \Phi_0 = \Omega^2 (x \hat{\bf x} 
+ y \hat{\bf y} )$.

We now consider perturbations to the surface density, velocity, 
pressure and potential,
\begin{eqnarray}
 \Sigma & = & \Sigma_0 + \Sigma_1 (x,y,t) \nonumber \\
 {\bf v} & = & {\bf v}_1 (x,y,t) \nonumber \\
 p & = & p_0 + p_1(x,y,t) \nonumber \\
 \Phi & = & \Phi_0 + \Phi_1 (x,y,z,t)
\end{eqnarray}
where it is assumed that $\Sigma_1 \ll \Sigma_0$ etc. Substituting 
these expressions into the fluid equations, and retaining only 
those terms that are linear in the perturbed quantities, we find,
\begin{eqnarray}
 \frac{\partial \Sigma_1}{\partial t} & + &  \Sigma_0 \nabla \cdot {\bf v}_1 = 0 \\
 \frac{\partial {\bf v}_1}{\partial t} & = & -\frac{c_s^2}{\Sigma_0} \nabla \Sigma_1 
 -\nabla \Phi_1 - 2 {\bf \Omega} \times {\bf v}_1 \\ 
 \nabla^2 \Phi_1 & = & 4 \pi G \Sigma_1 \delta(z)
\end{eqnarray}  
where we have made use of the fact that since $p$ is only a function 
of $\Sigma$, $\nabla p = ({\rm d}p / {\rm d}\Sigma) \nabla \Sigma$. Note 
that these equations {\em only} involve temporal or spatial derivatives 
of the perturbed quantities. Since the equations are (by construction) 
linear, the evolution of an arbitrary perturbation can be decomposed 
into fourier modes. Assuming a wavevector ${\bf k}$ that is parallel 
to $\hat{\bf x}$, we therefore write the perturbations in the form,
\begin{eqnarray}
 \Sigma_1 (x,y,t) & = & \Sigma_a e^{i(kx-\omega t)} \\
 {\bf v}_1 & = & (v_{ax} \hat{\bf x} + v_{ay} \hat{\bf y}) 
 e^{i(kx-\omega t)} \\
 \Phi_1 & = &  \Phi_a e^{i(kx-\omega t)} 
\end{eqnarray}
where the final expression describes the potential perturbations 
in the $z=0$ plane only. Substitution of these expressions into 
the perturbation equations will reduce them to algebraic expressions, 
which can be combined to yield the dispersion relation for the 
system.  

First though, we simplify the system by noting that perturbations in 
$\Sigma$ are the {\em source} of perturbations in $\Phi$. We can 
therefore write $\Phi_a$ in terms of $\Sigma_a$. To do this, let 
the general form for $\Phi_1$ (i.e. {\em not} just at $z=0$) be,
\begin{equation}
 \Phi_1 =  \Phi_a e^{i(kx-\omega t)} \times f(z)
\end{equation} 
where $f(z)$ is some function that needs to be determined. Requiring 
that $\nabla^2 \Phi_1 = 0$ for $z \neq 0$, we find,
\begin{equation}
 \frac{{\rm d}^2 f}{{\rm d} z^2} = k^2 f
\end{equation}
which has a general solution $f = Ae^{-kz} + B^{kz}$, with $A$ and 
$B$ arbitrary constants. Since $\Phi_1$ must remain finite as 
$z \rightarrow \pm \infty$, the general form of $\Phi_1$ is,
\begin{equation}
 \Phi_1 = \Phi_a e^{i(kx-\omega t) - \vert kz \vert}.  
\end{equation}
This is valid throughout all space.

\begin{figure}
\includegraphics[width=\columnwidth]{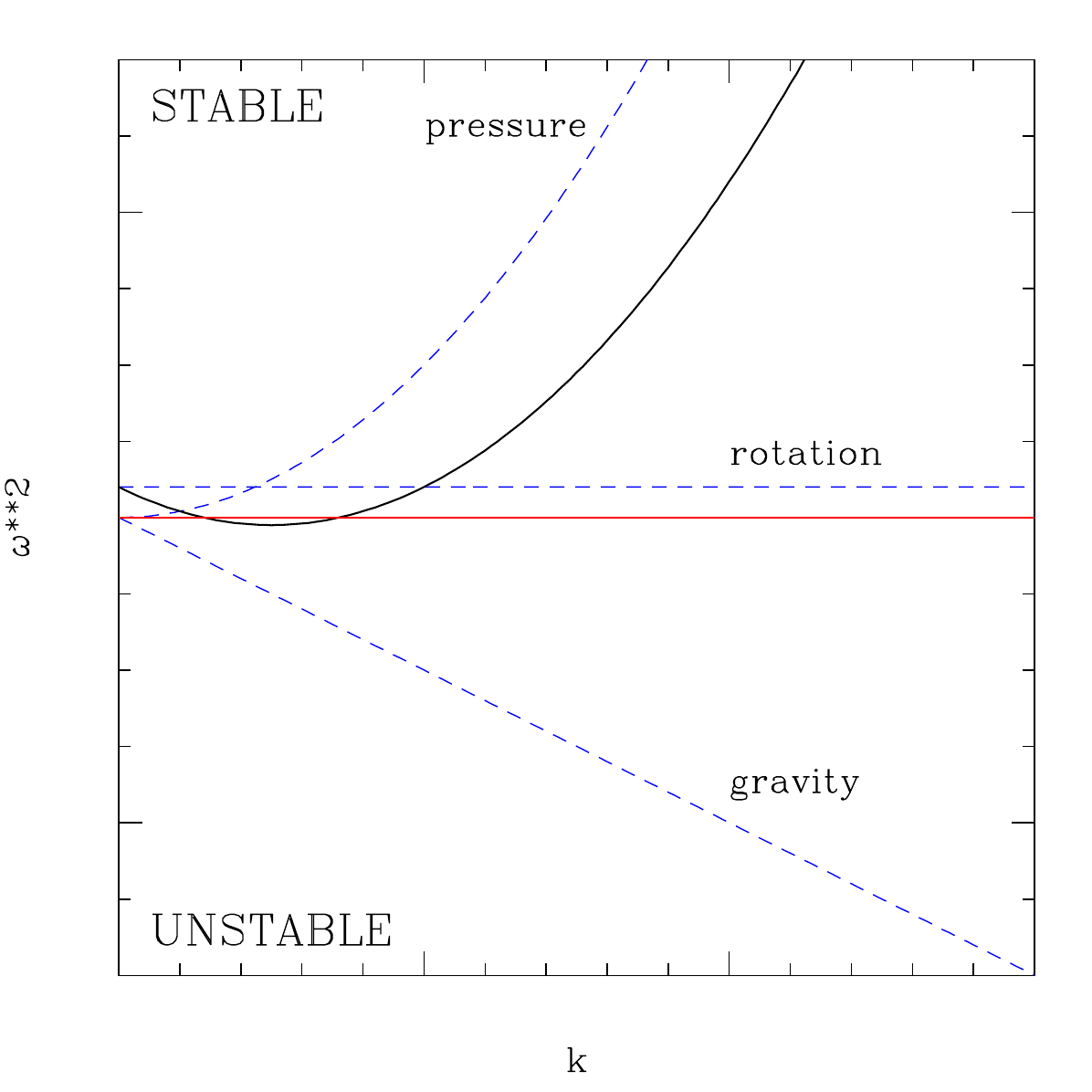}
\caption{The dispersion relation (solid black line) for a 
uniformly rotating sheet, illustrating the contributions 
from pressure, rotation, and self-gravity (dashed blue lines). 
The system is unstable if, at any value of the wavenumber $k$, 
$\omega^2$ falls below the red line and is negative. Pressure 
is a stabilizing influence that is most important at large 
$k$ (small spatial scales), while rotation acts to stabilize 
the system at small $k$ (large spatial scales).}
\label{fig_stability}
\end{figure}

To determine $\Phi_a$, we integrate the Poisson equation 
vertically between $z = -\epsilon$ and $z = + \epsilon$,
\begin{equation}
 \int_{-\epsilon}^{+\epsilon} \nabla^2 \Phi_1 dz = 
 \int_{-\epsilon}^{+\epsilon} 4  \pi G \Sigma_1 \delta(z) dz.
\end{equation}
Mathematically this requires a bit of care, since the integrand 
on the left hand side is zero everywhere except at $z=0$. However, noting 
that $\partial^2 \Phi_1 / \partial x^2$ and $\partial^2 \Phi_1 / \partial y^2$ 
are continuous at $z=0$, while $\partial^2 \Phi_1 / \partial z^2$ is {\em not}, 
we obtain,
\begin{equation} 
 \int_{-\epsilon}^{+\epsilon} \nabla^2 \Phi_1 dz = 
 \left. \frac{\partial \Phi_1}{\partial z} \right\vert_{-\epsilon}^{+\epsilon} =
 \int_{-\epsilon}^{+\epsilon} 4  \pi G \Sigma_1 \delta(z) dz.
\end{equation} 
Taking the limit as $\epsilon \rightarrow 0$,
\begin{equation}
 -2 \vert k \vert \Phi_a = 4 \pi G \Sigma_a
\end{equation}
and,
\begin{equation}
 \Phi_1 = - \frac{2 \pi G \Sigma_a}{\vert k \vert} e^{i(kx-\omega t) - \vert kz \vert}. 
\end{equation}
We are now in a position to substitute $\Sigma_1$, ${\bf v}_1$ and 
$\Phi_1$ into the remaining equations (continuity plus the $x$ and 
$y$ components of the momentum equation). The resulting algebraic 
equations are,
\begin{eqnarray} 
 -i \omega \Sigma_a & = & -i k \Sigma_0 v_{ax} \nonumber \\
 -i \omega v_{ax} & = & -\frac{c_s^2}{\Sigma_0} i k \Sigma_a 
 + \frac{2 \pi G i \Sigma_a k}{\vert k \vert} + 2 \Omega v_{ay} \\
 -i \omega v_{ay} & = & -2 \Omega v_{ax}.
\end{eqnarray} 
We seek a {\em dispersion relation} i.e. a formula for the 
growth rate $\omega = f(k)$ of modes of different scale $k$. 
Eliminating $v_{ax}$ and $v_{ay}$ in turn, we obtain,
\begin{equation}
 \omega^2 = c_s^2 k^2 - 2 \pi G \Sigma_0 \vert k \vert + 4 \Omega^2.
\label{eq_dispersion} 
\end{equation} 
This is the dispersion relation for a uniformly rotating thin sheet. 
The scale-dependence of the different terms is shown graphically in 
Figure \ref{fig_stability}.

Looking back to the form of the perturbations, we note that the 
sheet is:
\begin{itemize}
\item
STABLE if $\omega^2 \ge 0$, since in this case $\omega$ is real and 
the perturbations are oscillatory.
\item
UNSTABLE if $\omega^2 < 0$, for which case $\omega$ is imaginary and 
perturbations grow exponentially.
\end{itemize}
The rotational term ($4 \Omega^2$) is stabilizing at all scales, 
while the pressure term ($c_s^2 k^2$) has a strong stabilizing 
influence at large $k$ (i.e. small spatial scales). Self-gravity, 
represented by the $-2 \pi G \Sigma_0 \vert k \vert$ term, 
has a negative contribution to $\omega^2$ and so destabilizes the sheet.

The condition for marginal stability is that $\omega^2 \ge 0$ at all 
spatial scales. The most unstable scale $k_{\rm crit}$ can be found 
by setting ${\rm d}\omega^2 / {\rm d}k = 0$, which yields,
\begin{equation}
 k_{\rm crit} = \frac{\pi G \Sigma_0}{c_s^2}.
\end{equation} 
The sheet is marginally stable when $\omega^2 (k_{\rm crit}) = 0$, 
which gives the stability condition as,
\begin{equation}
 \frac{c_s \Omega}{G \Sigma_0} = \frac{\pi}{2}.
\end{equation} 
This analysis can be extended in several ways -- for example 
to include differential rotation or global rather than local 
stability. A generic way of expressing the results of such 
calculations is to define the Toomre $Q$ parameter,
\begin{equation} 
 Q \equiv \frac{c_s \Omega}{\pi G \Sigma}.
\end{equation}
In terms of $Q$, a disk is unstable\footnote{For 
a differentially rotating disk, it is easy to verify that stability 
depends upon the parameter combination $c_s \Omega / (G \Sigma_0)$ 
via a time scale argument. First derive the time scale for shear 
to separate two points that are initially $\Delta r$ apart, and 
equate this to the collapse time scale under gravity to find the 
maximum scale on which collapse can occur without being affected 
by shear. Taking the ratio of this scale to the Jeans scale (the 
smallest scale on which collapse can occur without being inhibited 
by pressure gradients) yields the correct functional form of $Q$.} 
to its own self-gravity if $Q < Q_{\rm crit}$, and stable if 
$Q > Q_{\rm crit}$. Typically $Q_{\rm crit} \simeq 1$ -- for the 
specific system we have investigated it would be $1/2$.  

We have derived the stability of a fluid disk in uniform rotation. 
Differential rotation and global effects alter the value of 
$Q_{\rm crit}$, but do not fundamentally change the result. For a 
collisionless disk (e.g. one made of stars or small solid particles) 
a comparable result applies if we replace the sound speed $c_s$ by 
the one-dimensional velocity dispersion $\sigma$.

The most unstable wavelength is,
\begin{equation} 
 \lambda_{\rm crit} = \frac{2 \pi}{k_{\rm crit}} = \frac{2 c_s^2}{G \Sigma_0}.
\end{equation}   
Comparing this to the scale height of the disk $h = c_s / \Omega$, 
we find that at marginal stability,
\begin{equation}
 \frac{\lambda_{\rm crit}}{h} \simeq 2 \pi
\end{equation}
i.e. the instability afflicts small-ish spatial scales within the disk. 

Let us apply this analysis to the problem of planetesimal formation. 
If we ignore radial drift, then at 1~AU $\Sigma_{\rm dust} \sim 10^{-2} 
\Sigma_{\rm gas}$, or about 10~g \ cm$^{-2}$ for a minimum mass Solar 
Nebula model (note that a gas to dust ratio of 100 is a commonly 
used approximation in protoplanetary disk theory). Setting 
$Q = \sigma \Omega / (\pi G \Sigma_{\rm dust}) = 1$, 
and taking $M_* = M_\odot$,  
we find that instability requires a critical velocity dispersion in the 
solid component,
\begin{equation}
 \sigma \simeq 10 \ {\rm cms}^{-1}.
\end{equation}
Since the {\em gas} sound speed at this radius is of the order of 
$10^5 \ {\rm cms}^{-1}$, and the scale heights of the gas and particle 
disks are respectively proportional to $c_s$ and $\sigma$, we see that 
an extremely thin disk is required before instability will set in! 

If instability occurs, the most unstable wavelength is predicted to be,
\begin{equation}
 \lambda_{\rm crit} \approx 3 \times 10^8 \ {\rm cm}.
\end{equation} 
The mass within an unstable patch is then,
\begin{equation}
 m \sim \pi \Sigma_{\rm dust} \lambda_{\rm crit}^2 \sim 3 \times 10^{18} \ {\rm g}
\end{equation}
which would correspond to a spherical body of size,
\begin{equation} 
 r = \left( \frac{3m}{4 \pi \rho_d} \right)^{1/3} \sim 6 \ {\rm km}
\end{equation}
for a material density of $\rho_d = 3 \ {\rm g \ cm}^{-3}$. The collapse time scale at 
distance $\lambda_{\rm crit}$ from mass $m$,
\begin{equation}
 t_{\rm ff} = \sqrt{\frac{\lambda_{\rm crit}^3}{2 G m}}
\end{equation}
is very short -- less than a year for the parameters adopted above. 
Even if we allow for the fact that angular momentum will preclude a 
prompt collapse, the derived time scale for planetesimal formation 
via gravitational instability remains extremely short -- perhaps 
of the order of 10$^3$~yr \cite{goldreich73}.

Formation of planetesimals via the Goldreich-Ward mechanism has 
several attractive features, most notably the short time scale 
and complete bypass of the size regime most vulnerable to radial 
drift. However in its simplest form, the mechanism fails to work. 
The problem lies in the fact that {\em even in an intrinsically 
non-turbulent} gas disk, the formation of a dense solid sub-disk 
leads to self-generated turbulence and associated vertical 
stirring {\em prior} to gravitational instability. As noted above, 
for gravitational instability to operate we require a thin 
sub-disk in which, for our choice of parameters,
\begin{equation}
 \frac{h_{\rm dust}}{h_{\rm gas}} \sim 10^{-4}.
\end{equation} 
Within this midplane layer, the volume density of solids 
would {\em exceed} the density of gas by a factor of the 
order of 100 -- i.e. the extreme thinness of the solid 
disk inverts the normal gas to dust ratio which favors 
{\em gas} by the same factor. Since the gas and dust are 
well coupled for small particle sizes, within the sub-disk 
(where the solid component dominates) we expect both the 
gas and the dust to orbit at the natural velocity for the 
solid component, which is the Kepler velocity. The gas 
just above the layer, on the other hand, will rotate 
slower due to the radial gas pressure gradient. There 
will therefore be a velocity gradient in the $z$ direction that 
is of the order of $(h_{\rm gas}/r)^2 v_K / h_{\rm dust}$. This 
shear will be Kelvin-Helmholtz unstable, leading to turbulence 
that prevents the layer ever getting thin enough to fragment 
into planetesimals \cite{cuzzi93}. The condition for 
Kelvin-Helmholtz instabilities to develop \cite{sekiya98,youdin02} is that the 
{\em Richardson number}, which measures the competition between 
vertical shear and buoyancy, is ${\rm Ri} < {\rm Ri}_{\rm crit}$, where,
\begin{equation} 
 {\rm Ri} \equiv \frac{N^2}{(\partial v_\phi / \partial z)^2}
\end{equation}
and $N$, the Brunt V\"ais\"al\"a frequency, is defined as,
\begin{equation}
 N^2 \equiv g_z \frac{\partial \ln \rho}{\partial z}.
\end{equation}  
The standard stability analysis obtains a critical Richardson 
number ${\rm Ri} = 0.25$, but both analytic calculations 
including the effect of Coriolis forces, and numerical 
simulations, favor a larger value of around unity 
\cite{gomez05,johansen06}.

\subsubsection{Streaming instabilities}
Although it is very hard to settle a particle layer to the point where it becomes gravitational 
unstable, coagulation plus settling can plausibly lead to a mid-plane layer in which 
the local particle density $\rho_p$ is comparable to that of the gas $\rho_g$. Current interest in gravitational 
instability as a mechanism for planetesimal formation is based upon the realization that 
the weaker condition $\rho_p \sim \rho_g$ suffices to trigger {\em non-gravitational} 
clumping of particles via the streaming instability, that can be strong enough to form clumps 
that will subsequently collapse under self-gravity.

The term ``streaming instability" is used generically to describe instability in aerodynamically 
coupled mixtures of particles and gas in Keplerian disks. The original analysis by \citet{youdin05} 
considered a vertically unstratified system in which an incompressible gas interacts with a 
compressible particle ``fluid" via two-way aerodynamic forces\footnote{Particle clustering 
in turbulence also occurs in the ``passive" limit where the aerodynamic back reaction on the 
gas is neglected. In this regime, particles concentrate in the regions {\em between} 
vortices. This type of aerodynamic clustering may have interesting consequences for 
protoplanetary disks \citep{cuzzi08,pan11}, but the natural scale is {\em much} smaller 
(of the order of the viscous scale in the turbulence) and the process is entirely 
distinct from the streaming instability.}. This system is described 
by a pair of continuity and momentum equations representing the two fluids,
\begin{eqnarray}
 \nabla \cdot {\bf v}_g &=& 0, \nonumber \\
 \frac{\partial \rho_p}{\partial t} + \nabla \cdot ( \rho_p {\bf v}_p ) &=& 0, \nonumber \\
 \frac{\partial {\bf v}_g}{\partial t} + {\bf v}_g \cdot \nabla {\bf v}_g &=&  
   - \Omega_K^2 {\bf r} + \frac{\rho_p}{\rho_g} \frac{ {\bf v}_p - {\bf v}_g }{t_{\rm fric}}  
   - \frac{\nabla P}{\rho_g}, \nonumber \\
 \frac{\partial {\bf v}_p}{\partial t} + {\bf v}_p \cdot \nabla {\bf v}_p &=&  
   - \Omega_K^2 {\bf r} - \frac{ {\bf v}_p - {\bf v}_g }{t_{\rm fric}}. 
\end{eqnarray} 
(The notation here ought to be self-explanatory.) The above equations are not a 
full description of the physical system found in protoplanetary disks. Vertical 
stratification is neglected, together with the effect of intrinsic turbulence which 
might loosely be supposed to lead to an effective diffusivity for the particles. 
Taking the gas to be incompressible and treating the particles as a fluid 
(with, necessarily, a single-valued velocity field at each point) are good 
approximations, but approximations nonetheless. 

The equilibrium state of steady radial drift defined by the above equations is the 
Nakagawa-Sekiya-Hayashi (NSH) equilibrium mentioned earlier \citep{nakagawa86}. 
\citet{youdin05} showed that the NSH equilibrium is linearly unstable for a broad 
range of system parameters, of which the important ones are the local solid to 
gas ratio, the dimensionless stopping time of the particles, and the amount of 
pressure support in the gas. The simplest unstable modes are axisymmetric, 
typically of small scale (substantially smaller than $h$), and have growth rates 
that can be as large as $\sim 0.1 \Omega_K^{-1}$ but which are often much 
smaller. The linear instability does not have any known explanation that is 
particularly compelling or intuitive.

Numerical simulations have established that the streaming instability provides a 
pathway to forming dense clumps that can collapse gravitationally to form 
planetesimals \citep{johansen07}, though the pre-requisite particle size and 
local metallicity are not trivially satisfied. Work by \citet{carrera15} and 
\citet{yang17} shows that the non-linear evolution of the streaming instability 
leads to strong clumping that would precipitate rapid planetesimal formation 
in a U-shaped region of local metallicity / stopping time space spanning 
$10^{-3} \lesssim \tau_{\rm fric} \lesssim 10$. The lowest required 
metallicity---which is still super-Solar \citep{johansen09}---occurs for $\tau_{\rm fric} \sim 0.1$, 
which corresponds to particles larger than those that obviously form 
from coagulation in the inner disk. Smaller values of $\tau_{\rm fric} \sim 10^{-3}$, 
which lead to strong clumping only for metallicities of $Z \gtrsim 0.04$, 
match up better with expected initial particle sizes. It appears plausible 
that planetesimal formation occurs, albeit at a slower rate, for local 
solid to gas ratios modestly lower than the currently established thresholds.

Once collapse of streaming-initiated over-densities occurs the outcome is a 
population of planetesimals. The initial mass function of the resultant 
planetesimals can be fit with a power-law that is cut-off above some 
characteristic mass. A recent determination by \citet{simon16} obtains 
a differential mass function,
\begin{equation}
 \frac{{\rm d}N}{{\rm d}M_p} \propto M_p^{-p},
\end{equation}
with $p = 1.6 \pm 0.1$. Results by \citet{schafer17} are consistent. 
This is a mass function in which the largest bodies in the population 
have most of the mass.

\begin{figure}
\includegraphics[width=0.9\columnwidth]{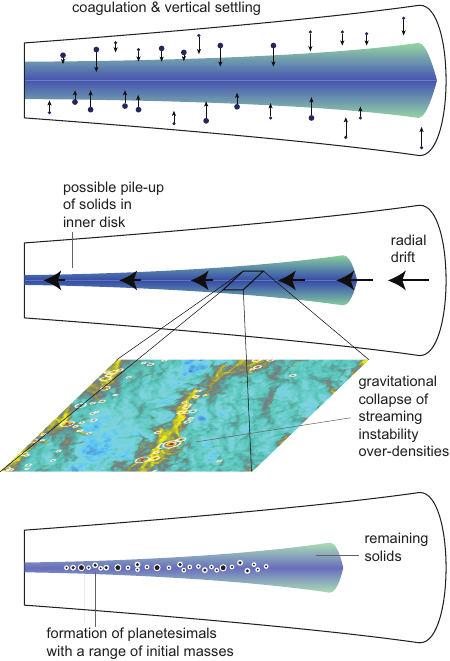}
\caption{Illustration of the physical processes that may be involved 
in the formation of planetesimals. Vertical settling and coagulation 
lead to the formation of a relatively high density particle layer 
near the disk mid-plane. Aerodynamics forces mean that particles 
drift radially as they grow, reducing the abundance of solids in the 
outer disk and (potentially) increasing it closer to the star. Where the 
stopping time and local dust-to-gas ratio satisfy the right conditions, the 
streaming instability \cite{youdin05} leads to strong clumping of 
solids. Some of these over-densities collapse gravitationally, 
forming a population of planetesimals with a range of masses. 
\cite[Simulation of planetesimal formation via the streaming instability from][{{\footnotesize \tt http://jila.colorado.edu/$\sim$jasi1566/3dstreaminginsta.html}}]{simon16}.}
\label{fig_streaming}
\end{figure}

Figure~\ref{fig_streaming} illustrates how the various processes that we 
have discussed---coagulation, vertical settling, radial drift, and the streaming instability---might 
conspire together to lead to planetesimal formation. It is not obvious what radial 
profile of planetesimal surface density ought to result, though barring some 
remarkable fluke it will not be the simple power-law envisaged in the 
Minimum Mass Solar Nebula model. There has been some discussion 
of how the hypothesis that the streaming instability forms planetesimals 
could be tested, either in the asteroid belt \citep{morbidelli09} or in the 
Kuiper Belt \citep{nesvorny10}, but more work is needed before drawing 
strong conclusions.

Other flavors of particle clustering in turbulence may also be important. 
Attaining the relatively high local dust to gas ratios needed to trigger the streaming 
instability may be aided and abetted by local pressure maxima in zonal flows 
\citep{johansen09b,simon14} or vortices \citep{barge95}, which may be 
formed at the edges of dead zones \citep{lyra09}. The loss of gas \citep{throop05} 
and formation of an inner hole via photoevaporative disk dispersal \citep{alexander07} could 
also enhance $Z$, potentially triggering a late episode of planetesimal formation. 

\subsection{Growth beyond planetesimals}
Once planetesimals have formed the gas in the disk will continue to influence 
their dynamics through two diametrically opposed effects. Residual 
{\em aerodynamic} interactions will act to damp planetesimal eccentricity 
and inclination, while surface density fluctuations produced by turbulence 
will exert fluctuating gravitational forces that excite eccentricity \citep{laughlin04,nelson05,okuzumi13}. 
These effects are significant, but overall further dynamical interaction between the 
solid and gaseous components of the disk is limited until bodies 
with sizes $> 10^3 \ {\rm km}$ form that are large enough to 
have a {\em gravitational} coupling to the gas. 
We will discuss the impact of gravitational coupling (``migration") 
later in the context of the early evolution of planetary systems. 

How do planetesimals grow to form planetary embryos, planets and giant planet cores? 
We will start by discussing the physics of the classical model for planet formation, 
in which the dominant dynamics is mutual gravitational interactions between the 
bodies and growth occurs from planetesimal-planetesimal and eventually 
planetesimal-protoplanet collisions. This is a well-posed problem that is 
usually studied using a combination of statistical and N-body methods. 
Later, we will describe a popular modern variant in which the dynamics 
remains largely gravitational, but where growth occurs due to the 
aerodynamically assisted accretion of small particles (``pebbles") 
that failed to form planetesimals.

\subsubsection{Gravitational focusing}

\begin{figure}
\includegraphics[width=\columnwidth]{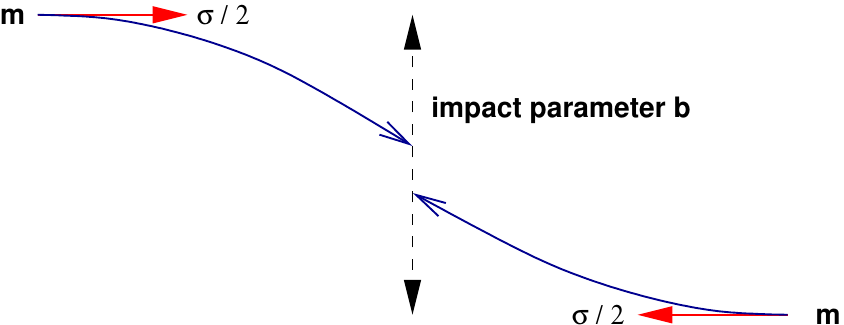}
\caption{Setup for calculation of gravitational focusing. Two 
bodies of mass $m$, moving on a trajectory with impact parameter $b$, 
have a velocity at infinity of $\sigma / 2$.}
\label{fig_focus}
\end{figure}

For sufficiently small bodies, the effects of gravity can be 
ignored for the purposes of determining whether they will physically 
collide. A massive planet, on the other hand, can {\em gravitationally 
focus} other bodies toward it, and as a result has a collision cross section that 
is much larger than its physical cross section. 

To evaluate the magnitude of this gravitational focusing, consider 
two bodies of mass $m$, moving on a trajectory with impact 
parameter $b$, as shown in Figure~\ref{fig_focus}. The relative velocity 
at infinity is $\sigma$. At closest approach, the bodies have 
separation $R_c$ and velocity $V_{\rm max}$. Equating energy in 
the initial (widely separated) and final (closest approach) states 
we have,
\begin{equation} 
 \frac{1}{4} m \sigma^2 = m V_{\rm max}^2 - \frac{Gm^2}{R_c}.
\end{equation}
Noting that there is no radial component to the velocity at the 
point of closest approach, angular momentum conservation gives,
\begin{equation}
 V_{\rm max} = \frac{1}{2} \frac{b}{R_c} \sigma.
\end{equation}
If the sum of the physical radii of the bodies is $R_s$, then 
for $R_c < R_s$ there will be a physical collision, while 
larger $R_c$ will result in a harmless flyby\footnote{This is 
true for solid bodies -- for giant planets or stars tidal 
effects can lead to significant dissipation of energy even when 
$R_c > R_s$ \cite{fabian75}.}. The {\em largest} value of the 
impact parameter that will lead to a physical collision is 
thus,
\begin{equation}
 b^2 = R_s^2 + \frac{4 G m R_s}{\sigma^2},
\end{equation}
which can be expressed in terms of the escape velocity from 
the point of contact, $v_{\rm esc}^2 = 4 Gm / R_s$ as,
\begin{equation}
 b^2 = R_s^2 \left( 1 + \frac{v_{\rm esc}^2}{\sigma^2} \right).
\end{equation}
The cross section for collisions is then,
\begin{equation}
 \Gamma = \pi R_s^2 \left( 1 + \frac{v_{\rm esc}^2}{\sigma^2} \right), 
\end{equation}
where the term in brackets represents the enhancement to the 
physical cross section due to gravitational focusing. Clearly 
a planet growing in a ``cold" planetesimal disk for which 
$\sigma \ll v_{\rm esc}$ will grow much more rapidly as a 
consequence of gravitational focusing. As a consequence, 
determining the velocity dispersion of bodies of different 
masses during the planet formation process is extremely 
important.      

\subsubsection{Growth versus fragmentation}
When two initially solid bodies physically collide the outcome can 
be divided broadly into three categories:
\begin{itemize}
\item
{\bf Accretion}. All or most of the mass of the impactor becomes part 
of the mass of the final body, which remains solid. Small fragments 
may be ejected, but overall there is net growth.
\item
{\bf Shattering}. The impact breaks up the target body into a number 
of pieces, but these pieces remain part of a single body (perhaps 
after reaccumulating gravitationally). The structure of the shattered 
object resembles that of a {\em rubble pile}.
\item
{\bf Dispersal}. The impact fragments the target into two or more 
pieces that do not remain bound.
\end{itemize}
To delineate the boundaries between these regimes quantitatively, we 
consider an impactor of mass $m$ colliding with a larger body of mass 
$M$ at velocity $v$. We define the specific energy $Q$ of the impact 
via,
\begin{equation}
 Q \equiv \frac{m v^2}{2 M},
\end{equation}
and postulate that this parameter largely controls the result. 
The thresholds for the various collision outcomes can then be expressed 
in terms of $Q$. Conventionally, we define the threshold for catastrophic 
disruption $Q_D^*$ as the minimum specific energy needed to disperse the 
target in two or more pieces, with the largest one having a mass $M/2$. 
Similarly $Q_S^*$ is the threshold for shattering the body. More work 
is required to disperse a body than to shatter it, so evidently $Q_D^* > Q_S^*$.  
It is worth keeping in mind that in detail the outcome of a particular collision 
will depend upon many factors, including the mass ratio between the target and 
the impactor, the angle of impact, and the shape and rotation rate of the bodies 
involved. Quoted values of $Q_D^*$ are often averaged over impact angles, but 
even when this is done the parameterization of collision outcomes in terms 
of $Q$ is only an approximation.

The estimated values of $Q_D^*$ for a target of a particular size vary by 
more than an order of magnitude depending upon the composition of the 
body, which can broadly be categorized into solid or shattered rock, and 
solid or porous ice. For any 
particular type of body, however, two distinct regimes can be identified:
\begin{itemize}
\item
{\bf Strength dominated regime}. The ability of small bodies to withstand 
impact without being disrupted depends upon the material strength of the 
object. In general, the material strength of bodies declines with increasing 
size, owing to the greater prevalence of defects that lead to cracks. In the 
strength dominated regime $Q_D^*$ decreases with increasing size.
\item
{\bf Gravity dominated regime}. Large bodies are held together primarily 
by gravitational forces. In this regime $Q_D^*$ must at the very least exceed the specific 
binding energy of the target, which scales with mass $M$ and radius $a$ 
as $Q_B \propto GM / a \propto \rho_d a^2$. In practice it requires a 
great deal more than this minimum amount of energy to disrupt the target -- so 
$Q_B$ is {\em not} a good estimate of $Q_D^*$ -- 
but nonetheless $Q_D^*$ does increase with increasing size.
\end{itemize}
Although the transition between these regimes is reasonably sharp there is 
{\em some} influence of the material properties (in particular the shear 
strength) on the catastrophic disruption threshold for smaller bodies 
within the gravity dominated regime.

Values of $Q_S^*$ and $Q_D^*$ can be determined experimentally for small 
targets \citep{arakawa02}. Experiments are not possible in the 
gravity dominated regime, but $Q_D^*$ can be estimated theoretically using 
numerical hydrodynamics \citep{benz99,leinhardt08} or (for rubble piles) 
rigid body dynamics simulations \citep{leinhardt02,korycansky06}. The 
simplest parameterization of the numerical results is as a broken 
power law that includes terms representing the strength and gravity 
regimes,
\begin{equation}
 Q_D^* = q_s \left( \frac{a}{1 \ {\rm cm}} \right)^c + 
 q_g \rho_d \left( \frac{a}{1 \ {\rm cm}} \right)^d.
\label{eq_C5_benzfit} 
\end{equation} 
Often (but not always) $Q_D^*$ is averaged over impact geometry, and $q_s$, $q_g$, $c$ and $d$ are 
all constants whose values are derived by fitting to the results of numerical 
simulations.

\begin{table}
\caption{Parameters for the catastrophic disruption threshold fitting formula 
(equation~\ref{eq_C5_benzfit}), which describes how $Q_D^*$ scales with 
the size of the target body. The quoted values were derived by Benz \& 
Asphaug (1999) and Leinhardt \& Stewart (2009) using numerical hydrodynamics 
simulations of collisions, which are supplemented in the strength dominated regime by experimental results.}
\begin{tabular}[t]{lccccc}
\hline \hline
 & $v$ / & $q_s$ / & $q_g$ / & $c$ & $d$ \\
 & km~s$^{-1}$ & erg~g$^{-1}$ & erg~cm$^3$~g$^{-2}$ &  &  \\
\hline 
Ice (weak) & 1.0 & $1.3 \times 10^6$ & 0.09 & -0.40 & 1.30 \\
Ice (strong) & 0.5 & $7.0 \times 10^7$ & 2.1 & -0.45 & 1.19 \\
Ice (strong) & 3.0 &  $1.6 \times 10^7$ & 1.2 & -0.39 & 1.26 \\
Basalt (strong) & 3.0 & $3.5 \times 10^7$ & 0.3 & -0.38 & 1.36 \\
Basalt (strong) & 5.0 & $9.0 \times 10^7$ & 0.5 & -0.36 & 1.36 \\
\hline \hline
\end{tabular}
\label{table_C5_benz}
\end{table}

\begin{figure}
\includegraphics[width=\columnwidth]{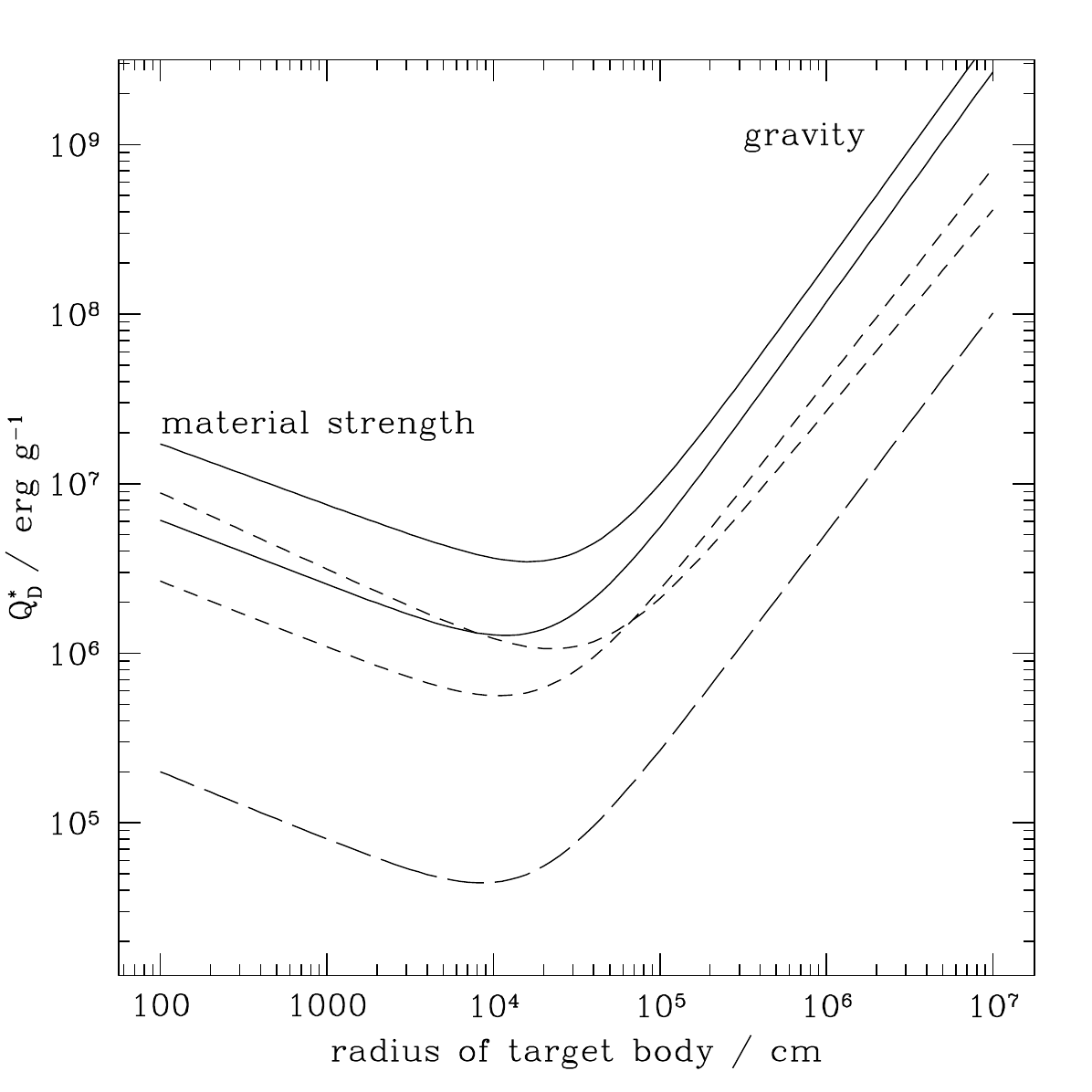}
\caption{The specific energy $Q_D^*$ for catastrophic disruption of solid bodies  
is plotted as a function of the body's radius. The solid and short dashed curves 
show results obtained using fits to theoretical 
calculations for impacts into ``strong" targets by Benz \& Asphaug (1999). The 
long dashed curve shows the recommended curve for impacts into ``weak" targets 
from Leinhardt \& Stewart (2008), derived from a combination of impact 
experiments and numerical simulations. 
In detail the solid curves show results for 
basalt at impact velocities of 5~km~s$^{-1}$ (upper curve) and 3~km~s$^{-1}$ 
(lower curve). The short dashed curves show results for water ice at 3~km~s$^{-1}$ 
(the lower curve for small target sizes) and 0.5~km~s$^{-1}$ (upper curve 
for small target sizes). The long dashed curve shows results for normal impacts into 
weak water ice targets at 1~km~s$^{-1}$.}
\label{figure_C5_benz}
\end{figure}

\citet{benz99} and \citet{leinhardt08} determined the 
values of the fitting parameters in 
equation~(\ref{eq_C5_benzfit}) from the results of an ensemble of simulations of 
impacts into icy or rocky targets. Their results are given in Table~\ref{table_C5_benz} 
and plotted as a function of target size in Figure~\ref{figure_C5_benz}. One observes 
immediately that the results for a particular target material vary with the 
impact velocity, and hence that $Q_D^*$ is {\em not} the sole determinant of 
the outcome of collisions. There is, however, a clear transition between 
the strength and gravity dominated regimes, with the weakest bodies being 
those whose size is comparable to the cross-over point. The most vulnerable 
bodies are generally those with radii in the 100~m to 1~km range. Just how 
vulnerable such bodies are to catastrophic disruption depends sensitively on 
their make-up, and it would be unwise to place too much trust in precise numbers. 
As a rough guide, however, the weakest icy bodies have minimum $Q_D^* \sim 
10^5 \ {\rm erg} \ {\rm g}^{-1}$, while the strongest conceivable planetesimals 
(unfractured rocky bodies) have minimum $Q_D^* > 
10^6 \ {\rm erg} \ {\rm g}^{-1}$.  

As a reality check, we may note that asteroids in the main belt with $e \simeq 0.1$ 
would be expected to collide today with typical velocities of the order of 2~km~s$^{-1}$. 
For a mass ratio $m / M = 0.1$ the specific energy of the collision is then around 
$Q = 2 \times 10^9 \ {\rm erg \ g}^{-1}$, which from Figure~\ref{figure_C5_benz} 
is sufficient to destroy even quite large solid bodies with $a \simeq 100 \ {\rm km}$. 
This is consistent with the observation of asteroid families, and the interpretation 
of such families as collisional debris. Evidently the random velocities that 
characterize collisions must have been {\em much} smaller during the epoch of 
planet formation if we are to successfully build large planets out of initially 
km-scale planetesimals.
  
\subsubsection{Shear versus dispersion dominated encounters}
\label{sec_shear_dispersion}
A more subtle distinction that nevertheless plays a crucial 
role in planet formation is whether encounters between bodies 
can be described via 2-body dynamics --- in which only the 
gravity of the two objects themselves matters --- or whether 
the tidal influence of the Sun also needs to be considered 
(3-body dynamics). \citet{goldreich04} summarize 
in simple terms why the distinction between 2 and 3-body 
dynamics matters at different stages of the planet formation 
process. We consider a 3-body system consisting of a large body 
(a ``planet") with mass $M$, a small body of negligible mass 
(described as a test particle), and the Sun, and define the 
{\em Hill radius} $r_H$ as the radius within which the gravity of 
the planet dominates (in astrophysical contexts, the same 
concept is referred to as the ``Roche lobe"). Roughly, this is 
obtained by equating the angular velocity for an orbit at distance $r_H$ from 
the planet with the angular velocity of the planet around the 
star. We find,
\begin{equation}
 r_H = \left( \frac{M_p}{3M_*} \right)^{1/3} a
\label{eq_rhill} 
\end{equation}
where the factor 3 is included for consistency with more 
detailed derivations. For circular orbits, collisions are 
forbidden for an orbital separation $\Delta a$ between the 
small body and the planet such that $\Delta a \lesssim r_H$ (c.f. the Trojan 
asteroids in the Solar System). If we define a characteristic 
velocity at the Hill radius,
\begin{equation}
 v_H \equiv \sqrt{\frac{GM_p}{r_H}}
\end{equation}
then for,
\begin{itemize}
\item
$\sigma > v_H$ 2-body dynamics describes collisions quite well. 
This regime is called {\bf dispersion dominated}.
\item
$\sigma < v_H$ 3-body effects are important. This regime is 
called {\bf shear dominated}.
\end{itemize}
When $\sigma < v_H$ and we are shear dominated, the collision 
rate is modified compared to expectations based on 2-body dynamics. 

\subsubsection{Accretion versus scattering}
A third general consideration is the balance between impacts (which will lead to 
accretion if the bodies are large enough) and gravitational scattering events. To 
estimate this, we consider a planetesimal orbiting just close enough to a growing 
planet that gravitational perturbations result in an encounter. As we will 
show later (Section \ref{sec_hill}), this condition implies that the orbital 
separation $\Delta a$ scales with the Hill radius $r_H$. The velocity 
difference between the planetesimal and the planet is then (ignoring factors of the 
order of unity),
\begin{equation} 
 \Delta v \sim \left\vert \frac{{\rm d}v_K}{{\rm d}a} \right\vert r_H \sim \sqrt{ \frac{G M_*^{1/3} M_p^{2/3}}{a} }.
\end{equation} 
A close encounter between the planetesimal and the planet will give the planetesimal a 
kick (a gravitational ``slingshot", akin to those used to modify the trajectory of spacecraft) 
whose magnitude depends upon the point of closest approach. The maximum kick will 
occur for a grazing counter, resulting in a kick whose size plausibly scales with the 
escape velocity $v_{\rm esc}$ from the planet. If $v_{\rm esc} \ll \Delta v$, it is then 
impossible for the planet to scatter the planetesimal into a significantly different orbit, 
and ultimately the two must collide. Conversely, for $v_{\rm esc} \gg \Delta v$, scattering 
will dominate over physical collisions. Taking the ratio, 
\begin{equation} 
 \frac{v_{\rm esc}}{\Delta v} \propto \left( \frac{M_p}{M_*} \right)^{1/6} \left( \frac{a}{R_s} \right)^{1/2},
\end{equation} 
we find that for fixed planetary properties scattering (favored at large values of $v_{\rm esc} / \Delta v$) 
becomes more important at large distances from the star, whereas collisions and accretion dominate 
close-in. This basic dynamical fact has two important consequences,
\begin{itemize}
\item
{\em Kepler's} super-Earth systems contain planets with masses greater than $M_\oplus$ and orbits 
substantially interior to 1~AU. This is well into the regime where accretion dominates over 
scattering. Accordingly --- in the absence of gas or other dissipative processes --- planets 
growing at small orbital radii end up accreting the large majority of the reservoir of solids 
dynamically accessible to them.
\item
A giant planet core, with a mass of the order of 10~$M_\oplus$, moves increasingly into the 
scattering regime for orbital radii substantially past 10~AU. In particular, at 50-100~AU,  
a core is much more efficient at scattering than accretion of planetesimals. The dominance of 
scattering means that there is a severe barrier to forming giant planets in situ at large 
orbital radii.
\end{itemize}
The Solar System's terrestrial planets fall into an intermediate regime. To a first approximation, 
the rocky bodies remaining near 1~AU after the gas disk disperses assemble ``in place" into 
the terrestrial planets, but some scattering and radial migration does take place. This is the 
basis, for example of the \citet{hansen09} model in which Mars is a planetary embryo 
scattered outward from a formation location closer to 1~AU.

\subsubsection{Growth rates}
We now proceed to derive an estimate for how fast a planet will 
grow due to accretion of planetesimals. We assume that the growing 
body, of mass $M$, radius $R_s$, and surface escape speed 
$v_{\rm esc}$ is embedded within a ``swarm" of planetesimals 
with local surface density $\Sigma_p$, velocity dispersion $\sigma$, 
and scale height $h_p$. The volume density of the planetesimal 
swarm is,
\begin{equation}
 \rho_{sw} = \frac{\Sigma_p}{2 h_p}.
\end{equation}
Then if 3-body effects can be ignored, the large body grows at 
a rate,
\begin{equation}
 \frac{{\rm d}M}{{\rm d}t} = \rho_{sw} \sigma \pi R_s^2 
 \left( 1 + \frac{v_{\rm esc}^2}{\sigma^2} \right). 
\end{equation}
This can be simplified since $h_p \sim \sigma / \Omega$ and 
hence $\rho_{sw}$ is inversely proportional 
to $\sigma$. We find,
\begin{equation}
 \frac{{\rm d}M}{{\rm d}t} = \frac{1}{2} \Sigma_p \Omega \pi R_s^2 
 \left( 1 + \frac{v_{\rm esc}^2}{\sigma^2} \right) 
\label{eq_growthrate} 
\end{equation}
where the numerical prefactor, which has not been derived accurately 
here, depends upon the assumed velocity 
distribution of the planetesimals. For an isotropic distribution 
the prefactor is $\sqrt{3} / 2$ \cite{lissauer93}.

This simple result is important. We note that:
\begin{itemize}
\item
The velocities of the planetesimals enter only via the gravitational 
focusing term, which can however by very large.
\item
The rate of mass growth scales linearly with $\Sigma_p$ --- we 
expect faster growth in disks that have more mass in planetesimals 
(due to a higher gas mass and / or a higher ratio of solids to 
gas).
\item
Other things being equal, growth will be slower at large radii, 
due to lower $\Sigma_p$ and smaller $\Omega$.
\end{itemize}
Complexity arises because as a planet grows, it stars to 
influence both the velocity dispersion and, eventually, the 
surface density of the planetesimal swarm in its vicinity.

Two simple solutions of the growth equation give an idea 
of the possibilities present in more sophisticated models. 
First, assume that the gravitational focusing term $F_g$ is 
constant. In this regime,
\begin{equation}
 \frac{{\rm d}M}{{\rm d}t} \propto R_s^2 \propto M^{2/3}
\end{equation}
which has solution,
\begin{equation}
 R_s \propto t.
\end{equation}  
The radius of the planet grows at a linear rate. Writing the 
planet mass $M = (4/3) \pi R_s^3 \rho_{\rm planet}$, where 
$\rho_{\rm planet}$ is the planet density,
\begin{equation}
 \frac{{\rm d}R_s}{{\rm d}t} = \frac{\Sigma_p \Omega}{8 \rho_{\rm planet}} F_g.
\end{equation} 
If we assume that at the orbital radius of Jupiter $\Sigma_p = 10 \ {\rm g \ cm}^{-2}$, 
then for $\rho_{\rm planet} = 3 \ {\rm g \ cm}^{-3}$,
\begin{equation} 
 \frac{{\rm d}R_s}{{\rm d}t} \simeq 0.2 F_g \ {\rm cm \ yr}^{-1}.
\end{equation} 
This initial growth rate is slow, which implies that to form the cores of the giant planets 
in a reasonable time, large gravitational focusing factors are needed. For 
example, to reach 1000~km in $10^5$~yr, we require $F_g \sim 5000$. The 
need for large gravitational enhancements to the collision rate is even 
more severe for the ice giants, but substantially easier in the terrestrial 
planet region.

Since empirically $F_g$ {\em must} be large, a second useful limit to 
consider is the case where $F_g \gg 1$. If we assume that $\sigma$ is 
constant (i.e. consider the regime where the growing planet has not 
yet managed to dominate the dynamical excitation of the planetesimal 
swarm) then,
\begin{eqnarray} 
 F_g & = & \left( 1 + \frac{v_{\rm esc}^2}{\sigma^2} \right) \nonumber \\
     & \simeq & \frac{v_{\rm esc}^2}{\sigma^2} \nonumber \\
     & \propto & \frac{M}{R_s}.
\end{eqnarray}     
The growth equation (\ref{eq_growthrate}) gives,
\begin{equation}
 \frac{{\rm d}M}{{\rm d}t} \propto M R_s
\end{equation}
with solution,
\begin{equation}
 M = \frac{1}{( M_0^{-1/3} - k t )^3}, 
\end{equation}
where $M_0$ is the initial mass at time $t=0$ and $k$ is a 
constant. In this regime the increasing gravitational focusing 
factor means that $M \rightarrow \infty$ in a finite 
time, allowing much more rapid growth.

\subsubsection{Isolation mass}
As noted above, rapid growth requires that $\sigma$ remain low --- 
i.e. that the planetesimals remain on roughly circular orbits. 
This means that there is a finite supply of planetesimals that 
have orbits that pass close enough to a growing planet to 
collide --- once these have all been consumed growth is 
bound to slow. The mass at which this slowdown occurs is 
described as the {\em isolation mass} $M_{\rm iso}$.

To estimate the isolation mass, we note that a planet grows 
by accreting planetesimals within a `feeding zone'. The 
size of the feeding zone $\Delta a_{\rm max}$ is set by 
the maximum distance over which the planet's gravity is 
able to perturb planetesimal orbits sufficiently to allow 
collisions, so it will scale with the Hill radius. Writing 
\begin{equation} 
 \Delta a_{\rm max} = C r_H
\end{equation}
with $C$ a constant of order unity, we have that the mass 
of planetesimals within the feeding zone is,
\begin{equation}
 2 \pi a \cdot 2 \Delta a_{\rm max} \cdot \Sigma_p \propto M^{1/3}.
\end{equation}
Note the $1/3$ power of the planet mass, which arises from the mass 
dependence of the Hill radius. As a planet grows, its feeding zone 
expands, but the mass of new planetesimals within the expanded 
feeding zone rises more slowly than linearly. We thus 
obtain the isolation mass by setting the planet mass equal to the 
mass of the planetesimals in the feeding zone of the original disk,
\begin{equation}
 M_{\rm iso} = 4 \pi a \cdot C \left( \frac{M_{\rm iso}}{3 M_*} \right)^{1/3} a
 \cdot \Sigma_p
\end{equation}
which gives,
\begin{equation}
 M_{\rm iso} =  \frac{8}{\sqrt{3}} \pi^{3/2} C^{3/2} M_*^{-1/2} \Sigma_p^{3/2} a^3.
\end{equation} 
Evaluating this expression in the terrestrial planet region, taking 
$a = 1 \ {\rm AU}$, $\Sigma_p = 10 \ {\rm g \ cm}^{-2}$, $M_* = M_\odot$ 
and $C = 2 \sqrt{3}$ \cite{lissauer93}, we obtain,
\begin{equation} 
 M_{\rm iso} \simeq 0.07 \ M_\oplus.
\end{equation}
Isolation is therefore likely to occur late in the formation of the 
terrestrial planets.
Repeating the estimate for the conditions appropriate to the formation 
of Jupiter's core, using $\Sigma_p = 10 \ {\rm g \ cm}^{-2}$ as adopted 
by \citet{pollack96}\footnote{Note that this is a factor of several 
enhanced above the minimum mass Solar Nebula value.}, gives,
\begin{equation}
 M_{\rm iso} \simeq 9 \ M_\oplus.
\end{equation} 
This estimate is comparable to, or larger than, the current best 
determinations for the mass of the Jovian core \cite{guillot05}. 
The concept of the isolation mass may or may not be relevant to the 
formation of Jupiter, depending upon the adopted disk model (and other 
issues, such as the importance of pebble accretion and migration).

\subsubsection{Pebble accretion}
\label{sec_pebble}
Observations show that mm and smaller-sized particles are retained within protoplanetary 
disks for most of their lifetimes, though the size of the dust disk may be substantially smaller 
than that of the gas \citep{cleeves16}. Protoplanets and planetesimals must therefore co-exist 
with smaller solids that remain aerodynamically coupled to the gas, and in some circumstances 
planetary growth may occur predominantly by accretion of the small solids rather than by 
collisions with planetesimals. This is the basic idea of aerodynamically assisted or ``pebble" 
accretion, which has been developed by \citet{ormel10} and \citet{lambrechts12}. Our 
account here leans heavily on the \citet{lambrechts12} treatment, though note that our 
definition of the radial pressure support parameter $\eta$ (equation~\ref{eq_dv}) differs 
by a factor of two from theirs. The calculation has three parts. First, we determine the 
speed with which aerodynamically coupled solids---henceforth pebbles---approach a 
protoplanet on a strictly Keplerian orbit. Second, we estimate the radius out to which 
the gravity of the protoplanet affects the trajectories of the pebbles. Finally, we obtain 
an effective cross-section by requiring that gravity acts fast enough to capture the 
pebbles before they are swept out of the planet's region of influence by the gas flow.

To begin, we compute the relative velocity between a radially drifting pebble and a 
protoplanet on a Keplerian orbit with zero eccentricity and inclination. The radial and 
azimuthal components of the pebble drift at the orbit of the planet are given by 
equations (\ref{eq_dv}), (\ref{eq_101}) and (\ref{eq_vrrate}). We have,
\begin{eqnarray}
 v_r & = & -\frac{\eta v_K}{\tau_{\rm fric} + \tau_{\rm fric}^{-1}} \\
v_\phi - v_K & = & -\frac{1}{2} \frac{\tau_{\rm fric}^{-1} \eta v_K}{\tau_{\rm fric} + \tau_{\rm fric}^{-1}}.
\end{eqnarray}
These expressions neglect radial flow of the {\em gas}, and are hence valid provided the 
radial drift speed of particles in the mid-plane exceeds that of the gas. Adding in 
quadrature the pebble approach speed is,
\begin{equation}
 \Delta v = \frac{1}{2} \frac{\sqrt{4 + \tau_{\rm fric}^{-2}}}{\tau_{\rm fric} + \tau_{\rm fric}^{-1}} \eta v_K.
\end{equation} 
The functional dependence on $\tau_{\rm fric}$ is within 15\% of unity for $10^{-3} \leq \tau_{\rm fric} \leq 1$, 
so for practical purposes it suffices to assume that,
\begin{equation}
 \Delta v \simeq \frac{1}{2} \eta v_K.
\label{eq_dv_pebble} 
\end{equation} 
The pebble approach speed is thus dependent on the disk properties (recall from equation~\ref{eq_eta_dependence} 
that $\eta \propto (h/r)^2$) but not on the particle size. Small particles approach the protoplanet on almost 
azimuthal trajectories, while large particles have a greater radial component but roughly the same total speed.

\begin{figure*}
\includegraphics[width=1.6\columnwidth]{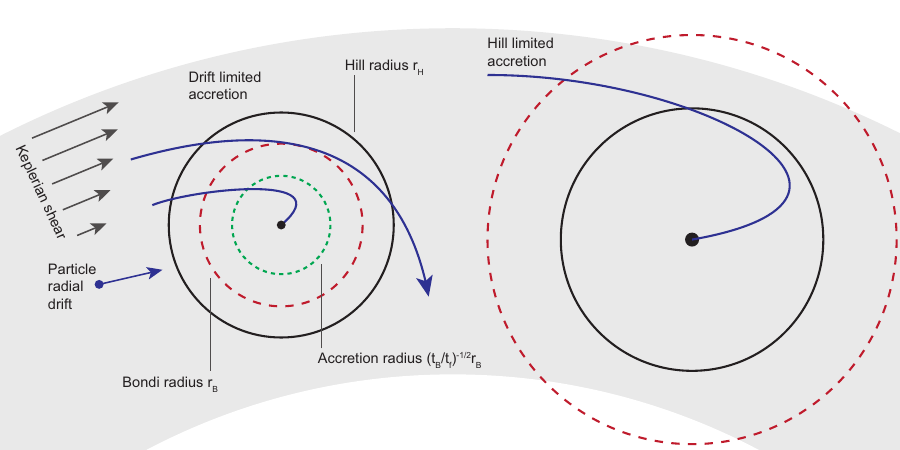}
\caption{Illustration of the two main regimes of aerodynamically assisted (``pebble") 
accretion. For lower mass protoplanets (left) the Bondi radius $r_B$, within which 
purely gravitational 2-body interactions would lead to strong scattering, is smaller 
than the Hill radius $r_H$. Aerodynamically coupled particles enter the Bondi 
radius with their radial drift velocity and are accreted if their stopping time is 
shorter than the time it takes to traverse 
the Bondi radius. For higher mass protoplanets (right) the Bondi radius exceeds the 
Hill radius, and the encounter velocity of particles approaching the planet is set by 
Keplerian shear rather than radial drift. Only those particles that enter the Hill 
sphere have a chance of being accreted.}
\label{fig_pebble}
\end{figure*}

In the same way that we distinguish between shear- and dispersion-dominated regimes of 
planetesimal accretion (\S\ref{sec_shear_dispersion}), how pebbles accrete depends upon 
whether the tidal field of the star needs to be taken into account. We first define two relevant 
radii. Ignoring both tidal (3-body) and gas drag effects a pebble approaching a protoplanet 
at velocity $\Delta v$ with impact parameter $b$ will suffer a {\em strong deflection} of the 
order of $\Delta \theta \sim 1 \ {\rm radian}$ if $b \leq r_B$, where 
\begin{equation}
 r_B = \frac{GM_p}{\Delta v^2}.
\end{equation} 
(This is justified to within factors of order unity in \S\ref{sec_torque_impulse}.) Following 
(rather confusing) convention we refer to this critical radius for large-angle scattering as 
the {\em Bondi radius}. A second characteristic scale is provided by the Hill radius 
(equation~\ref{eq_rhill}),
\begin{equation}
 r_H = \left( \frac{M_p}{3 M_*} \right)^{1/3} a,
\end{equation} 
 which defines the boundary beyond which tidal effects are 
important. At the Hill radius Keplerian shear introduces a velocity 
relative to the protoplanet $\Delta v_H = -(3/2) \Omega r_H$. Comparing 
this to the aerodynamic pebble approach speed (equation~\ref{eq_dv_pebble}) we find,
\begin{equation}
 \frac{\Delta v_H}{\Delta v} \approx \sqrt{ \frac{r_B}{r_H}}.
\end{equation} 
Under the assumption that strong 2-body deflection is a pre-requisite for aerodynamically 
assisted accretion we can then distinguish the two regimes of pebble accretion illustrated in 
Figure~\ref{fig_pebble}:
\begin{itemize}
\item
If $r_B < r_H$ pebbles that can potentially be accreted approach the protoplanet with 
a velocity set by radial drift. Accretion can be modeled without considering tidal effects. 
This is the regime of {\bf drift limited} accretion.
\item
If $r_B > r_H$ Keplerian shear dominates the approach velocity for large impact parameters, 
and limits the capture cross-section to roughly the Hill radius. This is {\bf Hill limited} accretion.
\end{itemize}
The transition between these regimes occurs when $r_B = r_H$. The transition mass is,
\begin{equation}
 M_p = \sqrt{\frac{1}{3}} \frac{\Delta v^3}{G \Omega},
\end{equation} 
with $\Delta v$ as given in eq.~(\ref{eq_dv_pebble}). There is a very strong dependence on 
the disk structure via the radial pressure support parameter $\eta$. Restricting ourselves to 
power-law disk models (equation~\ref{eq_def_n}) with mid-plane $P \propto r^{-n}$ we have 
that $\eta = n (h/r)^2$. A disk with $\Sigma \propto r^{-1}$ and $T_c \propto r^{-1/2}$ has 
$n=11/4$, and we estimate the transition mass as,
\begin{equation}
 M_p \approx 8 \times 10^{-3} \left( \frac{n}{11/4} \right)^3 
 \left( \frac{h/r}{0.05} \right)^6 \ M_\oplus.
\label{eq_pebble_mt} 
\end{equation} 
We should remember that the fundamental dependence is on ${\rm d}P / {\rm d}r$ rather 
than on $(h/r)$, and revert to the basic equations when considering locations such as 
planetary gap edges or pressure traps. With that caveat, however, we expect a modest 
increase in $(h/r)$ with radius in the disk, and a larger increase in the transition mass with 
distance from the star.

Requiring that particles suffer strong deflections due to the gravity of the planet is a 
necessary but not sufficient condition for them to be accreted in the presence of gas 
drag. We can determine the capture cross-section via a time scale argument. In the 
drift dominated case appropriate for small protoplanets there 
are three relevant time scales:
\begin{itemize}
\item
The friction time scale $t_{\rm fric}$.
\item
The time scale for a pebble to cross the Bondi radius $t_B = r_B / \Delta v$.  
\item
The time scale for the planet, with gravity $g$, to modify the velocity of a pebble 
with impact parameter $b$, $t_g = \Delta v / g = (\Delta v / GM_p) b^2$.
\end{itemize}
To be accreted we require that the planet's gravity be strong enough to ``pull" 
a pebble out of the aerodynamic flow, i.e. that $t_g < t_{\rm fric}$ \citep[this 
assertion is supported by numerical integrations;][]{ormel10,lambrechts12}. 
Taking $t_g = t_{\rm fric}$ to define the capture cross-section $r_d$ we 
obtain,
\begin{equation}
 r_d = \left( \frac{t_B}{t_{\rm fric}} \right)^{-1/2} r_B.
\end{equation}
The best situation for pebble accretion corresponds to $t_B \sim t_{\rm fric}$, in 
which case a large fraction of particles entering the Bondi radius end up getting 
accreted. With the same power-law disk model we used before this condition implies,
\begin{equation}
 \tau_{\rm fric} = \frac{8}{n^3}
 \left( \frac{M_p}{M_*} \right) 
 \left( \frac{h}{r} \right)^{-6}.
\end{equation} 
Once again there is a strong dependence on the geometric thickness of the 
gas disk. For nominal parameters appropriate to a protoplanet with a mass 
similar to the transition mass (equation~\ref{eq_pebble_mt}), $M_p = 10^{-2} \ M_\oplus$ 
and $(h/r) = 0.05$, the optimal dimensionless stopping time is $\tau_{\rm fric} \sim 1$.
 
Particles participating in pebble accretion are likely to have settled vertically, and 
depending upon the thickness of the settled layer (relative to $r_d$) the geometry 
of accretion may be either two- or three-dimensional. If the pebbles have surface 
density $\Sigma_p$ and mid-plane density $\rho_p$ simple geometry gives the 
appropriate accretion rate in the two regimes,
\begin{eqnarray}
 \dot{M}_{\rm pebble, 3D} & = &  \pi r_d^2 \rho_p \Delta v, \\
 \dot{M}_{\rm pebble, 2D} & = & 2 r_d \Sigma_p \Delta v. 
\end{eqnarray} 
What sort of growth is this? Substituting for $r_d$ in the expression 
above we find, in the three dimensional limit,
\begin{equation}
 \dot{M}_{\rm pebble, 3D} \propto \rho_p t_{\rm fric} M_p.
\end{equation}
Both $\rho_p$ and $t_{\rm fric}$ are properties of the particle disk rather than 
of the planet, so we have simply that $\dot{M}_p \propto M_p$ and a prediction 
of exponential growth. 

Identical considerations govern the expected accretion rate in the Hill limited 
regime \citep{lambrechts12}. In this case the crossing time of the Hill radius 
is $\Omega^{-1}$, and the requirement that $t_{\rm fric} \sim \Omega^{-1}$ 
implies immediately that the best particle size has $\tau_{\rm fric} \sim 1$. 
Most commonly the two dimensional accretion geometry is relevant, so we 
have,
\begin{equation}
 \dot{M}_{\rm pebble, Hill, 2D} = 2 r_H \Sigma_p \Delta v_H \propto r_H^2 \propto M_p^{2/3}.
\end{equation} 
We expect a transition in growth rates as protoplanets accreting pebbles via this 
aerodynamic mechanism pass from the drift to Hill limited regimes.

The sketch of the physics of pebble accretion given above is greatly simplified, and the 
reader is directed to the original papers of \citet{ormel10} and \citet{lambrechts12} both 
for more details and for justification of some of the assertions that we have made. Hydrodynamic 
calculations confirm that although the smallest particles get swept past embedded 
protoplanets by the aerodynamic flow, estimates based on three-body integrations 
of particle trajectories with drag are valid for most of the relevant particle sizes 
\citep{ormel13b,morbidelli12}. Within a pebble accretion scenario the growth rates of 
protoplanets depend critically upon the local abundance of particles of the right 
size to participate in the process. This is hard to determine, as the particle sizes of 
interest are precisely those that are subject to rapid radial drift with presumptively 
short residence times in the disk. Nonetheless, many estimates suggest that 
pebble accretion could be the dominant growth mechanism for at least some 
classes of planets. Examples of papers that discuss how planetary growth 
proceeds under pebble accretion include \citet{lambrechts14}, \citet{bitsch15}, 
\citet{levison15a} and \citet{levison15b}.
  
\subsubsection{Coagulation equation}
For both dust and planetary growth the basic mathematical question is how 
a size distribution evolves under the action of discrete collision events 
(possibly supplemented by a component of smooth accretion). 
The quantitative framework for addressing such questions is based on the 
{\em coagulation equation} \cite{smoluchowski16}. This allows us to drop 
our rather poorly defined descriptions of ``large" and ``small" bodies 
though at the expense of an enormous 
increase in complexity.

To write the coagulation equation in its simplest form\footnote{It is 
also possible to write the coagulation equation as an integro-differential 
equation for a continuous mass function $n(m,t)$ \cite{safronov69}, or 
as a discrete equation where bodies are binned into arbitrary mass 
intervals (typically logarithmic). \citet{kenyon98} provide a clear 
description of how the coagulation equation may be formulated and 
solved in the more general case.}, assume that 
the masses of bodies are integral multiples of some small mass $m_1$. 
At time $t$ there are $n_k$ bodies of mass $m_k = k m_1$. The 
coagulation equation in discrete form is,
\begin{equation}
 \frac{{\rm d}n_k}{{\rm d}t} = \frac{1}{2} 
 \sum_{i+j=k} A_{ij} n_i n_j - 
 n_k \sum_{i=1}^{\infty} A_{ki} n_i
\label{eq_coagulate} 
\end{equation}
where $A_{ij}$ is the rate of mergers between bodies of mass $m_i$ 
and $m_j$. The first term on the right-hand side of the equation 
describes the increase in the number of bodies of mass $m_k$ due 
to collisions of {\em all possible pairs} of bodies whose masses 
$m_i$ and $m_j$ sum to $m_k$. The second term describes the decrease 
due to bodies of mass $m_k$ being incorporated into even larger 
bodies. The possibility of fragmentation is here neglected. In 
this formulation of the problem of planetary growth, all of the 
physics --- such as gravitational focusing --- enters via the 
rate coefficients $A_{ij}$.

\begin{figure}
\includegraphics[width=\columnwidth]{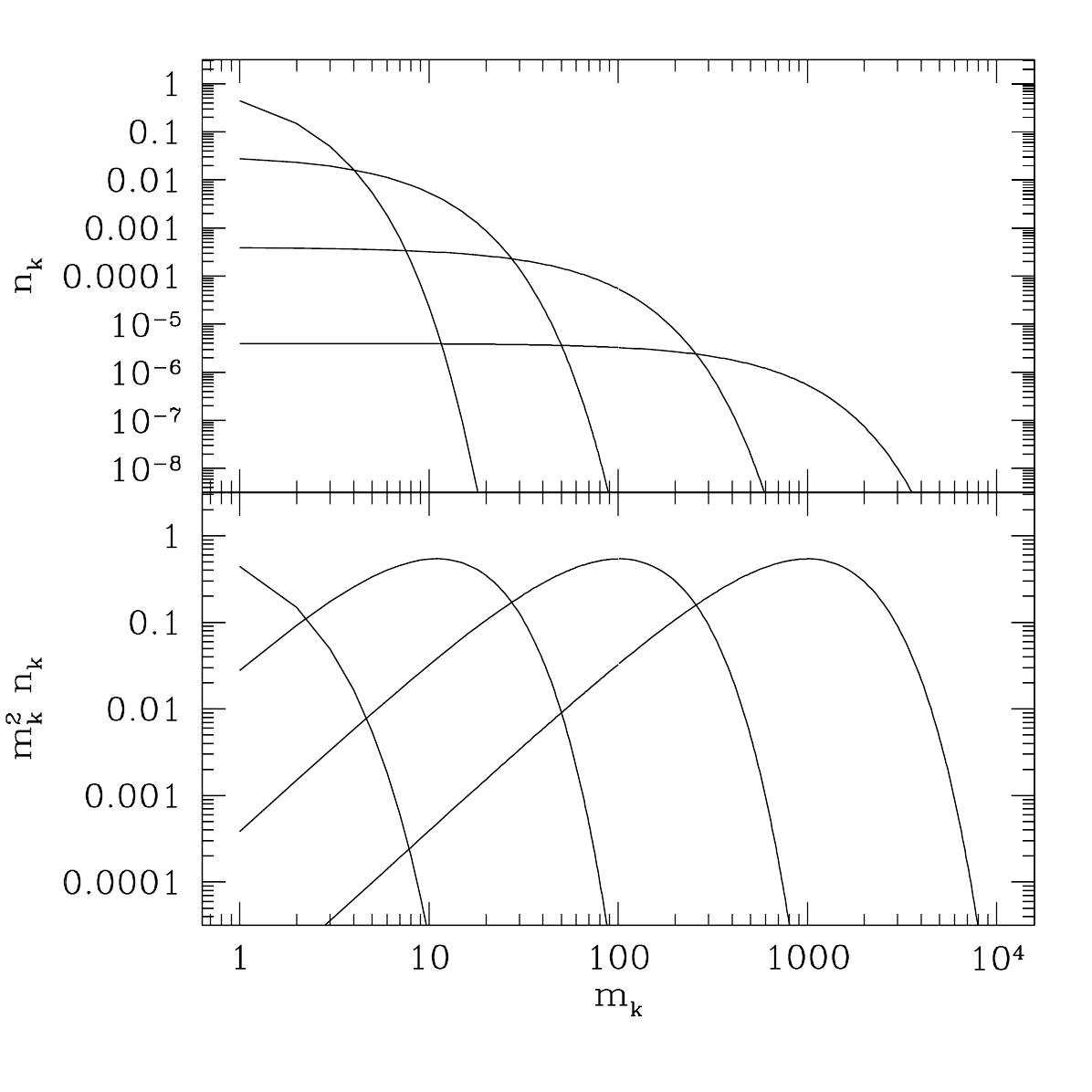}
\caption{Illustrative analytic solution to the coagulation equation for the 
simple case in which $A_{ij} = \alpha$, with $\alpha$ a constant. Initially 
all bodies have mass $m_1$. The solution is plotted for scaled times 
$t^\prime \equiv \alpha n_1 t$ equal to 1, 10, 100 and $10^3$. The upper 
panel shows the number of bodies $n_k$ of each mass (the vertical scale 
is arbitrary), while the lower panel shows how the mass distribution 
evolves. This solution is an example of orderly growth --- as time 
progresses the mean mass steadily increases while the shape of the 
mass spectrum remains fixed.}
\label{fig_coagulation}
\end{figure}

Equation (\ref{eq_coagulate}), or variants of it, has been used 
extensively to study planet formation \cite{safronov69,kenyon98,wetherill93,inaba01}, 
either on its own or in combination with direct N-body simulations 
\cite{bromley06}. Generally the coagulation equation needs to be 
supplemented with additional equations that describe the evolution 
of the velocity dispersion as a function of mass, as described for 
example in \citet{kenyon98}. Because of the fact that all $i$, $j$ such that 
$m_i + m_j = m_k$ contribute to the evolution of $n_k$, even the coagulation 
equation on its own is not  
a simple equation to deal with, and few analytic solutions are known. 
One (over)-simple case for which an analytic solution exists is for the 
case when,
\begin{equation}
 A_{ij} = \alpha
\end{equation}
with $\alpha$ a constant. Then,  if the initial state of the system 
comprises $n_1$ bodies all of mass $m_1$, the solution to equation 
(\ref{eq_coagulate}) is,
\begin{eqnarray}
 n_k &=& n_1 f^2 (1-f)^{k-1} \nonumber \\
 f &\equiv& \frac{1}{1 + \frac{1}{2} \alpha n_1 t}.
\end{eqnarray} 
This solution is shown as Figure~\ref{fig_coagulation}. The mass 
spectrum remains smooth and well-behaved as growth proceeds, and 
with increasing time the characteristic mass increases linearly 
while maintaining a fixed shape.

More generally, solutions to the coagulation equation fall into two 
classes \cite{lee00}:
\begin{itemize}
\item
Solutions that exhibit {\em orderly growth}, in which the mass distribution 
evolves smoothly with time toward higher mean masses. The analytic solution 
given above for the case $A_{ij} = {\rm constant}$ is an example of this 
type of evolution. Another analytic example is $A_{ij} \propto (m_i + m_j)$.
\item Solutions that show {\em runaway growth}. In this case the mass 
distribution develops a power-law tail toward high masses --- physically 
this corresponds to one or a handful of bodies growing rapidly at the 
expense of all the others. The long-term validity of the coagulation equation once 
runaway growth occurs is evidently limited. An analytic example occurs for a 
rate coefficient $A_{ij} \propto m_i m_j$.
\end{itemize}
Looking back to equation (\ref{eq_growthrate}), we note that the rate 
coefficient is expected to scale as $A \propto R_s^2 \propto m^{2/3}$ in the 
regime where gravitational focusing is unimportant, and $A \propto R_s^2 v_{\rm esc}^2 
\propto m^{4/3}$ once gravitational focusing is dominant. By comparison with the 
aforementioned analytic solutions, we expect that the initial growth of 
planetesimals will occur in the orderly regime, while runaway growth may 
occur once the largest bodies are massive enough for gravitational 
focusing to become significant.

\subsubsection{Overview of terrestrial planet formation}
We conclude the discussion of terrestrial planet formation by summarizing 
briefly the main stages of the process:
\begin{itemize}
\item[1.]
Dust particles agglomerate to form, eventually, planetesimals. Initially 
this occurs via pairwise collisions, though how (or whether) 
this process can continues to work for cm to meter scale particles remains 
somewhat murky. Gravitational instability, probably initiated by the streaming instability, 
may allow a bypass of these 
tricky sizes.
\item[2.]
Growth beyond planetesimals occurs via direct collisions, with an 
increasing role for gravitational focusing as masses become larger. 
Pebble accretion contributes to growth, by an amount that depends 
upon the abundance and size of surviving small solids. 
Dynamical friction keeps the velocity dispersion of the most 
massive bodies low. A phase of runaway growth may occur in which 
a few bodies grow rapidly at the expense of the rest.
\item[3.]
Runaway growth ceases once the largest bodies become massive 
enough to stir up the planetesimals in their vicinity. A 
phase of {\em oligarchic growth} ensues, in which the largest 
objects grow more slowly than they did during runaway growth, 
but still more rapidly than small bodies \cite{kokubo98,thommes03}. 
Growth continues in this mode until the isolation mass is 
approached, at which point growth slows further.
\item[4.]
Further evolution occurs as a result of collisions between 
the initially relatively isolated planetary embryos left 
over after oligarchic growth. The embryos are perturbed onto 
crossing orbits due to the influence of the giant planets 
and mutual secular resonances \cite{chambers98}. The final 
assembly of the terrestrial planets takes around 100~Myr, 
with the predicted configuration varying depending upon the 
assumed surface density of planetesimals and existence 
(or not) of giant planets \cite{kokubo06,levison03,raymond05}. 
In the Solar System, one of the final impacts on the Earth 
is widely considered to have given rise to the ejection of 
enough mass into orbit to subsequently form the Moon \cite{canup04}. 
\end{itemize}
The dominant uncertainties in theoretical models for terrestrial 
planet formation are arguably found during stage 1 --- the formation 
of planetesimals. It is also true that most simulations of the 
late stages of terrestrial planet formation lead to planetary 
properties (such as the eccentricity, and the mass of Mars compared 
to the other terrestrial planets) that differ somewhat from those observed in the 
Solar System. Thus, although there is general confidence that 
the basic physics of terrestrial planet formation is understood, 
it is clear that current models do not include all of the ingredients 
needed to accurately match Solar System constraints \citep{raymond09}. 

\subsection{Gas giant formation}
The formation of at least the vast majority of known gas giant planets 
is thought to occur as a consequence of {\em core accretion} \cite{mizuno80,bodenheimer86}. 
The core accretion model, which had a lengthy gestation period leading up 
to the landmark paper of \citet{pollack96}, postulates 
that the envelopes of gas giants are accreted subsequent to the 
formation of a large core, which is itself assembled in a manner 
analogous to terrestrial planet formation. 

Core accretion is the most widely accepted theory for massive planet formation. 
There is, however, an alternative model, based on the idea that 
a massive protoplanetary disk might collapse directly to form 
massive planets \cite{kuiper51,cameron78,boss97}. 
In this Section, we review the physics of these 
theories in turn. We also discuss the observational 
constraints on the different theories, which include 
inferences as to the core masses of the gas giants in 
the Solar System and the properties of extrasolar planetary systems.

\subsubsection{Core accretion model}
\begin{figure}
\includegraphics[width=\columnwidth,angle=90]{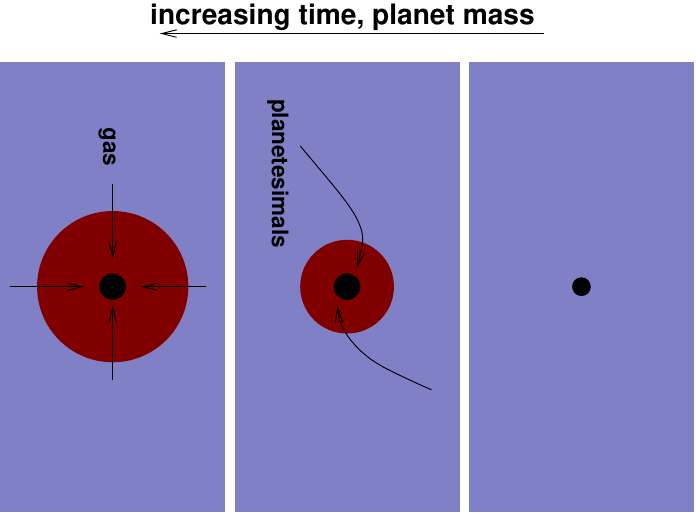}
\caption{Illustration of the main stages of the core accretion model 
for giant planet formation.}
\label{fig_coreaccretion}
\end{figure}

The main stages in the formation of a gas giant via core accretion 
are illustrated in Figure~\ref{fig_coreaccretion}. A core of rock 
and / or ice forms via the same mechanisms that we have previously 
outlined for terrestrial planet formation. Initially, there is 
either no atmosphere at all (because the potential is too shallow 
to hold on to a bound atmosphere), or any gas is dynamically 
insignificant. However, as the core grows, eventually it becomes 
massive enough to hold on to a significant envelope. At first, 
the envelope is able to maintain hydrostatic equilibrium. The core 
continues to grow via accretion of planetesimals (or pebbles), and the gravitational 
potential energy liberated as these solids rain down on the 
core provides the main source of luminosity. (In the limiting case 
where there is no planetesimal luminosity, analyzed in detail by 
\citet{piso14}, energy comes instead from Kelvin-Helmholtz contraction 
of the envelope.) This growth continues 
until the core reaches a {\em critical mass}. Once the critical 
mass is reached the envelope can no longer be maintained in 
hydrostatic equilibrium. The envelope contracts on its own 
Kelvin-Helmholtz time scale, and a phase of rapid gas accretion 
occurs. This process continues until (a) the planet becomes 
massive enough to open up a gap in the protoplanetary disk, 
thereby slowing down the rate of gas supply, or (b) the gas 
disk itself is dispersed.

The novel aspect of the core accretion  model is the existence 
of a critical core mass. \citet{mizuno80} used numerical models 
to demonstrate the existence of a maximum core mass, and showed 
that it depends only weakly on the local properties of the 
{\em gas} within the protoplanetary disk. A clear exposition 
of this type of calculation is given in, for example, 
\citet{papaloizou99}. The simplest toy model that 
exhibits the key property of a critical core mass is 
that due to \citet{stevenson82}, who considered the 
(unrealistic) case where energy transport is due 
solely to radiative diffusion. We reproduce his argument 
here. \citet{rafikov06} is a good place to start for understanding more 
realistic models in which convection also plays a role.

Consider a core of mass $M_{\rm core}$ and radius $R_{\rm core}$, 
surrounded by a gaseous envelope of mass $M_{\rm env}$. The 
total mass of the planet,
\begin{equation}
 M_t = M_{\rm core} + M_{\rm env}.
\end{equation} 
The envelope extends from $R_{\rm core}$ to some outer radius 
$R_{\rm out}$, which marks the boundary between the gas bound 
to the planet and the gas in the protoplanetary disk. $R_{\rm out}$ 
may be determined by thermal effects (in which case  $R_{\rm out} \sim 
GM_t / c_s^2$, with $c_s$ the disk sound speed) or by tidal 
considerations (giving an outer radius of $r_H$), whichever is 
the smaller. If the envelope is of low mass, then the largest 
contribution to the luminosity is from accretion of planetesimals 
onto the core. This yields a luminosity,
\begin{equation}
 L = \frac{GM_{\rm core} \dot{M}_{\rm core}}{R_{\rm core}}
\end{equation}
which is constant through the envelope.

If we assume that radiative diffusion dominates the energy transport, 
then the structure of the envelope is determined by the equations 
of hydrostatic equilibrium and radiative diffusion,
\begin{eqnarray}
 \frac{{\rm d}P}{{\rm d}r} & = & - \frac{GM(r)}{r^2} \rho \\
 \frac{L}{4 \pi r^2}       & = & - \frac{16}{3} \frac{\sigma T^3}{\kappa_R \rho} 
 \frac{{\rm d}T}{{\rm d}r}
\end{eqnarray}
where $\sigma$ is the Stefan-Boltzmann constant and $\kappa_R$ the 
Rosseland mean opacity (assumed constant). Eliminating the density 
between these equations we find that,
\begin{equation}
 \frac{{\rm d}T}{{\rm d}P} = \frac{3 \kappa_R L}{64 \pi \sigma G M T^3}.
\end{equation} 
We now integrate this equation inward from the outer boundary, making the 
approximation that $M(r) \approx M_t$ and taking $L$ and $\kappa_R$ to be constants,
\begin{equation}
 \int_{T_{\rm disk}}^T T^3 dT = 
 \frac{3 \kappa_R L}{64 \pi \sigma G M_t} \int_{P_{\rm disk}}^P dP.
\end{equation}
Once we are well inside the planet we expect that $T^4 \gg T_{\rm disk}^4$ and 
that $P \gg P_{\rm disk}$, so the integral yields, approximately,
\begin{equation}
 T^4 \simeq \frac{3}{16 \pi} \frac{\kappa_R L}{\sigma G M_t} P.
\end{equation}  
Substituting $P$ in this equation with an ideal gas equation of state,
\begin{equation}
 P = \frac{k_B}{\mu m_p} \rho T,
\end{equation}
we eliminate $T^3$ in favor of the expression involving ${\rm d}T / {\rm d}r$ 
and integrate once more with respect 
to radius to obtain,
\begin{eqnarray}
 T & \simeq & \left( \frac{\mu m_p}{k_B} \right) \frac{GM_t}{4 r} \\
 \rho & \simeq & \frac{64 \pi \sigma}{3 \kappa_R L} 
 \left( \frac{\mu m_p G M_t}{4 k_B} \right)^4 \frac{1}{r^3}.
\end{eqnarray} 
Having derived the density profile the mass of the envelope follows 
immediately,
\begin{eqnarray}
 M_{\rm env} & = & \int_{R_{\rm core}}^{R_{\rm out}} 4 \pi r^2 \rho (r) dr \nonumber \\
 & = & \frac{256 \pi^2 \sigma}{3 \kappa_R L} 
 \left( \frac{\mu m_p G M_t}{4 k_B} \right)^4
 \ln \left( \frac{R_{\rm out}}{R_{\rm core}} \right).
\end{eqnarray} 
The right-hand-side of this equation has a strong dependence on the total 
planet mass $M_t$ and a weaker dependence on the core mass $M_{\rm core}$ 
via the expression for the luminosity,
\begin{equation}
 L = \frac{G M_{\rm core} \dot{M}_{\rm core}}{R_{\rm core}} \propto M_{\rm core}^{2/3} \dot{M}_{\rm core}.
\label{eq_C6_lcore} 
\end{equation} 
In principle there are further dependencies to consider since $R_{\rm out}$ is a function 
of $M_t$ and $R_{\rm core}$ is a function of $M_{\rm core}$, but these enter only logarithmically 
and can be safely ignored. Noting that,
\begin{equation}
 M_{\rm core} = M_t - M_{\rm env},
\end{equation} 
we find that,
\begin{equation}
 M_{\rm core} = M_t - \left( \frac{C}{\kappa_R \dot{M}_{\rm core}} \right) 
 \frac{M_t^4}{M_{\rm core}^{2/3}},
\label{eq_C6_coreanalytic} 
\end{equation}
where we have shown explicitly the dependence on the envelope 
opacity and planetesimal accretion rate but have 
swept all the remaining constants 
(and near-constants) into a single constant $C$.       

\begin{figure}
\includegraphics[width=\columnwidth]{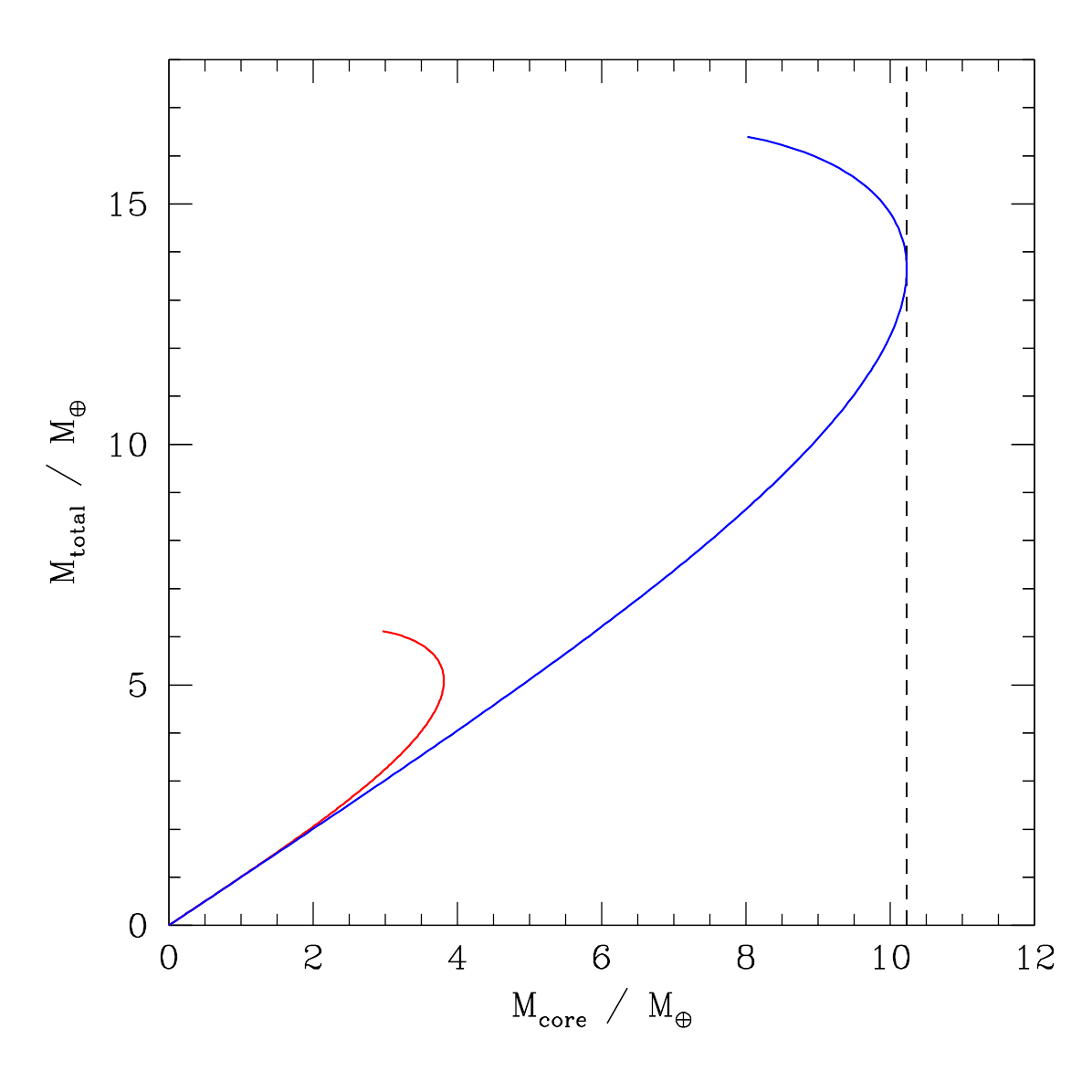}
\caption{Solutions to equation (\ref{eq_C6_coreanalytic}) for the core mass 
$M_{\rm core}$ and total mass $M_{\rm total}$. The blue curve is for a 
higher planetesimal accretion rate than for the red curve. The critical 
core mass is shown as the vertical dashed line. One should not take 
solutions to this toy model very seriously, but the numbers have been 
fixed here to correspond roughly to the values obtained from real 
calculations.}
\label{fig_coremass}
\end{figure}

Solutions to equation (\ref{eq_C6_coreanalytic}) are plotted as 
Figure~\ref{fig_coremass}. One sees that for fixed $\dot{M}_{\rm core}$, 
there exists a maximum or critical core mass $M_{\rm crit}$ beyond 
which no solution is possible. The physical interpretation of this 
result is that as the mass of the envelope increases relative to 
that of the core, ever higher pressures at the core-envelope 
boundary are needed to maintain the envelope in hydrostatic 
balance. At the critical mass, adding more gas to the envelope 
to raise the base pressure fails to help, because the added 
gas contributes more to the self-gravity tending to compress 
the envelope than to the pressure holding it up. Once the 
critical mass is exceeded hydrostatic equilibrium cannot be achieved 
in the envelope. Rather the envelope will contract, and further gas 
will fall in as fast as gravitational potential energy can be 
radiated. 

This toy model should not be taken too seriously. However, it 
does illustrate the most important result from more detailed 
calculations --- namely that the critical mass increases with larger 
$\dot{M}_{\rm core}$ and with enhanced opacity. \citet{ikoma00} 
derive an approximate fit to numerical results,
\begin{equation}
 \frac{M_{\rm crit}}{M_\oplus} \approx 12 \left( \frac{\dot{M}_{\rm core}}{10^{-6} \ M_\oplus 
 {\rm yr}^{-1}} \right)^{1/4} 
 \left( \frac{\kappa_R}{1 \ {\rm cm}^2{\rm g}^{-1}} \right)^{1/4}
\label{eq_corefit}  
\end{equation}
where the power-law indices are uncertain by around $\pm 0.05$. The 
weak dependence of the critical core mass on the planetesimal 
accretion rate means that, within a particular core accretion 
model, we can always speed up the approach to runaway gas accretion 
simply by increasing the assumed surface density of planetesimals 
in the vicinity of the growing core. Contrary to what is sometimes 
implied, there is no intrinsic difficulty in building planets quickly via  
core accretion. However, faster growth occurs {\em at the expense of} 
a larger final core mass. 

Although they appear very detailed, extant calculations of planet 
growth via core accretion should probably be regarded as illustrative 
rather than definitive. Three sources of particular uncertainty are present, 
\begin{itemize}
\item
{\bf What is the magnitude of the opacity?} Although $\kappa_R$ 
enters equation (\ref{eq_corefit}) as rather a weak power, its 
magnitude is highly uncertain. \citet{hubickyj05}, and more 
\citet{movshovitz10}, have computed core accretion models 
in which the opacity is either arbitrarily reduced or computed 
from a settling and coagulation model. These models suggest, 
first, that the appropriate value of the opacity in the 
envelope is greatly reduced (by a factor of the order of $10^2$) 
from the interstellar value \citep{podolak03}. Second, they indicate 
that the reduced opacity results in substantially faster growth 
of massive planets. Formation time scales as short as a Myr, 
or (for longer formation times) core masses as small as 5~$M_\oplus$, 
are achievable.
\item
{\bf The neglect of core migration}. Theoretical 
work, which we will discuss more fully in a subsequent Section, 
suggests that planets or planetary cores with masses exceeding 
$M_\oplus$ are highly vulnerable to radial migration as a 
consequence of gravitational torques exerted by the gas disk. 
This effect is not included in the calculations of \citet{pollack96} 
or \citet{hubickyj05}. \citet{papaloizou99} and \citet{alibert05} 
have studied the effect of steady inward migration on core formation, and 
have showed that it makes a large change to the time scale and 
outcome of the process. 
Radial migration could also be driven by dynamical interactions 
between growing cores and planetesimals \citep{levison10}.
\item
{\bf The relative role of planetesimal versus small particle accretion}. 
In the classic model of \citet{pollack96}, and its later refinements, the core 
grows exclusively by accreting planetesimals. However, as we noted in 
\S\ref{sec_pebble}, the accretion of smaller aerodynamically coupled 
solids could be equally or more important for core growth \citep{ormel10,lambrechts12,chambers14}. 
The size of primordial planetesimals is an important factor in determining 
the relative contribution of planetesimals versus pebbles. Small planetesimals---with 
radii of the order of a km---retain enough aerodynamic damping to stay dynamically 
cold, and have large gravitational focusing factors. Larger planetesimals, such as 
those produced to arise from simulations of the streaming instability \citep{simon16,schafer17}, 
are less favorable building blocks for rapid core formation, and if most of the mass 
in the planetesimal population is in $\sim 100 \ {\rm km}$ bodies pebble accretion 
may well be needed.
\end{itemize}
To summarize, the broad outlines of how core accretion works 
are well established, but even the most sophisticated models 
are probably lacking some essential physical elements.

\subsubsection{Gravitational instability model}
A sufficiently massive and / or cold gas disk is gravitationally 
unstable. Gravitational instability can lead to two possible 
outcomes: stable angular momentum transport or fragmentation. 
\citet{kratter16} provide a comprehensive review of both 
possibilities, here we focus on the idea that fragmentation 
could provide a mechanism for giant planet 
formation \cite{kuiper51,cameron78}. 

We have already derived the necessary conditions for gravitational 
instability to occur. We need the Toomre $Q$ parameter to be low 
enough, specifically,
\begin{equation}
 Q \equiv \frac{c_s \Omega}{\pi G \Sigma} < Q_{\rm crit} \simeq 1
\end{equation} 
where $c_s$ is the sound speed in a gas disk of local surface density 
$\Sigma$ and the disk mass is assumed to be small enough that the 
distinction between the orbital and epicyclic frequencies is of 
little import. If we consider a disk with $h/r = 0.05$ at 10~AU 
around a Solar mass star, then the relation $h/r = c_s / v_\phi$ 
yields a sound speed $c_s \simeq 0.5 \ {\rm km s}^{-1}$. To attain 
$Q=1$, we then require a surface density,
\begin{equation}  
 \Sigma \approx 1.5 \times 10^3 \ {\rm g \ cm}^{2}.
\end{equation} 
This is much larger than estimates based, for example, on the 
minimum mass Solar Nebula, from which we conclude that 
gravitational instability is most likely to occur at an early 
epoch when the disk mass is still high. Recalling that the 
characteristic wavelength for gravitational instability is 
$\lambda_{\rm crit} = 2 c_s^2 / (G \Sigma)$, we find that the 
mass of objects formed if such a disk fragmented would be,
\begin{equation} 
 M_p \sim \pi \Sigma \lambda_{\rm crit}^2 \sim 
 \frac{4 \pi c_s^4}{G^2 \Sigma} \sim 5 M_J
\end{equation}
where $M_J$ is the mass of Jupiter. At an order of magnitude  
level it appears that gravitational instability followed 
by fragmentation could form objects with masses in the 
giant planet range. For those objects to {\em remain} planets, 
however, requires that they accrete relatively modest amounts 
of gas from their young, gas-rich environment. Keeping fragments 
formed from gravitational collapse from growing into brown 
dwarfs or low mass stars is thus a significant challenge \citep{kratter10}.

It is also straightforward to derive where in the disk gravitational 
instability is most likely to occur. Noting that in a steady-state 
accretion disk $\nu \Sigma = \dot{M} / (3 \pi)$, we use the 
$\alpha$ prescription \cite{shakura73} and obtain,
\begin{equation} 
 Q \propto \frac{c_s^3}{\dot{M}}.
\end{equation} 
The sound speed in a protoplanetary disk decreases outward, so a 
steady-state disk becomes less stable at large radii. Indeed, 
unless the temperature becomes so low that external irradiation 
(not that from the central star) dominates the heating, a 
steady-state disk {\em will} become gravitational unstable 
provided only that it is large enough.

To derive sufficient conditions for fragmentation, we need to 
go beyond these elementary considerations and ask what happens 
to a massive disk as instability is approached. The critical 
point is that as $Q$ is reduced, {\em non-axisymmetric} 
instabilities set in which do not necessarily lead to fragmentation. 
Rather, the instabilities generate spiral arms which 
transport angular momentum and lead to dissipation and 
heating. The dissipation results in heating of the 
disk, which raises the sound speed and can lead to a stable 
self-gravitating state in which heating and cooling balance \citep{paczynski78}. 
On a longer time scale, angular momentum transport also 
leads to lower surface density and, again, enhanced stability \cite{lin90}.

Given these consideration, when will a disk fragment? \citet{gammie01} 
used both analytic arguments and local numerical simulations to identify 
the {\em cooling time} as the control parameter determining whether a 
gravitationally unstable disk will fragment. For an annulus of the 
disk we can define the equivalent of the Kelvin-Helmholtz time scale 
for a star,
\begin{equation} 
 t_{\rm cool} = \frac{U}{2 \sigma T_{\rm disk}^4}
\end{equation}
where $U$ is the thermal energy content of the disk per unit 
surface area. Then for an ideal gas equation of state with 
$\gamma = 5/3$ \citet{gammie01} found that the boundary for fragmentation is:
\begin{itemize}
\item
$t_{\rm cool} \lesssim 3 \Omega^{-1}$ --- the disk fragments.
\item
$t_{\rm cool} \gtrsim 3 \Omega^{-1}$ --- disk reaches a steady state 
in which heating due to dissipation of gravitational turbulence 
balances cooling.
\end{itemize}
This condition is intuitively reasonable. Spiral arms resulting from 
disk self-gravity compress patches of gas within the disk on a time scale 
that is to order of magnitude $\Omega^{-1}$. If cooling occurs on a 
time scale that is shorter that $\Omega^{-1}$, the heating due to 
adiabatic compression can be radiated away, and in the absence of 
extra pressure collapse is likely. The above condition was derived locally, but 
initial global simulations 
suggested that it provides a good approximation to the stability of 
protoplanetary disks more generally \cite{rice03b}. One can 
also express the fragmentation boundary in terms of a {\em maximum stress} 
that a self-gravitating disk can sustain without fragmenting 
\cite{gammie01}. Writing this in terms of an effective $\alpha$ 
parameter, $\alpha_{\rm max} \simeq 0.06$ \cite{rice05}.

In a real disk, the cooling time is determined by the opacity and 
by the mechanism of vertical energy transport: either radiative 
diffusion or convection. Using a disk model, one can then 
estimate analytically the conditions under which a disk will become unstable to 
fragmentation \citep{rafikov05,levin07,rafikov09,clarke09}. For 
standard opacities, the result is that fragmentation is expected 
only at quite large radii of the order of 50 or 100~AU. On these scales a 
large reservoir of mass is typically available locally and the 
likely outcome would be very massive planets 
or brown dwarfs \citep{stamatellos09}. At smaller 
radii the disk may still be gravitationally unstable, but the 
instability will saturate prior to fragmentation and, instead, 
contribute to angular momentum transport. 

The idea that the criterion for fragmentation can be written in terms of a threshold 
value of $t_{\rm cool} \Omega$, as discussed above, is useful but somewhat too simple.
On the 50-100~AU scales where fragmentation is most likely to occur stellar 
irradiation cannot be ignored, and modifies the criteria for fragmentation \citep{rice11}. 
Moreover, the numerical simulations used to study fragmentation exhibit quite 
subtle convergence properties \citep[as originally shown by][]{meru11}, and the 
detailed results can be sensitive both to stochastic effects \citep{paardekooper12} and 
to the treatment of cooling \citep{rice14}. Further high resolution simulations that 
include both radiative transport \citep{boley10} and MHD \citep{forgan17} are 
probably needed.

\subsubsection{Comparison with observations}
The architecture of the Solar System's giant planets provides 
qualified support for the core accretion model. The time scale 
for core accretion increases with orbital radius, which is 
qualitatively consistent with the general trend of planetary 
properties in the outer Solar System --- Jupiter is closest 
to Solar composition (reflecting a fully formed gas giant), 
while Saturn and the ice giants are relatively gas poor. 
Perhaps these outermost planets formed as the gas disk was 
in the process of being dispersed. 

The most direct test of core accretion in the Solar System involves 
measurement of the core masses of Jupiter and Saturn. Information 
as to the interior structure of the gas giants can be derived from 
exterior measurements of the gravitational multipole moments 
(for Jupiter, from the {\em Galileo} orbiter). When combined with 
interior structure models, which rely on knowledge of the equation of 
state at high pressures and densities, the measured multipole moments 
yield constraints on the core mass. Currently, the uncertainties on 
reasonable interior structure models appear to be large enough as to 
frustrate definitive conclusions. \citet{militzer08} have calculated Jovian models 
that include a substantial ($14-18 \ M_\oplus$) core, 
while \citet{nettelmann08} have computed similarly state-of-the-art 
models that support earlier suggestions \citep{guillot05} that 
any core must be small. The differences appear to result primarily 
from different assumptions made by the two groups as to the 
number of distinct layers within the interior of Jupiter. 
NASA's {\em JUNO} mission will return additional data that 
will yield new constraints on the interior structure of 
Jupiter, and it is to be hoped that this data will result in improved measurements 
of the planet's core mass.

Observations of extrasolar planets also yield constraints. Core accretion predicts that 
a greater surface density of planetesimals would lead to faster 
core growth and an increased chance of reaching runaway prior to 
disk dispersal. This is consistent with the observed correlation 
of massive planet frequency with host metallicity \cite{fischer05,ida04}. 
There are also known extrasolar planets whose small radii imply a 
large mass of heavy elements \citep[HD~149026 being one example;][]{sato05}, 
properties consistent with the outcome of core accretion.

This does not, of course, mean that disk instability does not occur. 
As we have emphasized fragmentation is expected to occur only at 
large disk radii, whereas almost all known exoplanets have been 
discovered via search techniques that are most sensitive to planets 
with small to intermediate separations. If disk instability does 
occur then we would expect a second population of 
massive planets in wide orbits \citep{boley09} with a different 
host metallicity distribution \citep{rice03c}. Among known systems 
HR~8799 \citep{marois08} is the closest to matching the expectations 
for a system formed via disk instability, but the evidence (for either 
fragmentation or core accretion) is inconclusive and to date there 
are no planets that clearly demand a disk instability origin.

\section{Evolution of Planetary Systems}
The story is not over once planets have managed to form. Theoretical 
models, which are now strongly supported by observations of the 
Solar System and of extrasolar planetary systems, suggest at least 
five mechanisms that can lead to substantial post-formation orbital 
evolution:
\begin{itemize}
\item
{\bf Interaction between planets and the gaseous protoplanetary 
disk}. This leads to orbital {\em migration} \cite{goldreich80} 
as a consequence of angular momentum exchange between the planet 
and the gas disk, and can be important for both terrestrial-mass 
planets and gas giants while the gas disk is still present. Gas 
disk migration was the first process suggested as an explanation for the  
existence of hot Jupiters \cite{lin96}.
\item
{\bf Interaction between planets and a remnant planetesimal disk}. 
Planets, especially giant planets, can also exchange angular 
momentum by interacting with and ejecting planetesimals left 
over from the planet formation process. This mechanism appears 
likely to have caused orbital migration of at least the ice 
giants, and possibly also Saturn, during the early history of the 
Solar System \cite{levison06}.
\item
{\bf Interaction within an initially unstable system of two or 
more massive planets}. There is no guarantee that the architecture 
of a newly formed planetary system will be stable over the long 
run. Instabilities can lead to planet-planet scattering, which 
usually results in the ejection of the lower mass planets, 
leaving the survivors on eccentric orbits. This could be the 
origin of the typically eccentric orbits seen in extrasolar 
planetary systems \cite{lin97,rasio96,weidenschilling96}. A 
fraction of scattered planets are typically injected on to highly eccentric 
and inclined orbits that may subsequently circularize to form hot Jupiters. 
\item
{\bf Interaction between orbiting planets and stellar binary companions}. 
A planet whose orbital plane is substantially misaligned with that of a 
binary companion has its eccentricity excited by the Kozai-Lidov mechanism \citep{kozai62,lidov62}. 
The eccentricity can become large enough that, at closest approach, 
energy is dissipated in tidal interactions with the star, forming hot 
Jupiters \citep{wu03}. 
\item
{\bf Tidal interactions between planets and their host stars}, 
which are of particular importance for hot Jupiters given that 
their orbital radii are, in some cases, just a handful of stellar 
radii.
\end{itemize}
This Section gives an elementary introduction to a selection of these 
mechanisms \citep[for a higher-level review, see][]{davies14}. 
The focus here is exclusively on {\em dynamical evolution} of 
newly formed planetary systems. 

\subsection{Gas disk migration}
The calculation of the torque experienced by a planet embedded 
within a viscous disk is highly technical, and although the basic 
principles are secure the details are anything but. Here, we 
first give a semi-quantitative treatment based on the impulse 
approximation \citep{lin79}. We then sketch some of the key 
ideas that underlie more detailed computations, which are 
based on summing the torque over a set of discrete resonances 
between the planet and the gaseous disk \citep{goldreich79}. Several 
excellent reviews are recommended for the reader seeking more 
details \citep{lubow10,kley12,baruteau14}.

\subsubsection{Torque in the impulse approximation}
\label{sec_torque_impulse}
Working in a frame of reference moving with the planet, we consider the 
gravitational interaction between the planet and gas flowing past with relative 
velocity $\Delta v$ and impact parameter $b$. The change to the perpendicular velocity that 
occurs during the encounter can be computed by summing the force along the unperturbed 
trajectory. It is,
\begin{equation}
 \vert \delta v_\perp \vert = \frac{2 G M_p}{b \Delta v},
\end{equation}
where $M_p$ is the planet mass. This velocity is directed radially, and 
hence does not correspond to any angular momentum change. The interaction 
in the two-body problem is however conservative, so the increase in the 
perpendicular velocity implies a reduction (in this frame) of the parallel 
component. Equating the kinetic energy of the gas particle well before and 
well after the interaction has taken place we have that,
\begin{equation}
 \Delta v^2 = \vert \delta v_\perp \vert^2 + (\Delta v - \delta v_\parallel )^2,
\end{equation}
which implies (for small deflection angles),
\begin{equation}
 \delta v_\parallel \simeq \frac{1}{2 \Delta v} 
 \left( \frac{2 G M_p}{b \Delta v} \right)^2.
\end{equation}
If the planet has a semi-major axis $a$ the implied angular momentum change 
per unit mass of the gas is,
\begin{equation}
 \Delta j = \frac{2 G^2 M_p^2 a}{b^2 \Delta v^3}.
\end{equation}  
It is worth pausing at this juncture to fix the {\em sign} of the angular 
momentum change experienced by the gas and by the planet firmly in our 
minds. Gas exterior to the planet's orbit orbits the star more slowly 
than the planet, and is therefore ``overtaken" by the planet. The 
decrease in the parallel component of the relative velocity of the gas 
therefore corresponds to an {\em increase} in the angular momentum of the 
gas exterior to the planet. Since the gravitational torque must be 
equal and opposite for the planet the sign is such 
that:
\begin{itemize}
\item
Interaction between the planet and gas exterior to the orbit increases 
the angular momentum of the gas, and decreases the angular momentum 
of the planet. The planet will tend to migrate inward, and the gas 
will be repelled from the planet.
\item
Interaction with gas interior to the orbit decreases the angular momentum 
of the gas and increases that of the planet. The interior gas is also 
repelled, but the planet tends to migrate outward.
\end{itemize}
In the common circumstance where there is gas both interior and exterior 
to the orbit of the planet the net torque (and sense of migration) will 
evidently depend upon which of the above effects dominates.

The total torque on the planet due to its gravitational interaction with 
the disk can be estimated by integrating the single particle torque over 
all the gas in the disk. For an annulus close to but exterior to the planet, 
the mass in the disk between $b$ and $b + db$ is,
\begin{equation}
 d m \approx 2 \pi a \Sigma db 
\end{equation}
where $\Sigma$ (assumed to be a constant) is some characteristic value 
of the gas surface density. If the gas in the annulus has angular 
velocity $\Omega$ and the planet has angular velocity $\Omega_p$ all 
of the gas within the annulus will encounter the planet in a time 
interval,
\begin{equation}
 \Delta t = \frac{2 \pi}{\vert \Omega - \Omega_p \vert}.
\label{eq_C7_dtenc} 
\end{equation}
Approximating $\vert \Omega - \Omega_p \vert$ as,
\begin{equation}
 \vert \Omega - \Omega_p \vert \simeq \frac{3 \Omega_p}{2 a} b,
\end{equation}
which is valid provided that $b \ll a$, we obtain the total torque 
on the planet due to its interaction with gas outside the orbit by 
multiplying $\Delta j$ by $dm$, dividing by $\Delta t$, and integrating 
over impact parameters. Formally we would have that,
\begin{equation}
 \frac{{\rm d}J}{{\rm d}t} = - \int_0^{\infty} 
 \frac{8 G^2 M_p^2 a \Sigma}{9 \Omega_p^2} \frac{db}{b^4},
\end{equation} 
but this integral is clearly divergent -- there is what must be 
an unphysically infinite contribution from gas passing very close 
to the planet. Sidestepping this problem for now, we replace the 
lower limit with a minimum impact parameter $b_{\rm min}$ and 
integrate. The result is,
\begin{equation}
 \frac{{\rm d}J}{{\rm d}t} = - \frac{8}{27} 
 \frac{G^2 M_p^2 a \Sigma}{\Omega_p^2 b_{\rm min}^3}.
\label{eq_C7_impulse} 
\end{equation}   
It is possible to tidy up this calculation somewhat, for example 
by taking proper account of the rotation of the planet frame around 
the star, and if this is done the result is that the expression 
derived above must be multiplied by a correction factor 
\citep{papaloizou06}.

Aside from the sign of the torque the most important result that 
we can glean from this calculation is that the torque on the planet 
due to its interaction with the gas scales as the {\em square} of the 
planet mass. This scaling can be compared to the orbital angular momentum of the planet, which is of course  
linear in the planet mass. We conclude that if all other factors 
are equal -- in particular if neither $\Sigma$ in the vicinity of the 
planet nor $b_{\rm min}$ vary with $M_p$ -- the time scale for the 
planet to change its orbital angular momentum significantly will 
scale inversely with the planet mass. The migration velocity in this limit will 
then be proportional to $M_p$ -- more massive planets will migrate faster.

Finally we can estimate the magnitude of the 
torque for parameters appropriate to different stages of giant 
planet formation. First, let us consider a rather low mass core 
($M_p = M_\oplus$) growing at 5~AU in a gas disk of surface density 
$\Sigma = 10^2 \ {\rm g \ cm}^{-2}$ around a Solar mass star. Our 
treatment of the interaction has assumed that the disk can be treated 
as a two-dimensional sheet, so it is arguably natural to take as a 
minimum impact parameter $b_{\rm min} = h \approx 0.05 a$. Using these 
numbers we find that the exterior torque would drive inward migration 
on a time scale,
\begin{equation}
 \tau = \frac{J}{\vert {\rm d}J / {\rm d}t \vert} \sim 1 \ {\rm Myr}.
\end{equation}
Of course this will be partly offset by the interior torque -- which 
has the opposite sign -- but absent some physical reason why these 
two torques should cancel to high precision we would predict changes 
to the semi-major axis on a time scale of the order of a Myr. This is 
already rapid enough to be a potentially important effect during 
giant planet formation via core accretion. Second, we can evaluate the 
torque for a fully-formed gas giant. A Jupiter mass planet has a 
Hill sphere $r_H > h$, so it seems reasonable to argue that the 
value of $b_{\rm min}$ that we adopted for an Earth mass core is too 
small in this case. Picking a modestly larger value $b_{\rm min} = 0.2 a$ 
results in a time scale,
\begin{equation}
 \tau \sim 2 \times 10^5 \ {\rm yr},
\end{equation} 
that is an order of magnitude shorter than typical protoplanetary disk 
lifetimes. If these estimates can be trusted to even an order of magnitude 
the conclusion is inescapable -- gas disk migration ought to be an 
important process for the early evolution of the orbits of giant planets.

\subsubsection{Torque at resonances}
\begin{figure}
\includegraphics[width=\columnwidth]{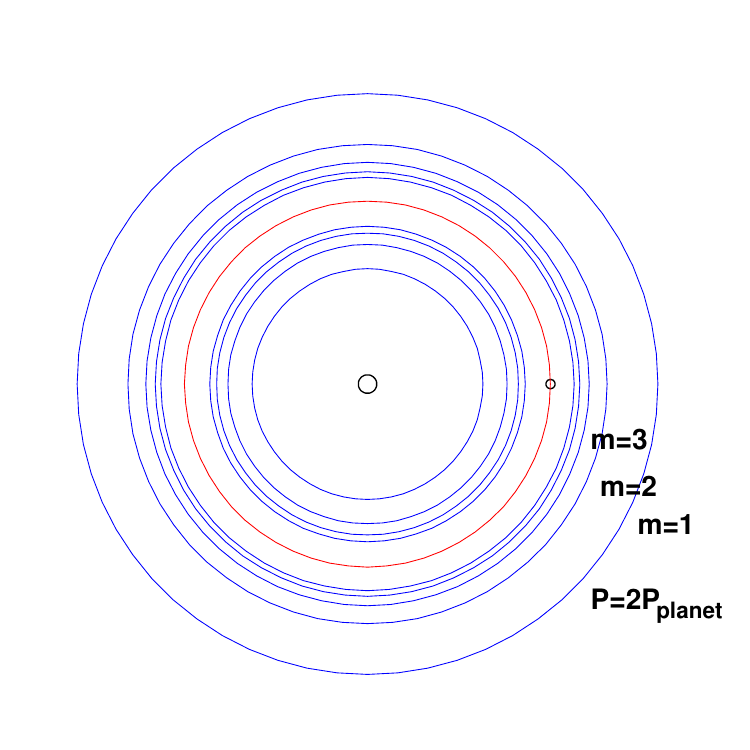}
\caption{Nominal locations of the corotation (red) and 
Lindblad resonances (blue) for a planet on a circular 
orbit. Only the low order Lindblad resonances are depicted --- 
there are many more closer to the planet.}
\label{fig_lindblad}
\end{figure}

A more involved -- but ultimately more powerful -- approach to calculate 
the torque is to decompose it into a sum over partial torques exerted 
at resonant locations with the disk \citep{goldreich79,tanaka02}. For 
simplicity, we start by considering a planet orbiting a star on a {\em circular} orbit with 
angular frequency $\Omega_p$. A standard exercise in dynamics 
(e.g. Binney \& Tremaine 1987) yields the conditions for 
resonances. A {\em corotation resonance} exists for radii in the 
disk where the angular frequency $\Omega$,
\begin{equation}
 \Omega = \Omega_p.
\label{eq_corotation} 
\end{equation} 
{\em Lindblad resonances} exist when,
\begin{equation}
 m ( \Omega - \Omega_p ) = \pm \kappa_0
\label{eq_lindblad}
\end{equation}
where $m$ is an integer and $\kappa_0$, the {\em epicyclic 
frequency}, is defined as,
\begin{equation}
 \kappa_0 \equiv \left( \frac{{\rm d}^2 \Phi_0}{{\rm d}r^2} + 
 3 \Omega^2 \right)
\end{equation}
with $\Phi_0$ the stellar gravitational potential. For a 
Keplerian potential $\kappa_0 = \Omega$. If we approximate the 
angular velocity of gas in the disk by the Keplerian angular 
velocity, the Lindlad resonances are located at,
\begin{equation}
 r_L = \left( 1 \pm \frac{1}{m} \right)^{2/3} r_p
\end{equation}
where $r_p$ is the planet orbital radius. The locations of 
some of the low order (small $m$) resonances are shown 
in Figure~\ref{fig_lindblad}. For an orbiting test particle, 
the resonances are locations where the planet can be a strong 
perturbation to the motion. For a gas disk, angular momentum 
exchange between the planet and the gas disk occurs at 
resonant locations. As we noted for the impulse approximation, the sense of the 
interaction is that the planet {\bf gains angular momentum} from interacting with the 
gas disk at the interior Lindblad resonances ($r_L < r_p$). This 
tends to drive the planet outward. The gas loses angular momentum, 
and moves inward. Conversely, the planet {\bf loses angular momentum} from interacting with the 
gas disk at exterior Lindblad resonances ($r_L > r_p$). This 
tends to drive the planet toward the star. The gas gains angular momentum, 
and moves outward. Notice that the gravitational interaction of a planet with a gas disk 
tends --- somewhat counter-intuitively --- to {\em repel} gas from 
the vicinity of the planet's orbit.

The flux of angular momentum exchanged at each Lindblad resonance 
can be written as,
\begin{equation}
 T_{LR}(m) \propto \Sigma M_p^2 f_c (\xi)
\label{eq_torque_lr} 
\end{equation}
where $\Sigma$ is the gas density and $M_p$ the planet mass. That the 
torque should scale with the {\em square} of the planet mass is 
intuitively reasonable --- the perturbation to the disk surface 
density scales as the planet mass in the linear regime so the 
torque scales as $M_p^2$. The factor $f_c(\xi)$ is the 
{\em torque cutoff function} \cite{artymowicz93}, which encodes 
the fact that resonances very close to the planet contribute 
little to the net torque. The torque cutoff function peaks at,
\begin{equation} 
 \xi \equiv m \left( \frac{c_s}{r \Omega} \right)_p \simeq 1
\end{equation}
i.e. at a radial location $r \simeq r_p \pm h$, where $h$ is the 
disk scale height (this result immediately implies that a 
three-dimensional treatment is necessary for the dominant 
resonances if the planet is completely embedded within a 
gas disk, as is the case for low mass planets). The strength 
of the torque exerted at each resonance is essentially 
independent of properties of the disk such as the disk 
viscosity, though of course the viscosity still matters  
inasmuch as it controls the value of the unperturbed disk surface 
density $\Sigma$.

\begin{figure}
\includegraphics[width=0.75\columnwidth,angle=90]{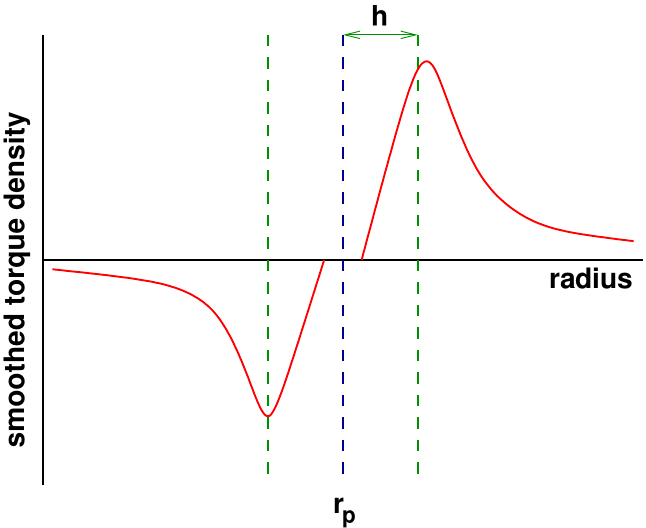}
\caption{Schematic illustration of the smoothed torque density 
due to angular momentum exchange between a planet and a gas disk 
at the location of Lindblad resonances, after \citet{ward97}. 
The peak torque occurs at $r \approx r_p \pm h$. The disk gains 
angular momentum from the planet as a result of the interaction 
for $r > r_p$, and loses angular momentum for $r < r_p$. The 
interaction is almost invariably asymmetric, such that when 
integrated over the entire disk the planet loses angular momentum 
and migrates inward.}
\label{fig_torquedensity}
\end{figure}

Figure~\ref{fig_torquedensity} illustrates the differential torque 
exerted on the disk by the planet, after smoothing over the Lindblad 
resonances \cite{ward97}. The flux of angular momentum is initially 
deposited in the disk as waves, which propagate radially before 
dissipating. The details of this dissipation matter little for the 
net rate of angular momentum exchange.

Angular momentum transfer at corotation requires 
additional consideration, and as we will see later thinking of 
these torques in terms of resonances is not as profitable as 
for the Lindblad torques. Formally though, 
in a two-dimensional disk the rate 
of angular momentum deposition at corotation is proportional 
to \cite{goldreich79,tanaka02},
\begin{equation} 
 T_{CR} \propto \frac{{\rm d}}{{\rm d}r} \left( \frac{\Sigma}{B} \right) 
\end{equation}
where $B$ is the Oort parameter,
\begin{equation} 
 B(r) = \Omega + \frac{r}{2} \frac{{\rm d}\Omega}{{\rm d}r}.
\end{equation}  
This implies that in a two-dimensional disk, the resonant corotation torque 
vanishes identically in the moderately interesting case of a 
disk with a surface density profile $\Sigma \propto r^{-3/2}$. 
This result does {\em not apply} to three-dimensional disks \cite{tanaka02}.

\subsubsection{Type I migration}
For low mass planets (generically $M_p \sim M_\oplus$, though the exact mass 
depends upon the disk properties) the angular momentum flux injected into the 
disk as a consequence of the planet-disk interaction is negligible when 
compared to the viscous transport of angular momentum. As a result, the 
gas surface density profile $\Sigma(r)$ remains approximately unperturbed, 
gas is present at the location of each of the resonances, and the net 
torque on the planet is obtained by summing up the torque exerted at 
each resonance. Schematically,
\begin{equation}
 T_{\rm planet} = \sum_{ILR} T_{LR} + \sum_{OLR} T_{LR} + T_{CR}\
\label{eq_resonance_sum} 
\end{equation}
where the planet gains angular momentum from the inner Lindblad 
resonances (ILR) and loses angular momentum to the outer Lindblad 
resonances (OLR). There is also a potentially important co-orbital 
torque $T_{CR}$. Changes to the planet's orbit as a result of 
this net torque are called {\bf Type~I migration} \cite{ward97}. 

As noted above (equation \ref{eq_torque_lr}) the torque exerted at 
each resonance scales as the planet mass squared. If the azimuthally 
averaged surface density profile of the gas disk remains unperturbed, 
it follows that the total torque will also scale as $M_p^2$ and the 
migration time scale,
\begin{equation} 
 \tau_I \propto \frac{M_p}{T_{\rm planet}} \propto M_p^{-1}.
\end{equation}
Type~I migration is therefore most rapid for the largest body for 
which the assumption that the gas disk remains unaffected by the 
planet remains valid. 

Actually evaluating the sum sketched out in equation (\ref{eq_resonance_sum}) 
is not easy, and there is no simple physical argument that I am aware of 
that gives even the {\em sign} of the net torque on the planet. However 
invariably it is found that the Lindblad resonances exterior to the planet 
are more powerful than those interior, so that the net torque due to 
Lindblad resonances leads to {\em inward} migration. Note that one might think (for 
example by looking at the surface density dependence of the torque 
in equation \ref{eq_torque_lr}) that the sense of migration ought to 
depend upon the surface density gradient --- i.e. that a steep 
surface density profile should strengthen the inner resonances relative 
to the outer ones and hence drive outward migration. This {\em is not 
true}. Pressure gradients (which depend upon the radial 
dependence of the surface density and temperature) alter the angular 
velocity in the disk from its Keplerian value (equation \ref{eq_dv}), 
and thereby shift the radial location of resonances from their 
nominal positions. A steep surface density profile implies a large 
pressure gradient, so that all the resonances move slightly inward. 
This weakens the inner Lindblad resonance relative to the outer ones, 
in such a way that the final dependence on the surface density profile 
is surprisingly weak \cite{ward97}.

\citet{tanaka02} compute the net torque on a planet in a three-dimensional 
but isothermal gas disk. For a disk in which,
\begin{equation}
 \Sigma(r) \propto r^{-\gamma}
\end{equation}
they obtain a net torque {\em due to Lindblad resonances only} of,
\begin{equation}
 T = -(2.34 -0.10 \gamma) 
 \left( \frac{M_p}{M_*} \frac{r_p \Omega_p}{c_s} \right)^2 
 \Sigma_p r_p^4 \Omega_p^2.
\label{eq_lindblad_torque} 
\end{equation}
This torque would result in migration on a time scale,
\begin{eqnarray}
 \tau & \equiv & -\frac{J}{\dot{J}} \nonumber \\
             & = & (2.34 - 0.1 \gamma)^{-1} 
	\frac{M_*}{M_p}
	\frac{M_*}{\Sigma_p r_p^2}
	\left( \frac{c_s}{r_p \Omega_p} \right)^2 
	\Omega_p^{-1},
\end{eqnarray}
where $\Sigma_p$, $c_s$ and $\Omega_p$ are respectively the gas 
surface density, gas sound speed, and angular velocity at the 
location of a planet orbiting at distance $r_p$ from a star of 
mass $M_*$. As expected based on the simple considerations discussed 
previously, the migration rate ($\propto \tau_I^{-1}$) scales linearly 
with both the planet mass and the local disk mass. The time scale 
becomes shorter for cooler, thinner disks --- provided that the 
interaction remains in the Type~I regime --- since for such disks
more resonances close to the planet contribute to the net torque.   

The most important thing to notice from this formula is that the 
predicted migration time scale is {\em very short}. If we consider a 
$5 \ M_\oplus$ core growing at 5~AU in a disk with typical parameters 
($\Sigma = 10^2 \ {\rm g \ cm}^{-2}$, $h/r=0.05$, $\alpha=1$) we find,
\begin{equation}
 \tau_{I,LR} \simeq 0.5 \ {\rm Myr}.
\end{equation}
One concludes that there is a strong argument that Type~I migration 
ought to play an important role in the formation of giant planet cores. 

\begin{figure}
\includegraphics[width=\columnwidth]{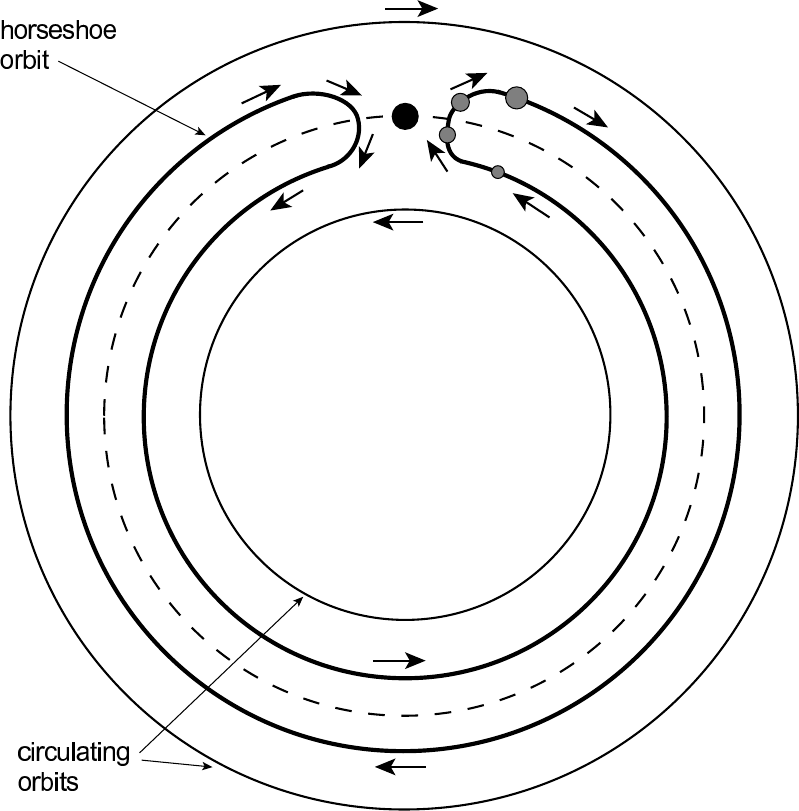}
\includegraphics[width=\columnwidth]{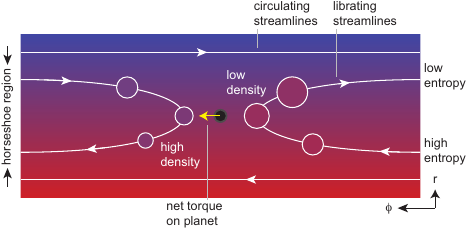}
\caption{A low mass planet embedded in a gas disk---shown in the upper panel in 
a global view and below in a local view---generically feels a torque from 
nearly co-orbital gas on close horseshoe orbits. In the example shown here we 
consider a disk with a negative entropy gradient and a cooling time scale that 
is comparable to the time scale for parcels of gas to execute horseshoe orbits. 
Changes in density as the gas executes the horseshoe turns results in an 
azimuthal asymmetry in the disk near the planet, creating a torque.}
\label{fig_horseshoe}
\end{figure}

The Lindblad torque is only part of the total torque exerted on the planet. What 
about the co-orbital torque? The physics behind the origin of torques from gas 
that is almost co-orbiting with the planet is illustrated in Figure~\ref{fig_horseshoe}. 
The key point is that gas that is almost, but not quite, co-orbital executes 
 {\em horseshoe orbits} when viewed in a frame co-rotating with the motion 
 of the planet. As the gas executes the U-shaped turns at the ends of the horseshoe, 
changes in the gas density occur that {\em are not symmetric} between the disk 
trailing and leading the planet. The density variations source a torque that, 
because it arises from quite close to the planet, can be as large or larger than 
the torque from the Lindblad resonances.

\citet{ward91} discussed how to calculate corotation torques from the physical 
perspective of gas on horseshoe orbits, but this way of thinking about the torque 
did not become widely known until simulations 
by \citet{paardekooper06} uncovered a dependence of the Type~I migration 
rate on the thermal properties of the disk. Subsequently, many authors 
have studied the co-orbital Type~I torque in both isothermal \citep{casoli09,paardekooper09a} 
and non-isothermal (radiative or adiabatic) disks \citep{kley08,masset09,kley09,paardekooper09b,paardekooper11}.  
The review by \citet{baruteau14} summarizes the results of these and other calculations.

The total torque on planets embedded in gaseous disks, and the corresponding Type~I migration 
rate, is given by the sum of the Lindblad and corotation torques. In addition to the planet 
mass, several properties of the disk enter into the result:
\begin{itemize}
\item
The surface density and temperature profiles. We have already noted the dependence of the 
Lindblad torque (in isothermal disks) on the surface density profile (equation~\ref{eq_lindblad_torque}), 
and the importance of the radial entropy profile for the horseshoe drag (Figure~\ref{fig_horseshoe}).
\item
The cooling time of parcels of gas in the horseshoe region (due to radiative diffusion in the 
optically thick limit or explicit heating / cooling processes otherwise). This matter because 
part of the corotation torque depends upon the presence of thermal gradients as gas moves 
around horseshoe orbits. The strength of those gradients depends upon how quickly the 
gas heats and cools relative to the time scale needed to execute a horseshoe orbit.
\item
The efficiency of angular momentum transport in the co-orbital region. The region of the 
disk that admits horseshoe orbits is closed and relatively small. It cannot absorb or give 
an arbitrary amount of angular momentum to a planet, unless it is ``connected" to the 
rest of the disk via viscous stresses. Large and persistent co-orbital torques 
are possible provided that the disk is viscous enough that the torque remains 
unsaturated; very low viscosity leads to saturation and vanishing torques.
\end{itemize}
With all these factors to consider the calculation of the Type~I migration rate is 
necessarily rather involved. One should remember, however, that both the Lindblad 
and corotation torques are large and would individually drive rapid migration. They 
may happen to sum to zero at special locations within the disk, but absent such 
happy coincidences migration of planets of roughly an Earth mass and higher 
will be significant. Type~I migration can rarely be ignored in any circumstance 
where substantial gas disks co-exist with planets.

\citet{bitsch13} provide an example of how the Type~I migration rate as a function 
of planet mass can be calculated given a specific disk model. They find that for 
sufficiently low mass planets ($M_p \lesssim 7 \ M_\oplus$ in their fiducial 
example) the Lindblad torque dominates, and migration is inward at all radii. 
For more massive planets there are radial zones of outward migration interspersed 
with radii where migration is inward. The boundaries where (as we move out) 
migration switches from outward to inward define convergence zones where 
planets might tend to accumulate.

\subsubsection{Type II migration}
For sufficiently large planet masses, the angular momentum flux from the 
planet locally dominates the viscous flux. As a consequence, gas is 
repelled from high-m resonances. The surface density drops near 
$r = r_p$, forming a {\em gap} --- an annular region in which the 
surface density is smaller than its unperturbed value.

Two conditions are necessary for gap formation. First, the Hill sphere 
(or Roche radius) of the planet needs to be comparable to the thickness 
of the gas disk,
\begin{equation} 
 r_H \equiv \left( \frac{M_p}{3 M_*} \right)^{1/3} r \gtrsim h
\end{equation} 
which requires a mass ratio $q \equiv M_p / M_*$,
\begin{equation}
 q \gtrsim 3 \left( \frac{h}{r} \right)_p^3.
\end{equation} 
This condition is satisfied for typical protoplanetary disk 
parameters for $q \sim 4 \times 10^{-4}$ --- i.e. for planet 
masses somewhere between that of Saturn and Jupiter.

\begin{figure}
\includegraphics[width=\columnwidth]{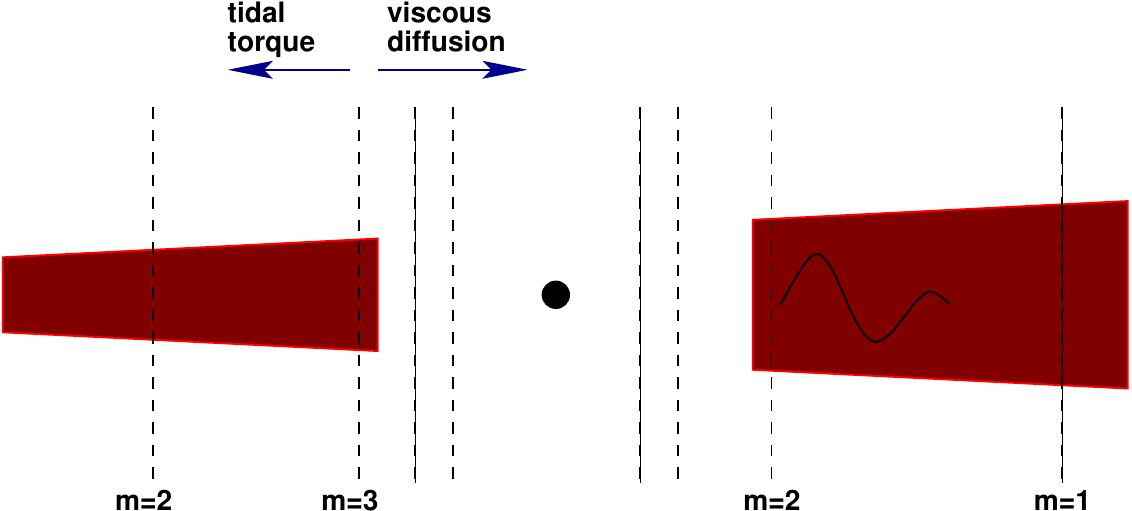}
\caption{Illustration of the viscous condition for gap opening. A gap can 
open when the time scale for opening a gap of width $\Delta r$ due to tidal 
torques becomes shorter than the time scale on which viscous diffusion can 
refill the gap.}
\label{fig_balance}
\end{figure}

\begin{figure}
\includegraphics[width=\columnwidth]{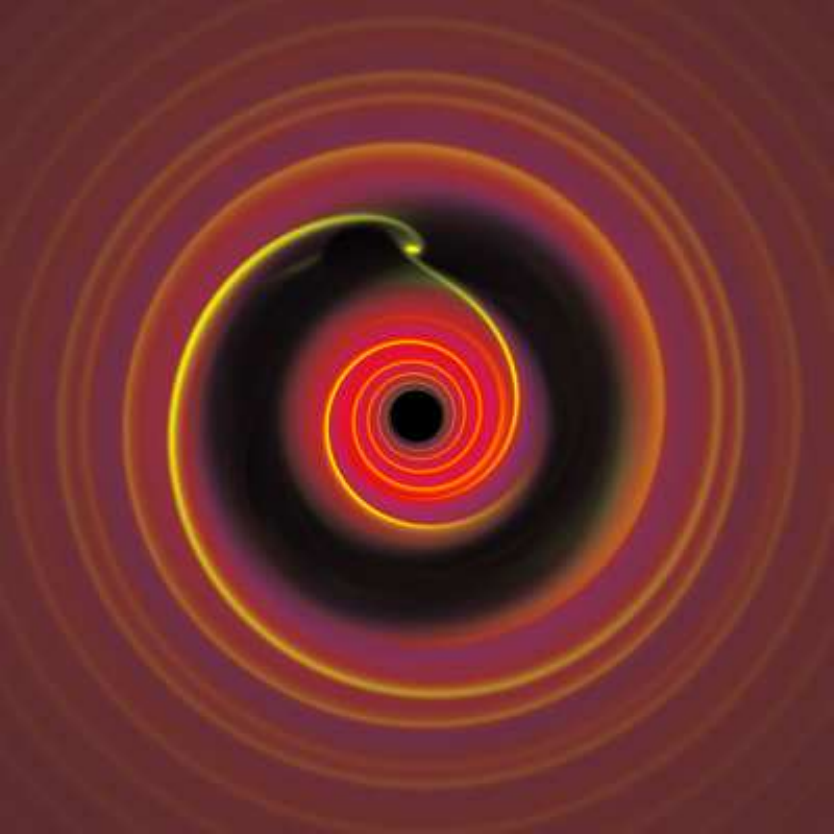}
\caption{Simulation of the planet-disk interaction in the Type~II regime 
in which the planet is sufficiently massive to open a gap in the gas 
disk. Note the presence of streams of gas that penetrate the gap region. A 
movie showing the interaction as a function of mass is available 
at { \footnotesize \tt http://jilawww.colorado.edu/$\sim$pja/planet{\underline{ }}migration.html. }}
\label{fig_type2}
\end{figure}

A second condition for gap opening arises due to the viscous considerations 
depicted in Figure~\ref{fig_balance}. To open a gap, we require that the 
tidal torques must be able to remove gas from the gap region faster than 
viscosity can fill the gap back in \cite{lin80,goldreich80,papaloizou84}. There are various ways 
to estimate the critical mass above which this condition is satisfied. 
Following \citet{takeuchi96}, we note that the time scale for viscous 
diffusion to close a gap of width $\Delta r$ is just,
\begin{equation}
 t_{\rm close} \sim \frac{(\Delta r)^2}{\nu}
\end{equation}
where $\nu = \alpha c_s h$ is the disk viscosity. The time scale to open 
a gap as a result of the tidal torque at an $m$-th order Lindblad 
resonance is,
\begin{equation}
 t_{\rm open} \sim \frac{1}{m^2 q^2 \Omega_p} \left( \frac{\Delta r}{r_p} \right)^2.
\end{equation} 
Setting $t_{\rm open} = t_{\rm close}$, and taking $m = r_p \Omega_p / c_s$ (since, 
as noted above, this value of $m$ is where the torque cutoff function peaks), 
we obtain,
\begin{equation}
 q \gtrsim \left( \frac{c_s}{r_p \Omega_p} \right)^2 \alpha^{1/2}.
\end{equation}
For typical disk parameters ($h/r=0.05$, $\alpha = 10^{-2}$), the viscous 
condition for gap opening is satisfied for $q$ modestly larger than $10^{-4}$. 
Combined with the thermal condition outlined above, we conclude that 
Jupiter mass planets ought to be massive enough to open a gap within 
the disk, whereas Saturn mass planets are close to the critical mass 
required for gap opening. Figure~\ref{fig_type2} from \citet{armitage05}, 
shows results from a two-dimensional simulation of the planet-disk 
interaction in the Type~II regime. Both the gap, and the presence of a 
prominent spiral wave excited within the gas disk, are obvious.

\subsubsection{The Type~II migration rate}
Once a planet becomes massive enough to open a gap, the orbital 
evolution becomes coupled to the viscous evolution of the gas within the disk. 
At small orbital radii the sense of gas motion will invariably be inward, and 
the planet will very probably follow suit \citep[by similar arguments, the planet 
can migrate outward in regions where the 
gas disk is expanding,][]{veras04}. 
The radial velocity of gas in the disk is,
\begin{equation}
 v_r = - \frac{\dot{M}}{2 \pi r \Sigma}, 
\end{equation}
which for a steady disk away from the boundaries can be written  
as,
\begin{equation}
 v_r = - \frac{3}{2} \frac{\nu}{r}.
\end{equation}
If the planet enforces a rigid tidal barrier at the outer edge 
of the gap (i.e. no gas is accreted by the planet, and no gas crosses the gap), 
then evolution of the disk will force the orbit 
to shrink at a rate $\dot{r}_p \simeq v_r$, provided that the 
local disk mass exceeds the planet mass, i.e. that $\pi r_p^2 \Sigma \gtrsim M_p$. 
This implies a nominal Type~II migration time scale, valid for 
{\em disk dominated migration} only,
\begin{equation}
 \tau_0 = \frac{2}{3 \alpha} \left( \frac{h}{r} \right)_p^{-2} 
 \Omega_p^{-1}.
\end{equation} 
For $h/r = 0.05$ and $\alpha = 10^{-2}$, the migration time scale 
at 5~AU is of the order of 0.5~Myr. 

In practice, the assumption that the local disk mass exceeds 
that of the planet often fails. For example, a plausible model 
of the protoplanetary disk with a mass of 
$0.01 \ M_\odot$ within 30~AU has a surface density profile,
\begin{equation}
 \Sigma = 470 \left( \frac{r}{{1 \ {\rm AU}} } \right)^{-1} \ 
 {\rm g \ cm}^{-2}.
\end{equation}
The condition that $\pi r_p^2 \Sigma = M_p$ gives an estimate of the 
radius within which disk domination ceases of,
\begin{equation}
 r = 6 \left( \frac{M_p}{M_J} \right) \ {\rm AU}.
\end{equation}
Interior to this radius, the planet acts as a slowly moving 
barrier which impedes the inflow of disk gas. If the 
barrier was assumed (unrealistically, as we will see) to 
be impermeable then gas would pile up outside it. 
This pile-up would increase the torque but would not 
continue without limit because the interaction also deposits 
angular momentum into the disk, causing it to expand 
\cite{pringle91}. The end result would be slower migration 
compared to the nominal rate quoted above. \citet{syer95} and 
\citet{ivanov99} provide explicit formulae for calculating the 
extent of the suppression given the assumption of an 
impermeable tidal barrier.

These simple estimates of the Type~II migration rate follow 
from the assumption that once a gap has been opened, the planet maintains an 
impermeable tidal barrier to gas inflow. In fact, simulations 
show that planets are able to accrete gas via tidal streams 
that bridge the gap \cite{lubow99}, and this breaks the link 
between the viscous evolution of the disk and migration.
\citet{duffell14}, \citet{durmann15} and \citet{durmann17} use 
numerical simulations to derive improved estimates of the 
Type~II migration rate. The Type~II migration rate 
can also be qualitatively altered---and even reversed---if 
two planets approach each other in the disk such that their 
gaps start to overlap or such that resonant interactions between 
the planets become important \citep{masset01}. Two planet migration may 
have played a role in the migration history of Jupiter and 
Saturn in the Solar System \citep{morbidelli07}, and is central 
to the ``Grand Tack" model in which an early intrusion of Jupiter 
into the inner Solar System reduces the inventory of solids in the 
region of Mars and the current asteroid belt \citep{walsh11}. 

\subsubsection{Stochastic migration}
In a turbulent disk migration (especially of low mass bodies) will 
not be perfectly smooth. 
Turbulence is accompanied by a spatially and temporally  
varying pattern of density fluctuations, which 
exert {\em random} torques on planets of any 
mass embedded within the disk. \citep[Analogously, transient 
spiral features in the Galactic disk increase the velocity 
dispersion of stellar populations;][]{carlberg85}. 
If we assume that the random torques 
are uncorrelated with the presence of a planet, then the random 
torques' linear scaling with planet mass will dominate over 
the usual Type~I torque (scaling as $M_p^2$) for sufficiently 
low masses. The turbulence will then act to increase the 
velocity dispersion of collisionless bodies, or, in the presence 
of damping, to drive a random walk in the semi-major axis 
of low mass planets.

To go beyond such generalities, and in particular to estimate the 
crossover mass between stochastic and Type~I migration, we need to 
specify the source of turbulence in the protoplanetary disk. MHD 
disk turbulence driven by the magnetorotational instability 
has been used as a model system for 
studying stochastic migration by several authors 
\cite{nelson04,nelson05,laughlin04,yang09}. Density fluctuations in MHD disk turbulence 
have a typical coherence time of approximately half an orbital period, 
and as a consequence are able to exchange angular momentum 
with an embedded planet across a range of disk radii (not 
only at narrow resonances). The study by \citet{nelson04} was based 
on both global ideal MHD disk simulations, with an aspect ratio of 
$h/r = 0.07$, and local shearing box calculations. For all masses 
considered in the range $3 \ M_\oplus \le M_p \le 30 \ M_\oplus$, 
the {\em instantaneous} torque on the planet from the MHD 
turbulent disk exhibited large fluctuations in both magnitude 
and sign. Averaging over $\approx 20$ orbital periods, the 
mean torque showed signs of converging to the Type~I rate, 
although the rate of convergence was slow, especially for 
the lowest mass planets in the global runs. Very roughly, 
the \citet{nelson04} simulations 
suggest that up to $M_p \sim 10 \ M_\oplus$ the random walk component 
dominates steady Type~I drift over time scales that substantially 
exceed the orbital period. 

How important stochastic (or diffusive) migration is for planet 
formation depends, first and foremost, on the strength and nature 
of the disk turbulence. Many existing studies are based on the 
properties of turbulence simulated under ideal MHD conditions, 
which as we noted earlier do not apply to protoplanetary disks. 
Nonetheless, turbulence even in more realistic disk model 
may be strong enough to pump the mean eccentricity (and perhaps inclination) of 
planetesimals, reducing the magnitude of gravitational focusing 
and leading to a greater likelihood of disruptive 
collisions \citep{ida08,ormel13}. A second situation in which 
fluctuating torques may play a critical role is in the breaking 
of mean-motion resonances between planets undergoing 
differential migration \citep{batygin17}.

\subsubsection{Eccentricity evolution during migration}
Most massive extrasolar planets are observed to be on significantly 
eccentric orbits. Since orbital migration due to planet-disk interactions 
is likely to have occurred in these systems, it is of interest to ask 
whether the same process---gravitational interactions between the 
gas disk and an orbiting planet in the Type~II regime---also leads to 
excitation of eccentricity. It seems likely that small eccentricities can 
be excited (or, at least, sustained) during migration, but that this is 
not the dominant process for most extrasolar planets.

The considerations relevant to this problem were set out in 
\citet{goldreich80}. As with the Type~I torque, the basic idea 
is to sum the contributions to $\dot{e}$ over resonances. The 
number of potentially important resonances is, however, much 
larger for an eccentric planet, and hence the calculation is 
harder. Eccentricity growth (or decay) depends upon 
the relative strength of:
\begin{itemize}
\item
External Lindblad resonances, which act to excite eccentricity.
\item
Non-co-orbital corotation resonances, which act to damp 
eccentricity. As noted above, the only corotation resonance 
that exists for a planet on a circular orbit is co-orbital, 
so a finite eccentricity is necessary for these resonances 
to be present.
\end{itemize}
Unfortunately, the effects leading to damping and excitation of 
eccentricity are finely balanced, making robust analytic assessment of the 
sign of the eccentricity evolution difficult. The simplest 
estimates favor damping, but only modest saturation of the corotation 
resonances would be needed to tilt the balance in favor of 
excitation \cite{ogilvie03,goldreich03,masset04}. Numerically, 
there is general agreement that substellar objects of brown 
dwarf mass and above suffer substantial eccentricity growth when embedded 
within a gas disk \cite{papaloizou01,artymowicz91,dunhill13}.
For Jovian mass planets the situation is more subtle, and the results of \citet{dangelo06} and 
\citet{duffell15} suggest that these planets are subject to a finite amplitude 
instability that can boost initially non-zero eccentricities up to 
$e \sim (h/r)$. Effects associated with the entropy gradient 
in the disk near gap edges may modify these results \citep{tsang14}. 

\subsubsection{Transition disks}
Gas giants must form while the protoplanetary disk remains gas-rich, 
and it is of considerable interest to try and identify candidate disk 
that may harbor newly formed (and presumably migrating) planets.
There are a number of T Tauri stars whose spectral energy 
distributions (SEDs) and sub-mm images exhibit characteristics 
broadly consistent with theoretical predictions for embedded planets. The SEDs 
show robust excesses in the mid-IR (indicative 
of gas and dust disks at AU scales) {\em without} matching excesses 
in the near-IR \cite{sicilia06}. Well-known examples of such {\em transition} disk 
sources include GM~Aur \cite{calvet05} and 
TW~Hya \cite{eisner06}, but many more such disks have now been 
identified via {\em Spitzer} observations \citep{muzerolle10}. 
By one common definition, 
these sources lack optically thick inner disks, from which one deduces 
that small grains are absent close to the star, though disks are 
unquestionably present at larger radii. \citet{espaillat14} review the 
observational properties of these sources.

What is going on in transition disks? Some may be stars 
caught in the act of dispersing their disks---perhaps as a result 
of the photoevaporative mechanism discussed earlier in these notes. 
Others, however, may be ``normal" Classical T Tauri stars 
around which an orbiting planet has created a tidal barrier to the inflow 
of gas and dust, thereby creating an inner hole. Theoretical models 
that invoke the presence of planets have the right basic morphology 
to explain the observed SEDs \cite{rice03d,quillen04}, though when the 
problem is examined in more detail additional processes, such as 
dust ``filtering" at the pressure maximum at the gap edge \citep{rice06}, 
and dust growth interior to the planet's orbit \citep{zhu12}, are needed 
to match observations. Planets can also (depending upon the disk 
model) induce strongly non-axisymmetric dust distributions similar to those 
seen in some recent {\em ALMA} observations \citep{vdM13}. Despite 
these encouraging signs, however, there is little or no direct evidence 
for the existence of planets in transition disks, some of which exhibit 
very large cavities that would need surprisingly massive planets at 
surprisingly large radius to explain. It is quite possible that the observed 
transition sources arise from a combination of physical mechanisms, 
including planets, photoevaporation, and possibly other processes 
\citep{alexander09}.
 
\subsection{Planetesimal disk migration}
It is unlikely that the formation of gas and ice giant planets consumes 
the entire inventory of planetesimals in their vicinity. The interaction 
of remnant planetesimals with planets, after the dispersal of the gas 
disk, can result in orbital migration of the planets. 

Here, we follow the simple discussion of \citet{malhotra95}\footnote{The 
treatment here is deliberately over-simplified. The 
reader interested in exploring more realistic analytic and numerical models is advised 
to consult \citet{ida00} and \citet{kirsh09}, and references therein.}. If we 
consider a single planetesimal of mass $\delta m$ interacting with a 
planet of mass $M_p$ at orbital radius $a$ there are two possible 
outcomes,
\begin{itemize}
\item
The planetesimal may be scattered outward---possibly sufficiently 
to be ejected---in which case the planet moves in by angular 
momentum conservation. Up to numerical factors,
\begin{equation}
 \frac{\delta a}{a} \simeq - \frac{\delta m}{M_p}.
\end{equation} 
\item
The scattering is inward, in which case ${\delta a} / {a} \simeq + {\delta m} / {M_p}$
\end{itemize}
Evidently for significant migration to occur we require that the total mass in 
planetesimals be comparable to the planet mass,
\begin{equation}
 \sum \delta m \sim M_p.
\end{equation}
This is a similar result to that obtained in the case of gas disk migration, 
though for planetesimals the restriction is more severe since while a low 
mass gas disk can still drive migration---albeit at a slower pace---ejected 
planetesimals are permanently removed from the system and cannot influence 
the planet further. We also note that for a single massive planet embedded 
within a sea of planetesimals, inward and outward scatterings will at least 
partially balance, leading to little net change in orbital radius.

The foregoing discussion suggests that planetesimal migration might be a 
negligible effect. However, 
\citet{fernandez84} showed that the architecture of the outer Solar System 
introduces an asymmetry in scattering that favors substantial {\em outward} migration of the ice giants. The key point 
is that Jupiter is able to eject planetesimals from the Solar System more 
easily that the other giant planets. Jupiter itself therefore tends to move inward 
by a relatively small amount due to the ejection of debris at initially 
larger orbital radii. The other outer planets scatter bodies inward, to locations 
from which they are removed by Jupiter. This depletion reduces the number of 
outward scatterings, and as a consequence the outer planets (minus Jupiter) 
migrate outward.

\subsubsection{Solar System evidence}
\citet{malhotra93} and \citet{malhotra95} considered the effect of the outward 
migration of Neptune on the origin of Pluto and dynamically similar Kuiper Belt 
Objects. When external forcing causes the semi-major axes of two bodies on 
Keplerian orbits to slowly converge, it is possible (and in some cases guaranteed) 
that they will be {\em captured} into mean motion resonance \citep{goldreich65,mustill11,batygin15}. 
Applying this concept to the Solar System, as Neptune migrated outward due to 
planetesimal scattering Pluto and smaller KBOs 
could have been captured into mean motion resonances. The eccentricities of captured 
bodies then increase as Neptune continues to move out. For a particle 
locked into a $j:j+1$ resonance, the eccentricity is \cite{malhotra95} 
\begin{equation}
 e^2 = e_0^2 + \frac{1}{j+1} \ln \left( \frac{a_{\rm Neptune}}{a_{\rm Neptune, init}} \right)
\end{equation}
where $e_0$ is the eccentricity on capture into the resonance, $a_{\rm Neptune, init}$ 
is the semi-major axis of Neptune when the particle was captured, and $a_{\rm Neptune}$ 
is the final semi-major axis. For example, if Pluto, then at 33~AU, was captured into 
3:2 resonance with Neptune when the latter was at 25~AU, then migration within the 
resonance out to Neptune's current location at 30.2~AU matches Pluto's current eccentricity 
of $e \approx 0.25$.

\begin{figure}
\includegraphics[width=\columnwidth]{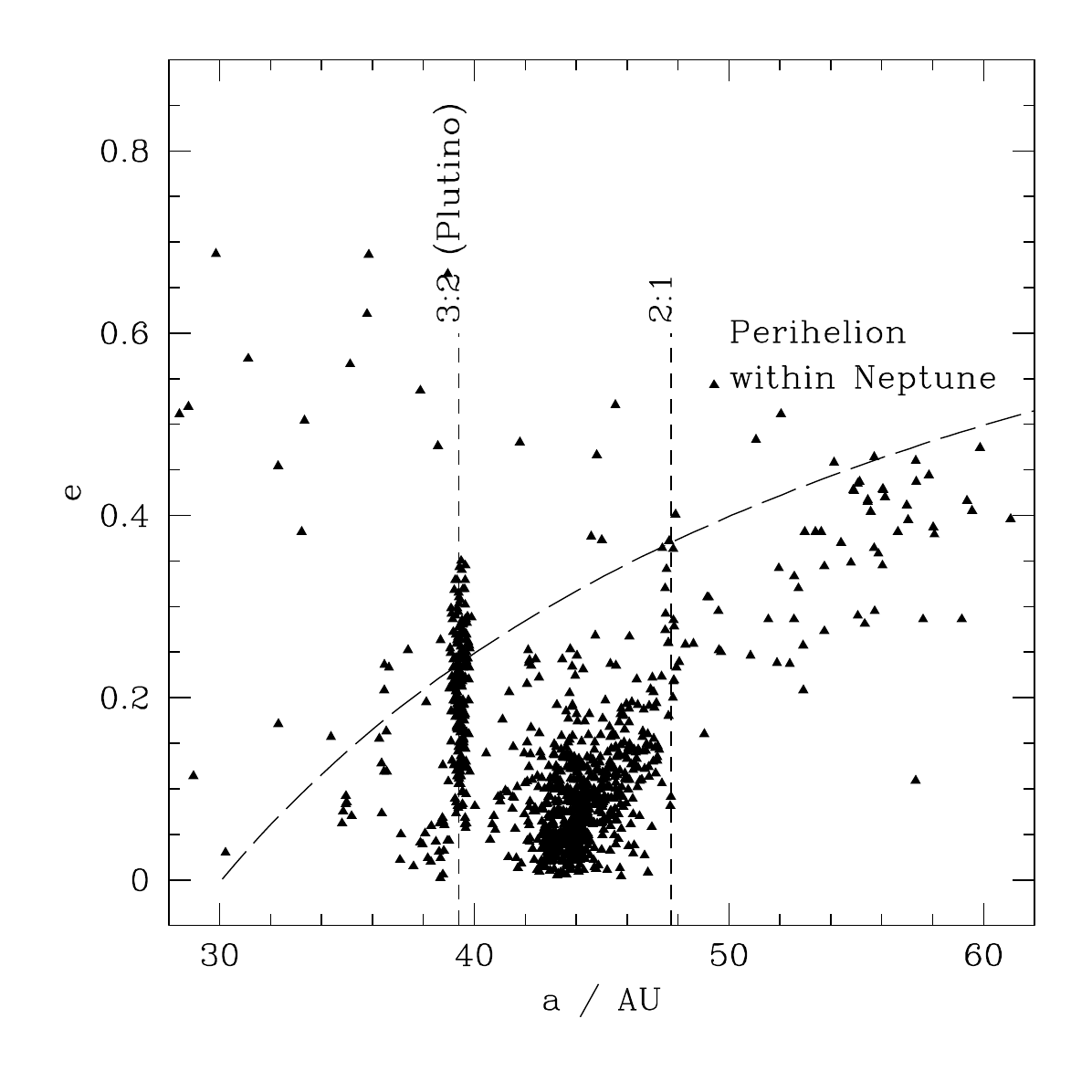}
\caption{The semimajor axes and eccentricities of known (as of 2008) transneptunian bodies. The vertical  
lines shows the location of the 3:2 and 2:1 resonances with Neptune, the dashed line 
shows the minimum eccentricity needed for a body to cross Neptune's orbit.}
\label{fig_kbo}
\end{figure}

This explanation for the origin of Pluto's peculiar orbit is attractive, but even 
more persuasive evidence for Neptune's migration comes from the existence of a 
large population of KBOs in 3:2 resonance (and smaller numbers in other major 
resonances) with Neptune. This population stands out in even the raw plot of $a$ 
vs $e$ for KBOs shown in Figure~\ref{fig_kbo}. In more detail, \citet{murray-clay05} 
and \citet{hahn05} have shown that the distribution of KBOs in resonance with Neptune 
(not just the 3:2 resonance) is broadly consistent with, 
and constrains the time scale of, outward migration of Neptune. 
Solar System evidence thus supports the hypothesis that 
substantial migration of Neptune captured a substantial disk of planetesimals 
and swept them into resonant configurations akin to that of Pluto.

\subsubsection{The Nice model}
The {\em Nice model} refers to any of several dynamical scenarios in which the Solar System's 
outer planets started in a more compact configuration, became unstable, and evolved into 
their current orbits with concomitant scattering of smaller bodies. An early model with this 
general flavor was proposed by \citet{thommes99}, and the basic idea is clearly a natural 
generalization of earlier models for Neptune's migration. \citet{tsiganis05} introduced the 
first version of what is now called the Nice model (named after the French city), which 
attracted immediate attention with the claim that the scattering of small bodies during 
a giant planet rearrangement could explain multiple puzzling features of the Solar System. 
The model has subsequently been revised and studied in great detail.

A variety of dynamical histories for the outer planets fall under the umbrella of the 
Nice model \citep{tsiganis05,morbi07,deienno17,levison11,batygin10b}. The ingredients 
we have to play with include:
\begin{itemize}
\item
The initial (at the time of gas disk dispersal) orbital configuration of the giant planets. 
It is typically assumed that the initial configuration was more compact than it is 
today, and probably featured most or all of the giant planets in a resonant chain. 
There may have been an additional ice giant present. From a broader planet 
formation perspective this class of initial configuration is quite plausible, 
giant planet cores may converge to roughly the Jupiter / Saturn region under 
Type~I migration, and become trapped in resonance due to Type~II migration.
\item
The initial configuration becomes destabilized through interaction with a disk 
of remnant planetesimals beyond Neptune (or beyond whatever planet occupies 
the most distant spot early on). The interaction could be direct planetesimal 
scattering, though it could also be scattering of dust produced collisionally in a 
belt slightly further out. The required planetesimal belt is massive, perhaps 
$35-50 \ M_\oplus$, but this is not an unreasonable mass of solid material 
at these radii.
\item
After the resonant chain is broken (or when resonances are crossed) there is 
some combination of (a) fast orbital evolution driven by close encounters between 
giant planets leading to scattering (or ejection, if there were initially more giant 
planets) and (b) slower orbital evolution driven by planetesimal scattering.
\item
Eventually the mass in the Kuiper Belt is reduced to a low enough level that 
planetesimal-driven migration becomes negligible, and the planets settle 
into their current orbits.
\end{itemize}
Assembling a single plausible history for the outer Solar System from this 
buffet of options is not easy, though there are a number of powerful 
constraints. \citet{deienno17} provide a summary. In general, constraints 
from the inner Solar System favor rapid evolution of the orbits of the 
Jupiter-Saturn subsystem after the onset of instability. This is because 
slow evolution would cause secular resonances---associated with the precession of these planets' 
orbits---to destabilize Mercury and the asteroid belt, contrary to observations 
\citep{roig15,kaib16,minton13}. The structure of the Kuiper Belt, on the 
other hand, is reproduced if Neptune undergoes a substantial stretch 
of relatively smooth planetesimal-driven migration \citep{nesvorny15}. 
The evidence thus suggests that a combination of planet-planet scattering and 
planetesimal-driven migration may have taken place.

The timing of the instability is uncertain. As part of the original Nice model, 
\citet{gomes05} associated the instability with the origin of the Late Heavy 
Bombardment on the Moon. The Late Heavy Bombardment \citep[LHB, for a 
review see][]{strom15} refers to cratering evidence for a spike in the 
rate of impacts on to the Moon (and other planets, including Mercury), 
nominally dated from an analysis of lunar samples to 600-700~Myr 
after the formation of the Solar System \citep{tera74}. Associating 
the LHB with the Nice model instability is highly constraining, as it 
is quite hard to find ultimately unstable configurations that survive 
that long. Both the chronology of the LHB \citep{norman14} and its association with 
the Nice model instability are, however, subject to debate, and an 
earlier timing of the instability remains a possibility.

A number of other Solar System properties have been suggested to result 
from Nice model dynamics. These include the capture of Jupiter's Trojan 
asteroids \citep{morbi05} and the capture of the giant planets' irregular 
satellites \citep{nesvorny14}, among others.

\subsection{Planet-planet scattering}
While the gas disk is present, gas damping can protect a multiple planet 
system against the development of crossing orbits from planet-planet 
gravitational interactions (at least if interactions with the gas disk 
actually damp eccentricity, which as noted above is somewhat uncertain). Once the 
gas is gone, gravity can go to work on what may be an unstable planetary system 
and change the orbital radii and eccentricities of the planets. This process---
gravitational scattering---is widely invoked as a mechanism to 
explain the large eccentricities of many extrasolar giant planets.

\subsubsection{Hill stability}
\label{sec_hill}
Let us begin with some analytic considerations. The general N-body problem 
of the motion of $N$ point masses interacting under Newtonian gravity is 
analytically insoluble for $N > 2$. Here, we start by considering a 
special case of $N=3$ in which two bodies of arbitrary mass have a 
circular orbit, while a third body of negligible mass orbits in the 
known gravitational field of the massive objects. This problem---called 
the {\em circular restricted 3-body problem}---still defies analytic 
solution, but it is possible to place useful limits on the motion of the 
third body (often described as a ``test particle"). The circular restricted 
3-body problem is a reasonable approximation to several situations of 
practical interest, including the motion of asteroids in the vicinity 
of Jupiter, and the evolution of planetesimals near a growing planet. 
A good description of the problem can be found in \citet{murray99}, 
whose treatment we largely mirror here. The more general 3-body problem is 
discussed (in both the planetary and multiple star contexts) in a book 
by \citet{valtonen06}.

\begin{figure}
\includegraphics[width=0.9\columnwidth]{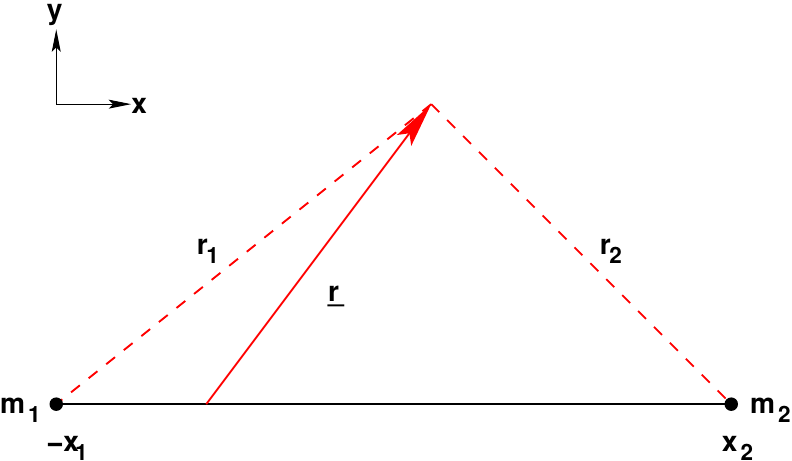}
\caption{Co-ordinate system for the restricted three body problem. 
We work in a co-rotating Cartesian co-ordinate system centered on the 
center of mass in which the star and planet are located at $(-x_1,0)$ 
and $(x_2,0)$ respectively. The test particle is at position ${\bf r}$.}
\label{fig_hill_setup}
\end{figure}

As shown in Figure~\ref{fig_hill_setup}, we consider a binary system 
in which the massive bodies have mass $m_1$ and $m_2$ respectively. We 
work in a corotating co-ordinate system centered on the center of 
mass. The orbital plane is $(x,y)$ in Cartesian co-ordinates, and the 
test particle is located at position ${\bf r}$.

If the angular velocity of the binary is $\Omega$, the equations of 
motion for the test particle are,
\begin{eqnarray}
 \ddot{\bf r} & = & - \nabla \Phi - 2 \left( {\bf \Omega} \times {\dot{\bf r}} \right) 
 - {\bf \Omega} \times \left( {\bf \Omega} \times {\bf r} \right), \\
 \Phi & = & -\frac{Gm_1}{r_1} -\frac{Gm_2}{r_2}.
\end{eqnarray} 
Expressed in components, we have,
\begin{eqnarray}
 \ddot{x} - 2 \Omega \dot{y} - \Omega^2 x & = & -G \left[ 
 \frac{m_1(x+x_1)}{r_1^3} + \frac{m_2(x-x_2)}{r_2^3} \right] \nonumber \\
 \ddot{y} + 2 \Omega \dot{x} - \Omega^2 y & = & -G \left[ 
 \frac{m_1}{r_1^3} + \frac{m_2}{r_2^3} \right] y \nonumber \\ 
 \ddot{z} & = & -G \left[ 
 \frac{m_1}{r_1^3} + \frac{m_2}{r_2^3} \right] z.
\end{eqnarray}
The acceleration due to the centrifugal force can be subsumed into a 
pseudo-potential. Defining,
\begin{equation}
 U \equiv \frac{\Omega^2}{2} \left( x^2 + y^2 \right) 
 + \frac{Gm_1}{r_1} + \frac{Gm_2}{r_2} 
\end{equation}
we obtain,
\begin{eqnarray} 
 \ddot{x} - 2 \Omega \dot{y} & = & \frac{\partial U}{\partial x} \nonumber \\ 
 \ddot{y} + 2 \Omega \dot{x} & = & \frac{\partial U}{\partial y} \nonumber \\
 \ddot{z} & = & \frac{\partial U}{\partial z}.
\label{eq_roche} 
\end{eqnarray}
Digressing briefly, we note that $U$ is (up to an arbitrary minus sign) the
``Roche potential". 
Two stars, or a star plus a planet, that rotate synchronously while on circular orbits occupy 
Roche equipotentials. If their size is comparable to the size of the 
{\em Roche lobe}---defined by the critical figure-of-eight shaped equipotential that 
passes through the inner Lagrange point $L_1$---then the bodies suffer 
significant tidal distortion. A useful approximation for the radius $R_{\rm RL}$ of a 
sphere with the same volume as the Roche lobe was provided by \citet{eggleton83}. 
For a binary with mass ratio $q \equiv m_2 / m_1$ and separation $a$,
\begin{equation}
 \frac{R_{\rm RL}}{a} \simeq \frac{0.49 q^{2/3}}{0.6 q^{2/3} + \ln (1 + q^{1/3})}.
\end{equation} 
This equation can be used to assess, for example, how close hot Jupiters are 
to overflowing their Roche lobes. For a Jupiter mass planet with $q = 10^{-3}$, 
\begin{equation}
 R_{\rm RL} \simeq 0.048 a.
\end{equation} 
A planet with the same radius as Jupiter ($7.14 \times 10^9$~cm) would then overflow 
its Roche lobe interior to $a = 0.01$~AU. A very short period hot Jupiter, such as  
OGLE-TR-56b \cite{torres04} with a period of 1.2~days, has a semi-major axis that 
is about 0.0225~AU. So this planet, and more securely other hot Jupiters that orbit 
modestly further out, is safe against mass transfer, though not by a large margin.

Returning to the general equations (\ref{eq_roche}), we eliminate the Coriolis 
terms by multiplying through by $\dot{x}$, $\dot{y}$ and $\dot{z}$ and adding. 
We then obtain,
\begin{eqnarray}
 \dot{x} \ddot{x} + \dot{y} \ddot{y} + \dot{z} \ddot{z} & = & 
 \dot{x} \frac{\partial U}{\partial x} + 
 \dot{y} \frac{\partial U}{\partial y} + 
 \dot{z} \frac{\partial U}{\partial z} \nonumber \\
 \frac{\rm d}{{\rm d}t} \left( \frac{1}{2} {\dot{x}}^2 + \frac{1}{2} {\dot{y}}^2
 + \frac{1}{2} {\dot{z}}^2 \right) & = & \frac{{\rm d}U}{{\rm d}t} \nonumber \\
 {\dot{x}}^2 + {\dot{y}}^2 + {\dot{z}}^2 & = & 2 U - C_J \nonumber \\
 C_J & = & 2U - v^2
\label{eq_cj} 
\end{eqnarray}
where $v$ is the velocity and $C_J$, called the {\em Jacobi constant}, 
is the arbitrary constant of integration. $C_J$ is an energy-like 
quantity that is a conserved quantity in the circular restricted 3-body 
problem.

The existence of this integral of motion is important because it places 
limits on the range of motion possible for the test particle. For a 
particle with a given initial position and velocity, we can use 
equation (\ref{eq_cj}) to compute $C_J$, and hence to specify {\em zero-velocity 
surfaces}, defined via,
\begin{equation}
 2U = C_J,
\end{equation}  
which the particle can never cross. If the volume enclosed by one of the 
zero-velocity surface is finite, then a particle initially within that 
region is guaranteed to remain there for all time. This concept is 
known as {\em Hill stability}. 
 
\begin{figure}
\includegraphics[width=0.8\columnwidth]{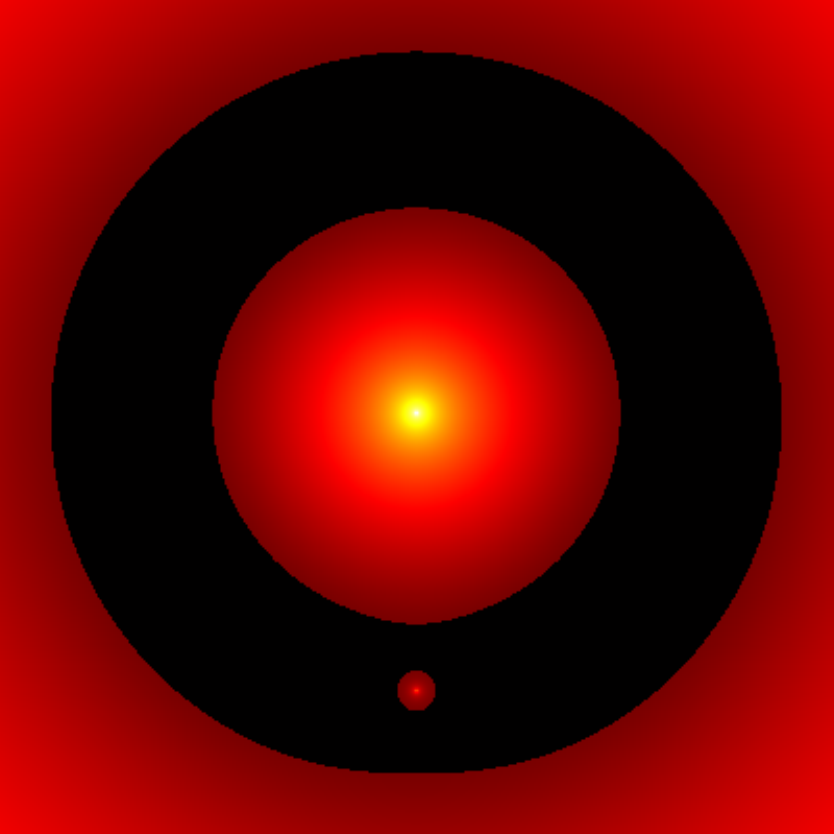}
\caption{Forbidden zones (dark regions) in an example of the restricted 
3-body problem. For this particular choice of the Jacobi constant $C_J$, particles 
can orbit (a) the star at small radii, (b) the planet in a tight orbit, or (c) the star-planet 
binary as a whole. The existence of zero-velocity surfaces, however, means that 
particles cannot be 
exchanged between these regions.}
\label{fig_hill}
\end{figure}

The topology of the zero-velocity surfaces in the restricted three-body problem 
varies according to the value of $C_J$. An example is shown in Figure~\ref{fig_hill}. 
In this instance the zero-velocity surfaces define three disjoint regions in the 
$(x,y)$ plane, one corresponding to orbits around the star, one corresponding to 
orbits around the planet, and one corresponding to orbits around the star-planet 
binary. A particle in any one of these states is stuck there -- it cannot 
cross the forbidden zone between the different regions to move into a different 
state.

\subsubsection{Scattering and exoplanet eccentricities}

\begin{figure}
\includegraphics[width=0.7\columnwidth]{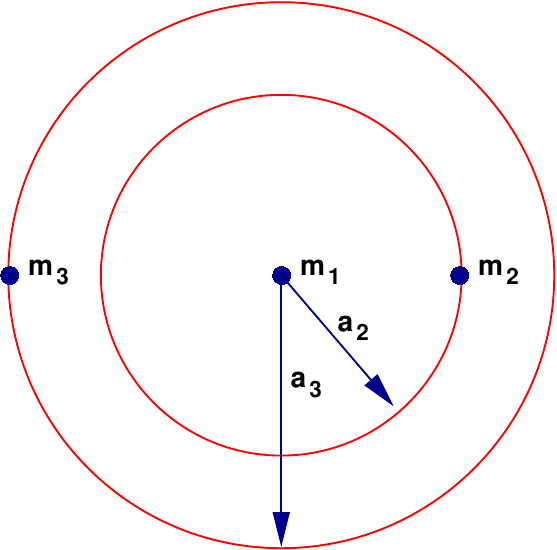}
\caption{Setup for the stability calculation of a two planet system in 
which both of the planets are on circular orbits. Unlike in the case 
of the Hill problem, here we strictly require that $m_2 \ll m_1$ and 
$m_3 \ll m_1$.}
\label{fig_gladman}
\end{figure}

Somewhat surprisingly, the result of the test particle analysis discussed above 
also applies in modified form to the much tougher problem of the stability of two 
planets orbiting a star. Consider the situation shown in Figure~\ref{fig_gladman}, 
in which planets of mass $m_2$ and $m_3$ orbit a star of mass $m_1$ in 
circular orbits with semi-major axes $a_2$ and $a_3$ respectively. The 
stability of the system evidently must depend upon the relative, rather 
than the absolute, spacing between the orbits. Accordingly we write,
\begin{equation}
 a_3 = a_2 (1 + \Delta)
\end{equation}
with $\Delta$ being a dimensionless measure of the orbital separation 
between the planets. We further define $\mu_2 = m_2 / m_1$ and 
$\mu_3 = m_3 / m_1$. Then for $\mu_2$,$\mu_3 \ll 1$, \citet{gladman93}, 
drawing on earlier results derived by \citet{marchal82} and others, showed 
that the system is guaranteed to be stable provided that the separation $\Delta$ 
exceeds a critical separation $\Delta_c$ given by,
\begin{equation}
 \Delta_c \simeq 2.40 \left( \mu_2 + \mu_3 \right)^{1/3}.
\end{equation} 
Note that analytic results leave open the question of whether systems with 
$\Delta < \Delta_c$ are actually unstable, all we know is that $\Delta > \Delta_c$ 
is sufficient for stability. This condition reduces to the test particle result 
if $\mu_3 \rightarrow 0$, as of course it should\footnote{Note, however, that the 
analysis for the restricted three-body problem applies for an arbitrary mass 
ratio of the massive bodies, whereas the result for two planets requires that 
{\em both} be much less massive than the star.}. As an example, if we compute 
the critical separation for planets of the mass of Jupiter and Saturn, 
we obtain $\Delta_c \simeq 0.26$. The actual separation of Jupiter and Saturn 
in these units is $\Delta \simeq 0.83$, so an isolated planetary system in 
which Jupiter and Saturn were on circular orbits would assuredly be stable 
for all time.

What about more complex systems? It is possible to include non-zero 
eccentricities into this analysis, but {\em not} more planets. For a 
multiple planet system one might plausibly reason that the system 
will be unstable if any pair substantially violates the critical two-planet 
separation for Hill stability. It is also true that the system will 
generally become more stable as the separations increase \citep{chambers96}. However, 
no absolute stability bound is known for any planetary system with $N > 3$.

If a two-planet system is unstable, the possible outcomes of the instability 
can be divided into four classes:
\begin{itemize}
\item[1.]
The separation evolves (increases) until the system achieves a state that is 
stable over the long term.
\item[2.]
One planet is ejected, while the other remains bound, generally with $e \neq 0$.
\item[3.]
The planets physically collide.
\item[4.]
One planet impacts the star, or is scattered into a short-period orbit for 
which tidal effects are important.
\end{itemize}
The last two channels are not possible in a model 3-body problem, in which the 
planets are represented by point masses, but can occur (especially planet-planet 
collisions, which become frequent at small radii) in real systems.

The idea that gravitational scattering and planetary ejections might account 
for the eccentricity of extrasolar planets was proposed as soon as it became 
clear that extrasolar planets were not typically on circular orbits 
\cite{rasio96,weidenschilling96,lin97}. Quantitative study of such models 
requires large-scale N-body integrations, first to derive the statistical 
distribution of outcomes of any given scenario (since the systems are 
typically chaotic, nothing can be said about any single run), and 
second to map out the large parameter space that results when one 
considers different numbers of planets with different initial 
separations, masses and so forth.

\citet{ford01} presented a comprehensive study of the dynamics of equal 
mass two planet systems. The planets were set up on circular orbits 
close to the stability boundary, and allowed to evolve under purely 
N-body forces until the system relaxed to a stable state. They found 
that the predicted fraction of collisions increases sharply for 
small orbital radii and / or larger planetary radii. For pairs of 
Jupiter mass and Jupiter radius planets initially located at 5~AU, 
the most common outcome is two planets (65\%), followed by ejections 
(35\%), with collisions (10\%) a distant third. If the same pair of 
planets starts at 1~AU, however, collisions occur roughly 30\% of the 
time. This conclusion is important for studies of extrasolar planet 
eccentricity, because collisions yield relatively low eccentricities 
for the merged planet. 

\begin{figure}
\includegraphics[width=\columnwidth]{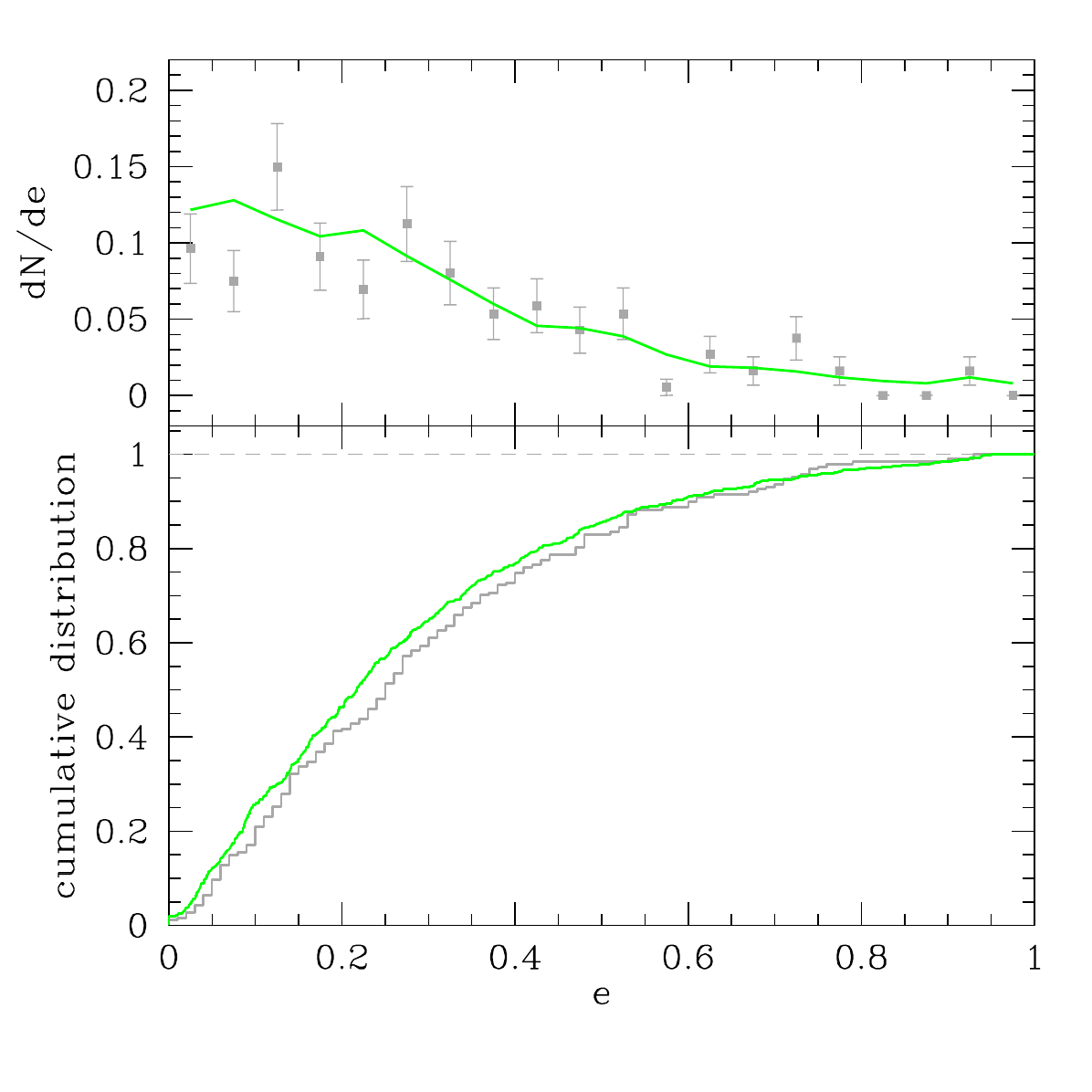}
\caption{The differential (upper panel) and cumulative (lower panel) 
eccentricity distribution of known extrasolar planets (grey curves) 
is compared to the predicted distribution that results from scattering 
in three planet systems (shown in green). The simulation results were derived by numerically 
evolving an ensemble of unstable planetary systems made up of three planets 
whose masses were drawn from the observed mass function for extrasolar 
planets in the range $0.3 \ M_J < M_p < 5 \ M_{J}$. The inner planet 
was initially at $a = 4.5 \ {\rm AU}$. Based on simulations by \citet{raymond08}, 
compared against data available in spring 2010.}
\label{fig_eccdist}
\end{figure}

There is only a rather small range of orbital separations which 
allows a two planet system to be unstable over the long term 
(greater than around $10^5$~yr, which is roughly the dispersal 
time for the gas disk), while not being violently unstable. This 
observation means that it is easier to set up an internally 
self-consistent scattering model with three or more planets, since 
a wider range of such systems eventually lead to interesting dynamics. 
Models starting with three or more planets have also been studied in 
some detail \cite{terquem02,marzari02,adams03}. Comprehensive 
studies, such as those by \citet{chatterjee08} and \citet{juric08}, 
find that scattering models yield a good quantitative match to the 
observed distribution of extrasolar planet eccentricities. Just 
how good this agreement is is illustrated in Figure~\ref{fig_eccdist}, 
which shows how the final simulated eccentricities compare to the 
data \citep{raymond08}. Although these results were derived from 
purely N-body calculations, hydrodynamic scattering simulations 
show that the persistence of modest amounts of disk gas does not 
ruin the agreement \citep{moeckel12}. Altogether, there is substantial 
but circumstantial evidence for a major role for scattering in establishing 
the broad distribution of 
exoplanet eccentricity.

\subsection{Predictions of migration theories}
In summary, there is persuasive circumstantial evidence for the 
action of at least three separate processes that lead to the 
early evolution of planetary systems:
\begin{itemize}
\item
{\bf Gas disk migration} in the Type~II regime appears to 
inevitable, since at least some gas must necessarily 
be present at the epoch when giant planets form. The presence of 
a reasonably high resonant fraction among systems with multiple 
giant planets also points to the existence of an early dissipative 
process in orbital evolution. 
\item
{\bf Planetesimal disk migration} provides a persuasive 
explanation for the origin of Pluto's odd orbit together 
with some of the more detailed properties of the Kuiper Belt. 
One can make a case that this process too ought to be common 
in the outer reaches of planetary systems. Gas giant formation 
almost certainly becomes more difficult further out in the 
disk, so it is quite plausible that the zone where gas and ice 
giants manage to form is often surrounded by a disk of 
planetesimal debris that has been unable to grow to large 
sizes.
\item
{\bf Planet-planet scattering} works well as an explanation 
for the eccentricity distribution of giant extrasolar 
planets. There is {\em no} straightforward independent argument that 
the unstable initial conditions needed for such models to 
work are generically realized in nature, but the empirical 
evidence seems to suggest that they are.
\end{itemize}
The presence of a large fraction of hot Jupiters on orbits inclined 
with respect to the stellar equator also yields constraints, though 
these are not unambiguous. Planet-planet scattering \citep{nagasawa08} 
or Kozai-Lidov excitation of eccentricity \citep{wu03,naoz16} are the mechanisms 
most obviously compatible with this observation, but it has also been 
suggested that the gas disk could be torqued out of the stellar 
equatorial plane \citep{bate10,batygin12}. If this process is common, 
some or all of the misaligned hot Jupiters could have arrived to 
their current orbits via Type~II migration.

Additional qualitatively different tests of these theoretical ideas are 
possible. 
Planet-planet scattering, for example, predicts the existence of a (small) 
population of very weakly bound, typically eccentric planets, 
with semi-major axes of $10^2$~AU and more \citep{veras09,scharf09}. 
Direct imaging surveys of young stars have the potential to 
detect this distinctive population.

The {\em combined} action of multiple evolutionary mechanisms 
may also give rise to new classes of planetary systems.  
At larger orbital radii (than those currently probed by observations 
of exoplanets) it seems likely that we ought to see planetary 
systems whose dynamics has been affected by both planet-planet 
scattering and planetesimal disk migration. N-body simulations 
suggest that the signature of this combination is a transition 
from generally eccentric to nearly circular planetary orbits 
as the mass of the planetary system is reduced \citep{raymond09b,raymond10}. 
If true, the near-circular orbits of the giant planets in the 
Solar System might in fact be typical of the architecture of 
relatively low-mass systems at large orbital radii. For higher 
mass systems the same simulations predict a high abundance of 
resonant configurations, including resonant chains that would 
be planetary analogs of the Laplace resonance in the Jovian 
satellite system.

\subsection{Tidal evolution}
Two bodies in a tight orbit experience tidal forces as a 
consequence of the gradient in the gravitational force across 
their finite radius. The tidal forces raise tidal bulges on the 
surface of the bodies, whose shape is approximately defined 
by the condition of hydrostatic equilibrium in the asymmetric 
gravitational potential. If the axis of the tidal bulge is 
offset with respect to the line joining the centers of the 
two bodies, the result is a tidal torque which modifies the 
semi-major axis and eccentricity of the system. 

Tides are dynamically important in the Solar System. Energy dissipation 
associated with oceanic tides, raised on the Earth by the Moon, is 
responsible for a slow but measureable increase in the Earth-Moon separation 
\citep{dickey94}. The basic framework for understanding these phenomena dates 
back to work by George Darwin (Charles' son) more than a century ago 
\citep{darwin79}, but even this classical theory involves many subtleties, 
while a general first-principles theory of tides remains 
elusive. \citet{ogilvie14} gives an excellent review of the state of 
theoretical knowledge of tidal phenomena. 

\subsubsection{The tidal bulge and tidal torque}
\begin{figure}
\includegraphics[width=\columnwidth]{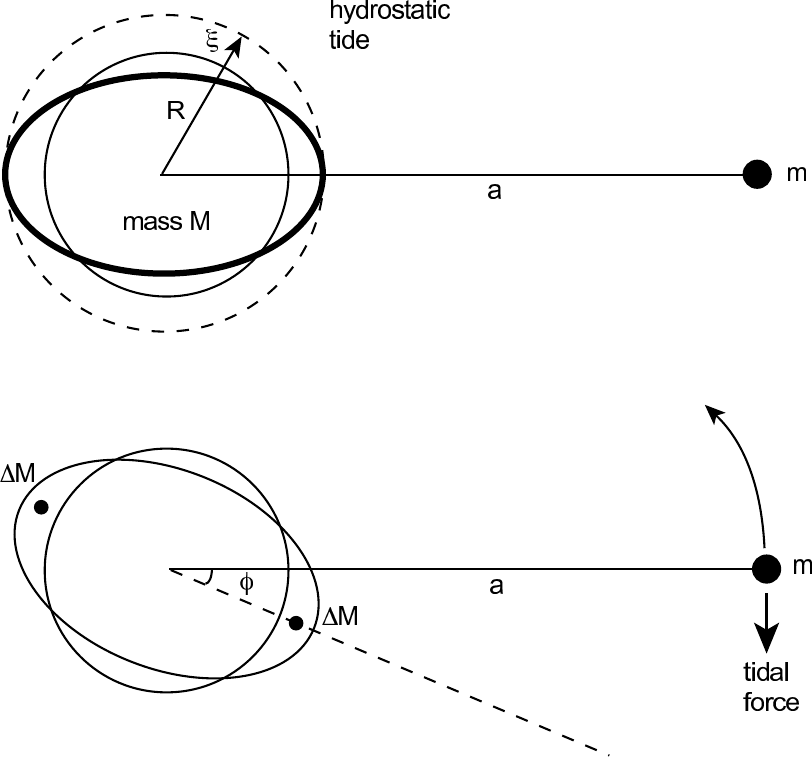}
\caption{Upper panel: illustration of the {\em hydrostatic} tidal response of a body 
(a star or a planet) of mass $M$ and radius $R$ to a point mass companion 
of mass $m$ orbiting at distance $a$. A tidal bulge of amplitude $\xi$ is raised 
on the surface, aligned with the separation vector between the two bodies. 
Lower panel: in the presence of dissipation, the tidal bulge lags (or leads, 
depending upon the spin) the motion of the companion by an angle $\phi$. As a 
consequence, the tidal bulge -- now idealized as two point masses of mass 
$\Delta M$ -- exerts a torque which modifies the orbit of the secondary.}
\label{fig_tides}
\end{figure}

Even the elementary theory of tides is quite intricate. We can gain considerable 
physical insight, however, from a simple ``back of the envelope" calculation that 
ignores order unity numerical factors and effects. Consider the tide 
raised on a fluid body of mass $M$ and radius $R$ by a companion, of mass $m$, 
orbiting in a circular orbit at distance $a$. The geometry is shown in 
Figure~\ref{fig_tides}. We seek to determine, first, the height of the tide $\xi$, 
and, second, the torque that results if the tidal bulge is misalinged by some 
angle $\phi$ with respect to the line joining the centers of the two bodies.
We will assume, throughout, that the tidal deformation corresponds closely 
to the hydrostatic response of the body in the gravitational field defined by 
both bodies (the ``static" tide). This is a reasonable approximation. Applying the 
virial theorem to the body on which the tide is raised, we find that the central 
sound speed ought to be comparable to the orbital velocity {\em around} the 
body at radius $R$. For $a \gg R$, it follows that one orbit of the companion 
corresponds to many sound crossing times of the fluid, and hydrostatic 
equilibrium has time to be established.

To estimate the height of the tidal bulge, $\xi$, we note that the gravitational 
force (per unit mass) exerted by the companion on the near side of the body differs 
from that exerted at the center by an amount,
\begin{equation}
 \Delta F = \frac{Gm}{(a-R)^2} - \frac{Gm}{a^2} \sim \frac{Gm}{a^3} R.
\end{equation}
This tidal force allows us to raise a tidal bulge up to a height where the 
self-gravity of the fluid body is reduced by the same amount,
\begin{eqnarray}
 \frac{{\rm d}}{{\rm d}R} \left( \frac{GM}{R^2} \right) \xi \sim - \frac{Gm}{a^3} R \\
 \frac{\xi}{R} \sim \left( \frac{m}{M} \right) \left( \frac{R}{a} \right)^3.
\end{eqnarray} 
The height of the bulge falls off rapidly for $a \gg R$. If we assume that the 
fluid body has a uniform density $\rho$ (and thereby ignore important structural 
factors) the mass associated with the bulge is given by,
\begin{equation}
 \Delta M = 4 \pi R^2 \xi \rho,
\end{equation}
which simplifies to yield,
\begin{equation}
 \Delta M \sim \left( \frac{R}{a} \right)^3 m.
\end{equation}
The work required to raise the bulge against the self-gravity of the fluid 
body is,
\begin{eqnarray}
 E_0 & \sim & \frac{GM \Delta M}{R^2} \xi \\
 & \sim & \frac{Gm^2}{R} \left( \frac{R}{a} \right)^6.
\end{eqnarray}  
This is the amount of energy associated with the tidal deformation of the 
body.

Let us now backtrack to revisit the assumption that the tidal response of 
the fluid is hydrostatic. If this were {\em exactly} true, the tidal 
bulge would line up precisely along the line joining the centers of the 
two bodies, and, by symmetry, there would be no tidal torque (as shown 
in the upper panel of Figure~\ref{fig_tides}). In reality, however, the 
tide represents the response of the fluid to forcing at some non-zero 
frequency $\Omega$, given (for a companion on a circular orbit around a 
non-rotating primary) simply by the orbital frequency. If the central 
body is non-rotating (or rotating slower than the orbital frequency), 
then the departure from hydrostatic equilibrium occasioned by the 
finite response time will cause the bulge to {\em lag} by some angle 
$\phi$. The bulge will lead if the spin of the central body is faster 
than the orbital frequency. A lagging or leading bulge will exert a 
torque on the companion that causes the orbit to decay or expand in 
radius, respectively (e.g. the Moon's orbital period exceeds the Earth's 
spin period, so the torque for the Earth-Moon system results in 
recession of the lunar orbit away from the Earth).

We now assume that the torque due to the tide can be represented as 
that due to two point masses $\Delta M$ offset from the line of centers 
by a small angle $\phi$ (Figure~\ref{fig_tides}, lower diagram). The 
tidal force is then, roughly, 
\begin{eqnarray}
 F_{\rm tidal} & \sim & \frac{G \Delta M m}{(a-R)^2} \frac{R}{a} \phi 
 - \frac{G \Delta M m}{(a+R)^2} \frac{R}{a} \phi \\
 & \sim & \frac{G \Delta M m}{a^4} R^2 \phi,
\end{eqnarray}
and the work done by the tidal force per orbit of the companion is,
\begin{equation}
 \Delta E \sim \frac{G \Delta M m}{a^3} R^2 \phi.
\end{equation}
If we knew $\phi$, this equation (with the 
missing numerical factors restored) would yield 
the rate of decay or rate of expansion of the orbit due to the tides.  

Calculating $\phi$ from first principles is a hard task. We can, 
however, at least gain a more transparent understanding of the 
physical problem of what {\em determines} $\phi$. To do so we 
define the tidal $Q$ as the ratio of the energy stored in the tidal 
deformation to the energy that is dissipated in one cycle,
\begin{equation}
 Q \equiv \frac{2 \pi E_0}{\int -\dot{E} dt}.
\end{equation}
Using our estimates, we have that,
\begin{equation}
 Q \sim \frac{E_0}{\Delta E} \sim \phi^{-1}.
\end{equation} 
This is an important result. We see that the magnitude of the tidal 
lag, and hence the strength of the tidal torque, is directly linked 
to the amount of {\em dissipation} within the tidally distorted 
body. 

Finally we can estimate the rate of decay of the orbital separation 
due to the (lagging) tidal bulge. For $m \ll M$ we have that,
\begin{equation}
 \frac{{\rm d}E_{\rm orbit}}{{\rm d}t} \sim - \frac{GMm}{a^2} 
 \frac{{\rm d}a}{{\rm d}t} = - \frac{\Delta E}{P},
\end{equation}
where $P = 2 \pi \sqrt{a^3/(GM)}$ is the orbital period. Substituting 
for the various quantities, we obtain,
\begin{equation}
 \frac{1}{a} \frac{{\rm d}a}{{\rm d}t} \sim 
 \left( \frac{G}{M} \right)^{1/2} \frac{m}{Q} \frac{R^5}{a^{13/2}}.
\end{equation}
Up to factors of the order of unity, this expression agrees with the 
standard tidal formula quoted, for example, by \citet{jackson08}, 
who also give the corresponding expressions for the change in orbital 
eccentricity. Standard references for the astrophysical theory of tides 
include \citet{goldreich66} and \citet{hut81}.

\subsubsection{Determining the tidal $Q$}
The above analysis suffices to illustrate two important points,
\begin{itemize}
\item
Tidal forces decline extremely rapidly with increasing orbital 
separation.
\item
The rate of tidal evolution depends upon the amount of dissipation 
present within the two bodies that are interacting tidally. 
\end{itemize}
The attentive reader may also have noticed the astrophysical sleight 
of hand by which all manner of intractable physics has been swept 
into a single unknown parameter, $Q$. Approaches to estimating $Q$ 
can be divided into those that rely on extrapolations from measured 
values in well-observed systems, and those that attempt to compute $Q$ 
by identifying dissipative processes within stars or planets.

For giant planets, the primary observational constraint on $Q$ comes 
from direct measurements of the tidal evolution in the orbits of the 
Galilean satellites \citep{yoder81,lainey09}\footnote{Historically, 
estimates of Jupiter's $Q$ have primarily been derived {\em indirectly}, 
by assuming that the excess heat flux from Io derives from tidal effects. 
\citet{greenberg10} gives a good review of these estimates.}. These 
empirical estimates suggest that $Q \simeq 10^5$, and based on this many 
workers adopt values for the tidal $Q$ of extrasolar planets that 
are similar (typically in the range $Q=10^5 - 10^6$). Considerable 
caution as to the validity of this extrapolation from Jupiter to 
extrasolar planets is, however, in order. As noted already, the $Q$ of a 
star or planet is {\em not} some fixed and immutable property of a body 
akin, say, to its mass. Rather, $Q$ describes the response of a body 
to forcing at one or more specific frequencies (in the case of the 
Jovian estimate, the frequency of relevance is the difference between 
the spin frequency of Jupiter and the orbital frequency of Io). To 
extrapolate correctly to extrasolar planets, we need to make some 
assumption as to how $Q$ varies as a function of frequency. One 
simple assumption is to postulate that the tidal lag {\em angle} remains 
constant---in which case $Q$ is indeed a constant---but one might 
equally assume that the tidal bulge has a constant {\em time} lag, 
in which case $Q=Q(\Omega)$. If one is interested in tidal eccentricity 
evolution, moreover, the forcing is not monochromatic but rather 
has a spread across a range of frequencies. As a result of these 
complications, empirical models of extrasolar tidal evolution are 
subject to substantial but unquantifiable uncertainties. 

Given the uncertainties in the empirical approach, a theoretical 
determination of $Q$ would evidently be extremely valuable. Achieving 
this goal requires first identifying, and then calculating, the primary 
source of dissipation that acts on the tide. Molecular viscosity is 
insufficient, so we are left with a variety of hard-to-calculate 
candidates that include non-linear dissipation of waves and interactions 
between the tide and turbulent processes within the body. Recent 
theoretical work \citep{ogilvie04,goodman09} in this area, although 
it still falls short of being able to predict $Q$ from first-principles, 
has nonetheless proven influential in identifying additional properties 
of planets that may influence the $Q$. The presence of a rigid, solid 
core, for example, can substantially alter the tidal response of a 
planet as compared to an observationally almost indistinguishable 
body lacking a core. Given the rapidly improving observations of 
extrasolar planets that are surely vulnerable to tidal evolution, 
one may hope that this is an area ripe for further theoretical and 
observational progress.

\section*{Acknowledgements}
These notes are based on a graduate 
course given at the University of Colorado, Boulder. 
My thanks to the students in those classes, and to my current and 
former postdoctoral collaborators, for stimulating my interest in this subject. 
My work on planet formation and accretion physics has been 
supported by NASA under the auspices of the Origins of Solar Systems, Exoplanet Research, 
Astrophysics Theory, Beyond Einstein Foundation Science and Hubble Theory programs, 
and by the National Science Foundation.


\begin{thebibliography}{}

\bibitem[Adams(2010)]{adams10}
 \textsc{Adams, F. C.} 2010, 
 \emph{ARA\&A}, 48, 47

\bibitem[Adams, Lada \& Shu(1987)]{adams87}	
 \textsc{Adams, F. C., Lada, C. J., \& Shu, F. H.} 1987, 
 \emph{ApJ}, 312, 788
 
\bibitem[Adams \& Laughlin(2003)]{adams03}
 \textsc{Adams, F. C., \& Laughlin, G.} 2003,
 \emph{Icarus}, 163, 290
 
\bibitem[Adams et al.(2006)]{adams06}
 \textsc{Adams, F. C., Proszkow, E. M., Fatuzzo, M., \& Myers, P. C.} 2006,
 \emph{ApJ}, 641, 504
 
\bibitem[Adams \& Shu(1986)]{adams86}
 \textsc{Adams, F. C., \& Shu, F. H.} 1986, 
 \emph{ApJ}, 308, 836
 
\bibitem[Afshordi, Mukhopadhyay \& Narayan(2005)]{afshordi05}
 \textsc{Afshordi, N., Mukhopadhyay, B., \& Narayan, R.} 2005, 
 \emph{ApJ}, 629, 373

\bibitem[Agol et al.(2005)]{agol05}
 \textsc{Agol, E., Steffen, J., Sari, R., \& Clarkson, W.} 2005, 
 \emph{MNRAS}, 359, 567
 
\bibitem[Agnor \& Hamilton(2006)]{agnor06}
 \textsc{Agnor, C. B., \& Hamilton, D. P.} 2006,
 \emph{Nature}, 441, 192
 
\bibitem[Alexander \& Armitage(2007)]{alexander07}
 \textsc{Alexander, R. D., \& Armitage, P. J.} 2007,
 \emph{MNRAS}, 375, 500
 
\bibitem[Alexander \& Armitage(2009)]{alexander09}
 \textsc{Alexander, R. D., \& Armitage, P. J.} 2009,
 \emph{ApJ}, 704, 989
  
\bibitem[Alexander, Clarke \& Pringle(2005)]{alexander05}	
 \textsc{Alexander, R. D., Clarke, C. J., \& Pringle, J. E.} 2005, 
 \emph{MNRAS}, 358, 283 
 
\bibitem[Alexander, Clarke \& Pringle(2006)]{alexander06b}	
 \textsc{Alexander, R. D., Clarke, C. J., \& Pringle, J. E.} 2006, 
 \emph{MNRAS}, 369, 229

\bibitem[Alexander et al.(2014)]{alexander14} 
 \textsc{Alexander, R., Pascucci, I., Andrews, S., Armitage, P., \& Cieza, L.} 2014, 
 \emph{Protostars \& Planets VI}, Henrik Beuther, Ralf S. Klessen, Cornelis P. Dullemond, and Thomas Henning (eds.), University of Arizona Press, Tucson, p.~475

\bibitem[Alibert et al.(2005)]{alibert05}
 \textsc{Alibert, Y., Mordasini, C., Benz, W., \& Winisdoerffer, C.} 2005,
 \emph{A\&A}, 434, 343
 
\bibitem[Andrews et al.(2009)]{andrews09}
 \textsc{Andrews, S. M., Wilner, D. J., Hughes, A. M., Qi, Chunhua, \& Dullemond, C. P.} 2009, 
 \emph{ApJ}, 700, 1502 
 
\bibitem[Arakawa, Leliwa-Kopystynski \& Maeno(2002)]{arakawa02}
 \textsc{Arakawa, M., Leliwa-Kopystynski, J., \& Maeno, N.} 2002, 
 \emph{Icarus}, 158, 516
 
\bibitem[Armitage(2002)]{armitage02b}
 \textsc{Armitage, P. J.} 2002, 
 \emph{MNRAS}, 330, 895 

\bibitem[Armitage(2010)]{armitage10}
 \textsc{Armitage, P. J.} 2010, 
 \emph{Astrophysics of Planet Formation}, Cambridge University Press (Cambridge: UK)
 
\bibitem[Armitage(2011)]{armitage11}
 \textsc{Armitage, P. J.} 2011,
 \emph{ARA\&A},  49, 195
 
\bibitem[Armitage(2015)]{armitage15}
 \textsc{Armitage, P. J.} 2015, 
 to appear in \emph{From Protoplanetary Disks to Planet Formation}, proceedings of the 45th Saas-Fee Advanced Course (arXiv:1509.06382)

\bibitem[Armitage \& Clarke(1996)]{armitage96}
 \textsc{Armitage, P. J., \& Clarke, C. J.} 1996, 
 \emph{MNRAS}, 280, 458
 
\bibitem[Armitage, Livio \& Pringle(2001)]{armitage01}
 \textsc{Armitage, P. J., Livio, M., \& Pringle, J. E.} 2001, 
 \emph{MNRAS}, 324, 705
 
\bibitem[Armitage, Simon \& Martin(2013)]{armitage13}
 \textsc{Armitage, P. J., Simon, J. B., \& Martin, R. G.} 2013, 
 \emph{ApJ}, 778, article id. L14
 
\bibitem[Armitage \& Rice(2005)]{armitage05} 
 \textsc{Armitage, P. J., \& Rice, W. K. M.} 2005, 
 in \emph{A Decade Of Extrasolar Planets Around Normal Stars}, 
 STScI May Symposium 2005, astro-ph/0507492
 
\bibitem[Artymowicz(1993)]{artymowicz93}
 \textsc{Artymowicz, P.} 1993, 
 \emph{ApJ}, 419, 155 
 
\bibitem[Artymowicz et al.(1991)]{artymowicz91}
 \textsc{Artymowicz, P., Clarke, C. J., Lubow, S. H., \& Pringle, J. E.} 1991,
 \emph{ApJ}, 370, L35
 
\bibitem[Bai(2014)]{bai14}
 \textsc{Bai, X.-N.} 2014, 
 \emph{ApJ}, 791, article id. 137
 
\bibitem[Bai(2016)]{bai16b}
 \textsc{Bai, X.-N.} 2016, 
 \emph{ApJ}, 821, article id. 80
 
\bibitem[Bai \& Stone(2011)]{bai11}
 \textsc{Bai, X.-N., \& Stone, J. M.} 2011, 
 \emph{ApJ}, 736, article id. 144

\bibitem[Bai \& Stone(2013)]{bai13}
 \textsc{Bai, X.-N., \& Stone, J. M.} 2013, 
 \emph{ApJ}, 769, article id. 76
 
\bibitem[Bai et al.(2016)]{bai16}
 \textsc{Bai, X.-N., Ye, J., Goodman, J., \& Yuan, F.} 2016, 
 \emph{ApJ}, 818, article id. 152

\bibitem[Balbus(2011)]{balbus09}
 \textsc{Balbus, S. A.} 2011, 
 \emph{Physical Processes in Circumstellar Disks Around Young Stars}, ed. Paulo J.V. Garcia, University of Chicago Press, p.~237 (arXiv:0906.0854)

\bibitem[Balbus \& Hawley(1991)]{balbus91}
 \textsc{Balbus, S. A., \& Hawley, J. F.} 1991,
 \emph{ApJ}, 376, 214
 
\bibitem[Balbus \& Hawley(1998)]{balbus98}
 \textsc{Balbus, S. A., \& Hawley, J. F.} 1998,
 \emph{Reviews of Modern Physics}, 70, 1 

\bibitem[Balbus \& Hawley(2006)]{balbus06}
 \textsc{Balbus, S. A., \& Hawley, J. F.} 2006,
 \emph{ApJ}, 652, 1020

\bibitem[Balbus, Hawley \& Stone(1996)]{balbus96}
 \textsc{Balbus, S. A., Hawley, J. F., \& Stone, J. M.} 1996,
 \emph{ApJ}, 467, 76
 
\bibitem[Balbus \& Terquem(2001)]{balbus01}
 \textsc{Balbus S. A., \& Terquem C.} 2001, 
 \emph{ApJ}, 552, 235  
 
\bibitem[Bally \& Scoville(1982)]{bally82}
 \textsc{Bally, J., \& Scoville, N. Z.} 1982,
 \emph{ApJ}, 255, 497 
 
\bibitem[Barge \& Sommeria(1995)]{barge95} 	
 \textsc{Barge, P., \& Sommeria, J.} 1995, 
 \emph{A\&A}, 295, L1
 
\bibitem[Barranco \& Marcus(2005)]{barranco05}
 \textsc{Barranco, J. A., \& Marcus, P. S.} 2005, 
 \emph{ApJ}, 623, 1157
 
\bibitem[Baruteau et al.(2014)]{baruteau14} 
 \textsc{Baruteau, C., et al.} 2014, 
 \emph{Protostars \& Planets VI}, Henrik Beuther, Ralf S. Klessen, Cornelis P. Dullemond, and Thomas Henning (eds.), University of Arizona Press, Tucson, p.~667
 
\bibitem[Basri, Walkowicz \& Reiners(2012)]{basri13}
 \textsc{Basri, G., Walkowicz, L. M., \& Reiners, A.} 2013, 
 \emph{ApJ}, 769, article id. 37
 
\bibitem[Bate, Lodato \& Pringle(2010)]{bate10}
 \textsc{Bate, M. R., Lodato, G., \& Pringle, J. E.} 2010, 
 \emph{MNRAS}, 401, 1505
 
\bibitem[Batygin(2012)]{batygin12}
 \textsc{Batygin, K.} 2012, 
 \emph{Nature},  491, 418
 
\bibitem[Batygin(2015)]{batygin15}
 \textsc{Batygin, K.} 2015, 
 \emph{MNRAS}, 451, 2589
 
\bibitem[Batygin \& Adams(2017)]{batygin17}
 \textsc{Batygin, K., \& Adams, F. C.} 2017, 
 \emph{AJ}, in press (arXiv:1701.07849)
 
\bibitem[Batygin \& Brown(2010)]{batygin10b}
 \textsc{Batygin, K., \& Brown, M. E.} 2010, 
 \emph{ApJ}, 716, 1323
 
\bibitem[Batygin \& Brown(2016)]{batygin16}
 \textsc{Batygin, K., \& Brown, M. E.} 2016, 
 \emph{AJ}, 151, article id. 22 
 
\bibitem[Batygin, Brown \& Fraser(2011)]{batygin11}
 \textsc{Batygin, K., Brown, M. E., \& Fraser, W. C.} 2011, 
 \emph{ApJ}, 738, 13 
 
\bibitem[Batygin \& Stevenson(2010)]{batygin10}
 \textsc{Batygin, K., \& Stevenson, D. J.} 2010, 
 \emph{ApJ}, 714, L238 
 
\bibitem[Beaulieu et al.(2006)]{beaulieu06}	
 \textsc{Beaulieu, J.-P., et al.} 2006, 
 \emph{Nature}, 439, 437 
 
\bibitem[Begelman, McKee \& Shields(1983)]{begelman83}
 \textsc{Begelman, M. C., McKee, C. F., \& Shields, G. A.} 1983, 
  \emph{ApJ}, 271, 70

\bibitem[Bell et al.(1997)]{bell97}
 \textsc{Bell, K. R., Cassen, P. M., Klahr, H. H., \& Henning, Th.} 1997, 
 \emph{ApJ}, 486, 372
 
\bibitem[Bell \& Lin(1994)]{bell94}
 \textsc{Bell, K. R., \& Lin, D. N. C.} 1994, 
 \emph{ApJ}, 427, 987
 
\bibitem[Belyaev, Rafikov \& Stone(2013)]{belyaev13}
 \textsc{Belyaev, M. A., Rafikov, R. R., \&  Stone, J. M.} 2013, 
 \emph{ApJ}, 770, article id. 68
 
\bibitem[Benz \& Asphaug(1999)]{benz99}
 \textsc{Benz, W., \& Asphaug, E.} 1999, 
 \emph{Icarus}, 142, 5
 
\bibitem[Benz, Slattery \& Cameron(1986)]{benz86}	
 \textsc{Benz, W., Slattery, W. L., \& Cameron, A. G. W.} 1986, 
 \emph{Icarus}, 66, 515
 
\bibitem[B\'ethune, Lesur \& Ferreira(2017)]{bethune17}
 \textsc{B\'ethune, W., Lesur, G., \& Ferreira, J.} 2017, 
 \emph{A\&A}, in press (arXiv:1612.00883)
 
\bibitem[Binney \& Tremaine(1987)]{binney87}
 \textsc{Binney, J., \& Tremaine, S.} 1987, 
 \emph{Galactic Dynamics}, Princeton University Press, p.~310 
 
\bibitem[Birnstiel, Klahr \& Ercolano(2012)]{birnstiel12}
 \textsc{Birnstiel, T., Klahr, H., \& Ercolano, B.} 2012, 
 \emph{A\&A},  539, id.~A148
 
\bibitem[Birnstiel, Ormel \& Dullemond(2011)]{birnstiel11}
 \textsc{Birnstiel, T., Ormel, C. W., \& Dullemond, C. P.} 2011, 
 \emph{A\&A}, 525, id. A11
 
\bibitem[Bitsch et al.(2013)]{bitsch13}
 \textsc{Bitsch, B., Crida, A., Morbidelli, A., Kley, W., \& Dobbs-Dixon, I.} 2013, 
 \emph{A\&A}, 549, id.~A124
 
\bibitem[Bitsch, Lambrechts \& Johansen(2015)]{bitsch15}
 \textsc{Bitsch, B., Lambrechts, M., \& Johansen, A.} 2015, 
 \emph{A\&A}, 582, id.~A112
 
\bibitem[Blaes \& Balbus(1994)]{blaes94}
 \textsc{Blaes, O. M., \& Balbus, S. A.} 1994,
 \emph{ApJ}, 421, 163
 
\bibitem[Blandford \& Payne(1982)]{blandford82}
 \textsc{Blandford, R. D., \& Payne, D. G.} 1982, 
 \emph{MNRAS}, 199, 883
 
\bibitem[Blum \& Wurm(2008)]{blum08}
 \textsc{Blum, J., \& Wurm, G.} 2008, 
 \emph{ARA\&A}, 46, 21
 
\bibitem[Bodenheimer \& Pollack(1986)]{bodenheimer86}
 \textsc{Bodenheimer, P., \& Pollack, J. B.} 1986,
 \emph{Icarus}, 67, 391

\bibitem[Boley et al.(2009)]{boley09}
 \textsc{Boley, A. C.} 2009,
 \emph{ApJ}, 695, L53
  
\bibitem[Boley et al.(2010)]{boley10}
 \textsc{Boley, A. C., Hayfield, T., Mayer, L., \& Durisen, R. H.} 2010,
 \emph{Icarus}, 207, 509 
 
\bibitem[Boss(1997)]{boss97}
 \textsc{Boss, A. P.} 1997, 
 \emph{Science}, 276, 1836

\bibitem[Bouvier et al.(2007)]{bouvier06}
 \textsc{Bouvier, J., Alencar, S. H. P., Harries, T. J., Johns-Krull, C. M., \& Romanova, M. M.} 2007, 
 \emph{Protostars and Planets V}, eds B. Reipurth, D. Jewitt, and K. Keil, University of 
 Arizona Press, Tucson, astro-ph/0603498

\bibitem[Brandenburg et al.(1995)]{axel95}
 \textsc{Brandenburg, A.,, Nordlund, A., Stein, R. F., \& Torkelsson, U.} 1995,
 \emph{ApJ}, 446, 741

\bibitem[Bromley \& Kenyon(2006)]{bromley06}
 \textsc{Bromley, B. C., \& Kenyon, S. J.} 2006,
 \emph{AJ}, 131, 2737

\bibitem[Brown, Trujillo \& Rabinowitz(2004)]{brown04}
 \textsc{Brown, M. E., Trujillo, C., \& Rabinowitz, D.} 2004, 
 \emph{ApJ}, 617, 645
 
\bibitem[e.g. Burrows et al.(1996)]{burrows96}
 \textsc{Burrows, C. J.,  et al.} 1996, 
 \emph{ApJ}, 473, 437 

\bibitem[Burrows et al.(2007)]{burrows07} 
 \textsc{Burrows, A., Hubeny, I, Budaj, J., \& Hubbard, W. B.} 2007, 
 \emph{ApJ}, 661, 502
 
\bibitem[Butler et al.(1996)]{butler96}
 \textsc{Butler, R. P., Marcy, G. W., Williams, E., McCarthy, C., Dosanjh, P., \& Vogt, S. S.} 1996, 
 \emph{PASP}, 108, 500
  
\bibitem[Calvet et al.(2005)]{calvet05}
 \textsc{Calvet, N., et al.} 2005, 
 \emph{ApJ}, 630, L185 
  
\bibitem[Cameron(1978)]{cameron78}
 \textsc{Cameron, A. G. W.} 1978, 
 \emph{Moon and the Planets}, 18, 5  
  
\bibitem[Cannizzo(1993)]{cannizzo93}
 \textsc{Cannizzo, J. K.} 1993, 
 \emph{ApJ}, 419, 318  
 
\bibitem[Canup(2004)]{canup04}
 \textsc{Canup, R. M.} 2004, 
 \emph{ARA\&A}, 42, 441 
 
\bibitem[Canup(2012)]{canup12}
 \textsc{Canup, R. M.} 2012, 
 \emph{Science}, 338, 1052 
  
\bibitem[Canup \& Ward(2002)]{canup02} 
 \textsc{Canup, R. M., \& Ward, W. R.} 2002, 
 \emph{AJ}, 124, 3404
 
\bibitem[Canup \& Ward(2008)]{canup08} 
 \textsc{Canup, R. M., \& Ward, W. R.} 2009, 
 in \emph{Europa}, Edited by Robert T. Pappalardo, William B. McKinnon, Krishan K. Khurana. 
 University of Arizona Press, Tucson,  p.~59
 
\bibitem[Carlberg \& Sellwood(1985)]{carlberg85}
 \textsc{Carlberg, R. G., \& Sellwood, J. A.} 1985. 
 \emph{ApJ}, 292, 79
 
\bibitem[Carrera, Johansen \& Davies(2015)]{carrera15}
 \textsc{Carrera, D., Johansen, A., \& Davies, M. B.} 2015, 
 \emph{A\&A}, 579, id.~A43
 
\bibitem[Carter et al.(2012)]{carter12}
 \textsc{Carter,  J. A., et al.} 2012,
 \emph{Science}, 337, 556
 
\bibitem[Casoli \& Masset(2009)]{casoli09}
 \textsc{Casoli, J., \& Masset, F. S.} 2009, 
 \emph{ApJ}, 703, 845
 
\bibitem[Chambers(2014)]{chambers14}
 \textsc{Chambers, J. E.} 2014, 
 \emph{Icarus}, 233, 83
 
\bibitem[Chambers \& Wetherill(1998)]{chambers98}
 \textsc{Chambers, J. E., \& Wetherill, G. W.} 1998,  
 \emph{Icarus}, 136, 304
 
\bibitem[Chambers, Wetherill \& Boss(1996)]{chambers96}	
 \textsc{Chambers, J. E., Wetherill, G. W., \& Boss, A. P.} 2006,
 \emph{Icarus}, 119, 261
 
\bibitem[Chandrasekhar(1961)]{chandra61}
 \textsc{Chandrasekhar, S.} 1961, 
 \emph{Hydrodynamic and hydromagnetic stability}, International 
 Series of Monographs on Physics, Oxford: Clarendon
 
\bibitem[Chapman \& Cowling(1970)]{chapman70}
 \textsc{Chapman, S., \& Cowling, T. G.} 1970, 
 \emph{The mathematical theory of non-uniform gases}, 
 Cambridge University Press
 
\bibitem[Charbonneau et al.(2009)]{charbonneau09}
 \textsc{Charbonneau, D., et al.} 2009, 
 \emph{Nature}, 462, 891
 
\bibitem[Chatterjee \& Ford(2015)]{chatterjee15} 
 \textsc{Chatterjee, S., \& Ford, E. B.} 2015, 
 \emph{ApJ}, 803, article id. 33
 
\bibitem[Chatterjee et al.(2008)]{chatterjee08}
 \textsc{Chatterjee, S., Ford, E. B., Matsumura, S., \& Rasio, F. A.} 2008, 
 \emph{ApJ}, 686, 580	

\bibitem[Chiang \& Goldreich(1997)]{chiang97}
 \textsc{Chiang, E. I., \& Goldreich, P.} 1997,
 \emph{ApJ}, 490, 368

\bibitem[Chiang et al.(2007)]{chiang06}	
 \textsc{Chiang, E., Lithwick, Y., Murray-Clay, R., Buie, M., Grundy, W., \& Holman, M.} 2007, 
 \emph{Protostars and Planets V}, eds B. Reipurth, D. Jewitt, and K. Keil, University of 
 Arizona Press, Tucson, astro-ph/0601654
    
\bibitem[Clarke(2009)]{clarke09}
 \textsc{Clarke, C. J.} 2009, 
 \emph{MNRAS}, 396, 1066   
   
\bibitem[Clarke, Gendrin \& Sotomayor(2001)]{clarke01}
 \textsc{Clarke, C. J., Gendrin, A., \& Sotomayor, M.} 2001, 
 \emph{MNRAS}, 328, 485
 
\bibitem[Cleeves et al.(2015)]{cleeves15} 
 \textsc{Cleeves, L. I., Bergin, E. A., Qi, C., Adams, F. C., \& \"Oberg, K. I.} 2015, 
 \emph{ApJ}, 799, article id. 204
 
\bibitem[Cleeves et al.(2016)]{cleeves16}
 \textsc{Cleeves, L. I., \"Oberg, K. I., Wilner, D. J., Huang, J., Loomis, R. A., Andrews, S. M., \& Czekala, I.} 2016, 
 \emph{ApJ}, 832, article id.~110
 
\bibitem[Coleman \& Nelson(2016)]{coleman16}
 \textsc{Coleman, G. A. L., \& Nelson, R. P.} 2016, 
 \emph{MNRAS}, 457, 2480
 
\bibitem[Collier Cameron \& Campbell(1993)]{acc93}
 \textsc{Collier Cameron, A., \& Campbell, C. G.} 1993, 
 \emph{A\&A}, 274, 309
 
\bibitem[\'Cuk \& Stewart(2012)]{cuk12}
 \textsc{\'Cuk, M., \& Stewart, S. T.} 2012, 
 \emph{Science}, 338, 1047 
 
\bibitem[\'Cuk et al.(2016)]{cuk16} 
 \textsc{\'Cuk, M., Hamilton, D. P., Lock, S. J., \& Stewart, S. T.} 2016,
 \emph{Nature}, 539, 402

\bibitem[Cumming et al.(2008)]{cumming08}
 \textsc{Cumming, A., et al.} 2008,
 \emph{PASP}, 120, 531
 
\bibitem[Cuzzi, Dobrovolskis \& Champney(1993)]{cuzzi93}
 \textsc{Cuzzi, J. N., Dobrovolskis, A. R., \& Champney, J. M.} 1993, 
 \emph{Icarus}, 106, 102
 
\bibitem[Cuzzi, Hogan \& Shariff(2008)]{cuzzi08}
 \textsc{Cuzzi, J. N., Hogan, R. C., \& Shariff, K.} 2008,  
 \emph{ApJ}, 687, 1432
 
\bibitem[D'Angelo, Lubow \& Bate(2006)]{dangelo06}
 \textsc{D'Angelo, G., Lubow, S. H., \& Bate, M. R.} 2006, 
 \emph{ApJ}, 652, 1698
 
\bibitem[Darwin(1879)]{darwin79}
 \textsc{Darwin, G. H.} 1879, 
 \emph{Philos. Trans. of Roy. Soc.}, 170, 447
 
\bibitem[Davies et al.(2014)]{davies14}
 \textsc{Davies, M. B., Adams, F. C., Armitage, P., Chambers, J., Ford, E., Morbidelli, A., Raymond, S. N., \& Veras, D.} 2014, 
 \emph{to appear in in Protostars \& Planets VI}, eds. H. Beuther, C. Dullemond, Th. Henning, R. Klessen, University of Arizona Press  
 
\bibitem[Davis, Stone \& Pessah(2010)]{davis10}
 \textsc{Davis, S. W., Stone, J. M., \& Pessah, M. E.} 2010,   
 \emph{ApJ}, 713, 52
 
\bibitem[Dawson \& Murray-Clay(2012)]{dawson12}
 \textsc{Dawson, R. I., \& Murray-Clay, R.} 2012, 
 \emph{ApJ}, 750, 43
 
\bibitem[Deienno et al.(2017)]{deienno17} 
 \textsc{Deienno, R., Morbidelli, A., Gomes, R. S., \& Nesvorny, D.} 2017, 
 \emph{AJ}, in press (arXiv:1702.02094)
 
\bibitem[Demory \& Seager(2011)]{demory11}
 \textsc{Demory, B.-O., \& Seager, S.} 2011, 
 \emph{ApJS}, 197, article id. 12
 
\bibitem[Desch(2004)]{desch04}
 \textsc{Desch, S. J.} 2004, 
 \emph{ApJ}, 608, 509 
 
\bibitem[Dickey et al.(1994)]{dickey94}
 \textsc{Dickey, J. O., et al.} 1994,
 \emph{Science}, 265, 482
 
\bibitem[Doyle et al.(2011)]{doyle11}
 \textsc{Doyle,  L. R., et al.} 2011, 
 \emph{Science}, 333, 1602
 
\bibitem[Dubrulle, Morfill \& Sterzik(1995)]{dubrulle95}
 \textsc{Dubrulle, B., Morfill, G., \& Sterzik, M.} 1995, 
 \emph{Icarus}, 114, 237
 
\bibitem[Duffell et al.(2014)]{duffell14}  
 \textsc{Duffell, P. C., Haiman, Z., MacFadyen, A. I., D'Orazio, D. J., \& Farris, B. D.} 2014, 
 \emph{ApJ}, 792, article id. L10
 
\bibitem[Duffell \& Chiang(2015)]{duffell15}
 \textsc{Duffell, P. C., \& Chiang, E.} 2015, 
 \emph{ApJ}, 812, article id.~94 

\bibitem[Dullemond \& Dominik(2005)]{dullemond05}
 \textsc{Dullemond, C. P., \& Dominik, C.} 2005, 
 \emph{A\&A}, 434, 971

\bibitem[Dullemond et al.(2007)]{dullemond06}
 \textsc{Dullemond, C. P., Hollenbach, D., Kamp, I., \& D'Alessio, P.} 2007, 
 \emph{Protostars and Planets V}, eds B. Reipurth, D. Jewitt, and K. Keil, University of 
 Arizona Press, Tucson, astro-ph/0602619
 
\bibitem[Dunhill, Alexander \& Armitage(2013)]{dunhill13}
 \textsc{Dunhill, A. C.; Alexander, R. D., \& Armitage, P. J.} 2013, 
 \emph{MNRAS}, 428, 3072
  
\bibitem[Durisen et al.(2005)]{durisen05}
 \textsc{Durisen, R. H., Cai, K., Mej\'ia, A. C., \& Pickett, M. K.} 2005, 
 \emph{Icarus}, 173, 417
 
\bibitem[D\"urmann \& Kley(2015)]{durmann15} 
 \textsc{D\"urmann, C., \& Kley, W.} 2015, 
 \emph{A\&A}, 574, id.~A52
 
\bibitem[D\"urmann \& Kley(2017)]{durmann17} 
 \textsc{D\"urmann, C., \& Kley, W.} 2017, 
 \emph{A\&A}, 598, id.~A80

\bibitem[Eggleton(1983)]{eggleton83}
 \textsc{Eggleton, P. P.} 1983, 
 \emph{ApJ}, 268, 368

\bibitem[Eisner, Chiang \& Hillenbrand(2006)]{eisner06}
 \textsc{Eisner, J. A., Chiang, E. I., \& Hillenbrand, L. A.} 2006,
 \emph{ApJ}, 637, L133

\bibitem[Eisner et al.(2005)]{eisner05}
 \textsc{Eisner, J. A., Hillenbrand, L. A., Carpenter, J. M., \& Wolf, S.} 2005, 
 \emph{ApJ}, 635, 396
 
\bibitem[Elser et al.(2011)]{elser11}
 \textsc{Elser, S., Moore, B., Stadel, J., \& Morishima, R.} 2011, 
 \emph{Icarus}, 214, 357 

\bibitem[Ercolano, Clarke \& Drake(2009)]{ercolano09}
 \textsc{Ercolano, B., Clarke, C. J., \& Drake, J. J.} 2009,
 \emph{ApJ}, 699, 1639

\bibitem[Ercolano \& Glassgold(2013)]{ercolano13}
 \textsc{Ercolano, B., \& Glassgold, A. E.} 2013, 
 \emph{MNRAS}, 436, 3446
 
\bibitem[Espaillat et al.(2014)]{espaillat14}
 \textsc{Espaillat, C., et al.} 2014, 
 \emph{Protostars and Planets VI}, Henrik Beuther, Ralf S. Klessen, Cornelis P. Dullemond, and Thomas Henning (eds.), University of Arizona Press, Tucson, p.~497

\bibitem[Estrada et al.(2009)]{estrada09}
 \textsc{Estrada, P. R., Mosqueira, I., Lissauer, J. J., D'Angelo, G., \& Cruikshank, D. P.} 2009, 
 in \emph{Europa}, Edited by Robert T. Pappalardo, William B. McKinnon, Krishan K. Khurana. 
 University of Arizona Press, Tucson,  p.~27
 
\bibitem[Evans \& Tabachnik(1999)]{evans99}
 \textsc{Evans, N. W., \& Tabachnik, S.} 1999, 
 \emph{Nature}, 399, 41 
 
\bibitem[Everett et al.(2013)]{everett13}
 \textsc{Everett, M. E., Howell, S. B., Silva, D. R., \& Szkody, P.} 2013, 
 \emph{ApJ}, 771, article id. 107
 
\bibitem[Fabian, Pringle \& Rees(1975)]{fabian75}
 \textsc{Fabian, A. C., Pringle, J. E., \& Rees, M. J.} 1975,
 \emph{MNRAS}, 172, 15
 
\bibitem[Fabrycky et al.(2014)]{fabrycky14}
 \textsc{Fabrycky, D. C., et al.} 2014,
 \emph{ApJ}, 790, article id. 146

\bibitem[Feigelson et al.(2007)]{feigelson07}
 \textsc{Feigelson, E., Townsley, L., G\"udel, M., \& Stassun, K.} 2007, 
 \emph{Protostars and Planets V}, eds B. Reipurth, D. Jewitt, and K. Keil, University of 
 Arizona Press, Tucson, p.~313

\bibitem[Fernandez \& Ip(1984)]{fernandez84}
 \textsc{Fernandez, J. A., \& Ip, W.-H.} 1984,
 \emph{Icarus}, 58, 109
 
\bibitem[Fienga et al.(2016)]{fienga16}
 \textsc{Fienga, A., Laskar, J., Manche, H., \& Gastineau, M.} 2016, 
 \emph{A\&A}, 587, id.~L8

\bibitem[Fischer \& Valenti(2005)]{fischer05}
 \textsc{Fischer, D. A., \& Valenti, J.} 2005, 
 \emph{ApJ}, 622, 1102
 
\bibitem[Flaherty et al.(2015)]{flaherty15}
 \textsc{Flaherty, K. M., Hughes, A. M., Rosenfeld, K. A., Andrews, S. M., Chiang, E., Simon, J. B., Kerzner, S., \& Wilner, D. J.} 2015, 
 \emph{ApJ}, 813, article id. 99 

\bibitem[Font et al.(2004)]{font04}
 \textsc{Font, A. S., McCarthy, I. G., Johnstone, D., \& Ballantyne, D. R.} 2004,
 \emph{ApJ}, 607, 890
 
\bibitem[Ford, Havlickova \& Rasio(2001)]{ford01}
 \textsc{Ford, E. B., Havlickova, M., \& Rasio, F. A.} 2001,
 \emph{Icarus}, 150, 303
 
\bibitem[Ford et al.(2012)]{ford12}
 \textsc{Ford, E. B., et al.} 2012, 
 \emph{ApJ}, 756, article id. 185
 
\bibitem[Forgan, Price \& Bonnell(2017)]{forgan17} 
 \textsc{Forgan, D., Price, D. J., \& Bonnell, I.} 2017, 
 \emph{MNRAS}, 466, 3406

\bibitem[Frank, King \& Raine(2002)]{frank02}
 \textsc{Frank, J., King, A., \& Raine, D. J.} 2002, 
 \emph{Accretion Power in Astrophysics}, (3rd edition, Cambridge University Press)

\bibitem[Fraser et al.(2014)]{fraser14}
 \textsc{Fraser, W. C., Brown, M. E., Morbidelli, A., Parker, A., \& Batygin, K.} 2014, 
 \emph{ApJ}, 782, 100

\bibitem[Fraser \& Kavelaars(2009)]{fraser09}
 \textsc{Fraser, W. C., \& Kavelaars, J. J.} 2009, 
 \emph{AJ}, 137, 71

\bibitem[Fricke(1968)]{fricke68}
 \textsc{Fricke, K.} 1968, 
 \emph{Zeitschrift f\"ur Astrophysik}, 68, 317

\bibitem[Fromang \& Papaloizou(2006)]{fromang06}
 \textsc{Fromang, S., \& Papaloizou, J.} 2006,
 \emph{A\&A}, 452, 751
 
\bibitem[Galicher(2016)]{galicher16}
 \textsc{Galicher, R., et al.} 2016,
 \emph{A\&A}, 594, id.~A63

\bibitem[Gammie(1996)]{gammie96}
 \textsc{Gammie, C. F.} 1996, 
 \emph{ApJ}, 457, 355
 
\bibitem[Gammie(2001)]{gammie01}
 \textsc{Gammie, C. F.} 2001, 
 \emph{ApJ}, 553, 174

\bibitem[Garaud \& Lin(2007)]{garaud06}
 \textsc{Garaud, P., \& Lin, D. N. C.} 2007,
 \emph{ApJ}, 654, 606

\bibitem[Gaudi(2010)]{gaudi10}
 \textsc{Gaudi, B. S.} 2010,
 \emph{in Exoplanets}, ed. S. Seager, University of Arizona Press (arXiv:1002.0332)
 
\bibitem[Gilliland et al.(2012)]{gilliland12}
 \textsc{Gilliland, R. L., et al.} 2012, 
 \emph{ApJ}, 197, article id. 6
 
\bibitem[Ginsburg \& Sari(2016)]{ginsburg16}
 \textsc{Ginzburg, S., \& Sari, R.} 2016, 
 \emph{ApJ}, 819, 116

\bibitem[Gladman(1993)]{gladman93}
 \textsc{Gladman, B.} 1993,
 \emph{Icarus}, 106, 247

\bibitem[Godon \& Livio(1999)]{godon99}
 \textsc{Godon, P., \& Livio, M.} 1999, 
 \emph{ApJ}, 523, 350
 
\bibitem[Goldreich(1965)]{goldreich65}
 \textsc{Goldreich, P.} 1965, 
 \emph{MNRAS}, 130, 159 
 
\bibitem[Goldreich, Lithwick \& Sari(2004)]{goldreich04}
 \textsc{Goldreich, P., Lithwick, Y., \& Sari, R.} 2004, 
 \emph{ARA\&A}, 42, 549

\bibitem[Goldreich \& Sari(2003)]{goldreich03}
 \textsc{Goldreich, P., \& Sari, R.} 2003,
 \emph{ApJ}, 585, 1024
 
\bibitem[Goldreich \& Schlichting(2014)]{goldreich14} 
 \textsc{Goldreich, P., \& Schlichting, H. E.} 2014,
 \emph{AJ}, 147, article id. 32

\bibitem[Goldreich \& Schubert(1967)]{goldreich67} 
 \textsc{Goldreich, P., \& Schubert, G.} 1967, 
 \emph{ApJ}, 150, 571
 
\bibitem[Goldreich \& Soter(1966)]{goldreich66} 
 \textsc{Goldreich, P., \& Soter, S.} 1966,
 \emph{Icarus}, 5, 375

\bibitem[Goldreich \& Tremaine(1979)]{goldreich79}
 \textsc{Goldreich, P., \& Tremaine, S.} 1979, 
 \emph{ApJ}, 233, 857
 
\bibitem[Goldreich \& Tremaine(1980)]{goldreich80}
 \textsc{Goldreich, P., \& Tremaine, S.} 1980, 
 \emph{ApJ}, 241, 425
 
\bibitem[Goldreich \& Ward(1973)]{goldreich73}
 \textsc{Goldreich, P., \& Ward, W. R.} 1973, 
 \emph{ApJ}, 183, 1051
 
\bibitem[Gomes et al.(2005)]{gomes05}
 \textsc{Gomes, R., Levison, H. F., Tsiganis, K., \& Morbidelli, A.} 2005, 
 \emph{Nature}, 435, 466
 
\bibitem[Gomez \& Ostriker(2005)]{gomez05}
 \textsc{G\'omez, G. C., \& Ostriker, E. C.} 2005, 
 \emph{ApJ}, 630, 1093
 
\bibitem[Goodman \& Lackner(2009)]{goodman09} 
 \textsc{Goodman, J., \& Lackner, C.} 2009,
 \emph{ApJ}, 696, 2054

\bibitem[Gorti \& Hollenbach(2009)]{gorti09}
 \textsc{Gorti, U., \& Hollenbach, D.} 2009, 
 \emph{ApJ}, 690, 1539
 
\bibitem[Greenberg(2010)]{greenberg10}
 \textsc{Greenberg, R.} 2010,
 \emph{Reports on Progress in Physics}, 73, 036801
 
\bibitem[Gressel et al.(2015)]{gressel15}
 \textsc{Gressel, O., Turner, N. J., Nelson, R. P., \& McNally, C. P.} 2015, 
 \emph{ApJ}, 801, article id. 84
 
\bibitem[G\"udel et al.(2014)]{gudel14}
 \textsc{G\"udel, M., et al.} 2014, 
 \emph{Protostars and Planets VI}, Henrik Beuther, Ralf S. Klessen, Cornelis P. Dullemond, and Thomas Henning (eds.), University of Arizona Press, Tucson, p.~883

\bibitem[Guillot(2005)]{guillot05}
 \textsc{Guillot, T.} 2005, 
 \emph{Annual Review of Earth and Planetary Sciences}, 33, 493
 
\bibitem[Guillot \& Showman(2002)]{guillot02}
 \textsc{Guillot, T., \& Showman, A. P.} 2002, 
 \emph{A\&A}, 385, 156 

\bibitem[Gullbring et al.(1998)]{gullbring98}	
 \textsc{Gullbring, E., Hartmann, L., Briceno, C., \& Calvet, N.} 1998,
 \emph{ApJ}, 492, 323
 
\bibitem[Gundlach \& Blum(2015)]{gundlach15} 
 \textsc{Gundlach, B., \& Blum, J.} 2015, 
 \emph{ApJ}, 798, article id. 34
 
\bibitem[G\"uttler et al.(2010)]{guttler10}
 \textsc{G\"uttler, C., Blum, J., Zsom, A., Ormel, C. W., \& Dullemond, C. P.} 2010, 
 \emph{A\&A}, 513, 56
 
\bibitem[Hahn \& Malhotra(2005)]{hahn05}
 \textsc{Hahn, J. M., \& Malhotra, R.} 2005, 
 \emph{AJ}, 130, 2392
 
\bibitem[Haisch, Lada \& Lada(2001)]{haisch01}
 \textsc{Haisch, K. E., Lada, E. A., \& Lada, C. J.} 2001,
 \emph{ApJ}, 553, L153
 
\bibitem[Hands, Alexander \& Dehnen(2014)]{hands14}
 \textsc{Hands, T. O., Alexander, R. D., \& Dehnen, W.} 2014, 
 \emph{MNRAS}, 445, 749

\bibitem[Hansen(2009)]{hansen09}
 \textsc{Hansen, B. M. S.} 2009,
 \emph{ApJ},  703, 1131

\bibitem[Hansen(2010)]{hansen10}
 \textsc{Hansen, B. M. S.} 2010,
 \emph{ApJ},  723, 285
 
\bibitem[Hartman et al.(2014)]{hartman14}
 \textsc{Hartman, J. D., et al.} 2014,
 \emph{AJ}, 147, article id. 128

\bibitem[Hartmann et al.(1998)]{hartmann98}
 \textsc{Hartmann, L., Calvet, N., Gullbring, E., \& D'Alessio, P.} 1998,
 \emph{ApJ}, 495, 385
 
\bibitem[Hartmann \& Davis(1975)]{hartmann75}
 \textsc{Hartmann, W. K., \& Davis, D. R.} 1975, 
 \emph{Icarus}, 24, 504
 
\bibitem[Hartmann \& Kenyon(1995)]{hartmann95}
 \textsc{Hartmann, L., \& Kenyon, S. J.} 1995, 
 \emph{ARA\&A}, 34, 207
 
\bibitem[Hayashi(1981)]{hayashi81}
 \textsc{Hayashi, C.} 1981, 
 \emph{Progress of Theoretical Physics Supplement}, 70, 35 
 
\bibitem[Herbst \& Mundt(2005)]{herbst05}
 \textsc{Herbst, W., \& Mundt, R.} 2005,
 \emph{ApJ}, 633, 967
 
\bibitem[Hirose \& Turner(2011)]{hirose11}
 \textsc{Hirose, S., \& Turner, N. J.} 2011, 
 \emph{ApJ},  732, article id. L30
 
\bibitem[Hollenbach et al.(1994)]{hollenbach94}
 \textsc{Hollenbach, D., Johnstone, D., Lizano, S., \& Shu, F.} 1994, 
 \emph{ApJ}, 428, 654
 
\bibitem[Holman \& Murray(2005)]{holman05}
 \textsc{Holman, M. J., \& Murray, N. W.} 2005, 
 \emph{Science}, 307, 1288
 
\bibitem[Holman \& Wisdom(1993)]{holman93}
 \textsc{Holman, M. J., \& Wisdom, J.} 1993, 
 \emph{AJ}, 105, 1987
 
\bibitem[Howard et al.(2012)]{howard12}
 \textsc{Howard, A. W., et al.} 2012, 
 \emph{ApJS}, 201, article id. 15
 
\bibitem[Hubickyj, Bodenheimer \& Lissauer(2005)]{hubickyj05}
 \textsc{Hubickyj, O., Bodenheimer, P., \& Lissauer, J. J.} 2005, 
 \emph{Icarus}, 179, 415
 
\bibitem[Hut(1981)]{hut81}
 \textsc{Hut, P.} 1981, 
 \emph{A\&A}, 99, 126 
 
\bibitem[Ida et al.(2000)]{ida00}
 \textsc{Ida, S., Bryden, G., Lin, D. N. C., \& Tanaka, H.} 2000, 
 \emph{ApJ}, 534, 428
 
\bibitem[Ida, Guillot \& Morbidelli(2008)]{ida08}
 \textsc{Ida, S., Guillot, T., \& Morbidelli, A.} 2008, 
 \emph{ApJ}, 686, 1292 
 
\bibitem[Ida \& Lin(2004)]{ida04}
 \textsc{Ida, S., Lin, D. N. C.} 2004, 
 \emph{ApJ}, 616, 567
 
\bibitem[Ikoma, Nakazawa \& Emori(2000)]{ikoma00}
 \textsc{Ikoma, M., Nakazawa, K., \& Emori, H.} 2000,  
 \emph{ApJ}, 537, 1013
 
\bibitem[Ioannou \& Kakouris(2001)]{ioannou01}	
 \textsc{Ioannou, P. J., \& Kakouris, A.} 2001, 
 \emph{ApJ}, 550, 931
 
\bibitem[Inaba et al.(2001)]{inaba01}
 \textsc{Inaba, S., Tanaka, H., Nakazawa, K., Wetherill, G. W., \& Kokubo, E.} 2001, 
 \emph{Icarus}, 149, 235
 
\bibitem[Ivanov, Papaloizou \& Polnarev(1999)]{ivanov99}
 \textsc{Ivanov, P. B., Papaloizou, J. C. B., \& Polnarev, A. G.} 1999, 
 \emph{MNRAS}, 307, 79
 
\bibitem[Jackson, Greenberg \& Barnes(2008)]{jackson08}
 \textsc{Jackson, B., Greenberg, R., \& Barnes, R.} 2008, 
 \emph{ApJ}, 678, 1396
 
\bibitem[Jewitt \& Haghighipour(2007)]{jewitt07}	
 \textsc{Jewitt, D., \& Haghighipour, N.} 2003, 
 \emph{ARA\&A}, 45, 261
 
\bibitem[Jewitt \& Luu(1993)]{jewitt93}
 \textsc{Jewitt, D., \& Luu, J.} 1993, 
 \emph{Nature}, 362, 730  
 
\bibitem[Ji et al.(2006)]{ji06}
 \textsc{Ji, H., Burin, M., Schartman, E., \& Goodman, J.} 2006,
 \emph{Nature}, 444, 343
 
\bibitem[Johansen, Henning \& Klahr(2006)]{johansen06}
 \textsc{Johansen, A., Henning, T., \& Klahr, H.} 2006,  
 \emph{ApJ}, 643, 1219
 
\bibitem[Johansen et al.(2007)]{johansen07}
 \textsc{Johansen, A., Oishi, J. S., Mac Low, M.-M., Klahr, H., Henning, T., \& Youdin, A.} 2007, 
 \emph{Nature}, 448, 1022
 
\bibitem[Johansen, Youdin \& Klahr(2009b)]{johansen09b} 
 \textsc{Johansen, A., Youdin, A., \& Klahr, H.} 2009,
 \emph{ApJ}, 697, 1269
 
\bibitem[Johansen, Youdin \& Mac Low(2009)]{johansen09}
 \textsc{Johansen, A., Youdin, A., \& Mac Low, M.-M.} 2009, 
 \emph{ApJ}, 704, L75
 
\bibitem[Johnson \& Gammie(2005)]{johnson05}
 \textsc{Johnson, B. M., \& Gammie, C. F.} 2005, 
 \emph{ApJ}, 635, 149
 
\bibitem[Johnstone, Hollenbach \& Bally(1998)]{johnstone98}
 \textsc{Johnstone, D., Hollenbach, D., \& Bally, J.} 1998,
 \emph{ApJ}, 499, 758
 
\bibitem[Juri\'c \& Tremaine(2008)]{juric08}
 \textsc{Juri\'c, M., \& Tremaine, S.} 2008, 
 \emph{ApJ}, 686, 603 
 
\bibitem[Kaib \& Chambers(2016)]{kaib16}
 \textsc{Kaib, N. A., \& Chambers, J. E.} 2016, 
 \emph{MNRAS},  455, 3561
 
\bibitem[Kasting, Whitmire \& Reynolds(1993)]{kasting93}
 \textsc{Kasting, J. F., Whitmire, D. P., \& Reynolds, R. T.} 1993, 
 \emph{Icarus}, 101, 108
 
\bibitem[Kenyon \& Hartmann(1987)]{kenyon87}
 \textsc{Kenyon, S. J., \& Hartmann, L.} 1987, 
 \emph{ApJ}, 323, 714
 
\bibitem[Kenyon \& Luu(1998)]{kenyon98}
 \textsc{Kenyon, S. J., \& Luu, J. X.} 1998,
 \emph{AJ}, 115, 2136 
 
\bibitem[King, Pringle \& Livio(2007)]{king07}
 \textsc{King, A. R., Pringle, J. E., \& Livio, M.} 2007, 
 \emph{MNRAS}, 376, 1740
 
\bibitem[Kirsh et al.(2009)]{kirsh09}
 \textsc{Kirsh, D. R., Duncan, M., Brasser, R., \& Levison, H. F.} 2009, 
 \emph{Icarus}, 199, 197 
 
\bibitem[Klahr \& Bodenheimer(2003)]{klahr03}
 \textsc{Klahr, H. H., \& Bodenheimer, P.} 2003,
 \emph{ApJ}, 582, 869 
 
\bibitem[Kley, Bitsch \& Klahr(2009)]{kley09}
 \textsc{Kley, W., Bitsch, B., \& Klahr, H.} 2009, 
 \emph{A\&A}, 506, 971 
 
\bibitem[Kley \& Crida(2008)]{kley08}
 \textsc{Kley, W., \& Crida, A.} 2008, 
 \emph{A\&A}, 487, L9
 
\bibitem[Kley \& Nelson(2012)]{kley12}
 \textsc{Kley, W., \& Nelson, R. P.} 2012, 
 \emph{ARA\&A}, 50, 211
 
\bibitem[Kobayashi et al.(2012)]{kobayashi12}
 \textsc{Kobayashi, S., Hainick, Y., Sari, R., \& Rossi, E. M.} 2012, 
 \emph{ApJ}, 748, 105  
 
\bibitem[Kokubo \& Ida(1998)]{kokubo98}
 \textsc{Kokubo, E., \& Ida, S.} 1998, 
 \emph{Icarus}, 131, 171
 
\bibitem[Kokubo, Ida \& Makino(2000)]{kokubo00}
 \textsc{Kokubo, E., Ida, S., \& Makino, J.} 2000, 
 \emph{Icarus}, 148, 419 
 
\bibitem[Kokubo, Kominami \& Ida(2006)]{kokubo06}
 \textsc{Kokubo, E., Kominami, J., \& Ida, S.} 2006, 
 \emph{ApJ}, 642, 1131
 
\bibitem[K\"onigl(1991)]{konigl91}
 \textsc{K\"onigl, A.} 1991, 
 \emph{ApJ}, 370, L39 
 
\bibitem[K\"onigl \& Salmeron(2011)]{konigl10}	
 \textsc{K\"onigl, A., \& Salmeron, R.} 2011, 
 \emph{Physical Processes in Circumstellar Disks around Young Stars}, ed. Paulo J.V. Garcia, University of Chicago Press, p.~283
 
 \bibitem[Kopparapu et al.(2013)]{kopparapu13}
 \textsc{Kopparapu, R. K., et al.} 2013, 
 \emph{ApJ}, 765, article id. 131
 
\bibitem[Kopparapu et al.(2014)]{kopparapu14}
 \textsc{Kopparapu, R. K., Ramirez, R. M., SchottelKotte, J., Kasting, J. F., Domagal-Goldman, S., \& Eymet, V.} 2014,  
 \emph{ApJL}, 787, article id. L29
 
\bibitem[Korycansky \& Asphaug(2006)]{korycansky06}	
 \textsc{Korycansky, D. G., \& Asphaug, E.} 2006, 
 \emph{Icarus}, 181, 605
 
\bibitem[Kozai(1962)]{kozai62}
 \textsc{Kozai, Y.} 1962,
 \emph{AJ}, 67, 591
 
\bibitem[Kratter \& Lodato(2016)]{kratter16} 
 \textsc{Kratter, K., \& Lodato, G.} 2016, 
 \emph{ARA\&A}, 54, 271
 
\bibitem[Kratter, Murray-Clay \& Youdin(2010)]{kratter10}
 \textsc{Kratter, K. M., Murray-Clay, R. A., \& Youdin, A. N.} 2010, 
 \emph{ApJ}, 710, 1375
 
\bibitem[Kuiper(1951)]{kuiper51} 
 \textsc{Kuiper, G. P.} 1951, 
 \emph{Proc. Natl. Acad. Sci.}, 37, 1
  
 \bibitem[Kunz(2008)]{kunz08}
  \textsc{Kunz M. W.} 2008, 
  \emph{MNRAS}, 385, 1494 
  
 \bibitem[Kunz \& Balbus(2004)]{kunz04}
  \textsc{Kunz M. W., \& Balbus S. A.} 2004, 
  \emph{MNRAS}, 348, 355   
  
\bibitem[Lada \& Lada(2003)]{lada03}
 \textsc{Lada, C. J., \& Lada, E. A.} 2003, 
 \emph{ARA\&A}, 41, 57
 
\bibitem[Lainey et al.(2009)]{lainey09}
 \textsc{Lainey, V., Arlot, J.-E., Karatekin, O., \& van Hoolst, T.} 2009, 
 \emph{Nature}, 459, 957

\bibitem[Lambrechts \& Johansen(2012)]{lambrechts12} 
 \textsc{Lambrechts, M., \& Johansen, A.} 2012, 
 \emph{A\&A}, 544, id.~A32
 
\bibitem[Lambrechts \& Johansen(2014)]{lambrechts14} 
 \textsc{Lambrechts, M., \& Johansen, A.} 2014, 
 \emph{A\&A}, 572, id.~A107 
  
\bibitem[Laughlin, Steinacker \& Adams(2004)]{laughlin04}
 \textsc{Laughlin, G., Steinacker, A., Adams, F. C.} 2004, 
 \emph{ApJ}, 608, 489
 
\bibitem[Lecar et al.(2006)]{lecar06}
 \textsc{Lecar, M., Podolak, M., Sasselov, D., \& Chiang, E.} 2006, 
 \emph{ApJ}, 640, 1115

\bibitem[e.g. Lee(2000)]{lee00}
 \textsc{Lee, M. H.} 2000, 
 \emph{Icarus}, 143, 74
 
\bibitem[Lee \& Peale(2002)]{lee02} 
 \textsc{Lee, M. H., \& Peale, S. J.} 2002, 
 \emph{ApJ}, 567, 596
 
\bibitem[Leinhardt \& Richardson(2002)]{leinhardt02}
 \textsc{Leinhardt, Z. M., \& Richardson, D. C.} 2002, 
 \emph{Icarus}, 159, 306

\bibitem[Leinhardt \& Stewart(2009)]{leinhardt08}
 \textsc{Leinhardt, Z. M., \& Stewart, S. T.} 2009, 
 {\emph Icarus}, 199, 542
 
\bibitem[Lesur \& Latter(2016)]{lesur16}
 \textsc{Lesur, G. R. J., \& Latter, H.} 2016, 
 \emph{MNRAS}, 462, 4549

\bibitem[Lesur, Kunz \& Fromang(2014)]{lesur14} 
 \textsc{Lesur, G., Kunz, M. W., \& Fromang, S.} 2014, 
 \emph{A\&A}, 566, id.A56
 
\bibitem[Lesur \& Ogilvie(2010)]{lesur10}	
 \textsc{Lesur, G., \& Ogilvie, G. I.} 2010,
 \emph{MNRAS}, 404, L64
 
\bibitem[Lesur \& Papaloizou(2009)]{lesur09}
 \textsc{Lesur, G., \& Papaloizou, J. C. B.} 2009, 
 \emph{A\&A}, 498, 1
 
\bibitem[Lesur \& Papaloizou(2010)]{lesur10b}
 \textsc{Lesur, G., \& Papaloizou, J. C. B.} 2010, 
 \emph{A\&A}, 513, 60 
 
\bibitem[Levin(2007)]{levin07}
 \textsc{Levin, Y.} 2007, 
 \emph{MNRAS}, 374, 515 
 
\bibitem[Levison \& Agnor(2003)]{levison03}
 \textsc{Levison, H. F., \& Agnor, C.} 2003,
 \emph{AJ}, 125, 2692
 
\bibitem[Levison, Kretke \& Duncan(2015a)]{levison15a} 
 \textsc{Levison, H. F., Kretke, K. A., \& Duncan, M. J.} 2015, 
 \emph{Nature},  524, 322
 
\bibitem[Levison et al.(2015b)]{levison15b}
 \textsc{Levison, H. F., Kretke, K. A., Walsh, K. J., \& Bottke, W. F.} 2015, 
 \emph{PNAS}, 112, 14180

\bibitem[Levison et al.(2007)]{levison06}
 \textsc{Levison, H. E., Morbidelli, A., Gomes, R., \& Backman, D.} 2007,
 \emph{Protostars and Planets V}, eds B. Reipurth, D. Jewitt, and K. Keil, University of 
 Arizona Press, Tucson
 
\bibitem[Levison et al.(2011)]{levison11}
 \textsc{Levison, H. F., Morbidelli, A., Tsiganis, K., Nesvorn\'y, D., \& Gomes, R.} 2011, 
 \emph{AJ}, 142, article id.~152

\bibitem[Levison et al.(2008)]{levison08}
 \textsc{Levison, H. F., Morbidelli, A., Van Laerhoven, C., Gomes, R., \& Tsiganis, K.} 2008, 
 \emph{Icarus}, 196, 258
 
\bibitem[Levison, Thommes \& Duncan(2010)]{levison10}
 \textsc{Levison, H. F., Thommes, E., \& Duncan, M. J.} 2010,
 \emph{AJ}, 139, 1297
 
\bibitem[Li et al.(2014)]{li14}
 \textsc{Li, Z.-Y.; Banerjee, R.; Pudritz, R. E.; Jorgensen, J. K.; Shang, H.; Krasnopolsky, R.; Maury, A.} 2014, 
 \emph{Protostars and Planets VI}, Henrik Beuther, Ralf S. Klessen, Cornelis P. Dullemond, and Thomas Henning (eds.), University of Arizona Press, Tucson, p.~173
 
\bibitem[Lidov(1962)]{lidov62}
 \textsc{Lidov, M. L.} 1962, 
 \emph{Planetary and Space Science}, 9, 719

\bibitem[Lin, Bodenheimer \& Richardson(1996)]{lin96}
 \textsc{Lin, D. N. C., Bodenheimer, P., \& Richardson, D. C.} 1996,
 \emph{Nature}, 380, 606

\bibitem[Lin \& Ida(1997)]{lin97}
 \textsc{Lin, D. N. C., \& Ida, S.} 1997,
 \emph{ApJ}, 477, 781

\bibitem[Lin \& Papaloizou(1979)]{lin79}
 \textsc{Lin, D. N. C., \& Papaloizou, J.} 1979, 
 \emph{MNRAS}, 186, 799

\bibitem[Lin \& Papaloizou(1980)]{lin80}
 \textsc{Lin, D. N. C., \& Papaloizou, J.} 1980, 
 \emph{MNRAS}, 191, 37

\bibitem[Lin \& Pringle(1990)]{lin90}
 \textsc{Lin, D. N. C., \& Pringle, J. E.} 1990,
 \emph{ApJ}, 358, 515
 
\bibitem[Lin \& Youdin(2015)]{lin15}
 \textsc{Lin, M.-K., \& Youdin, A. N.} 2015, 
 \emph{ApJ}, 811, article id. 17
 
\bibitem[Lines et al.(2014)]{lines14}
 \textsc{Lines, S., Leinhardt, Z. M., Paardekooper, S., Baruteau, C., \& Thebault, P.} 2014, 
 \emph{ApJ}, 782, article id. L11

\bibitem[Lissauer(1993)]{lissauer93}
 \textsc{Lissauer, J. J.} 1993,
 \emph{ARA\&A}, 31, 129 
 
\bibitem[Lissauer et al.(2012)]{lissauer12}
 \textsc{Lissauer, J. J., et al.} 2012,
 \emph{ApJ},  750, article id. 112
 
\bibitem[Lithwick(2009)]{lithwick09}
 \textsc{Lithwick, Y.} 2009, 
 \emph{ApJ}, 693, 85 
 
\bibitem[Lodders(2003)]{lodders03} 
 \textsc{Lodders, K.} 2003, 
 \emph{ApJ}, 591, 1220
 
\bibitem[Lovett(1895)]{lovett95}
 \textsc{Lovett, E. O.} 1895, 
 \emph{AJ}, 15, 113  
 
\bibitem[Lubow \& Ida(2010)]{lubow10}
 \textsc{Lubow, S. H., \& Ida, S.} 2010, 
 in \emph{Exoplanets}, editor S. Seager, University of Arizona Press, arXiv:1004.4137v1
 
\bibitem[Lubow, Siebert \& Artymowicz(1999)]{lubow99}
 \textsc{Lubow, S. H., Seibert, M., \& Artymowicz, P.} 1999, 
 \emph{ApJ}, 526, 1001
 
\bibitem[Lunine \& Stevenson(1982)]{lunine82} 
 \textsc{Lunine, J. I., \& Stevenson, D. J.} 1982, 
 \emph{Icarus}, 52, 14
 
\bibitem[Lynden-Bell(1969)]{lyndenbell69}
 \textsc{Lynden-Bell, D.} 1969,
 \emph{Nature}, 223, 690 
 
\bibitem[Lynden-Bell \& Pringle(1974)]{lyndenbell74}
 \textsc{Lynden-Bell, D., \& Pringle, J. E.} 1974, 
 \emph{MNRAS}, 168, 603
 
\bibitem[Lyra et al.(2009)]{lyra09}
 \textsc{Lyra, W., Johansen, A., Zsom, A., Klahr, H., \& Piskunov, N.} 2009, 
 \emph{A\&A}, 497, 869 
 
\bibitem[Madigan \& McCourt(2016)]{madigan16}
 \textsc{Madigan, A.-M., \& McCourt, M.} 2016, 
 \emph{MNRAS}, 457, L89
 
\bibitem[Malhotra(1993)]{malhotra93}
 \textsc{Malhotra, R.} 1993, 
 \emph{Nature}, 365, 819
 
\bibitem[Malhotra(1995)]{malhotra95}
 \textsc{Malhotra, R.} 1995, 
 \emph{AJ}, 110, 420
 
\bibitem[Marchal \& Bozis(1982)]{marchal82}
 \textsc{Marchal, C., \& Bozis, G.} 1982, 
 \emph{Celestial Mechanics}, 26, 311
 
\bibitem[Marcus et al.(2015)]{marcus15}
 \textsc{Marcus, P. S., Pei, S., Jiang, C.-H., Barranco, J. A., Hassanzadeh, P., \& Lecoanet, D.} 2015,
 \emph{ApJ}, 808, article id. 87
 
\bibitem[Marcy et al.(2008)]{marcy08}
 \textsc{Marcy,  G. W., et al.} 2008, 
 \emph{Physica Scripta}, 130, id. 014001
 
\bibitem[Marcy et al.(2014)]{marcy14}
 \textsc{Marcy, G. W., et al.} 2014,
 \emph{ApJS}, 210, article id. 20
 
\bibitem[Marois et al.(2008)]{marois08}
 \textsc{Marois, C., et al.} 2008, 
 \emph{Science}, 322, 1348 
 
\bibitem[Marois et al.(2010)]{marois10}
 \textsc{Marois, C., Zuckerman, B., Konopacky, Q. M., Macintosh, B., \& Barman, T.} 2010, 
 \emph{Nature}, 468, 1080
 
\bibitem[Martin \& Lubow(2011)]{martin11}
 \textsc{Martin, R. G., \& Lubow, S. H.} 2011,  
 \emph{MNRAS}, 413, 1447
 
\bibitem[Marzari \& Weidenschilling(2002)]{marzari02}	
 \textsc{Marzari, F., \& Weidenschilling, S. J.} 2002, 
 \emph{Icarus}, 156, 570
 
\bibitem[Masset \& Casoli(2009)]{masset09}
 \textsc{Masset, F. S., \& Casoli, J.} 2009, 
 \emph{ApJ}, 703, 857
 
\bibitem[Masset \& Ogilvie(2004)]{masset04}
 \textsc{Masset, F. S., \& Ogilvie, G. I.} 2004, 
 \emph{ApJ}, 615, 1000
 
\bibitem[Masset \& Snellgrove(2001)]{masset01}
 \textsc{Masset, F., \& Snellgrove, M.} 2001, 
 \emph{MNRAS}, 320, L55
 
\bibitem[Mathis, Rumpl \& Nordsieck(1977)]{mathis77}
 \textsc{Mathis, J. S., Rumpl, W., \& Nordsieck, K. H.} 1977, 
 \emph{ApJ}, 217, 425 

\bibitem[Matt \& Pudritz(2005)]{matt05}
 \textsc{Matt, S., \& Pudritz, R. E.} 2005, 
 \emph{ApJ}, 632, L135
 
\bibitem[Mayor \& Queloz(1995)]{mayor95}
 \textsc{Mayor, M., \& Queloz, D.} 1995, 
 \emph{Nature}, 378, 355 
 
\bibitem[McLaughlin(1924)]{mclaughlin24}	
 \textsc{McLaughlin, D. B.} 1924, 
 \emph{ApJ}, 60, 22
 
\bibitem[Meru \& Bate(2011)]{meru11}
 \textsc{Meru, F., \& Bate, M. R.} 2011, 
 \emph{MNRAS}, 411, L1
 
\bibitem[Militzer et al.(2008)]{militzer08}
 \textsc{Militzer, B., Hubbard, W. B., Vorberger, J., Tamblyn, I., \& Bonev, S. A.} 2008, 
 \emph{ApJ}, 688, L45
  
\bibitem[Miller \& Stone(2000)]{miller00}
 \textsc{Miller, K. A., \& Stone, J. M.} 2000, 
 \emph{ApJ}, 534, 398
 
\bibitem[Minton \& Malhotra(2013)]{minton13}
 \textsc{Minton, D. A., Malhotra, R.} 2013, 
 \emph{ApJ}, 732, article id. 53

\bibitem[Mizuno(1980)]{mizuno80}
 \textsc{Mizuno, H.} 1980, 
 \emph{Progress of Theoretical Physics}, 64, 544
 
\bibitem[Moeckel \& Armitage(2012)]{moeckel12}
 \textsc{Moeckel, N., \& Armitage, P. J.} 2012, 
 \emph{MNRAS}, 419, 366 
 
\bibitem[Morbidelli et al.(2009)]{morbidelli09}
 \textsc{Morbidelli, A., Bottke, W., Nesvorny, D., \& Levison, H. F.} 2009, 
 \emph{Icarus}, 204, 558
 
\bibitem[Morbidelli et al.(2000)]{morbi00} 
 \textsc{Morbidelli, A., Chambers, J., Lunine, J. I., Petit, J. M., Robert, F., Valsecchi, G. B., \& Cyr, K. E.} 2000,
 \emph{Meteoritics \& Planetary Science}, 35, 1309
 
\bibitem[Morbidelli \& Crida(2007)]{morbidelli07}	
 \textsc{Morbidelli, A., Crida, A.} 2007, 
 \emph{Icarus}, 191, 158
 
\bibitem[Morbidelli et al.(2005)]{morbi05}
 \textsc{Morbidelli, A., Levison, H. F., Tsiganis, K., \& Gomes, R.} 2005, 
 \emph{Nature}, 435, 462
 
\bibitem[Morbidelli \& Nesvorny(2012)]{morbidelli12}
 \textsc{Morbidelli, A., \& Nesvorny, D.} 2012, 
 \emph{A\&A}, 546, id.~A18
 
\bibitem[Morbidelli et al.(2007)]{morbi07}
 \textsc{Morbidelli, A., Tsiganis, K., Crida, A., Levison, H. F., \& Gomes, R.} 2007, 
 \emph{Icarus}, 134, 1790

\bibitem[Mosqueira \& Estrada(2003a)]{estrada03a} 
 \textsc{Mosqueira, I., \& Estrada, P. R.} 2003, 
 \emph{Icarus}, 163, 198
 
\bibitem[Mosqueira \& Estrada(2003b)]{estrada03b} 
 \textsc{Mosqueira, I., \& Estrada, P. R.} 2003, 
 \emph{Icarus}, 163, 232 
  
\bibitem[Movshovitz et al.(2010)]{movshovitz10}
 \textsc{Movshovitz, N., Bodenheimer, P., Podolak, M., \& Lissauer, J. J.} 2010,
 \emph{Icarus}, 209, 616
 
\bibitem[Murray \& Dermott(1999)]{murray99}
 \textsc{Murray, C. D., \& Dermott, S. F.} 1999,
 \emph{Solar System Dynamics}, Cambridge University Press
 
\bibitem[Murray-Clay \& Chiang(2005)]{murray-clay05}
 \textsc{Murray-Clay, R. A., \& Chiang, E. I.} 2005,
 \emph{ApJ}, 619, 623
 
\bibitem[Mustill \& Wyatt(2011)]{mustill11}
 \textsc{Mustill, A. J., \& Wyatt, M. C.} 2011,
 \emph{MNRAS},  413, 554
 
\bibitem[Muzerolle et al.(2010)]{muzerolle10} 
 \textsc{Muzerolle, J., Allen, L. E., Megeath, S. T., Hernandez, J., \& Gutermuth, R. A.} 2010,  
 \emph{ApJ}, 708, 1107
 
\bibitem[Nagasawa, Ida \& Bessho(2008)]{nagasawa08}
 \textsc{Nagasawa, M., Ida, S., \& Bessho, T.} 2008,
 \emph{ApJ}, 678, 498
 
\bibitem[Nakagawa, Sekiya \& Hayashi(1986)]{nakagawa86} 
 \textsc{Nakagawa, Y., Sekiya, M., \& Hayashi, C.} 1986, 
 \emph{Icarus}, 67, 375
 
\bibitem[Naoz(2016)]{naoz16}
 \textsc{Naoz, S.} 2016, 
 \emph{ARA\&A},  54, 441

\bibitem[Nelson(2005)]{nelson05}
 \textsc{Nelson, R. P.} 2005,
 \emph{A\&A}, 443, 1067
 
\bibitem[Nelson, Gressel \& Umurhan(2013)]{nelson13} 
 \textsc{Nelson, R. P., Gressel, O., \& Umurhan, O. M.} 2013, 
 \emph{MNRAS}, 435, 2610

\bibitem[Nelson \& Papaloizou(2004)]{nelson04}
 \textsc{Nelson, R. P., \& Papaloizou, J. C. B.} 2004,
 \emph{MNRAS}, 350, 849
 
\bibitem[Nesvorn\'y(2015)]{nesvorny15}
 \textsc{Nesvorn\'y, D.} 2015, 
 \emph{AJ}, 150, article id. 73
 
\bibitem[Nesvorn\'y, Vokrouhlick\'y \& Deienno(2014)]{nesvorny14}
 \textsc{Nesvorn\'y, D., Vokrouhlick\'y, D., \& Deienno, R.} 2014, 
 \emph{ApJ},  784, article id. 22
 
\bibitem[Nesvorn\'y, Vokrouhlick\'y \& Morbidelli(2007)]{nesvorny07}
 \textsc{Nesvorn\'y, D., Vokrouhlick\'y, D., \& Morbidelli, A.} 2007, 
 \emph{AJ}, 133, 1962 
 
\bibitem[Nesvorn\'y, Youdin \& Richardson(2010)]{nesvorny10}
 \textsc{Nesvorn\'y, D., Youdin, A. N., \& Richardson, D. C.} 2010, 
 \emph{AJ}, 140, 785  
 
\bibitem[Nettelmann et al.(2008)]{nettelmann08}
 \textsc{Nettelmann, N., Holst, B., Kietzmann, A., French, M., Redmer, R., \& Blaschke, D.} 2008,  
 \emph{ApJ}, 683, 1217
 
\bibitem[Norman \& Nemchin(2014)]{norman14} 
 \textsc{Norman, M. D., \& Nemchin, A. A.} 2014, 
 \emph{Earth and Planetary Science Letters}, 388, 387

\bibitem[O'Dell, Wen \& Hu(1993)]{odell93}
 \textsc{O'Dell, C. R., Wen, Z., \& Hu, X.} 1993,
 \emph{ApJ}, 410, 696

\bibitem[Ogihara \& Ida(2012)]{ogihara12}
 \textsc{Ogihara, M., \& Ida, S.} 2012,  
 \emph{ApJ}, 753, 60
 
\bibitem[Ogilvie(2014)]{ogilvie14}
 \textsc{Ogilvie, G. I.} 2014, 
 \emph{ARA\&A}, 52, 171
  
\bibitem[Ogilvie \& Lin(2004)]{ogilvie04}
 \textsc{Ogilvie, G. I., \& Lin, D. N. C.} 2004, 
 \emph{ApJ}, 610, 477
 
\bibitem[Ogilvie \& Lubow(2003)]{ogilvie03}
 \textsc{Ogilvie, G. I., \& Lubow, S. H.} 2003,
 \emph{ApJ}, 587, 398
 
\bibitem[Okuzumi \& Ormel(2013)]{okuzumi13}
 \textsc{Okuzumi, S., \& Ormel, C. W.} 2013, 
 \emph{ApJ}, 771, article id. 43
 
\bibitem[Okuzumi et al.(2012)]{okuzumi12}
 \textsc{Okuzumi, S., Tanaka, H., Kobayashi, H., \& Wada, K.} 2012, 
 \emph{ApJ}, 752, article id. 106 
 
\bibitem[Ormel(2013)]{ormel13b}
 \textsc{Ormel, C. W.} 2013,
 \emph{MNRAS}, 428, 3526
 
\bibitem[Ormel \& Cuzzi(2007)]{ormel07}
 \textsc{Ormel, C. W., \& Cuzzi, J. N.} 2007, 
 \emph{A\&A}, 466, 413
 
\bibitem[Ormel \& Klahr(2010)]{ormel10}
 \textsc{Ormel, C. W., \& Klahr, H. H.} 2010, 
 \emph{A\&A}, 520, id. A43

\bibitem[Ormel \& Okuzumi(2013)]{ormel13}
 \textsc{Ormel, C. W., \& Okuzumi, S.} 2013, 
 \emph{ApJ}, 771, article id. 44
 
\bibitem[Pardekooper(2012)]{paardekooper12}
 \textsc{Paardekooper, S.-J.} 2012, 
 \emph{MNRAS}, 421, 3286
 
\bibitem[Paardekooper et al.(2010)]{paardekooper09b}
 \textsc{Paardekooper, S.-J., Baruteau, C., Crida, A., \& Kley, W.} 2010, 
 \emph{MNRAS}, 401, 1950
 
\bibitem[Paardekooper, Baruteau \& Kley(2011)]{paardekooper11} 
 \textsc{Paardekooper, S.-J., Baruteau, C., \& Kley, W.} 2011, 
 \emph{MNRAS}, 410, 293
  
\bibitem[Paardekooper \& Mellema(2006)]{paardekooper06}
 \textsc{Paardekooper, S.-J., \& Mellema, G.} 2006, 
 \emph{A\&A}, 459, L17
 
\bibitem[Paardekooper \& Papaloizou(2009)]{paardekooper09a}
 \textsc{Paardekooper, S.-J., \& Papaloizou, J. C. B.} 2009,
 \emph{MNRAS}, 394, 2283
 
\bibitem[Paczynski(1978)]{paczynski78}
 \textsc{Paczynski, B.} 1978,
 \emph{Acta Astronomica}, 28, 91
 
\bibitem[Pahlevan \& Stevenson(2007)]{pahlevan07}
 \textsc{Pahlevan, K., \& Stevenson, D. J.} 2007, 
 \emph{Earth and Planetary Science Letters}, 262, 438 
 
\bibitem[Pan et al.(2011)]{pan11}
 \textsc{Pan, L., Padoan, P., Scalo, J., Kritsuk, A. G., Norman, M. L.} 2011, 
 \emph{ApJ},  740, article id. 6
 
\bibitem[Papaloizou \& Lin(1984)]{papaloizou84}
 \textsc{Papaloizou, J. C. B., \& Lin, D. N. C.} 1984, 
 \emph{ApJ}, 285, 818

\bibitem[Papaloizou, Nelson \& Masset(2001)]{papaloizou01}
 \textsc{Papaloizou, J. C. B., Nelson, R. P., \& Masset, F.} 2001,
 \emph{A\&A}, 366, 263

\bibitem[Papaloizou \& Terquem(1999)]{papaloizou99}
 \textsc{Papaloizou, J. C. B., \& Terquem, C.} 1999,
 \emph{ApJ}, 521, 823
 
\bibitem[Papaloizou \& Terquem(2006)]{papaloizou06}
 \textsc{Papaloizou, J. C. B., \& Terquem, C.} 2006,
 \emph{Reports on Progress in Physics}, 69, 119  
 
\bibitem[Pepe et al.(2011)]{pepe11}
 \textsc{Pepe, F., et al.} 2011, 
 \emph{A\&A}, 534, id.~A58

\bibitem[Perez-Becker \& Chiang(2011)]{perezbecker11} 
 \textsc{Perez-Becker, D., \& Chiang, E.} 2011, 
 \emph{ApJ}, 735, article id. 8,
 
\bibitem[Perri \& Cameron(1973)]{perri73}
 \textsc{Perri, F., \& Cameron, A. G. W.} 1973, 
 \emph{Icarus}, 22, 416 
 
\bibitem[Petersen, Stewart \& Julien(2007)]{petersen07}
 \textsc{Petersen, M. R., Stewart, G. R., \& Julien, K.} 2007, 
 \emph{ApJ}, 658, 1252
 
\bibitem[Petigura, Marcy \& Howard(2013)]{petigura13}
 \textsc{Petigura, E. A., Marcy, G. W., \& Howard, A. W.} 2013, 
 \emph{ApJ}, 770, article id. 69
 
\bibitem[Petrovich, Malhotra \& Tremaine(2013)]{petrovich13}
 \textsc{Petrovich, C., Malhotra, R., \& Tremaine, S.} 2013,
 \emph{ApJ}, 770, article id. 24
 
\bibitem[Pinilla et al.(2012)]{pinilla12}
 \textsc{Pinilla, P., Birnstiel, T., Ricci, L., Dullemond, C. P., Uribe, A. L., Testi, L., \& Natta, A.} 2012, 
 \emph{A\&A}, 538, id. A114
 
\bibitem[Piso \& Youdin(2014)]{piso14}
 \textsc{Piso, A.-M. A., \& Youdin, A. N.} 2014, 
 \emph{ApJ},  786, article id. 21

\bibitem[Podolak(2003)]{podolak03}
 \textsc{Podolak, M.} 2003, 
 \emph{Icarus}, 165, 428

\bibitem[Pollack et al.(1996)]{pollack96}
 \textsc{Pollack, J. B., Hubickyj, O., Bodenheimer, P., Lissauer, J. J., Podolak, M., \&
 Greenzweig, Y.} 1996, 
 \emph{Icarus}, 124, 62

\bibitem[Popham et al.(1993)]{popham93}
 \textsc{Popham, R., Narayan, R., Hartmann, L., \& Kenyon, S.} 1993, 
 \emph{ApJ}, 415, L127
 
\bibitem[Poppe, Blum \& Henning(2000)]{poppe00}
 \textsc{Poppe, T., Blum, J., \& Henning, T.} 2000, 
 \emph{ApJ}, 533, 454
 
\bibitem[Pringle(1977)]{pringle77}
 \textsc{Pringle, J. E.} 1977, 
 \emph{MNRAS}, 178, 195 
 
\bibitem[Pringle(1981)]{pringle81}
 \textsc{Pringle, J. E.} 1981, 
 \emph{ARA\&A}, 19, 137
 
\bibitem[Pringle(1989)]{pringle89}
 \textsc{Pringle, J. E.} 1989, 
 \emph{MNRAS}, 236, 107
 
\bibitem[Pringle(1991)]{pringle91}
 \textsc{Pringle, J. E.} 1991, 
 \emph{MNRAS}, 248, 754
 
\bibitem[Pringle, Verbunt \& Wade(1986)]{pringle86}
 \textsc{Pringle, J. E., Verbunt, F., \& Wade, R. A.} 1986, 
 \emph{MNRAS}, 221, 169
 
\bibitem[Pu \& Wu(2015)]{pu15}
 \textsc{Pu, B., \& Wu, Y.} 2015, 
 \emph{ApJ}, 807, article id. 44
 
\bibitem[Quillen et al.(2004)]{quillen04}
 \textsc{Quillen, A. C., Blackman, E. G., Frank, A., \& Varni\'ere, P.} 2004, 
 \emph{ApJ}, 612, L137
 
\bibitem[Rafikov(2005)]{rafikov05}
 \textsc{Rafikov, R. R.} 2005,
 \emph{ApJ}, 621, L69

\bibitem[Rafikov(2006)]{rafikov06}
 \textsc{Rafikov, R. R.} 2006,
 \emph{ApJ}, 648, 666
 
\bibitem[Rafikov(2009)]{rafikov09}
 \textsc{Rafikov, R. R.} 2009,
 \emph{ApJ}, 704, 281
 
\bibitem[Rasio \& Ford(1996)]{rasio96}	
 \textsc{Rasio, F. A., \& Ford, E. B.} 1996,
 \emph{Science}, 274, 954 
 
\bibitem[Raymond, Armitage \& Gorelick(2009)]{raymond09b}
 \textsc{Raymond, S. N., Armitage, P. J., \& Gorelick, N.} 2009, 
 \emph{ApJ}, 699, L88

\bibitem[Raymond, Armitage \& Gorelick(2010)]{raymond10}
 \textsc{Raymond, S. N., Armitage, P. J., \& Gorelick, N.} 2010, 
 \emph{ApJ}, 711, 772
 
\bibitem[Raymond et al.(2008)]{raymond08}
 \textsc{Raymond, S. N., Barnes, R., Armitage, P. J., \& Gorelick, N.} 2008,
 \emph{ApJ}, 687, L107
 
\bibitem[Raymond et al.(2009)]{raymond09}
 \textsc{Raymond, S. N., O'Brien, D. P., Morbidelli, A., \& Kaib, N. A.} 2009, 
 \emph{Icarus}, 203, 644
 
\bibitem[Raymond, Quinn \& Lunine(2005)]{raymond05}
 \textsc{Raymond, S. N., Quinn, T., \& Lunine, J. I.} 2005,
 \emph{ApJ}, 632, 670
 
\bibitem[Rebull et al.(2006)]{rebull06}
 \textsc{Rebull, L. M., Stauffer, J. R., Megeath, S. T., Hora, J. L., \& Hartmann, L.} 2006, 
 \emph{ApJ}, 646, 297
 
\bibitem[Rice \& Armitage(2009)]{rice09}
 \textsc{Rice, W. K. M., \& Armitage, P. J.} 2009, 
 \emph{MNRAS}, 396, 228
 
\bibitem[Rice et al.(2003b)]{rice03b}
 \textsc{Rice, W. K. M., Armitage, P. J., Bate, M. R., \& Bonnell, I. A.} 2003, 
 \emph{MNRAS}, 339, 1025
 
\bibitem[Rice et al.(2003c)]{rice03c}
 \textsc{Rice, W. K. M., Armitage, P. J., Bonnell, I. A., Bate, M. R., Jeffers, S. V., \& Vine, S. G.} 2003, 
 \emph{MNRAS}, 346, L36 
 
\bibitem[Rice et al.(2011)]{rice11}
 \textsc{Rice, W. K. M., Armitage, P. J., Mamatsashvili, G. R., Lodato, G., \& Clarke, C. J.} 2011, 
 \emph{MNRAS}, 418, 1356
 
\bibitem[Rice et al.(2006)]{rice06}
 \textsc{Rice, W. K. M., Armitage, P. J., Wood, K., \& Lodato, G.} 2006, 
 \emph{MNRAS}, 373, 1619 
 
\bibitem[Rice, Lodato \& Armitage(2005)]{rice05}
 \textsc{Rice, W. K. M., Lodato, G., \& Armitage, P. J.} 2005, 
 \emph{MNRAS}, 364, L56
 
\bibitem[Rice et al.(2004)]{rice04}
 \textsc{Rice, W. K. M., Lodato, G., Pringle, J. E., Armitage, P. J., \& Bonnell, I. A.} 2004, 
 \emph{MNRAS}, 355, 543
 
\bibitem[Rice et al.(2014)]{rice14}
 \textsc{Rice, W. K. M., Paardekooper, S.-J., Forgan, D. H., \& Armitage, P. J.} 2014, 
 \emph{MNRAS}, 438, 1593
 
\bibitem[Rice et al.(2003d)]{rice03d}
 \textsc{Rice, W. K. M., Wood, K., Armitage, P. J., Whitney, B. A., \& Bjorkman, J. E.} 2003, 
 \emph{MNRAS}, 342, 79
 
\bibitem[Rogers(2015)]{rogers15}
 \textsc{Rogers, L. A.} 2015, 
 \emph{ApJ}, 801, article id. 41

\bibitem[Roig \& Nesvorn\'y(2015)]{roig15}
 \textsc{AJ}, 150, article id. 186
 
\bibitem[Ros \& Johansen(2013)]{ros13}
 \textsc{Ros, K., \& Johansen, A.} 2013, 
 \emph{A\&A}, 552, id.~A137
 
\bibitem[Rossiter(1924)]{rossiter24}
 \textsc{Rossiter, R. A.} 1924, 
 \emph{ApJ}, 60, 15 
 
\bibitem[Rybicki \& Lightman(1979)]{rybicki79}
 \textsc{Rybicki, G. B., \& Lightman, A. P.} 1979, 
 \emph{Radiative Processes in Astrophysics}, (Wiley) 

\bibitem[Safronov(1969)]{safronov69}
 \textsc{Safronov, V. S.} 1969, 
 \emph{Evolution of the Protoplanetary Cloud and Formation of the Earth and 
 the Planets}, English translation NASA TT F-677 (1972)

\bibitem[Sagan \& Mullen(1972)]{sagan72}
 \textsc{Sagan, C., \& Mullen, G.} 1972, 
 \emph{Science}, 177, 52 
 
\bibitem[Salmeron \& Wardle(2005)]{salmeron05}
 \textsc{Salmeron, R., \& Wardle, M.} 2005,
 \emph{MNRAS}, 361, 45

\bibitem[Sasaki, Stewart \& Ida(2010)]{sasaki10} 
 \textsc{Sasaki, T., Stewart, G. R., \& Ida, S.} 2010, 
 \emph{ApJ}, 714, 1052

\bibitem[Sato et al.(2005)]{sato05}
 \textsc{Sato, B., et al.} 2005, 
 \emph{ApJ}, 633, 465
 
\bibitem[Sch\"afer, Yang \& Johansen(2017)]{schafer17} 
 \textsc{Sch\"afer, U., Yang, C.-C., \& Johansen, A.} 2017,
 \emph{A\&A}, 597, id.~A69
 
\bibitem[Scharf \& Menou(2009)]{scharf09}
 \textsc{Scharf, C., \& Menou, K.} 2009,
 \emph{ApJ}, 693, L113
 
\bibitem[Schartman et al.(2012)]{schartman12} 
 \textsc{Schartman, E., Ji, H., Burin, M. J., \& Goodman, J.} 2012, 
 \emph{A\&A}, 543, id.A94
 
\bibitem[Scholl \& Froeschle(1991)]{scholl91}
 \textsc{Scholl, H., \& Froeschle, Ch.} 1991, 
 \emph{A\&A}, 245, 316  
 
\bibitem[Sekiya(1998)]{sekiya98}
 \textsc{Sekiya, M.} 1998, 
 \emph{Icarus}, 133, 298
 
\bibitem[Shakura \& Sunyaev(1973)]{shakura73}
 \textsc{Shakura, N. I., \& Sunyaev, R. A.} 1973,
 \emph{A\&A}, 24, 337
 
\bibitem[Shen, Stone \& Gardiner(2006)]{shen06}
 \textsc{Shen, Y., Stone, J. M., \& Gardiner, T. A.} 2006, 
 \emph{ApJ}, 653, 513  
 
\bibitem[Shlosman \& Begelman(1989)]{shlosman89}
 \textsc{Shlosman, I., \& Begelman, M. C.} 1989,
 \emph{ApJ}, 341, 685
 
\bibitem[Shu, Johnstone \& Hollenbach(1993)]{shu93}
 \textsc{Shu, F. H., Johnstone, D., \& Hollenbach, D.} 1993,
 \emph{Icarus}, 106, 92
 
\bibitem[Sicilia-Aguilar et al.(2006)]{sicilia06} 	
 \textsc{Sicilia-Aguilar, A., Hartmann, L. W., F\"ur\'esz, G., Henning, T., Dullemond, C., \& Brandner,
 W.} 2006, \emph{AJ}, 132, 2135
 
\bibitem[Simon \& Armitage(2014)]{simon14} 
 \textsc{Simon, J. B., \& Armitage, P. J.} 2014, 
 \emph{ApJ}, 784, article id. 15
 
\bibitem[Simon et al.(2016)]{simon16}
 \textsc{Simon, J. B., Armitage, P. J., Li, R., \& Youdin, A. N.} 2016, 
 \emph{ApJ},  822, article id. 55
 
\bibitem[Simon et al.(2013a)]{simon13} 
 \textsc{Simon, J. B., Bai, X.-N., Stone, J. M., Armitage, P. J., \& Beckwith, K.} 2013, 
 \emph{ApJ}, 764, article id. 66
 
\bibitem[Simon et al.(2013b)]{simon13b}
 \textsc{Simon, J. B., Bai, X.-N., Armitage, P. J., Stone, J. M., \& Beckwith, K.} 2013, 
 \emph{ApJ}, 775, article id. 73
 
\bibitem[Simon, Beckwith \& Armitage(2012)]{simon12}
 \textsc{Simon, J. B., Beckwith, K., \& Armitage, P. J.} 2012, 
 \emph{MNRAS}, 422, 2685
 
\bibitem[Simon et al.(2015b)]{simon15b}
 \textsc{Simon, J. B., Hughes, A. M., Flaherty, K. M., Bai, X.-N., \& Armitage, P. J.} 2015, 
 \emph{ApJ},  808, article id. 180
 
\bibitem[Simon et al.(2015)]{simon15}
 \textsc{Simon, J. B., Lesur, G., Kunz, M. W., \& Armitage, P. J.} 2015,
 \emph{MNRAS},  454, 1117
 
\bibitem[Smoluchowski(1916)]{smoluchowski16}
 \textsc{Smoluchowski, M. V.} 1916, 
 \emph{Physik. Zeit.}, 17, 557
 
\bibitem[Sousa et al.(2008)]{sousa08}
 \textsc{Sousa, S. G., et al.} 2008, 
 \emph{A\&A}, 487, 373
 
\bibitem[Spiegel \& Burrows(2013)]{spiegel13}
 \textsc{Spiegel, D. S., \& Burrows, A.} 2013, 
 \emph{ApJ}, 772, article id. 76
 
\bibitem[Stamatellos \& Whitworth(2009)]{stamatellos09}
 \textsc{Stamatellos, D., \& Whitworth, A. P.} 2009,
 \emph{MNRAS}, 392, 413
 
\bibitem[Steffen et al.(2012)]{steffen12}
 \textsc{Steffen, J. H., et al.} 2012,
 \emph{Proceedings of the National Academy of Sciences}, 109, 7982
 
\bibitem[Stern \& Durda(2000)]{stern00}
 \textsc{Stern, S. A., \& Durda, D. D.} 2000, 
 \emph{Icarus}, 143, 360
 
\bibitem[Stevenson(1982)]{stevenson82}
 \textsc{Stevenson, D. J.} 1982, 
 \emph{Planetary and Space Science}, 30, 755
 
\bibitem[Stevenson \& Lunine(1988)]{stevenson88}
 \textsc{Stevenson, D. J., \& Lunine, J. I.} 1988, 
 \emph{Icarus}, 75, 146
 
\bibitem[Stone et al.(1996)]{stone96b}
 \textsc{Stone, J. M., Hawley, J. F., Gammie, C. F., \& Balbus, S. A.} 1996, 
 \emph{ApJ}, 463, 656
 
\bibitem[Strom et al.(2015)]{strom15}
 \textsc{Strom, R. G., Renu, M., Zhi-Yong, X., Takashi, I., Fumi, Y., \& Ostrach Lillian, R.} 2015, 
 \emph{Research in Astronomy and Astrophysics}, 15, article id. 407
 
\bibitem[Syer \& Clarke(1995)]{syer95}
 \textsc{Syer, D., \& Clarke, C. J.} 1995, 
 \emph{MNRAS}, 277, 758

\bibitem[Tabachnik \& Tremaine(2002)]{tabachnik02}
 \textsc{Tabachnik, S., \& Tremaine, S.} 2002, 
 \emph{MNRAS}, 335, 151

\bibitem[Takeuchi, Clarke \& Lin(2005)]{takeuchi05} 
 \textsc{Takeuchi, T., Clarke, C. J., \& Lin, D. N. C.} 2005, 
 \emph{ApJ}, 627, 286
 
\bibitem[Takeuchi \& Lin(2002)]{takeuchi02}
 \textsc{Takeuchi, T., \& Lin, D. N. C.} 2002,
 \emph{ApJ}, 581, 1344 
 
\bibitem[Takeuchi, Miyama \& Lin(1996)]{takeuchi96}
 \textsc{Takeuchi, T., Miyama, S. M., \& Lin, D. N. C.} 1996,
 \emph{ApJ}, 460, 832

\bibitem[Tanigawa, Ohtsuki \& Machida(2012)]{tanigawa12}
 \textsc{Tanigawa, T., Ohtsuki, K., \& Machida, M. N.} 2012, 
 \emph{ApJ}, 747, 47
 
\bibitem[Tanaka, Takeuchi \& Ward(2002)]{tanaka02}
 \textsc{Tanaka, H.,. Takeuchi, T., \& Ward, W. R.} 2002, 
 \emph{ApJ}, 565, 1257
 
\bibitem[Teague et al.(2016)]{teague16}
 \textsc{Teague, R., Guilloteau, S., Semenov, D., Henning, Th., Dutrey, A., Pi\'etu, V., Birnstiel, T., Chapillon, E., Hollenbach, D., \& Gorti, U.} 2016, 
 \emph{A\&A}, 592, id.~A49
 
\bibitem[Tera, Papanastassiou \& Wasserburg(1974)]{tera74} 
 \textsc{Tera, F., Papanastassiou, D. A., \& Wasserburg, G. J.} 1974, 
 \emph{Earth and Planetary Science Letters}, 22, 1
 
\bibitem[Terquem \& Papaloizou(2002)]{terquem02}
 \textsc{Terquem, C., \& Papaloizou, J. C. B.} 2002, 
 \emph{MNRAS}, 332, L39 
 
\bibitem[Testi et al.(2014)]{testi14}
  \textsc{Testi, L., et al.} 2014, 
  \emph{Protostars and Planets VI}, Henrik Beuther, Ralf S. Klessen, Cornelis P. Dullemond, and Thomas Henning (eds.), University of Arizona Press, Tucson, 339
 
\bibitem[Thommes, Duncan \& Levison(1999)]{thommes99}
 \textsc{Thommes, E. W., Duncan, M. J., \& Levison, H. F.} 1999, 
 \emph{Nature}, 402, 635
 
\bibitem[Thommes, Duncan \& Levison(2003)]{thommes03}
 \textsc{Thommes, E. W., Duncan, M. J., \& Levison, H. F.} 2003, 
 \emph{Icarus}, 161, 431

\bibitem[Throop \& Bally(2005)]{throop05}
 \textsc{Throop, H. B., \& Bally, J.} 2005,
 \emph{ApJ}, 623, L149

\bibitem[Toomre(1964)]{toomre64}
 \textsc{Toomre, A.} 1964, 
 \emph{ApJ}, 139, 1217
 
\bibitem[Torres et al.(2004)]{torres04} 
 \textsc{Torres, G., Konacki, M., Sasselov, D. D., \& Jha, S.} 2004,
 \emph{ApJ}, 609, 1071

\bibitem[Trujillo, Jewitt \& Luu(2001)]{trujillo01}
 \textsc{Trujillo, C. A., Jewitt, D. C., \& Luu, J. X.} 2001, 
 \emph{AJ}, 122, 457
 
\bibitem[Trujillo \& Sheppard(2014)]{trujillo14} 
 \textsc{Trujillo, C. A., \& Sheppard, S. S.} 2014, 
 \emph{Nature}, 507, 471
 
\bibitem[Tsang, Turner \& Cumming(2014)]{tsang14}
 \textsc{Tsang, D.,Turner, N. J., \& Cumming, A.} 2014, 
 \emph{ApJ}, 782, article id. 113
 
\bibitem[Tsiganis et al.(2005)]{tsiganis05}
 \textsc{Tsiganis, K., Gomes, R., Morbidelli, A., \& Levison, H. F.} 2005, 
 \emph{Nature}, 435, 459
 
\bibitem[Turner et al.(2014)]{turner14} 
 \textsc{Turner, N. J., Fromang, S., Gammie, C., Klahr, H., Lesur, G., Wardle, M., \& Bai, X.-N.} 2014, 
 \emph{Protostars and Planets VI}, Henrik Beuther, Ralf S. Klessen, Cornelis P. Dullemond, and Thomas Henning (eds.), University of Arizona Press, Tucson, p.~411
 
\bibitem[Umebayashi(1983)]{umebayashi83}	
 \textsc{Umebayashi, T.} 1983, 
 \emph{Progress of Theoretical Physics}, 69, 480

\bibitem[Umebayashi \& Nakano(1981)]{umebayashi81}
 \textsc{Umebayashi, T., \& Nakano, T.} 1981,
 \emph{PASJ}, 33, 617
 
\bibitem[Valtonen \& Karttunen(2006)]{valtonen06}
 \textsc{Valtonen, M., \& Karttunen, H.} 2006, 
 \emph{The Three-Body Problem}, Cambridge University Press 
 
\bibitem[van der Marel et al.(2013)]{vdM13}
 \textsc{van der Marel, N., et al.} 2013, 
 \emph{Science}, 340, 1199 

\bibitem[Velikhov(1959)]{velikhov59}
 \textsc{Velikhov, E. T.} 1959, 
 \emph{Sov. Phys. JETP}, 36, 995

\bibitem[Veras \& Armitage(2004)]{veras04}
 \textsc{Veras, D., \& Armitage, P. J.} 2004,
 \emph{MNRAS}, 347, 613
 
\bibitem[Veras, Crepp \& Ford(2009)]{veras09}	
 \textsc{Veras, D., Crepp, J. R., \& Ford, E. B.} 2009, 
 \emph{ApJ}, 696, 1600

\bibitem[Wadhwa et al.(2007)]{wadhwa06}
 \textsc{Wadhwa, M., Amelin, Y., Davis, A. M., Lugmair, G. W., Meyer, B., Gounelle, M., \& 
 Desch, S.} 2007, \emph{Protostars and Planets V}, eds B. Reipurth, D. Jewitt, and K. Keil, University of 
 Arizona Press, Tucson
 
\bibitem[Walker, Hays \& Kasting(1981)]{walker81}
 \textsc{Walker, J. C. G., Hays, P. B., \& Kasting, J. F.} 1981, 
 \emph{Journal of Geophysical Research}, 86, 9776
 
\bibitem[Walsh et al.(2011)]{walsh11}
 \textsc{Walsh, K. J., Morbidelli, A., Raymond, S. N., O'Brien, D. P., \& Mandell, A. M.} 2011, 
 \emph{Nature}, 475, 206

\bibitem[Ward(1991)]{ward91}
 \textsc{Ward, W. R.} 1991
 \emph{Abstracts of the Lunar and Planetary Science Conference}, 22, 1463

\bibitem[Ward(1997)]{ward97}
 \textsc{Ward, W. R.} 1997
 \emph{Icarus}, 126, 261

\bibitem[Weidenschilling(1977)Weidenschilling]{weidenschilling77}
 \textsc{Weidenschilling, S. J.} 1977, 
 \emph{Astrophysics and Space Science}, 51, 153
 
\bibitem[Weidenschilling(1977b)Weidenschilling]{weidenschilling77b}
 \textsc{Weidenschilling, S. J.} 1977, 
 \emph{MNRAS}, 180, 57
 
\bibitem[Weidenschilling(1980)]{weidenschilling80}
 \textsc{Weidenschilling, S. J.} 1980, 
 \emph{Icarus},  44, 172
 
\bibitem[Weidenschilling \& Cuzzi(1993)]{weidenschilling93}
 \textsc{Weidenschilling, S. J., \& Cuzzi, J. N.} 1993, 
 \emph{in Protostars and Planets III}, University of Arizona Press, p.~1031
 
\bibitem[Weidenschilling \& Marzari(1996)]{weidenschilling96}
 \textsc{Weidenschilling, S. J., \& Marzari, F.} 1996, 
 \emph{Nature}, 384, 619
 
\bibitem[Weiss \& Marcy(2014)]{weiss14}
 \textsc{Weiss, L. M., \& Marcy, G. W.} 2014, 
 \emph{ApJ}, 783, article id. L6
 
\bibitem[Welsh et al.(2012)]{welsh12}
 \textsc{Welsh, W. F., et al.} 2012, 
 \emph{Nature}, 481, 475 
 
\bibitem[Wetherill \& Stewart(1993)]{wetherill93}
 \textsc{Wetherill, G. W., \& Stewart, G. R.} 1993,  
 \emph{Icarus}, 106, 190
 
\bibitem[Whipple(1972)]{whipple72}
 \textsc{Whipple, F. L.} 1972, in 
 \emph{From Plasma to Planet, Proceedings of the Twenty-First Nobel Symposium}, 
 editor Aina Evlius. Wiley Interscience Division (New York), p.~211
  
\bibitem[Winn et al.(2012)]{winn12}
 \textsc{Winn, J. N., Fabrycky, D., Albrecht, S., \& Johnson, J. A.} 2012,
 \emph{ApJ}, 718, L145
 
\bibitem[Wolszczan \& Frail(1992)]{frail92}
 \textsc{Wolszczan, A., \& Frail, D. A.} 1992, 
 \emph{Nature}, 355, 145 
 
\bibitem[Wright et al.(2011)]{wright11}
 \textsc{Wright, J. T., et al.} 2011, 
 \emph{ApJ},  730, article id. 93 
 
\bibitem[Wu \& Murray(2003)]{wu03}
 \textsc{Wu, Y.,\&  Murray, N.} 2003, 
 \emph{ApJ},  589, 605
 
\bibitem[Yang, Johansen \& Carrera(2017)]{yang17}
 \textsc{Yang, C.-C., Johansen, A., \& Carrera, D.} 2017, 
 \emph{A\&A}, submitted (arXiv:1611.07014)
 
\bibitem[Yang, Mac Low \& Menou(2009)]{yang09}
 \textsc{Yang, C.-C., Mac Low, M.-M., \& Menou, K.} 2009, 
 \emph{ApJ}, 707, 1233
 
\bibitem[Yoder \& Peale(1981)]{yoder81}
 \textsc{Yoder, C. F., \& Peale, S. J.} 1981, 
 \emph{Icarus}, 47, 1 
 
\bibitem[Youdin(2011)]{youdin11}
 \textsc{Youdin, A. N.} 2011,
 \emph{ApJ}, 742, article id. 38
 
\bibitem[Youdin \& Chiang(2004)]{youdin04}
 \textsc{Youdin, A. N., \& Chiang, E. I.} 2004, 
 \emph{ApJ}, 601, 1109
 
\bibitem[Youdin \& Goodman(2005)]{youdin05}
 \textsc{Youdin, A. N., \& Goodman, J.} 2005, 
 \emph{ApJ}, 620, 459
 
\bibitem[Youdin \& Lithwick(2007)]{youdin07}
 \textsc{Youdin, A. N., \& Lithwick, Y.} 2007, 
 \emph{Icarus}, 192, 588
 
\bibitem[Youdin \& Shu(2002)]{youdin02}
 \textsc{Youdin, A. N., \& Shu, F. H.} 2002, 
 \emph{ApJ}, 580, 494
 
\bibitem[Zhu, Hartmann \& Gammie(2009)]{zhu09}
 \textsc{Zhu, Z., Hartmann, L., \& Gammie, C.} 2009, 
 \emph{ApJ}, 694, 1045
 
\bibitem[Zhu et al.(2012)]{zhu12}
 \textsc{Zhu, Z., Nelson, R. P., Dong, R., Espaillat, C., \& Hartmann, L.} 2012, 
 \emph{ApJ}, 755, article id.~6
 
\bibitem[Zhu, Stone \& Bai(2015)]{zhu15}
 \textsc{Zhu, Z., Stone, J. M., \& Bai, X.-N.} 2015, 
 \emph{ApJ}, 801, article id.~81
 
\bibitem[Zsom et al.(2010)]{zsom10} 
 \textsc{Zsom, A., Ormel, C. W., Guettler, C., Blum, J., \& Dullemond, C. P.} 2010, 
 \emph{A\&A}, 513, id. A57

\end{thebibliography}
\end{document}